%% file: thesis1.tex
\documentclass[12pt]{uvic}

\newcommand{\babar}{\mbox{\sc{BaBar}}\index{BaBar}}

\newcommand{\thesis}{dissertation}
\newcommand{\we}{we }

\newcommand{\lp}{
}

\newcommand{\EXDM}{5}      

\newcommand{\sca}{\phi_1}
\newcommand{\scb}{\phi_2}

\newcommand{\hef}{$^4$He}
\newcommand{\het}{$^3$He}
\newcommand{\lisx}{$^6$Li\index{lithium!$^6$Li}}
\newcommand{\lisv}{$^7$Li\index{lithium!$^7$Li}}
\newcommand{\bes}{$^7$Be\index{beryllium!$^7$Be}}
\newcommand{\beet}{$^8$Be\index{beryllium!$^8$Be}}

\newcommand{\hefm}{{\rm ^4He}}
\newcommand{\hetm}{{\rm ^3He}}
\newcommand{\lisxm}{{\rm ^6Li}}
\newcommand{\lisvm}{{\rm ^7Li}}
\newcommand{\besm}{{\rm ^7Be}}
\newcommand{\beetm}{{\rm ^8Be}}

\newcommand{\bex}{(\bes $X^-$)}

\newcommand{\ddme}{}

\usepackage{index}

\usepackage{amsfonts}
\usepackage{amssymb}
\usepackage{amsmath}

\usepackage{epsfig}
\usepackage{graphicx}
\usepackage{cancel}
\usepackage{fancyhdr}

\usepackage{type1cm}
\usepackage{courier}
\usepackage{mathrsfs}
\DeclareMathAlphabet{\mathpzc}{OT1}{pzc}{m}{it}

\usepackage{setspace}
\title{The Early Universe as a Probe of New Physics}
\author{Christopher Shane Bird}
\thesistype{Dissertation}
\prevdegree{B.Sc. University of Victoria 1999}
\prevdegree{M.Sc. University of Victoria 2001}
\date{2007}
\degree{Doctor of Philosophy}
\department{Department of Physics \& Astronomy}
\supervisor{Dr. Maxim Pospelov}
\committeemember{Dr.\ Maxim\ Pospelov, Supervisor %
(Department of Physics \& Astronomy)}
\committeemember{Dr.\ Charles\ Picciotto, Member %
(Department of Physics \& Astronomy)}
\committeemember{Dr.\ Richard\ Keeler, Member %
(Department of Physics \& Astronomy)}
\committeemember{Dr.\ Alexandre\ Brolo, Outside Member %
(Department of Chemistry)}

\copyrightyear{2008}
\copyrightpagedate{November 24, 2008}

\makeindex

\begin{document}

\pagenumbering{roman}
\maketitlepage
\lp
\makesupervisorycommittee
\lp
\begin{abstract}
\par
The Standard Model of Particle Physics has been verified to unprecedented precision in the last few decades. However there are
still phenomena in nature which cannot be explained, and as such
new theories will be required. Since terrestrial experiments are
limited in both the energy and precision that can be probed, new methods are required to search for signs of physics beyond the Standard Model. In this \thesis, I demonstrate how these theories can be probed by searching for remnants of their effects in the early Universe. In particular I focus on three possible extensions of the Standard Model: the addition of massive neutral particles as dark matter, the addition of charged massive particles, and the existence of higher dimensions. For each new model, I review the existing experimental bounds and the potential for discovering new physics in the next generation of experiments.
\par
For dark matter, I introduce six simple models which I have developed, and which involve a minimum amount of new physics, as well as reviewing one existing model of dark matter. For each model I calculate the latest constraints from astrophysics experiments, nuclear recoil experiments, and collider experiments. I also provide motivations for studying sub-GeV mass dark matter, and propose the possibility of searching for light WIMPs in the decay of B-mesons and other heavy particles.
\par
For charged massive relics, I introduce and review the recently proposed model of catalyzed Big Bang nucleosynthesis. In particular I review the production of \lisx\ by this mechanism, and calculate the abundance of \lisv\ after destruction of \bes\ by charged relics. The result is that for certain natural relics CBBN is capable of removing tensions between the predicted and observed \lisx\ and \lisv\ abundances which are present in the standard model of BBN.
\par
For extra dimensions, I review the constraints on the ADD model from both astrophysics and collider experiments. I then calculate the constraints on this model from Big Bang nucleosynthesis in the early Universe. I also calculate the bounds on this model from Kaluza-Klein gravitons trapped in the galaxy which decay to electron-positron pairs, using the measured $511 \; keV$ $\gamma$-ray flux.
\par
For each example of new physics, I find that remnants of the early Universe provide constraints on the models which are complimentary to the existing constraints from colliders and other terrestrial experiments. 
\end{abstract}


\maketableofcontents
\lp
\makelistoftables
\makelistoffigures

\begin{acknowledge} \input{ack.tex} \end{acknowledge}

\lp
\afterpage{\pagenumbering{arabic}}

\pagestyle{fancy}
\fancyhead[RO,LE]{\thepage}
\fancyhead[LO,RE]{\slshape \leftmark}
\fancyfoot[LC,RC]{ }

\input{chapter0.tex}

\newpage
\lp
\input{chapter1b.tex}

\newpage
\input{chapter2.tex}

\newpage
\input{chapter3.tex}

\newpage
\input{conclusions.tex}
\lp 



\pagestyle{plain}
\bibliographystyle{apsrev2}
\bibliography{thesis1}

\lp
\lp
\end{document}

%% file: ack.tex
\vspace{-5mm}
\par
As with any significant endeavor, this dissertation represents the cumulative result of the influence, input, and guidance of many people. It is only through the generous support of these individuals that this work could be completed.

\par
First and foremost I must thank my adviser, Dr. Maxim Pospelov, for guiding my research and letting me explore many interesting topics, and for always being willing and able to discuss new ideas.  I also thank Dr. Charles Picciotto for always being supportive as both a teacher and adviser, and for helping me at every stage of my university education,Dr. Fred Cooperstock for many interesting discussions, and Dr. Harold Fearing for guiding me through the initial stages of my graduate work.

\par
It is also a pleasure to thank all of the members of the Department of Physics \& Astronomy at the University of Victoria, who taught and supported me for these many years, and the support staff who through the years have helped me with the administrative side of graduate school. 

\par
In addition, I would like to thank all of my collaborators for many informative conversations and interesting research projects. I also thank all of those who came before me, for building the theories that I have now modified and hopefully improved. 

\par
Finally, I thank all of the friends and family who at one time or another supported my dreams and ambitions, and those individuals whose passion for physics first inspired me to pursue this course of study. 

\begin{flushright}
Christopher S. Bird
\end{flushright}

%% file: chapter0.tex
\chapter{Introduction}
\thispagestyle{empty}

\par
The Standard Model has been very successful for the last three decades, with numerous experiments confirming the existence of several particles and measuring the fundamental parameters to increasing precision. In spite of many dedicated searches for new physics at high energy colliders, as yet there has been no confirmed data which is inconsistent with with Standard Model.

\par
However this success does not extend to explaining cosmological data. For example, the WMAP\index{WMAP} satellite \cite{Spergel:2003cb,Dunkley:2008ie} which measured anisotropy in the cosmic microwave background, and experiments studying both supernovae\index{supernovae} and large scale structure\index{large scale structure} in the Universe, have provided strong evidence for the existence of at least two new forms of energy confirming the results of previous astrophysics experiments. The first of these is an electrically neutral form of matter referred to as {\it dark matter}\index{dark matter} comprising 23\% of the energy content of the Universe, whose existence had been previously inferred from the discrepancy between luminous mass and gravitation masses of galaxies and more recently in the observed gravitational lensing\index{gravitational lensing} of the bullet cluster\index{dark matter!bullet cluster} \cite{Clowe:2006eq}. The other new form of energy, which is referred to as {\it dark energy}\index{dark energy}, has a negative pressure and comprises 73\% of the energy content of the universe, and was originally detected in supernovae\index{supernovae} surveys \cite{Perlmutter:1998np,Riess:1998cb}. Astrophysics experiments have also indicated excess positrons\index{galactic positron excess} in the galaxy \cite{Jean:2003ci}, ultra-high energy cosmic rays\index{cosmic rays} \footnote{Recent preliminary results from the Auger observatory\index{Auger observatory} have suggested that the source of these ultra-high energy cosmic rays\index{cosmic rays} are not isotropic, and may instead be produced in active galactic nuclei, which would indicate that new physics may not be required to explain them\cite{Abraham:2007bb}.} \cite{Hayashida:1994hb,Hayashida:2000zr}, and a net baryon number in the Universe (which can be measured both in the CMB and in comparisons of the predictions of Big Bang Nucleosynthesis\index{Big Bang Nucleosynthesis}\index{nucleosynthesis} to observed abundances of light elements). None of these phenomena can currently be resolved within the context of the Standard Model.
\par
In addition to these experimental anomalies,the Standard Model also fails to explain why gravity is fifteen orders of magnitude weaker than the other forces\index{hierarchy problem}, why there exists three generations of particles, and several other problems related to the underlying theory. There are numerous proposals which try to solve these problems, but as yet none have been confirmed by experiments. Furthermore economical and technological constraints restrict both the energy and precision which can be probed directly in either current or next generation of collider experiments.

\par
However nature has provided an alternate laboratory in the search for new physics in the form of the early Universe. Moments after the big bang, the energy scales involved in typical particle reactions were well in excess of those accessible to terrestrial experiments, allowing for previously undetected physical phenomena to have an effect on the evolution and particle content of the Universe. If these effects leave a signature which can be studied in modern times, then they may provide evidence for the existence and nature of new physics.

\par
In this dissertation, I present several examples of physics beyond the Standard Model that will have an effect on the early universe, including many models which my collaborators and I have developed in previous published papers. I also present several new methods of searching for the effects of these models in the modern Universe, including both methods which my collaborators and I originally published and methods which I have developed for this \thesis, which are previously unpublished. As will be demonstrated, each of these new methods provides either stronger constraints on new physics models than previously existed, or allows existing experiments to probe new regions of the parameter space for each model.

\par
In Chapter \ref{Chapter:DarkMatter}, I review the motivation for dark matter and present several simple models. Although the models presented involve minimal extensions of the Standard Model, they also serve as effective theories for more complicated models and the bounds presented can be applied to other dark matter candidates. In Section \ref{Section::LDM}, I present the motivations for the special case of light dark matter\index{dark matter!light}, involving sub-GeV dark matter, and the possibility of detection in B-meson\index{B mesons} experiments as originally published in:
\newline
\vspace{14pt}
\noindent
\begin{itemize}{\small
\item  {\bf C.Bird}, P.~Jackson, R.~Kowalewski and M.~Pospelov,
  ``Search for dark matter in b $\to$ s transitions with missing energy'', Phys. Rev. Lett.  {\bf 93}, 201803 (2004), [arXiv:hep-ph/0401195]. \vspace{14pt}
\item    {\bf C.Bird}, R.~Kowalewski and M.~Pospelov,
  ``Dark matter pair-production in b $\to$ s transitions'',
  Mod. Phys. Lett. A {\bf 21}, 457 (2006)
  [arXiv:hep-ph/0601090].
  }
\end{itemize}
\vspace{14pt}
\noindent
With the exception of the Minimal Model of Dark Matter (MDM) presented in Section \ref{Section::MDM}, which was previously published in Ref \cite{Silveira:1985rk,McDonald:1993ex,Burgess:2000yq}, all of the models represent original research. The constraints on each model, which are derived from existing experimental data as well as updated bounds on the MDM, also constitute original research.

\par
In Chapter \ref{Chapter:ChargedRelics} , I review the existing bounds on long lived charged relics\index{CHAMPs}\index{charged relics} which may exist in the Universe, and  how the presence of metastable charged particles\index{CHAMPs}\index{charged relics} could affect the predictions of Big Bang nucleosynthesis\index{Big Bang Nucleosynthesis}\index{nucleosynthesis}. The possibility that charged particles could catalyzed the standard reactions in Big Bang nucleosynthesis (BBN) was originally published in Ref \cite{Pospelov:2006sc}, and the resulting constraints from Catalyzed BBN\index{Big Bang Nucleosynthesis!catalyzed} on charged relics\index{charged relics}\index{CHAMPs} are reviewed. In particular, I calculate the effect of charged particles on the primordial Lithium-7 and Beryllium-7 abundances, and demonstrate how the presence of charged particles during nucleosynthesis could  catalyze the destruction of these elements. This work was originally published in:

\vspace{14pt}

\begin{itemize}
\item {\bf C.~Bird}, K.~Koopmanns, M.~Pospelov, ''Primordial Lithium Abundance in Catalyzed Big Bang Nucleosynthesis'', Phys. Rev.  D {\bf 78}, 083010 (2008)
,[  arXiv:hep-ph/0703096v3]
\end{itemize}

\vspace{14pt}

\noindent
Using the measured \lisv\ abundance, which is known to be smaller than the abundance predicted in the standard BBN, constraints on the lifetime and abundance of the charged relic are derived and compared with previously published constraints derived from catalyzed production of \lisx\ .

\par
In Chapter \ref{Chapter:ExtraDimensions}, I review the motivations for introducing extra dimensions\index{extra dimensions} into spacetime as well as the existing constraints on higher dimensions from both collider experiments and astrophysics experiments. I derive new bounds on the size of nonwarped extra dimensions by calculating the abundance of Kaluza-Klein gravitons\index{Kaluza-Klein gravitons} in such models, and comparing this result to limits derived from the comparison of BBN predictions to the observed abundance of primordial \lisx\ . This calculation and constraints were originally published in:
\vspace{14pt}
\noindent
{\small
\begin{itemize}\item
   R.~Allahverdi, {\bf C.~Bird}, S.~Groot Nibbelink and M.~Pospelov,
  ``Cosmological bounds on large extra dimensions from non-thermal  production
  of Kaluza-Klein modes'',
  Phys.\ Rev.\ D {\bf 69}, 045004 (2004)
  [arXiv:hep-ph/0305010].

\end{itemize}}
\vspace{14pt}
\noindent
In addition, I demonstrate the Kaluza-Klein gravitons produced in the early Universe could become trapped in the galaxy, and decay in the present. These decays produce both $\gamma$-rays and positrons, with the positrons annihilating to produce an observable flux of $511 \; keV$ $\gamma$-rays\index{$\gamma$-rays!511 keV}. By comparison with the $511 \; keV$ flux observed by the INTEGRAL\index{INTEGRAL} satellite, I derive new constraints on the size of the extra dimensions. These calculations represent original research which is previously unpublished.

\par
Through these three classes of physics beyond the Standard Model, I will introduce and demonstrate a variety of methods in which new theories can be probed by examining both their effects on the early Universe and the remnant signatures that they have left in the modern Universe. As will be shown throughout this \thesis, the effects of new physics in the early Universe can be used to probe phenomena that are beyond the reach of terrestrial experiments.

%% file: chapter1b.tex
\chapter{Dark Matter \label{Chapter:DarkMatter}}\index{dark matter|(}

\section{Overview}

\par
One of the oldest and most important problems in modern cosmology is the missing mass of the Universe. Baryonic matter\index{baryonic matter}, such as luminous matter in the form of stars and nebulae\index{nebulae} and non-luminous matter in the form of dust and planets, accounts for less than 5\% of the total energy content \cite{Spergel:2003cb} of the Universe. The remaining matter, which forms 23 \% of the energy density of the Universe, is believed to be in the form of dark matter\index{dark matter}, and cannot be explained by the Standard Model\footnote{It is possible to explain dark matter using massive neutrinos, however limits on the mass of the neutrinos\index{neutrinos} in the Standard Model exclude them as the primary form of dark matter.}. 

\par
The direct detection of dark matter\index{dark matter} and the determination of its properties is inhibited by the apparent weakness of its interactions. At present dark matter can only be detected through its gravitational effects , and therefore the nature of dark matter\index{dark matter} is still undetermined. Most models require dark matter to interact with the Standard Model through other forces as well, however studying dark matter with these other forces requires either collider experiments or nuclear scattering experiments, neither of which has yet detected a clear signal of dark matter.  

\par
There are several candidates for dark matter. The most common are in the form of weakly interacting massive particles (WIMPs)\index{dark matter!WIMPs}. These as yet undetected particles are expected to by thermally produced in the early universe. If the particles were strongly interacting, then all WIMPs\index{dark matter!WIMPs} would have annihilated early in the history of the Universe, while WIMPs\index{dark matter!WIMPs} with no interactions are overproduced. The evolution of the Universe with WIMPs\index{dark matter!WIMPs} is well understood, and standard methods from cosmology (see for example Ref. \cite{Kolb:1990vq}) can be used to precisely calculate the dark matter abundance.

\par
In this chapter the properties of several dark matter candidates will be derived and presented. In Section \ref{Section:DarkMatter}, seven minimal models will be developed which rely on a minimum amount of new physics. Although these models are minimal, they represent effective models for more complicated dark matter models and the properties and constraints derived are generic. In Section \ref{Section:Abundance}, I calculate the dark matter abundance predicted by each model, and use this result and the observed dark matter abundance to constrain the parameter space of the model. In Section \ref{Section:DDMS}, I further constrain the parameter space of each model by calculating the cross-section for scattering of the WIMP\index{dark matter!WIMPs} from a nucleon, and comparing this result to the limits set by dedicated dark matter searches. In Section \ref{Section:ColliderS} I review the methods used at high energy particle colliders to search for invisible Higgs\index{Higgs!invisible} decays, which is the signal expected for each of the models presented in this \thesis. These results provide the region of parameter space for each model which can be probed by experiments such as the LHC\index{Large Hadron Collider} and the Tevatron\index{Tevatron}. Finally, in Section \ref{Section::LDM} I review both the motivations and limitations of sub-GeV WIMPs\index{dark matter!WIMPs}\index{dark matter!light}, and then derive constraints on light dark matter in each model using the abundance constraints and the current limits on the invisible decays of the B-meson\index{B mesons}. As will be demonstrated in this chapter, each of the minimal models has unique and interesting properties.

\input{chapter1e.tex}

%% file: chapter1e.tex
\section{Minimal Models of Dark Matter \label{Section:DarkMatter}}

\par
There are many interesting candidates for dark matter. In many of these models, the dark matter candidate is motivated by another, often more complicated  theory, such as the lightest supersymmetric\index{supersymmetry} particle and Kaluza-Klein\index{Kaluza-Klein gravitons} gravitons (motivated by the possible existence of higher dimensions). However it is also possible that dark matter is unrelated to any other theory, and is just a single new particle or a few new particles. 

\par 
In this section, I present several minimal models of dark matter in which only a few new particles are added to the Standard Model \footnote{A complete review of all minimal models is beyond the scope of this dissertation, and as such I will only include models in which the interaction with the Standard Model is provided by a Higgs\index{Higgs} or Higgs-like boson.}. In addition, in each model the WIMP is made stable by only allowing interactions containing an even number of WIMPs\index{dark matter!WIMPs}. 
\par
These models are simple, yet provide an explanation for the effects of dark matter, and can also be used as effective theories for more complicated dark matter models.  The models considered in this \thesis\ are:

\begin{itemize}
\item Model 1: {\it Minimal Model of Dark Matter (MDM)}\index{dark matter!Minimal Model}\index{dark matter!scalar}
\newline - In this model, a single scalar field is added to the Standard Model, which couples only to the Standard Model Higgs boson. This represents the simplest model of dark matter that can produce the observed abundance. 
\item Model 1b: {\it Next-to-Minimal Model of Dark Matter} \nopagebreak
\newline - This model is identical to MDM, but introduces a second scalar field which couples to the scalar WIMP and mixes with the Higgs boson. 
\item Model 2: {\it Minimal Model of Dark Matter with Two Higgs Doublet} \nopagebreak
\newline - The simplest Higgs model involves a single Higgs boson, but this may not be the correct model of nature. There exist several models which include two Higgs bosons\index{Higgs!two higgs doublet}, with one coupled to up-type quarks and one coupled to down-type quarks and leptons. This model of dark matter introduces a scalar WIMP\index{dark matter!scalar} which can couple to one or both of these Higgs bosons.

\item Model 3: {\it Minimal Model of Fermionic Dark Matter (MFDM)} \index{dark matter!fermionic} \nopagebreak
\newline- In this model, a Majorana fermion\index{Majorana fermions} WIMP is added to the Standard Model. However the fermion in this model cannot couple directly to the Higgs boson, and so an additional scalar field is required to mediate the interactions between dark matter and the Standard Model.
\item Model 4: {\it Minimal Model of Fermionic Dark Matter with Two Higgs Doublets} \nopagebreak
\newline - As with Model 2, it is possible that there are two different Higgs bosons. In this model a Majorana fermion WIMP is the dark matter candidate, which can couple to one or both of these Higgs fields.
\item Model 4b: {\it Higgs-Higgsino Model} \nopagebreak \index{dark matter!higgsino}\index{higgsino}
\newline- In supersymmetric models\index{supersymmetry}, each Higgs boson is accompanied by a fermionic partner, the Higgsino. In this model a fermionic WIMP is coupled to both the Higgs and the Higgsino. 
\item Model 5: {\it Dark Matter in Warped Extra Dimensions}\index{extra dimensions!warped} \newline \nopagebreak
 - In models with warped extra dimensions, there is an additional field known as the radion\index{Kaluza-Klein gravitons!radion} which has similar properties to the Higgs boson. In this model either a scalar or fermion WIMP is added to the Standard Model, but with no Standard Model interactions. Instead the gravitational forces mediate interactions with the Standard Model through via the radion.\index{dark matter!scalar}\index{dark matter!fermionic}
\end{itemize}

\noindent
In this section each model will be developed, with constraints and experimental sensitivities given in the following sections.

\subsection{Model 1: Minimal Model of Dark Matter \label{Section::MDM}}\index{dark matter!Minimal Model|(}\index{dark matter!scalar|(}
\nopagebreak
\par
The minimal model of dark matter introduces a singlet scalar to the Standard Model \cite{Silveira:1985rk,McDonald:1993ex,Burgess:2000yq}, which interacts with the Standard Model through the exchange of a Higgs boson. This represents the simplest model which can explain the properties of dark matter. 
\par
The Lagrangian for this model is given by

\begin{equation}
\begin{split}
-L_S &= \frac{m_0^2}{2}S^2 + \frac{\lambda_S}{4} + \lambda S^2 H^{\dag}H \\ &= \frac{m_0^2 + \lambda v_{ew}^2}{2}S^2 + \frac{\lambda_S}{4} + \lambda v_{ew} h S^2  + \frac{ \lambda}{2} S^2 h^2
\end{split}
\end{equation}

\noindent
where H is the Standard Model Higgs doublet\index{Higgs!Standard Model}, $v_{ew} = 246\; GeV$ is the Higgs vacuum expectation value, and h is the corresponding Higgs boson, with $H = (0 , (v_{ew}+h)/\sqrt{2})$. The physical mass of the scalar is $m_S^2 = m_0^2 + \lambda v_{ew}^2 $.

\par
As will be demonstrated in later sections, the coupling constant and the Higgs mass appear together in each calculation. As such, the model is reparameterized using

\begin{equation}
\kappa^2 \equiv \lambda^2 \left( \frac{100 GeV}{m_h} \right)^4
\end{equation}

\noindent
where $m_h$ is the Higgs boson mass.

\index{dark matter!Minimal Model|)}
\subsection{Model 1b: Next to Minimal Model of Dark Matter}
\nopagebreak
\par
It is also possible that the scalar WIMP has no interactions with the Standard Model particles. In this case, a next-to-minimal model of dark matter can be constructed in which the scalars are coupled to a second singlet scalar,U. Since the WIMPs\index{dark matter!WIMPs} must annihilate to Standard Model particles, this new intermediate scalar must couple to the Standard Model. However existing experimental bounds restrict a direct coupling of U to Standard Model fermions or gauge bosons. Therefore in this model the U-boson is taken to mix with the Standard Model Higgs field.
\par
The Lagrangian for this model is

\begin{equation}
\begin{split}
-{\cal L}_S &= \frac{\lambda_S}{4}S^4+\frac{m_0^2}{2} S^2+
(\mu_1 U + \mu_2 U^2) S^2   + V(U)+ \eta' U^2 H^{\dag} H 
\\&\\
&= \frac{m_S^2}{2} S^2+\frac{m_u^2}{2} u^2+ 
\mu u S^2   + \eta v_{EW} u h +...,
\end{split}
\label{tsdm}
\end{equation}

\noindent
where in the second line only the mass terms and relevant interaction terms are listed, and where $u$ is the excitation of the U-boson field above its vacuum expectation value,
\begin{displaymath}
U = <U> + u
\end{displaymath}

\noindent
 The final term in the second line of Eq. \ref{tsdm}
gives the mixing between $u$ and $h$. 

\par
If this mixing is significant,
the existing bounds on the higgs mass would also place a lower bound on the mass of the u-boson. If the u-boson is light, then it is possible that it could violate existing experimental bounds. It is also possible that a light u-boson could contribute as a second component of dark matter. Therefore in this dissertation it is assumed that $m_u \gg m_h, m_S$. However this region of parameter space is identical to the MDM, with the redefinition 

\begin{equation}
\kappa^2 \equiv \frac{\mu^2 \eta^2 }{m_u^4} \left(\frac{100 \;{\rm GeV}}{m_h} \right)^4
\end{equation}

\noindent
and as a result, all of the experimental bounds and searches for the scalar in the minimal model also apply to the scalar WIMP in the next-to-minimal model.

\subsection{Model 2: Minimal Model of Dark Matter with 2HDM \label{Section:2HDMMDM}}
\nopagebreak
\par
Another possible extension of the minimal model of dark matter is the addition of another Higgs particle. One motivation for this model is to allow more freedom in the properties of the Higgs mechanism. Although the Standard Model can be viable with a single Higgs field, there is no evidence from experiment or from theoretical predictions for there to exist only one type of Higgs. Furthermore, the existence of a second Higgs doublet is required in supersymmetric models to avoid both a gauge anomaly and to allow both the up and down type quarks to have Higgs couplings. 
\par
There are a few different common two-higgs doublet models, with different Standard Model particles coupled to each of the Higgs bosons. In this \thesis, the Type-II model is used in which one Higgs is coupled to the up type quarks and the second one is coupled to down type quarks and leptons, and which is the model required for the minimal model of supersymmetry\footnote{For a more detailed review of the Type-II model see, eg. Ref\cite{Gunion:1989we}}. 
\par
In contrast to the Standard Model Higgs, the vacuum expectation values for the 2HDM are not known, with the only constraint being $v_u^2 + v_d^2 = v_{ew}^2$. Due to this constraint, it is common to use the parameter $\tan \beta \equiv v_u/v_d$. Furthermore, since the mass ratio of top and bottom quarks is proportional to $v_u/v_d$, it is also common to take $\tan \beta$ to be large \cite{Carena:1994bv,Hall:1993gn,Rattazzi:1995gk} so that the Yukawa couplings for top and bottom type quarks are similar in magnitude. 
\par
The other motivation for this model of dark matter is in the possibility of light WIMPs\index{dark matter!WIMPs}.  As will be outlined in Section \ref{Section::LDM}, there are several experiments whose results could be interpreted as evidence of lighter WIMPs\index{dark matter!WIMPs}, with masses in the O($1 \; GeV$) range. In the minimal model of dark matter, the mass of the scalar WIMP receives a contribution from the Higgs vev of O(200\; GeV), $m_{DM}^2 = m_0^2 + \lambda_S v_{ew}^2$ , and therefore a sub-GeV WIMP requires significant fine-tuning of $m_0$ and $\lambda_S$ to reduce the mass by the required two orders of magnitude. In the 2HDM model, the corresponding correction to the WIMP mass can be of order $v_d \sim O(1 \; GeV)$ and therefore it may require very little fine-tuning. 

\par
In this section I will introduce three special cases of the minimal model of dark matter with 2HDM. In general the dark matter couplings to the Higgs bosons will be of the form

\begin{equation}
-{\cal L} = \frac{m_0^2}{2}S^2 + \lambda_1 S^2 (|H_d^0|^2 + |H_d^-|^2) + 
\lambda_2 S^2 (|H_u^0|^2 + |H_u^+|^2) + \lambda_3 S^2 (H_d^-H_u^+ - H_d^0 H_u^0)
\end{equation}

\noindent
However unlike the minimal model of dark matter, this model has too many unknown parameters to be fully constrained by the dark matter abundance. However the most interesting properties of this model are observable in certain special cases. The three special cases which will be studied in this \thesis are those in which a single $\lambda_i$ is taken to be non-zero. In particular, the special cases are:

\begin{list}{}
\item {\it Case 1} corresponds to $\lambda_1 \gg \lambda_2,\lambda_3$ or a scalar WIMP which interacts with down-type quarks and leptons through Higgs mediation.
\item {\it Case 2} corresponds to $\lambda_2 \gg \lambda_1,\lambda_3$ or a scalar WIMP which interacts with up-type quarks through Higgs mediation. In this case, the Higgs vev which appears in all the calculations in close to $v_{SM}$, and for most of the WIMP mass range $H_u$ decays predominantly to the weak bosons as in the single Higgs model. As a result, this case is almost identical to the minimal model of dark matter presented in Section \ref{Section::MDM}. 
\item {\it Case 3} corresponds to $\lambda_3 \gg \lambda_1,\lambda_2$ or a scalar WIMP which interacts with both up and down-type quarks and leptons through Higgs mediation.
\end{list}

\par
In the general model, the physical mass of the scalar is given by

\begin{equation}
m_S^2 = m_0^2 + \lambda_1 v_d^2 + \lambda_2 v_u^2 - \lambda_3 v_u v_d
\end{equation}

\noindent
In the special case of $\tan \beta$ large, and $\lambda_1 \gg \lambda_2,\lambda_3$ ({\it Case 1}), the scalar mass is

\begin{equation}
m_S^2 = m_0^2 + \frac{\lambda_1 v_{ew}^2 }{\tan^2 \beta}
\end{equation}

\noindent
which, unlike the previous models, can be of order $O(1 \; GeV)$ without significant fine-tuning. For the case of $\lambda_3 \gg \lambda_1,\lambda_2$({\it Case 3}) and large $\tan \beta$, the mass is

\begin{equation}
m_S^2 = m_0^2 + \frac{\lambda_3 v_{ew}^2}{\tan \beta}
\end{equation}

\noindent
which can also be small without requiring significant fine-tuning. The third case, in which $\lambda_2$ dominates ({\it Case 3}), is nearly identical to the MDM and cannot contain sub-GeV WIMPs\index{dark matter!WIMPs} without significant fine-tuning.

\input{chapter1f.tex}
\index{dark matter!Minimal Fermionic Model|)}

\subsection{Model 4: Fermionic Dark Matter with 2HDM \label{Section::2HDM+f}}

\par
The model presented in this section is similar to the model presented in Section \ref{Section:2HDMMDM}, but in this model the WIMPs\index{dark matter!WIMPs} are Majorana fermions. As in that section, it is assumed in this model the there exist two Higgs doublets, with one Higgs coupled only to up-type quarks and one coupled only to down-type quarks and leptons.
\par
As in the previous section, the fermions cannot couple directly to the Higgs but must instead couple through an intermediate boson,

\begin{equation}
\begin{split}
- {\cal L} =&\frac{m_0^2}{2} \overline{\chi}\chi+\frac{m_U^2}{2} U^2 + 
\mu U \overline{\chi}\chi + \eta_1 U^2 (|H_d^0|^2 + |H_d^-|^2) \\& + 
\eta_2 U^2 (|H_u^0|^2 + |H_u^+|^2)  + \eta_3 U^2 (H_d^-H_u^+ - H_d^0 H_u^0)+ \eta_U U^4
\end{split}
\end{equation}

\noindent
After symmetry breaking, the relevant terms reduce to

\begin{equation}
\begin{split}
- {\cal L} =
&\frac{m_{\chi}^2}{2} \overline{\chi}\chi + \lambda_1 v_d H_d \overline{\chi}\chi + \lambda_2 v_u H_u \overline{\chi}\chi + \lambda_3 v_u H_d \overline{\chi} \chi
\end{split}
\end{equation}

\noindent
assuming that $M_U >> m_H, m_{\chi}$. 

\par
As in Section \ref{Section:2HDMMDM}, three special cases of this model will be considered corresponding to a single $\lambda_i$ dominant.  
 In this \thesis, only the special cases of $\lambda_1$ and $\lambda_3$ dominant will be studied. As will be seen in Section \ref{Section::LDM}, the special case of $\lambda_3$ dominant is particularly interesting as it produces sub-GeV fermionic dark matter. The special case of $\lambda_2$ dominant is similar to the minimal model of fermionic dark matter presented in Section \ref{Section::MFDM}, and will not be studied further.

\subsection{Model 4b: Higgs-Higgsino Model \label{Section::Higgsino}}
\index{higgsino|(}\index{dark matter!higgsino|(}

\par
Another simple form of fermionic dark matter is a Majorana fermion coupled to a Higgs-Higgsino pair. This model is inspired by supersymmetry, in which each Higgs boson is accompanied by a fermion field known as a {\it Higgsino}. However in this model the Higgsino is only assumed to be a fermion field with an SU(2)$\times$U(1) charge, with the quantum numbers of a Higgs, without requiring the presence of supersymmetry. In this model, the dark matter is the Majorana fermion, which is analogous to the neutralino in supersymmetric models. This model exhibits the basic properties of many supersymmetric models of dark matter, without the additional complications that are present in such models.

\par
The terms of the Lagrangian for this model which are relevant for these calculations are 

\begin{equation}\label{Eq::Higgsino}  
- L_f = \frac{1}{2} M \overline{\psi} \psi + \mu \overline{\bar{H}_d} \bar{H}_u + \lambda_d \overline{\psi} \bar{H}_d H_d + \lambda_u \overline{\psi} \bar{H_u} H_u
\end{equation}

\noindent
where $\bar{H}_d, \bar{H}_u$ are the Higgsino fields. In this model it is also assumed that $M \ll \mu,\lambda_u v_u$, and as before $\tan \beta$ is assumed to be large.

\par
The physical fields in this model are linear combinations of the fields given in Eq \ref{Eq::Higgsino}. The dark matter candidate is

\begin{equation}
\chi = - \psi \cos \theta + \bar{H}_d \sin \theta \quad \quad \sin^2 \theta \equiv \frac{\lambda_u^2 v_u^2}{\lambda_u^2 v_u^2 + \mu^2}
\end{equation}

\begin{displaymath}
m_1 = M \cos^2 \theta
\end{displaymath}

\noindent
which is the lightest mass eigenstate. The terms in the effective Lagrangian which describe the mass and interactions of this state are

\begin{equation}
L_{eff} = \frac{1}{2} m_1 \overline{\chi} \chi - \lambda_d \sin \theta \cos \theta H_d \overline{\chi} \chi 
\end{equation}

\noindent
 At energy scales significantly smaller than $m_U$, which is taken to be large, this model then reduces to the model in Section \ref{Section::2HDM+f}, with $\lambda_d \sin \theta \cos \theta$ corresponding to $\lambda_1$. The 2HDM+fermion model does include an additional effective two Higgs - two fermion coupling which is not significant in the tree-level annihilation cross-sections for WIMPs\index{dark matter!WIMPs} lighter than $m_h$, but which will become important in searching for light dark matter in B-meson decays, in which Higgs loops are present, as shown in Section \ref{Section::LightHiggsino}. As a result, the constraints from abundance calculations, dedicated dark matter searches, and collider searches for the 2HDM+fermion model also apply to this model with the reparameterization
 
 \begin{equation} \label{Eq:higgsino-kappa}
\kappa \equiv \left( \frac{\lambda_d \lambda_u v_u \mu}{\lambda_u^2 v_u^2 + \mu^2} \right) \left(\frac{100 \; GeV}{M_H} \right)^2 \left( \frac{\tan \beta}{100} \right)
\end{equation}
 
\noindent
while the constraints on light dark matter from B-meson decays will be different for the two models. 

\index{dark matter!fermionic|)}
\index{higgsino|)}\index{dark matter!higgsino|)}

\subsection{Model 5: Dark Matter \& Warped Extra Dimensions} \label{Section::EXDM}
\index{extra dimensions!Randall-Sundrum model|(}
\index{extra dimensions!warped|(}

\par
The models presented in the previous sections have used the Higgs boson to provide an interaction between the dark matter candidate and the Standard Model, as is required to produce the correct dark matter abundance. In this section I will introduce an alternative method, in which warped extra dimensions can effectively mediate WIMP annihilations.

\par
Since WIMPs\index{dark matter!WIMPs} cannot interact through electromagnetic or strong nuclear forces\index{strong nuclear forces}, and since interactions through weak nuclear forces\index{weak nuclear forces} are tightly constrained by experiments, it is tempting to consider WIMPs\index{dark matter!WIMPs} which only interact through gravity. However gravity is too weak to produce a significant abundance of thermally produced dark matter. If dark matter is produced in decays of heavy relics, then the gravitational interactions are too weak to produce efficient annihilation, and the result is an overabundance of dark matter. One possible exception is to produce dark matter in regions where gravity is stronger, such as in warped extra dimensions.

\par
The possible existence of extra dimensions\footnote{A more complete review of the motivations for extra dimensions are presented in Chapter \ref{Chapter:ExtraDimensions}, along with several of the common models and constraints. } has become very popular in recent years \cite{Arkani-Hamed:1998rs,Randall:1999ee,Randall:1999vf}, with the primary motivation for such models being a resolution of the {\it hierarchy problem}. The electroweak forces have couplings of the order $O(TeV^{-1})$, while gravitational couplings are of order $M_{PL}^{-1} = \sqrt{G_N} = 0.82 \times 10^{-16} \; TeV^{-1} $. However the Standard Model cannot explain this large difference in the strengths of the forces.

\par
One explanation is that gravity exists in higher dimensions, effectively diluting the gravitational field relative to the other Standard Model fields. 
In these models, the Standard Model fields are trapped on a four-dimensional spacetime brane\index{extra dimensions!branes} while gravity can propagate in higher dimensions as well. Gravitation experiments can probe these higher dimensions, and currently restrict the size of the non-warped extra dimensions to be less than $\sim O(0.1 \;mm)$ \cite{Adelberger:2003zx,Hoyle:2004cw}

\par
The Randall-Sundrum model\index{extra dimensions!Randall-Sundrum model} avoids these constraints by introducing a single extra dimension which is strongly warped \cite{Randall:1999ee,Randall:1999vf}. The spacetime metric for this model is

\begin{equation}
ds_{RS}^2 = e^{- 2 k \phi |y|} \eta_{\mu \nu} dx^{\mu}dx^{\nu} - \phi(x_{\mu})^2 dy^2
\end{equation}

\noindent
where $\phi(x_{\mu})$ behaves in the same manner as a scalar field trapped on the brane, and is referred to as the {\it radion}. As a result of this exponential warping, the extra dimension could be large or non-compact without violating constraints from gravitation experiments. In addition, the effective Planck mass $M_{PL}$, which  determines the gravitation couplings on the brane, is reduced relative to the true Planck mass, $M_*$, by the relation

\begin{equation}
M_* \approx M_{PL} e^{- k \pi r_c}
\end{equation}

\noindent
where $r_c \equiv <\phi>$ is the vacuum expectation value of the radion field. In this model, $M_*$ can be as small as 1 TeV while $M_{PL} = 1.22 \times 10^{16} \; TeV$.

\par
There are a number of possible candidates for dark matter which are naturally contained in extra dimensional models.
 For example, when the gravitational field propagates in the higher dimensions, it can only have certain energy levels or modes due to the boundary conditions on the extra dimension. Each of these modes has the same properties as a massive particle trapped on the brane, and this effective particle is referred to as a {\it Kaluza-Klein graviton} or a {\it Kaluza-Klein mode}. Another possibility is that the brane on which the SM fields are trapped can fluctuate in the higher dimensions, forming bumps in the brane. These fluctuations can also behave like particles trapped on the brane, referred to as {\it branons}\index{extra dimensions!branons}. In the early Universe, the KK gravitons  and the branons can be formed both in the decay of other particles and in the annihilations of Standard Model particles. In the same manner that WIMPs\index{dark matter!WIMPs} freeze-out of thermal equilibrium to form a dark matter abundance, these effective particles can also freeze-out and replicate the effects of dark matter. These models have been studied extensively in Ref \cite{Servant:2002aq,Cheng:2002ej} and Ref \cite{Cembranos:2003mr}.
\par
 In this model, it is only assumed that the dark matter candidate is a new particle and not necessarily an effect of the extra dimensions. It is also assumed that this new particle accounts for the entire dark matter abundance, although it is possible that the observed abundance is a combination of WIMPs\index{dark matter!WIMPs} and Kaluza-Klein gravitons\index{Kaluza-Klein gravitons} or branons\index{extra dimensions!branons}.  

\par
In the previous sections, a minimal number of new particles were introduced, which were then coupled to the Standard Model through the exchange of a Higgs boson. In this section, I again introduce a single new particle \footnote{In this section both a scalar and a fermion are added to the Standard Model, however these are to be considered as two separate models for dark matter}, but now couple it to the Standard Model through the exchange of a Randall-Sundrum radion. 
\par
Since gravitons and radions naturally couple to the energy-momentum tensor, the WIMPs\index{dark matter!WIMPs} naturally interact with the Standard Model without requiring additional interactions. This has the additional benefit of removing one parameter from the model, as the WIMP-gravity coupling is proportional to the WIMP mass instead of an arbitrary coupling constant. Although these properties are also present in models without extra dimensions, in those cases the gravitational interaction is too weak to efficiently annihilate WIMPs\index{dark matter!WIMPs} in the early Universe, with typical annihilation cross sections being of order $\sigma_{ann} \sim O(m_{dm}^2 M_{PL}^{-4})$. 
 Since the Planck mass is several orders of magnitude lower in the Randall-Sundrum model, the annihilation cross-section is much larger in the presence of warped extra dimensions and the WIMPs\index{dark matter!WIMPs} can annihilate efficiently. 

\par
In this section I introduce two models. The first model is a singlet scalar WIMP, with no non-gravitational interactions, and with Lagrangian 

\begin{equation}
L = \frac{1}{2} (\partial_{\mu} S)^2 - \frac{1}{2} m_S^2 S^2
\end{equation}

\noindent
The second model is similar, except the WIMP is a Majorana fermion. The Lagrangian for the second model is,

\begin{equation}
L = \frac{1}{2} \bar{\chi} \cancel{\partial} \chi - \frac{m_{\chi}}{2} \bar{\chi}\chi
\end{equation}

\noindent
As outlined in Ref \cite{Bae:2001id}, in the Randall-Sundrum model, the radion\index{extra dimensions!radions} couples to the trace of the energy-momentum tensor, denoted by $\Theta_{\mu}^{\mu} $, 

\begin{equation}
L_{int} = \frac{\phi}{\Lambda_{\phi}} \Theta_{\mu}^{\mu} 
\end{equation}

\noindent
where $\Lambda_{\phi}$ is the vacuum expectation value of the radion. The couplings of the radion\index{Kaluza-Klein gravitons!radion} to the Standard Model fields was derived in Ref \cite{Bae:2001id}, and for the case of strongly warped extra dimensions are similar to the Higgs couplings.

\par
It should be noted that in the figures for this model, it is assumed that $\Lambda_{\phi} = v_{EW}$. While solving the hierarchy problem does require the size of the extra dimensions to be stabilized with $\Lambda_{\phi} \sim O(TeV)$ \cite{Goldberger:1999un}, there is no further restriction on its size. For comparison with the previous models which rely on a Higgs coupling, and following the examples in Ref \cite{Bae:2001id}, it will be assumed that $\Lambda_{\phi} = v_{EW}$ for the purpose of each calculation. The actual $\Lambda_{\phi}$ dependence included in an effective coupling constant,

\begin{displaymath}
\kappa \equiv \left( \frac{m_{S,f}}{1 \; TeV}^2 \right)\left( \frac{v_{EW}}{\Lambda_{\phi}} \right)^2 \left(  \frac{1\; TeV}{M_{\phi}}\right)^2
\end{displaymath} 

\noindent
where $M_{\phi}$ is the mass of the radion\index{Kaluza-Klein gravitons!radion}. It should also be noted that in the range of $m_{S,f} \gg \Lambda_{\phi}$ the couplings can become non-perturbative and therefore such heavy WIMPs\index{dark matter!WIMPs} are not considered in this model.

\index{extra dimensions!Randall-Sundrum model|)}
\index{extra dimensions!warped|)}

\vspace{16pt}
\section{Abundance Constraints \label{Section:Abundance}}

\par
The primary constraint on any proposed dark matter candidate is that it not overclose the Universe, so that the predicted energy density of dark matter should not exceed the energy density of the Universe. Furthermore, the dark matter density predicted by each model should be consistent with the observed value of $\Omega_{DM} h^2 = 0.1099 \pm 0.0062$ \cite{Dunkley:2008ie} measured by the WMAP\index{WMAP} satellite. 

\par
The most common mechanism for production of dark matter in the early Universe is through thermal production. The early Universe contained high energy fields in hot thermal equilibrium, with all species of particles being created and annihilating. As the Universe expanded and the temperature dropped, the density of each particle species decreased (due to dilution in an expanding universe) and the  production and annihilation reaction rates lowered. At a certain temperature, referred to as the {\it freeze out temperature} and taken to be the temperature where $H \approx \Gamma_{ann} = <\sigma_{ann} v> \Omega_{DM} \rho_{cr}$ for each species, the WIMPs\index{dark matter!WIMPs} became too diffuse to effectively annihilate and the dark matter density froze out\index{freeze out temperature}. 

\par
Using standard methods(see for example Ref. \cite{Kolb:1990vq}), the dark matter abundance at freeze-out can be derived, 

\begin{equation}\label{Eq::Abundance}
\Omega_{DM} h^2 = \frac{1.07 \times 10^9 x_f}{g_*^{1/2} M_{PL} \; GeV \; <\sigma_{ann} v>}
\end{equation}

\noindent
where $x_f = m/T_f$ is the inverse freeze out temperature in units of the WIMP mass, and $g_*$ is the number of degrees of freedom available at freeze out. The annihilation cross-section term $<\sigma_{ann} v>$ in this equation represents the thermal average of the cross-section and the relative velocity of the WIMPs\index{dark matter!WIMPs} at the time of freeze-out. From Eq \ref{Eq::Abundance} and the observed dark matter abundance, it follows that the annihilation cross section has to be $\sigma_{ann} \approx 0.7 \; pb$. 
\par
For most of the parameter space, the thermal average can be related to the cross-section by the formula

\begin{equation}\label{Eq:SCS}
\sigma_{ann} = a + b v^2  \to <\sigma_{ann} v> = a + \frac{6 b T}{m_{DM}}
\end{equation}

\noindent
where a and b represent the s-wave and p-wave parts of the cross-section. However near the resonances, such as occurs at $m_{DM} \sim m_h/2$ in the minimal model of dark matter, this formula fails because the cross-section cannot be written in the form given in Eq \ref{Eq:SCS} due to the presence of the resonance. This formula also fails close to thresholds, where a particle with a slightly higher energy can annihilate to additional particles.
In those mass ranges, the thermal average is given by \cite{Griest:1990kh}

\begin{equation}\label{Eq:TCS}
<\sigma_{ann} v> = \frac{m^{3/2}}{2 \sqrt{ \pi T^{3}}} \int_0^{\infty} e^{- m v^2 / 4T} \sigma_{ann} v^3 dv
\end{equation}

\noindent
This equation provides corrections to account for the highest energy particles in the thermal equilibrium which can annihilate either through a resonance or the particles heavier than the WIMPs\index{dark matter!WIMPs}. These effects widen the resonances in the annihilation cross-section, with the largest correct occurring for WIMPs\index{dark matter!WIMPs} whose masses are slightly below the resonance, and reduce the sharp increase in the cross section at the threshold for production of heavier particles.

\begin{figure}
\begin{center}
\psfig{file=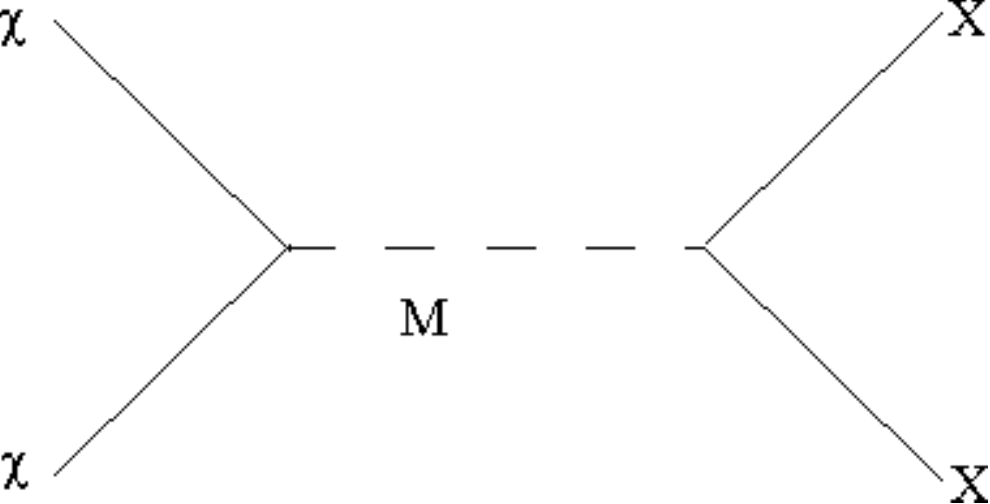,width=0.65\textwidth,angle=0}
\end{center}
\caption{\label{Figure::Annihilation} Generic Feynman diagram for annihilation of WIMPs\index{dark matter!WIMPs}, denoted here by $\chi$. In this diagram, M is a mediator particle and X represents any Standard Model field.}
\end{figure}

\par
It should also be noted that in general, the abundance must be calculated separately for two mass ranges. For WIMPs\index{dark matter!WIMPs} in the range $m_{DM} \gtrsim 2 \;GeV$ the abundance freezes out before hadronization\index{hadronization temperature}, meaning that the annihilation produces unbound quarks, leptons, and (for sufficiently heavy WIMPs\index{dark matter!WIMPs}) gauge bosons and Higgs pairs. For lighter dark matter, with $m_{DM} \lesssim 2 \; GeV$, the WIMPs\index{dark matter!WIMPs}\index{dark matter!light} freeze out after hadronization, and therefore the annihilation produces hadrons as well as leptons, but not unbound quarks.  

\par
In addition, for each model there exists a lower bound on the WIMP mass that results from requiring the model to have perturbative couplings. This bound is called the {\it Lee-Weinberg limit}\index{dark matter!Lee-Weinberg limit} \cite{Lee:1977ua,Vysotsky:1977pe}. As a result, it was originally believed that WIMPs\index{dark matter!WIMPs} could not be lighter than $m_{DM} \sim O(10 \; GeV)$. Since the annihilation cross-sections for fermions are usually suppressed by a factor of $m_{DM}^2/M^4$, where M is the mass of a mediator particle, light fermionic WIMPs\index{dark matter!WIMPs} would require new forces below the electroweak scale. However several recent papers have demonstrated that it is possible to produce $O(GeV)$ mass WIMPs\index{dark matter!WIMPs} with the correct abundance using either scalar WIMPs\index{dark matter!WIMPs}\index{dark matter!scalar} \cite{Boehm:2003bt,Fayet:2004bw,Fayet:2004kz, Bird:2006jd}, or using certain models of fermionic WIMPs\index{dark matter!WIMPs} with enhanced annihilation cross-sections \cite{Bird:2006jd}.

\par
In this section, I derive abundance constraints for each of the minimal models presented in the last section. In each case the abundance is plotted separately for light dark matter\index{dark matter!light}, with the exception of the minimal model of fermionic dark matter in which light WIMPs\index{dark matter!WIMPs} are not possible and in the model of dark matter with warped extra dimensions\index{extra dimensions!warped}, in which case light WIMPs\index{dark matter!WIMPs}\index{dark matter!light} are already excluded.

\subsection{Model 1: Minimal Model of Dark Matter \label{Section::MDM-AC}}\index{dark matter!Minimal Model|(}
\index{dark matter!scalar|(}

\par
\begin{figure}
\begin{center}
$\begin{array}{c}
\psfig{file=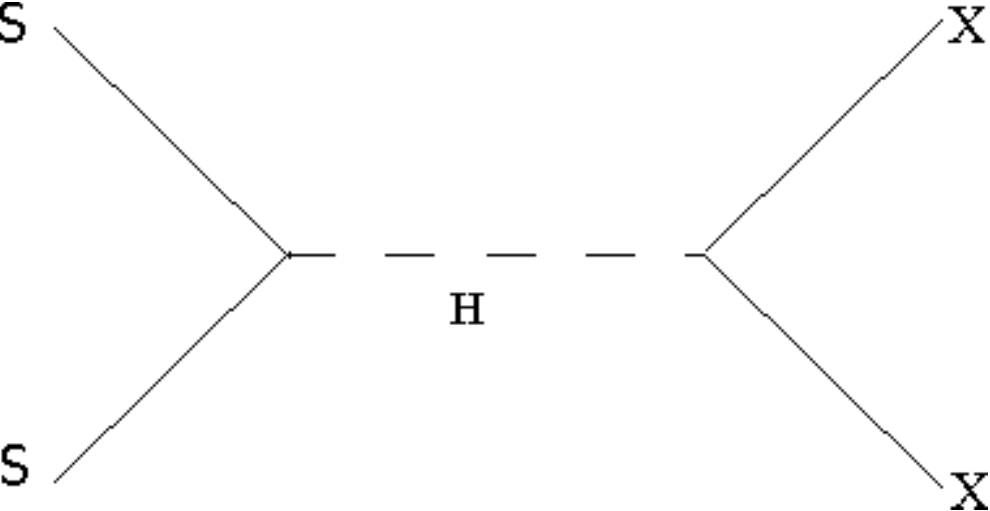,width=0.7\textwidth,angle=0} \\ (a) \\ \psfig{file=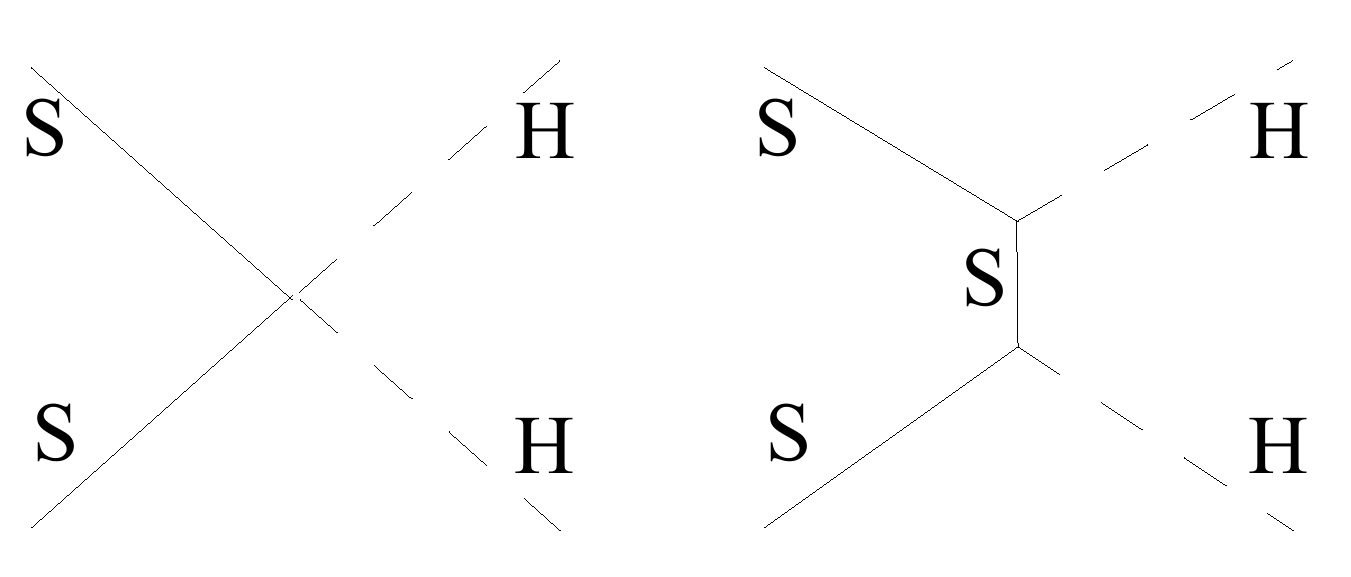,width=0.7\textwidth,angle=0} \\ (b) \quad \quad \quad \quad \quad \quad \quad \quad \quad \quad (c)
\end{array}
$
\end{center}
\caption{\label{Figure::MDM-ann} The Feynman diagram for the annihilation of scalar WIMPs\index{dark matter!WIMPs}\index{dark matter!scalar} in the Minimal Model of Dark Matter\index{dark matter!Minimal Model}.In (a), the scalars annihilate via an intermediate Higgs boson\index{Higgs!Standard Model} to produce any Standard Model fields. For sufficiently heavy scalars, diagrams (b) and (c) also contribute to the annihilation of scalars into Higgs boson pairs\index{Higgs!Standard Model}.}
\end{figure}

For the minimal model of dark matter, the annihilation cross section is calculated using the diagrams in Figure \ref{Figure::MDM-ann}. 
The cross section can then be written in terms of the decay width of a virtual Higgs boson\index{Higgs},

\begin{equation}
\sigma_{ann} v_{rel} = \frac{8 v_{EW}^2 \lambda^2}{(4 m_S^2 - m_h^2)^2+m_h^2 \Gamma_h^2} \; \; \lim_{m_h \to 2 m_s} \frac{\Gamma_{hX}}{m_h}
\end{equation}

\noindent
The Higgs decay width has been studied extensively in searches for the Higgs boson\index{Higgs!Standard Model} (for a review, see \cite{Gunion:1989we}), and writing the cross-section in this form then simplifies the abundance calculation.

\par
\begin{figure}
\label{Figure:MDMabundance}
\begin{center}
$\begin{array}{c}
\psfig{file=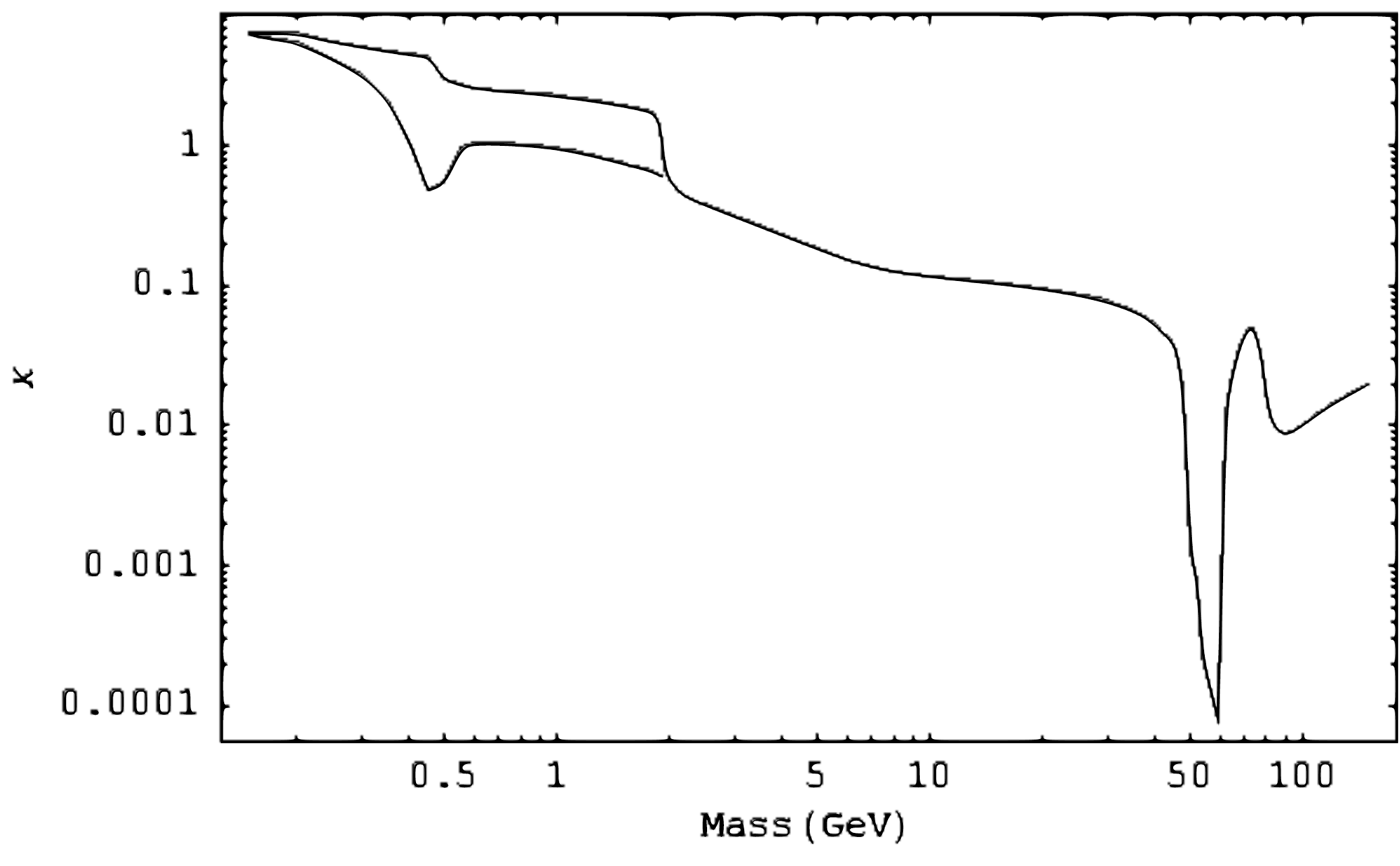,width=0.9 \textwidth,angle=0} \\ (a)\\
\psfig{file=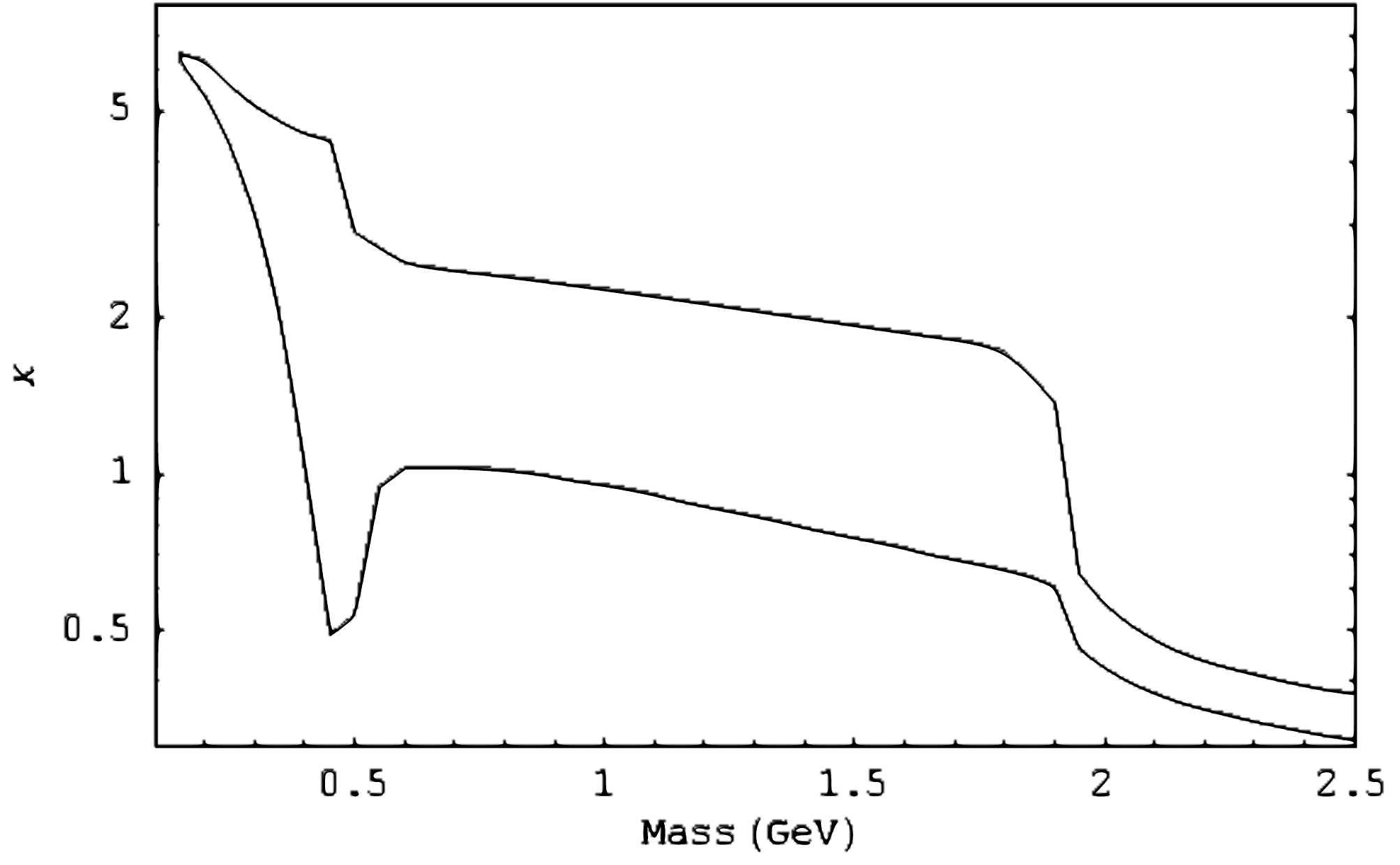,width=0.9 \textwidth,angle=0} \\(b) 
\end{array}$
\end{center}
\caption{Abundance constraints on the coupling and mass of the scalar in the minimal model of dark matter. The first plot gives the constraints for heavy WIMPs\index{dark matter!WIMPs}, while the second plot gives the approximate constraints for GeV scale WIMPs\index{dark matter!WIMPs}. 
For sub-GeV WIMPs\index{dark matter!WIMPs}, there is some uncertainty in the annihilation cross section related to the effects of non-zero temperature on resonant annihilation modes and the effects of annihilations during hadronization. The region above the curves corresponds to abundances below the observed dark matter abundance, but are not excluded.}
\end{figure}

For WIMPs\index{dark matter!WIMPs} in the range of $m_s \lesssim 60 \; GeV$ the annihilation cross-section is dominated by production of b-quarks and $\tau^+ \tau^-$ pairs, while heavier WIMPs\index{dark matter!WIMPs} in the range of $m_s \gtrsim 85 \; GeV$ annihilate efficiently to $W^+W^-$ and $Z^0Z^0$ pairs. It should also be noted that the peak in the annihilation cross-section corresponding to the production of an on-shell Higgs is located at the Higgs mass\index{Higgs}, which is currently unknown but is constrained to $m_h \geq 114 \; GeV$\cite{Abbiendi:2006gd,Abbiendi:2007jk,Nagai:2008zz} and $m_h \leq 182 \; GeV$ \cite{Grunewald:2007pm}, while data from the ALEPH detector may indicate $m_h \sim 115 \;GeV$\cite{Heister:2001kr}. In this calculation the Higgs mass will be taken to be $m_h = 120 \; GeV$. If the Higgs mass is different from this, the peak will be located in a different region and the corresponding lowering of the coupling constant, illustrated in Figure \ref{Figure:MDMabundance}(a), will also move.

\par
For sub-GeV WIMPs\index{dark matter!WIMPs}, this cross section depends on the decay width of a light Higgs, which was previously studied two decades ago \cite{Voloshin:1985tc,Raby:1988qf,Truong:1989my}. However there exist uncertainties in the annihilation cross section due to the fact that previous calculations were done at zero-temperature, while the decay width used here is properly calculated at a finite temperature. In particular, it is unclear whether the resonances in the Higgs decay width will have an effect, since the thermal bath in the early Universe may significantly broaden the hadronic resonances. There also exist some uncertainty as to the temperature at which hadronization becomes important\index{hadronization temperature}. Therefore in the  abundance calculation, \we introduce a range of decay widths corresponding to the zero temperature case and the high temperature case, with the true decay falling somewhere between these two extremes. The result is plotted in Figure \ref{Figure:MDMabundance}b.

\par
For scalars lighter than $\sim 150 \; MeV$, the main annihilation channel is to electrons and muons. In the range $150 \; MeV \lesssim m_S \lesssim 350 \; MeV$ the annihilation cross-section is dominated by annihilation to pion pairs. The Higgs-pion coupling is calculated using the standard low-energy theorems \cite{Voloshin:1985tc}. It should also be noted that the requirement that the scalar abundance be equal to the observed dark matter abundance requires the coupling to be large, with $\kappa \gtrsim \sqrt{4 \pi}$, and as a result the theory would become non-perturbative.

\par
In the range $350 \; MeV \lesssim m_S \lesssim 650 \; MeV$ kaons and other bound strange quarks will begin to be produced, as well as several resonances. The most important of these is the $f_0(980)$ resonance, which creates an enhancement in the annihilation cross section at $m_S \sim 490 \; MeV$. However the width of this resonance is only known at zero temperature, whereas in the early Universe this resonance is important at $T \sim (0.05-0.1) m_S \approx (25 - 50) \; MeV$. The result of this higher temperature is to destroy a fraction of the resonances during the annihilation, which results in a weakening of the effect. For this reason, \we have taken one extreme to be the narrowest resonance consistent with experimental bounds, which results in the largest cross-section, and the other extreme to be complete destruction of the resonance and no effect on the cross-section.

\par
For WIMPs\index{dark matter!WIMPs} in the range $650 \; MeV \lesssim m_S \lesssim 1 \; GeV$ the annihilation cross-section includes several resonances and numerous decay channels. Although the calculation cannot be done precisely in this range, it is reasonable to assume that there will be no significant source of suppression or enhancement of the cross-section in this range, and as such \we extrapolate the cross-section in this region. 

\par
Above $m_S \sim 1 \; GeV$, the freeze-out temperature\index{freeze-out temperature} of the WIMPs\index{dark matter!WIMPs} is sufficiently high that hadronization has not occurred and the annihilation cross-section can be calculated using unbound quarks. However as before there is still some uncertainty in this calculation. At the threshold for charm quark production the temperature is just below the hadronization temperature\index{hadronization temperature}, while at the threshold for D-meson\index{D mesons} production (the lightest bound state of a charm quark) the temperature is high enough to destroy these states. Therefore \we take one extreme for the cross-section to be introduction of charm quarks at the lower threshold , and one extreme to be introduction of charm quarks only at the higher threshold. When the scalars are taken to heavier still, annihilation to $\tau$-leptons also becomes important.

\par
The total cross-section has been calculated, and using Eq \ref{Eq::Abundance}, the abundance has been calculated. The results are plotted in Figure \ref{Figure:MDMabundance} in terms of the parameter,

\begin{equation} \label{Eq:KappaMDM}
\kappa \equiv \lambda \left( \frac{100 \; GeV) }{m_h} \right)^2
\end{equation} 

\noindent
Using the requirement of perturbative couplings, with $\kappa \lesssim \sqrt{4 \pi}$, the range of $m_S \lesssim 300 \; MeV$ is excluded. As already mentioned in this section, there is uncertainty in the decay width of a virtual Higgs boson at low energies and non-zero temperatures, resulting in uncertainties in the constraint on $\kappa$ for $m_S \lesssim 2 \; GeV$. For $m_S \sim m_h/2$, the scalars annihilate through the Higgs resonance, resulting in a larger cross-section, which then requires $\kappa$ to be smaller in this region. It should also be noted that in most of the models in this section, the abundance constraints are only given for $m_S \lesssim 100 \; GeV$. The WIMPs\index{dark matter!WIMPs} could be heavier than this, with masses as high as a few TeV still being viable candidates for dark matter, however such WIMPs\index{dark matter!WIMPs} would be difficult to detect and are not expected to be well constrained by present experiments. Also in each of these plots, the region of parameter space above the lines corresponds to models which have an abundance lower than the observed dark matter abundance, although the scalar could still be one component of dark matter.
\index{dark matter!Minimal Model|)}
\subsection{Model 2: Minimal Model of Dark Matter with 2HDM \label{Section:MDM2HDM-AC}}

\par
The abundance calculation in this model proceeds in the same manner as in Section \ref{Section::MDM-AC}, with the Standard Model Higgs decay width replaced with the appropriate decay width for one of the higgses in the two higgs doublet\index{Higgs!two higgs doublet}. For the purpose of comparison with other models in this \thesis, it is assumed that each of the Higgs bosons has a mass of $m_H = 120 \; GeV$, although this assumption is not required. As with the minimal model of dark matter\index{dark matter!Minimal Model}, if the mass of the Higgs is changed the constraints on the parameter $\kappa$ remain the same, except for the position of the Higgs resonance (which appears as a dip located at $m_S \sim m_h/2$ in the plots below).

\begin{figure}
$\begin{array}{c}
\psfig{file=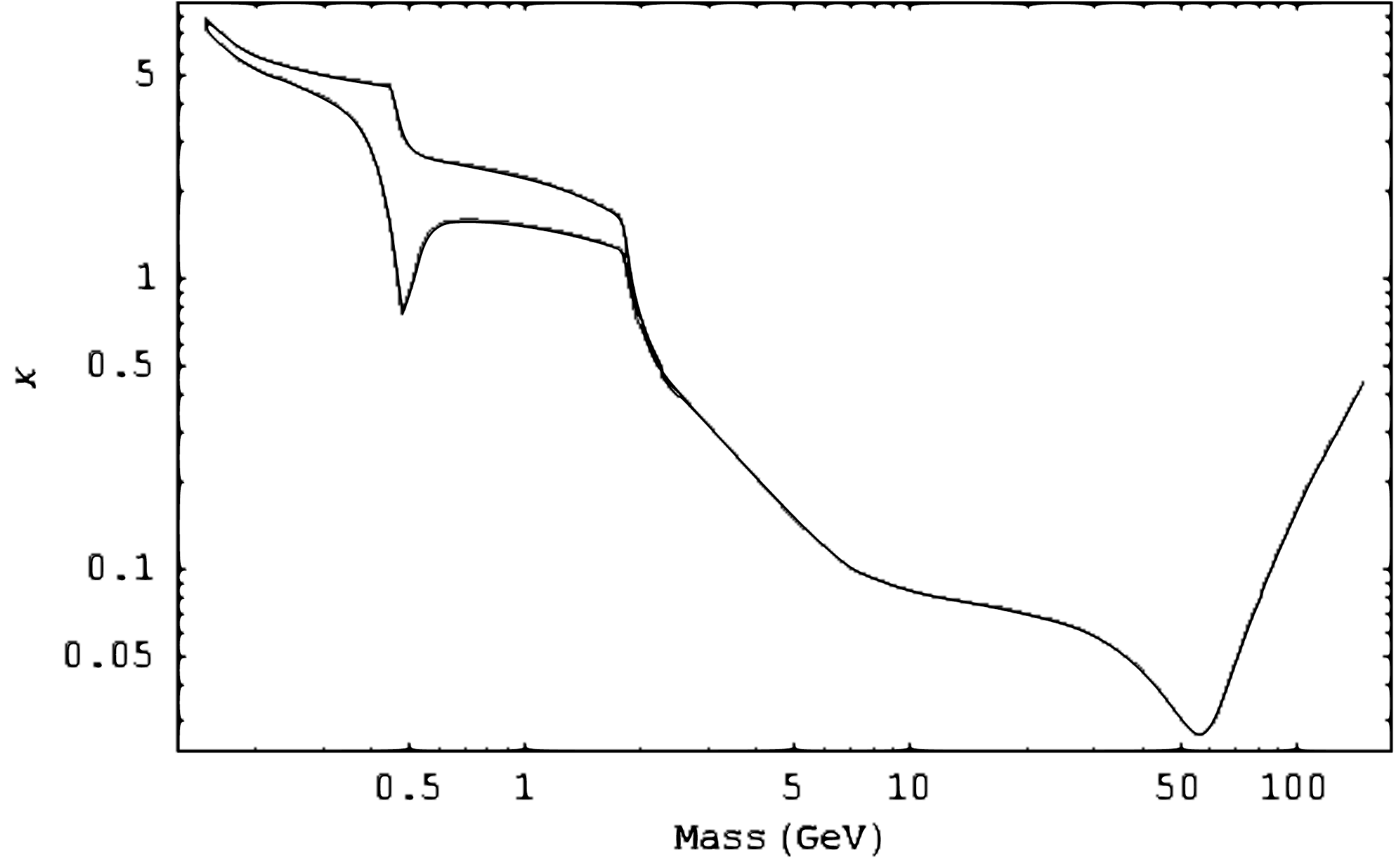,width=\textwidth,angle=0}\\(a) \\
\psfig{file=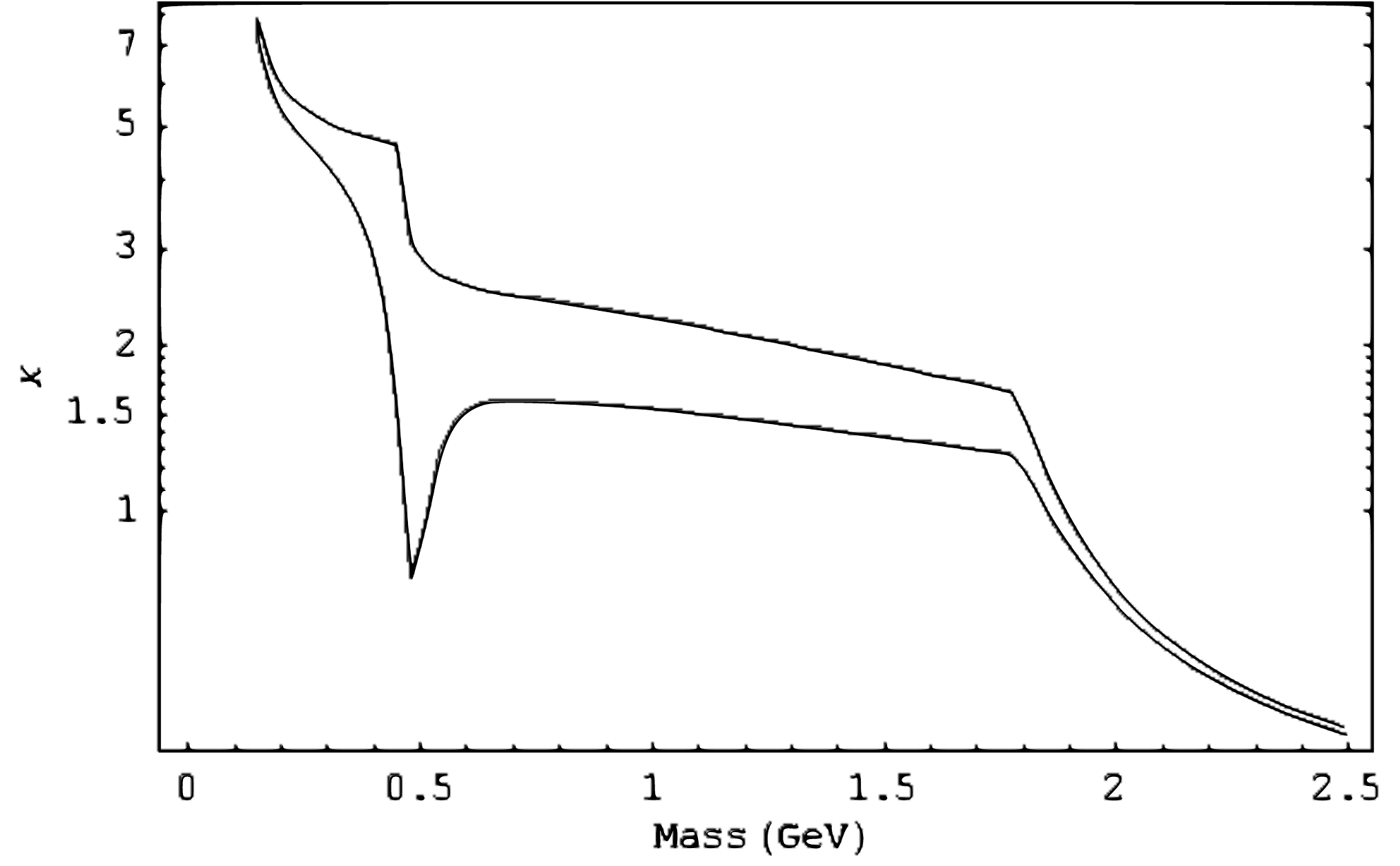,width=\textwidth,angle=0} \\ (b)
\end{array}$
\caption{ \label{Figure::2HDM1-AC} Abundance constraints for the minimal dark matter + 2 HDM, with $\lambda_1 \gg \lambda_2,\lambda_3$ and with $M_{H_d} = 120 {\rm GeV}$.}
\end{figure}

\begin{figure}
\psfig{file=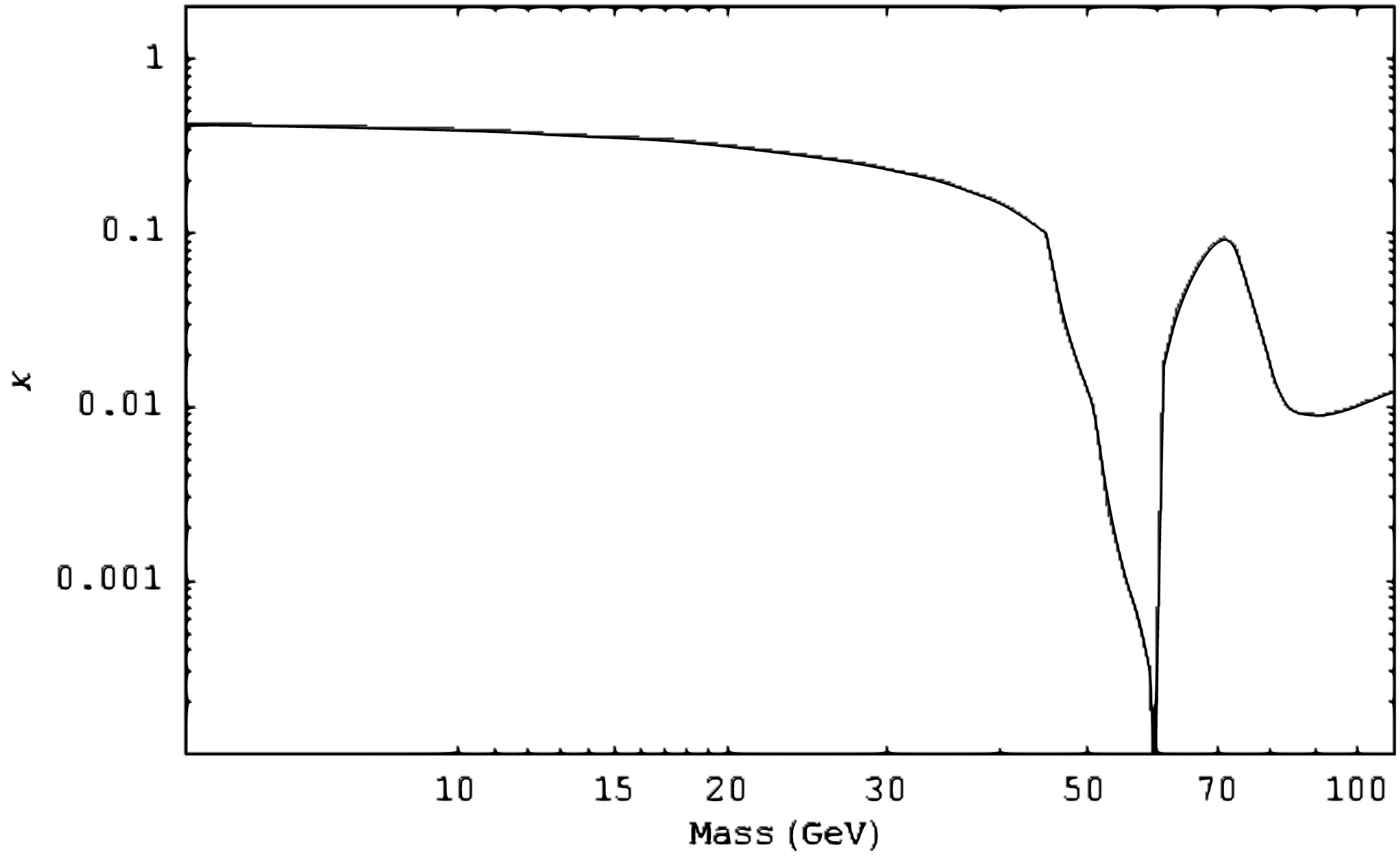,width=\textwidth,angle=0}
\caption{ \label{Figure::2HDM2-AC} Abundance constraints for the minimal dark matter + 2 HDM, with $\lambda_2 \gg \lambda_1,\lambda_3$ and with $M_{H_u} = 120 {\rm GeV}$}
\end{figure}

\par
For the first special case, with $\lambda_1 \gg \lambda_2,\lambda_3$, the scalars annihilate via the $H_d$ boson, which decays to leptons and down-type quarks. The abundance constraints are given in Figure \ref{Figure::2HDM1-AC}. Since $v_d \ll v_{EW}$, the width of the Higgs resonance is increased resulting in a less apparent dip in the allowed value of $\kappa$ when compared with the MDM results. 

\par
As mentioned before,when $\tan \beta$ is taken to be large the case of $\lambda_2$ dominant is very similar to the Minimal Model of Dark Matter\index{dark matter!Minimal Model}, presented in Section \ref{Section::MDM}. The difference is in the lack of annihilation to strange and bottom quarks when the WIMPs\index{dark matter!WIMPs} have masses of a few GeV. As a result, in this mass range the abundance constraints require $\kappa$ to be significantly larger than in the MDM. Although there are uncertainties in the decay width of the virtual Higgs in this case, most of the parameter space which gives the correct dark matter abundance also requires non-perturbative couplings. For WIMPs\index{dark matter!WIMPs} heavier than a few GeV, the abundance constraints are identical to the MDM.

\par
The third case, with $\lambda_3$ dominant, is more interesting. In the large $\tan \beta$ limit, the scalars annihilate through $H_d$ as in the first model, except the effective coupling of the WIMP is enhanced by a factor of $v_u/v_d = \tan \beta$. Therefore, using 

\begin{equation}
\kappa = \lambda_3 \left(\frac{100  \; {\rm GeV}}{M_{H_d}} \right)^2 \left( \frac{\tan \beta}{100} \right)^2
\end{equation}

\noindent
the abundance constraints are identical to those plotted in Figure \ref{Figure::2HDM1-AC}, but two orders of magnitude smaller.

\par
In summary, the three special cases considered are defined in a similar manner, but provide very different abundances. The case of $\lambda_1$ dominant allows for WIMPs\index{dark matter!WIMPs} with $m_S > 400 \; MeV$ to have perturbative couplings, and does not display as large a variation in the allowed values of $\kappa$ over the entire mass range when compared to the other cases. When $\lambda_2$ is dominant, the range of $m_S \gtrsim 5 \; GeV$ is very similar to the minimal model of dark matter\index{dark matter!Minimal Model}, but the lack of annihilations to leptons and strange quarks reduces the possibility of light WIMPs\index{dark matter!WIMPs} by requiring non-perturbative couplings for most of the parameter space.  The final case of $\lambda_3$ dominant has a much smaller value of $\kappa$ due to the $\tan \beta$ enhancement of the annihilation cross section, but otherwise has features identical to the case of $\lambda_1$ dominant. In the general case, where all three $\lambda_i$ are of comparable magnitude, it is expected that the abundance constraint will resemble the third case, since the total cross-section is dominated by the $\lambda_3 v_u/v_d$ terms.

\vspace{18pt}
\index{dark matter!scalar|)}
\subsection{Model 3: Minimal Model of Fermionic Dark Matter \label{Section:MFDM-AC}}
\index{dark matter!fermionic|(}\index{dark matter!Minimal Fermionic Model|(}
\par
As in the previous models, the abundance of dark matter in this model is calculated using the annihilation cross-section. In the MFDM, this cross section is

\begin{equation}
\sigma_{ann} = \frac{\eta_1^2 \sin^2 \theta \cos^2 \theta   \sqrt{s-4m_{\chi}^2}}{2 } \left( \frac{(m_1^2-m_2^2)^2+(m_2 \Gamma_2 - m_1 \Gamma_1)^2}{((s-m_1^2)^2+m_1^2 \Gamma_1^2)((s-m_2^2)^2+m_2^2  \Gamma_2^2)} \right) \Gamma_{h \to X}
\end{equation}

\begin{figure}
\begin{center}
\psfig{file=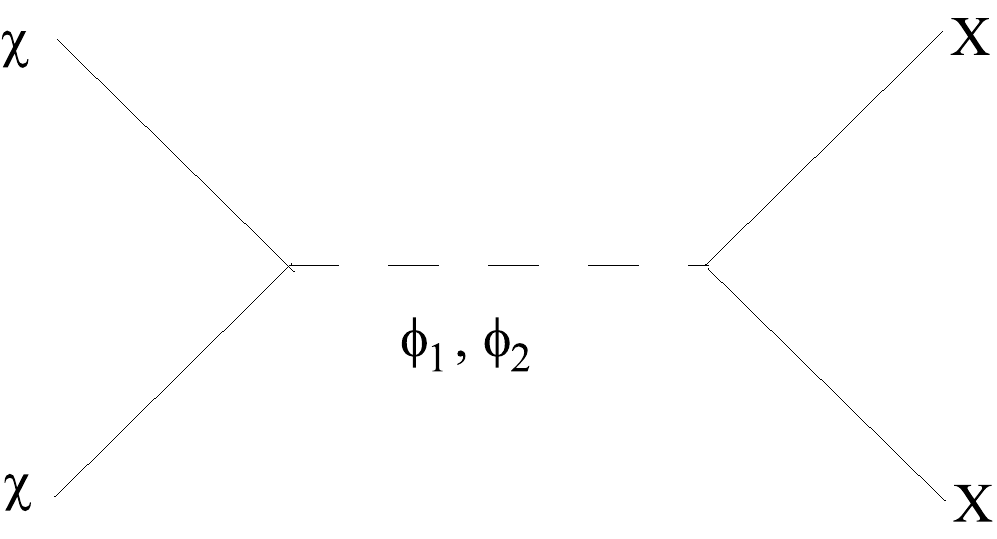,width=0.7\textwidth,angle=0}
\caption{\label{Figure:MFDM-diagrams}Feynman diagram for the annihilation of WIMPs\index{dark matter!WIMPs} in the Minimal Model of Fermionic Dark Matter. In this diagram, $\phi_1,\phi_2$ represent the higgs and higgs-like scalars of the theory and X represents any Standard Model particles.}
\end{center}
\end{figure}

\noindent
where $\Gamma_1$ and $\Gamma_2$ represent the decay widths for $\sca$ and $\scb$ respectively, and $\Gamma_{h \to X}$ represents to decay width for a virtual Higgs with mass $m_h = 2 m_{\chi}$. The thermal average for the annihilation cross-section can be approximated as

\begin{equation}
<\sigma_{ann} v> \approx 6 T \eta_1^2 \sin^2 \theta \cos^2 \theta    \left( \frac{(m_1^2-m_2^2)^2+(m_2 \Gamma_2 - m_1 \Gamma_1)^2}{((s-m_1^2)^2+m_1^2 \Gamma_1^2)((s-m_2^2)^2+m_2^2  \Gamma_2^2)} \right) \Gamma_{h \to X} 
\end{equation}

\noindent
Although this approximation is valid for most of the parameter space, it is not accurate if $m_{\chi} \sim m_1/2$, $m_{\chi}\sim m_2/2$, or if the fermion mass is close to the threshold for annihilations to heavier particles \cite{Griest:1990kh}. For these special cases, the thermal average has been calculated numerically using Eq \ref{Eq:TCS}.

\par
\begin{figure}
\begin{center}
\psfig{file=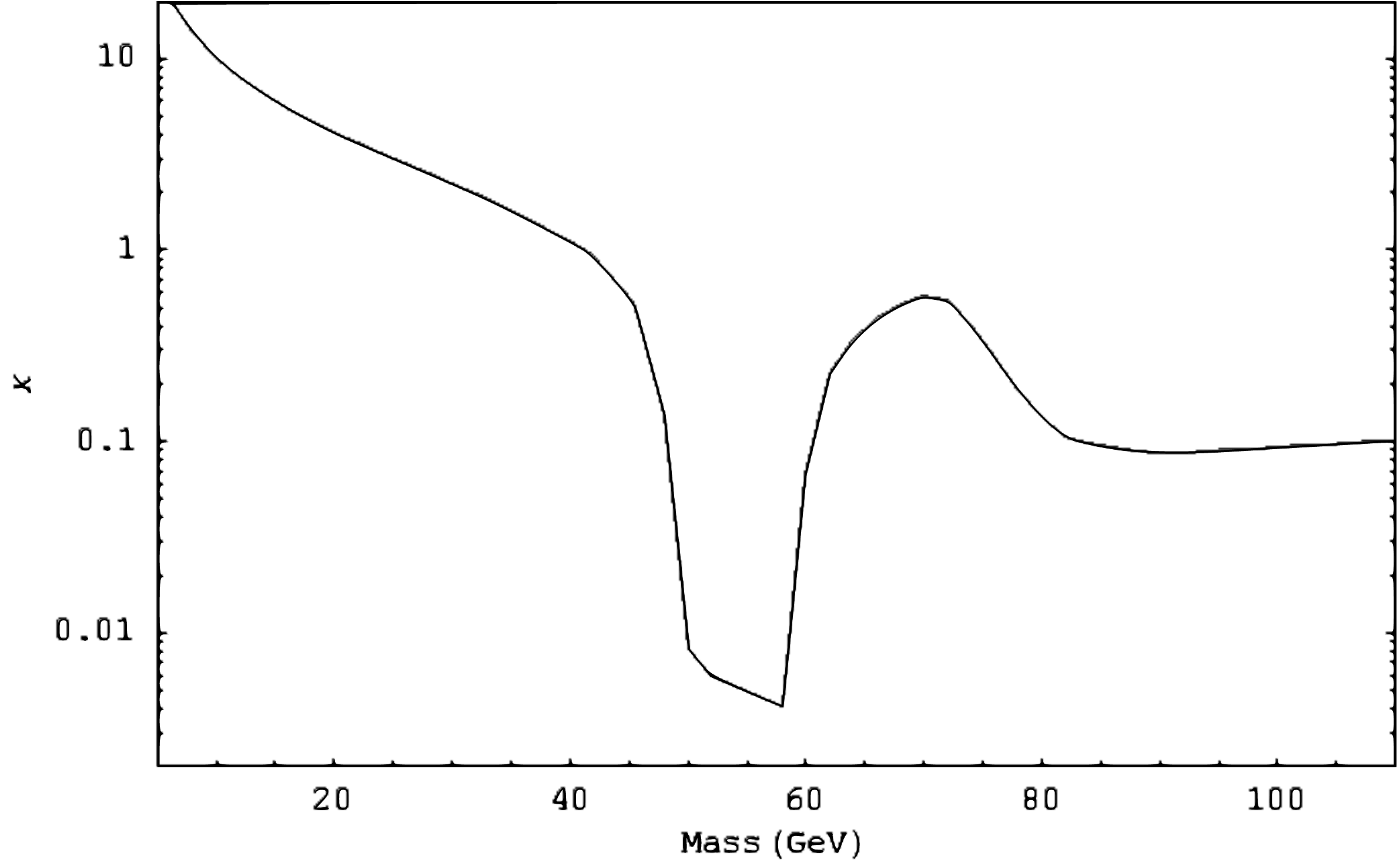,width=\textwidth,angle=0}
\end{center}
\caption{\label{Figure:MFDM-kappa} Abundance constraints on the minimal model of fermionic dark matter. In contrast to the scalar dark matter models, fermionic dark matter requires non-perturbative couplings for masses of $O(1 \; GeV)$, and therefore this range is omitted from the plot.}
\end{figure}

The resulting bounds on the coupling constants are plotted in Figure \ref{Figure:MFDM-kappa} , with the region of parameter space above the line allowed, but leading to an abundance lower than the observed dark matter abundance. The bounds are written in terms of the parameter 

\begin{equation}
\kappa \equiv \eta_1 \cos \theta \sin \theta \left( \frac{100 \; GeV}{m_1} \right)^2
\end{equation}

\noindent
as this combination appears in all of the relevant cross-sections for production, annihilation, and scattering of WIMPs\index{dark matter!WIMPs}.

\par
It should also be noted that light dark matter\index{dark matter!light} is not possible in this model. Compared to the model presented in Section \ref{Section::MDM}, the annihilation cross-section is suppressed by a factor of

\begin{equation} \label{EQ:MFDMenh}
\frac{\sigma_{fermion}}{\sigma_{scalar}} \sim \frac{m_{\chi}^2 v_{rel}^2}{m_u^2} \sim O(10^{-5})
\end{equation}

\noindent
which results in a coupling strength of $\kappa \sim O(10^{2}-10^3)$, and is therefore non-perturbative.\footnote{It is possible to produce light fermionic dark matter in this model when $m_u \ll v_{ew}$ and $\eta \ll \mu < 1$, but this region of parameter space has been extensively explored in searches for lighter higgs bosons and such a model would need significant fine-tuning.} Requiring the coupling to be perturbative excludes $m_{\chi} \lesssim 25 \; GeV$. 
\index{dark matter!Minimal Fermionic Model|)}

\subsection{Model 4: Fermionic Dark Matter with 2HDM \label{Section:2HDMf-ab}}

\par
As with the case of scalar WIMPs\index{dark matter!WIMPs} coupled to two Higgs doublets, this model contains multiple free parameters and as such the general model cannot be easily studied or plotted. Instead it can be examined through three special cases, corresponding to a single non-zero or dominant coupling constant $\lambda_i$ for each case.
\par
For the case of $\lambda_2$ dominant, the model reduces to the minimal model from the previous section, and the constraint is very similar to Figure \ref{Figure:MFDM-kappa}. The difference between these two models is that in this model the WIMPs\index{dark matter!WIMPs} cannot annihilate to b-quarks pairs or to $\tau$-leptons. However these effects are only significant for lighter WIMPs\index{dark matter!WIMPs} for which $\kappa$ is already required to be non-perturbative. 

\begin{figure}
\begin{center}
$\begin{array}{c}
\psfig{file=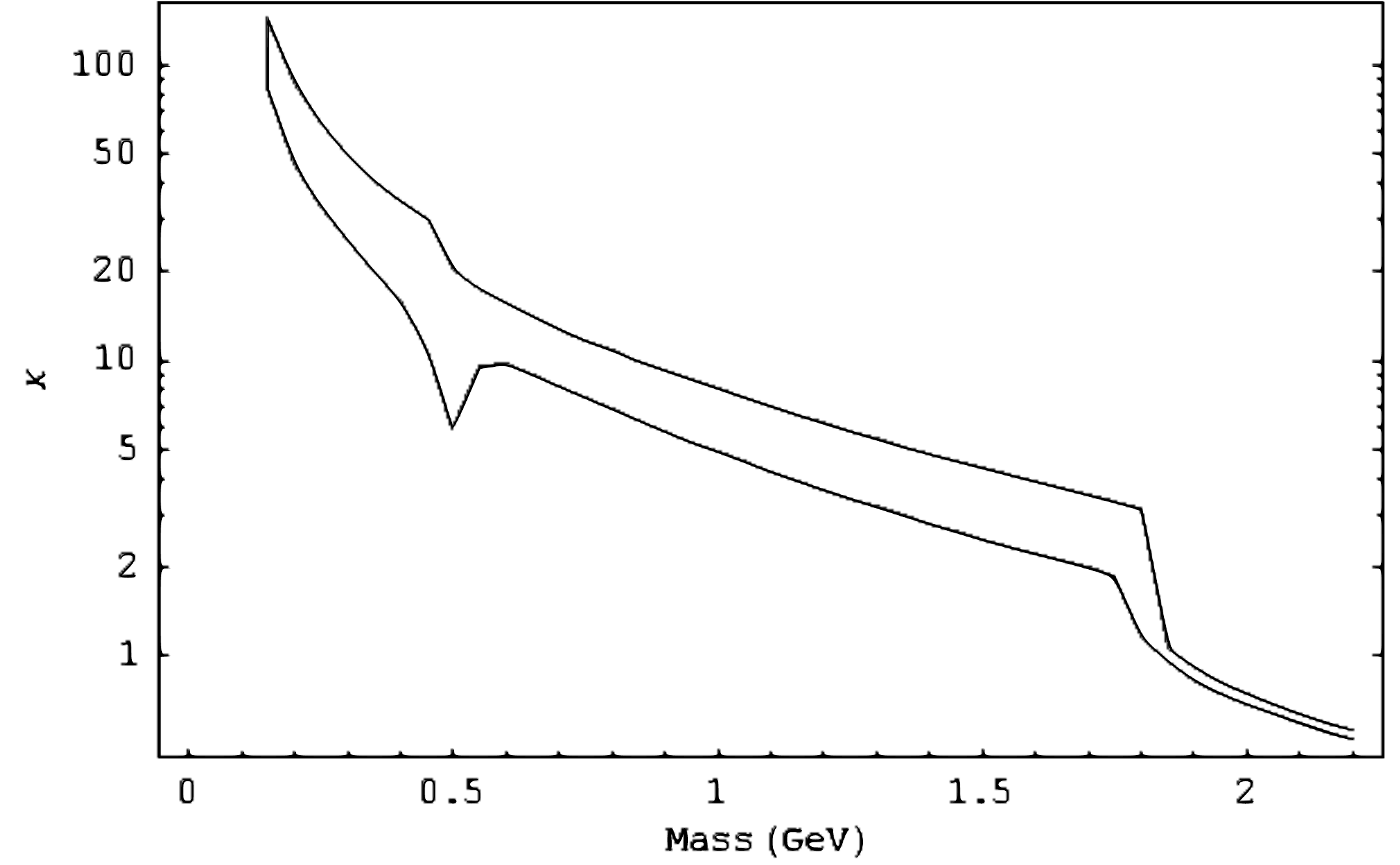,width=\textwidth,angle=0} \\(a)\\
\psfig{file=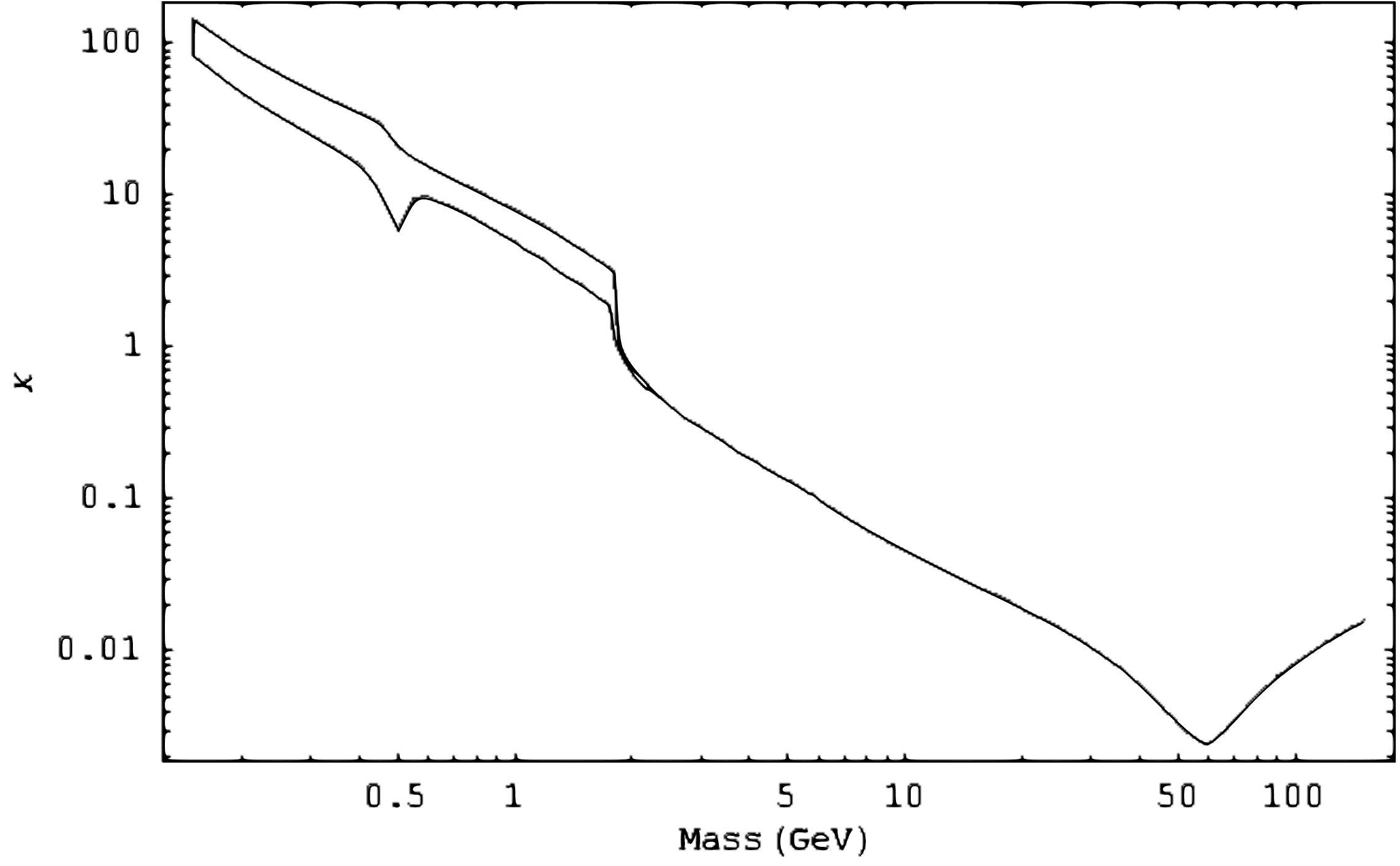,width=\textwidth,angle=0} \\(b)
\end{array}$
\caption{\label{figure:2HDMchi} Abundance constraints on $\kappa$ in the 2HDM plus fermionic WIMP model for the ranges (a) $m_{\chi} < 2 \; GeV$  and (b) $m_{\chi} \lesssim 100 \; GeV$, with $\lambda_1$ dominant. These abundance constraints also apply to the Higgs-Higgsino model. }
\end{center}
\end{figure}

\begin{figure}
\begin{center}
$\begin{array}{c}
\psfig{file=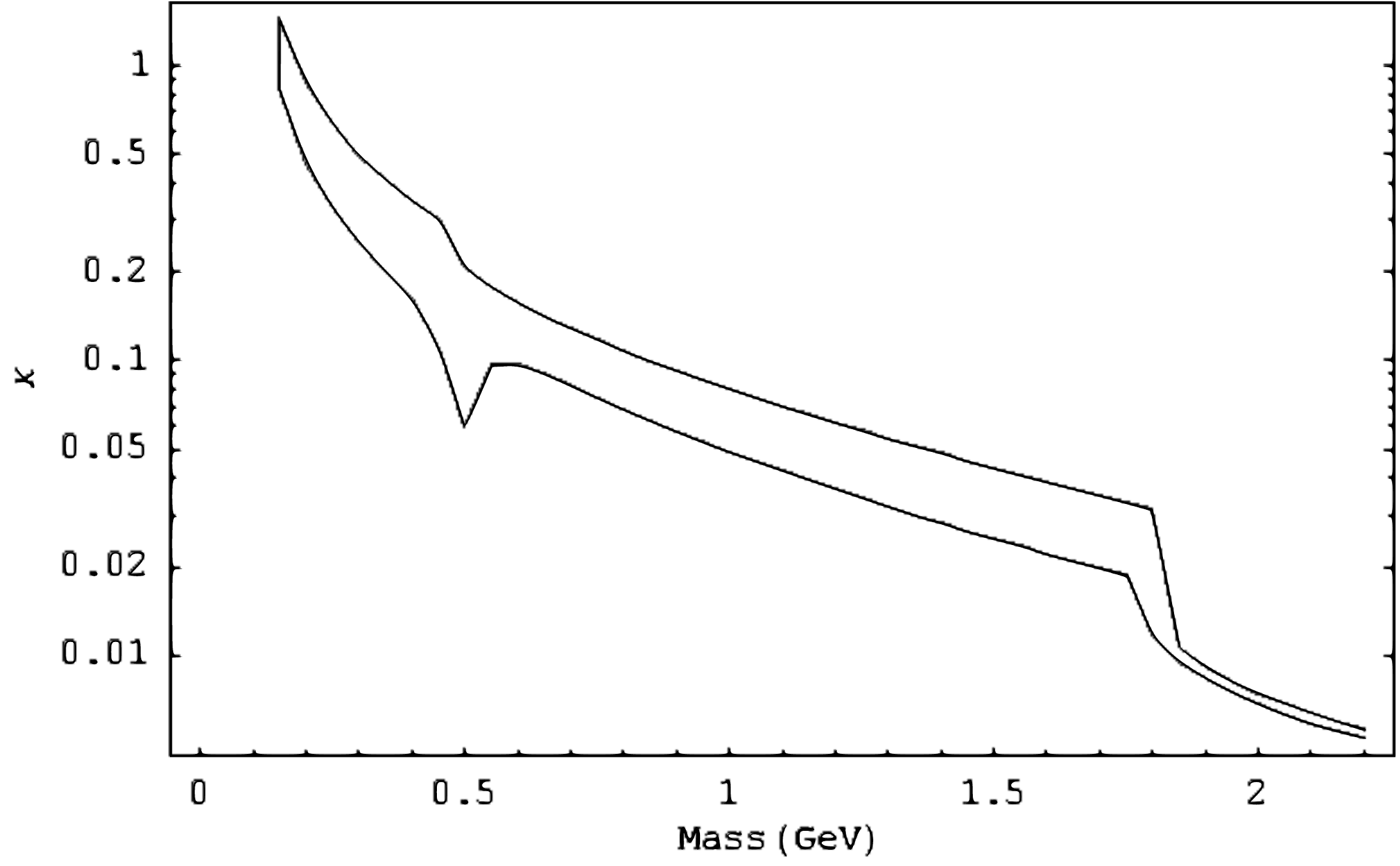,width=\textwidth,angle=0} \\(a)\\
\psfig{file=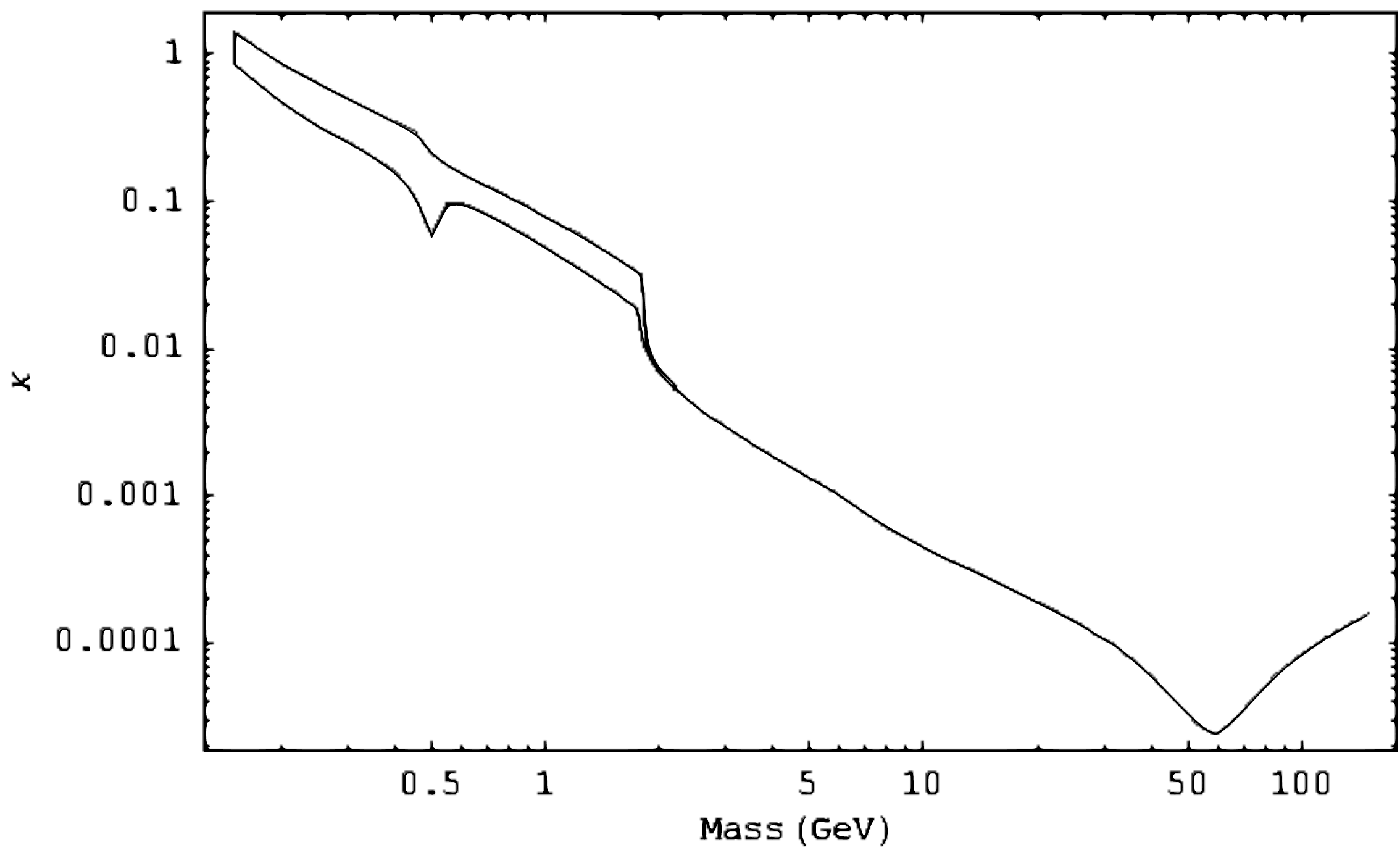,width=\textwidth,angle=0} \\(b)\end{array}$
\caption{\label{figure:2HDMchi2} Abundance constraints on $\kappa$ in the 2HDM plus fermionic WIMP model for the ranges (a) $m_{\chi} < 2 \; GeV$  and (b) $m_{\chi} \lesssim 100 \; GeV$, with $\lambda_3$ dominant. }
\end{center}
\end{figure}

\par
The abundance constraints for the case of $\lambda_1$ dominant are given in Figure \ref{figure:2HDMchi}.
As in the minimal model of fermionic dark matter, the coupling constant must be larger than in the analogous scalar model by a factor of $\sim m_u^2/(m_{\chi}<U>)$ and therefore sub-GeV WIMPs\index{dark matter!WIMPs} are not possible for the cases of $\lambda_1$ and $\lambda_2$ dominant, with the coupling constant only perturbative for $m_{\chi} \gtrsim 1.3 \;GeV$ in the first case and $m_{\chi} \gtrsim 25 \;GeV$ in the second case. 
\par
However the special case of $\lambda_3 \gg \lambda_1, \lambda_2$ results in a suppression of the coupling constants by a factor of $\tan^{-1} \beta \sim 0.01$. This suppression of the coupling constant by a factor of $v_d/v_u$ allows for sub-GeV fermionic WIMPs\index{dark matter!WIMPs} without requiring $ \kappa$ to be non-perturbative. The abundance constraints are given in Figure \ref{figure:2HDMchi}, using the parameterization, 

\begin{equation}
\kappa \equiv 2 \lambda_3 \mu \left( \frac{v_{sm} w}{m_u^2} \right) \left( \frac{100 \; GeV}{M_H}\right)^2 \left( \frac{\tan \beta}{100} \right)
\end{equation}


\par
For the general case in which all of the coupling constants are of similar magnitude, the annihilation cross-section is still dominated by the $\lambda_3$ term due to the $(v_u/v_d)^2 \sim \tan^2 \beta$ enhancements, arising from the $\lambda_3 v_u$ coupling of the WIMP\index{dark matter!WIMPs} to $H_d$ and the $m_f/v_d$ couplings of $H_d$ to the Standard Model fields.





\index{dark matter!fermionic|)}

\subsection{Model 5: Dark Matter \& Warped Extra Dimensions \label{Section:RSDM}}

\par
The WIMPs\index{dark matter!WIMPs} in this model annihilate via a virtual radion, which subsequently decays into Standard Model fields. The annihilation cross-section can be written in terms of the radion decay width, given in Ref \cite{Bae:2001id},

\begin{equation}
<\sigma_{s} v> = \frac{8 M_S^4}{\Lambda_{\phi}^2} \frac{1}{(4 M_S^2 - M_{\phi}^2)^2 + M_{\phi^2}\Gamma_{\phi}^2}  \left( \frac{\Gamma_{\phi \to X}}{M_{\phi}} \right)_{M_{\phi} \to 2 M_S}
\end{equation}

\begin{equation}
<\sigma_{f} v> = \frac{12 m_f^3 T}{\Lambda_{\phi}^2} \frac{1}{(4 m_f^2 - M_{\phi}^2)^2 + M_{\phi}^2\Gamma_{\phi}^2}  \left( \frac{\Gamma_{\phi \to X}}{M_{\phi}} \right)_{M_{\phi} \to 2 m_f}
\end{equation}



\begin{figure}
\begin{center}
\psfig{file=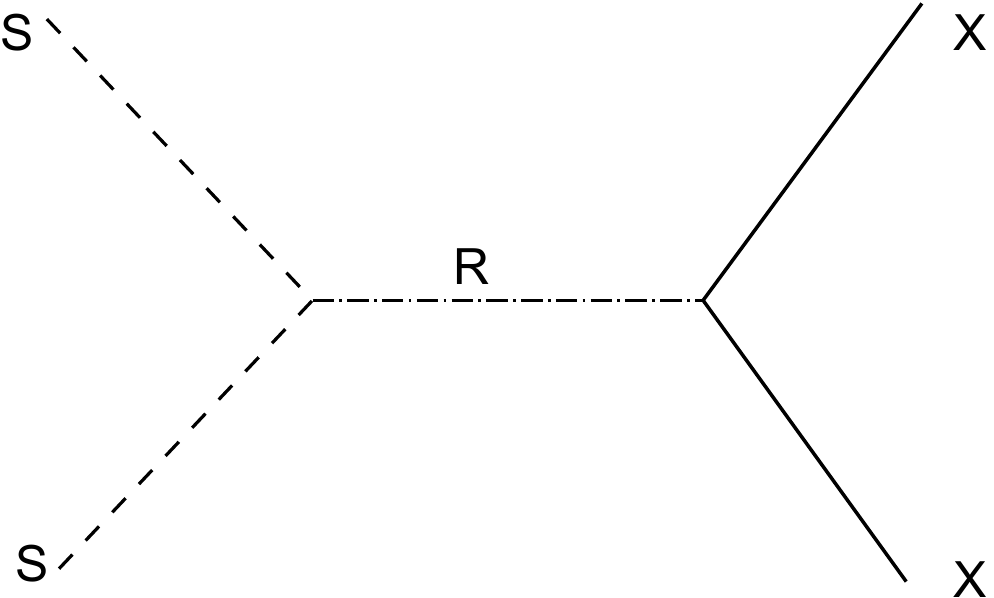,width=0.65\textwidth,angle=0}
\end{center}
\caption{\label{Figure:EXDM-ann} Feynman diagram for the annihilation cross section of WIMPs\index{dark matter!WIMPs} in the presence of warped extra dimensions. In this diagram, R represents the radion which acts as a mediator for the annihilation and S represents either scalar\index{dark matter!scalar} or fermionic\index{dark matter!fermionic} WIMPs\index{dark matter!WIMPs}.}
\end{figure}

\noindent
where the first equation corresponds to scalar WIMPs\index{dark matter!WIMPs} and  the second to fermionic WIMPs\index{dark matter!WIMPs}, and $\Lambda_{\phi}$ is the vacuum expectation for the radion field. As discussed in the introduction to this section, this form of the thermally average cross-sections is only valid when the WIMP mass is not close to the resonance in the radion propagator, and not close to a threshold for producing heavier Standard Model fields. For these regions the thermal average of the cross-sections are calculated numerically using Eq \ref{Eq:TCS}.

\begin{figure}
\psfig{file=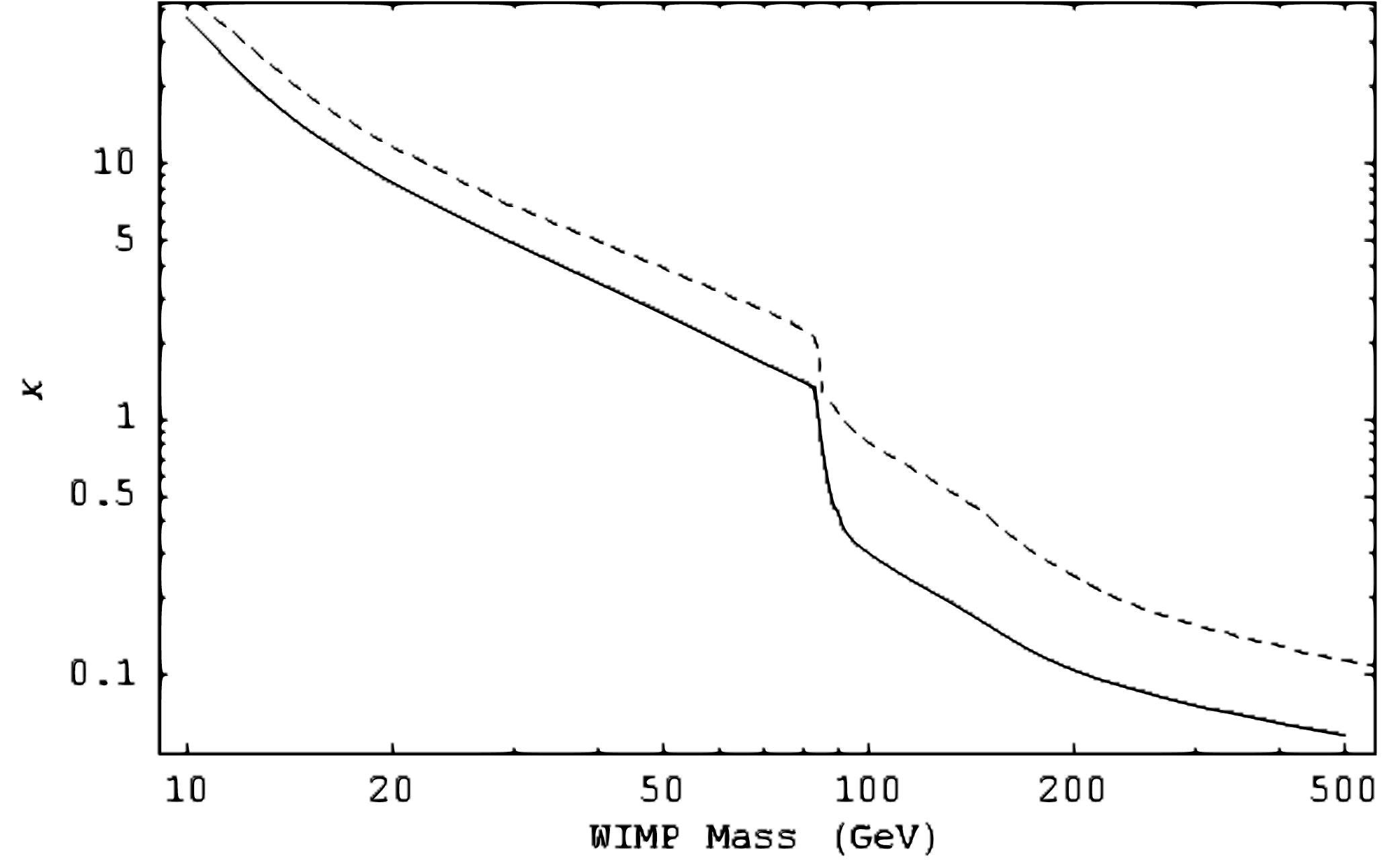,width=\textwidth,angle=0}
\caption{\label{Figure:RS-AC} Abundance constraints on scalar (solid line) and fermionic (dashed line) WIMPs\index{dark matter!WIMPs} in the presence of warped extra dimensions. 
}
\end{figure}

\par
The dark matter abundance is calculated using Eq \ref{Eq::Abundance}, and the results are plotted in Figure \ref{Figure:RS-AC} in terms of the effective coupling constant

\begin{equation}\label{Eq::EXDMkappa}
\kappa \equiv\left( \frac{v_{EW}}{\Lambda_{\phi}} \right)^2 \left(  \frac{m_{S,f}}{M_{\phi}}\right)^2
\end{equation}

\noindent
For both scalars\index{dark matter!scalar} and fermions\index{dark matter!fermionic} there is a lowering of $\kappa$ at $m_{S,f} \sim 85 \;GeV$, due to the availability of annihilations to gauge bosons. This decay channel is efficient, leading to a larger cross-section and requires smaller values of $\kappa$ to produce the correct dark matter abundance. It should be noted that, unlike the previous models, the coupling of the WIMPs\index{dark matter!WIMPs} to the radion is determined by the mass of the WIMP, and therefore the abundance constraints leave only the radion mass\index{Kaluza-Klein gravitons!radions} as a free parameter analogous to the Higgs mass in previous models. 
\par
Also unlike the previous models, the value of $\Lambda_{\phi}$ is unknown while the corresponding parameter in the previous models, $v_{EW}$ is known from electroweak measurements. However in the reactions relevant to this model, $\Lambda_{\phi}$ only appears in the parameter $\kappa$ and therefore variations in $\Lambda_{\phi}$ do not affect the constraints given for this model. 

\par
It should also be noted that, although one of these model includes scalar WIMPs\index{dark matter!WIMPs}, sub-GeV WIMPs\index{dark matter!WIMPs} are not possible. Requiring $\kappa$ to be perturbative sets $m_S \gtrsim 35 \;GeV$ and $m_f \gtrsim 50 \; GeV$. 

\par
The calculations and results of this section demonstrate that the presence of warped extra dimensions can allow a WIMP to have no gauge or Yukawa interactions with other particles, but still annihilate efficiently through gravitational forces to provide the correct dark matter abundance. 

\vspace{16pt}
\section{Dedicated Dark Matter Searches \label{Section:DDMS}}

\par
At present, the primary method of searching for WIMPs\index{dark matter!WIMPs} is with dedicated dark matter detectors which search for the recoil of nuclei which results from from collisions with WIMPs\index{dark matter!WIMPs}. In each experiment, an array of semiconductor detectors is located in a shielded location, usually underground, and surrounded by detectors which measure either ionization, phonons, or photons which result from the scattering of WIMPs\index{dark matter!WIMPs} in the solar system with nuclei in the detectors. 
\par
Using various methods, experiments such as DAMA\index{\ddme DAMA} \cite{Bernabei:2000qi} , CDMS\index{\ddme CDMS} \cite{Abusaidi:2000wg,Akerib:2005kh,Ahmed:2008eu}, and XENON10\index{\ddme XENON10} \cite{Angle:2007uj} have already reported upper bounds on the WIMP-nucleon elastic scattering cross section. Furthermore, the DAMA\index{\ddme DAMA} collaboration has claimed a positive signal of dark matter, although this result conflicts with exclusions set by other experiments.

\begin{figure}
\begin{center}
\psfig{file=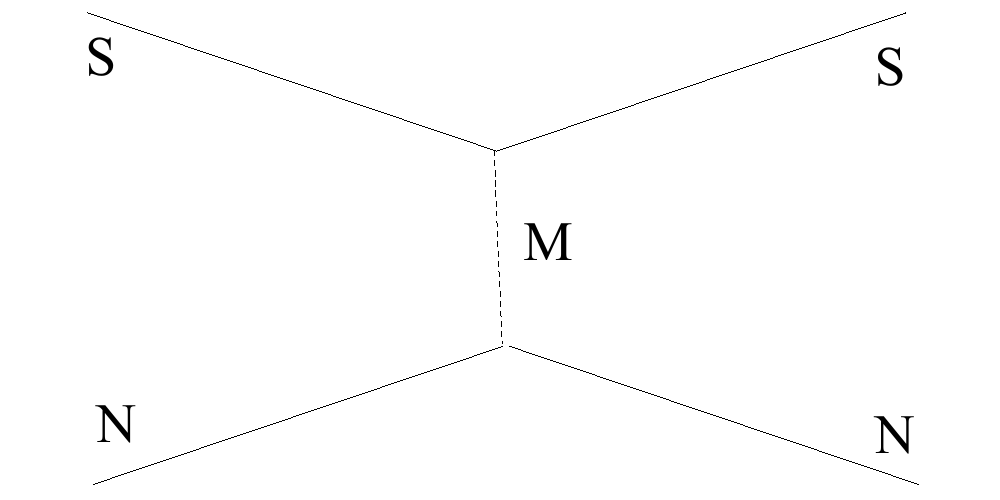,width=0.8\textwidth,angle=0}
\caption{\label{Figure:FeynmanRecoil} A generic Feynman diagrams for elastic WIMP-nucleon scattering.}
\end{center}
\end{figure}

\par
For the models presented in this \thesis, the WIMP-nucleon scattering is mediated by a Higgs or Higgs-like particle.  However the Higgs-nucleon coupling\index{Higgs!effective couplings} is not well known, and as such low energy theorems have to be used \cite{Shifman:1978zn,Gunion:1989we}. 
\par
The coupling of the Higgs\index{Higgs!effective couplings} to the quarks is defined as

\begin{equation}
L = - \sum_i \frac{m_i}{v_{EW}} h \bar{q}_i q_i
\end{equation}

\noindent
although the effect of the lightest quarks is negligible and can be omitted. The coupling to gluons, via heavy quark loops, is given by

\begin{equation}
L = \frac{\alpha_s N_H}{12 \pi v_{EW}} G_{\mu \nu}^a G^{\mu \nu}_a h
\end{equation}

\noindent
where $N_H$ is the number of heavy quark flavours that the Higgs can couple to. By equating the terms in this interaction to the trace of the QCD energy-momentum tensor,

\begin{equation}
\Theta_{\mu}^{\mu} = -\frac{\alpha_s (11 - 2/3 N_L)}{8 \pi} G_{\mu \nu}^a G^{\mu \nu}_a + \sum_{i=1}^{N_L} m_i \bar{q}_i q_i
\end{equation}

\noindent
and using the known expectation value of the energy momentum-tensor for nucleons,

\begin{equation}
<N| \Theta_{\mu}^{\nu} | N> = m_N <N|\bar{\psi}_N \psi_N |N>
\end{equation}

\noindent
the Higgs-Nucleon coupling\index{Higgs!effective couplings} can be expressed in the form

\begin{equation}
<N|h|N> = \frac{2 N_H m_N}{3(11 - 2/3 N_L) v_{EW} } <N|\bar{\psi}_N \psi_N |N> h
\end{equation}

\noindent
It should be noted however that this interaction fails to take into account the direct coupling of the Higgs to the small strange quark component in the nucleon\footnote{In some of the models presented in this \thesis, the type of Higgs boson involved does not couple to the strange quark, and therefore this correction is dropped}. The heavier top, bottom and charm quarks are more strongly coupled to the Higgs boson, however their abundance in the nucleon is almost non-existent. However virtual strange quarks in the nucleon, while having a smaller mass and therefore a weaker Higgs coupling, do have a significant effect on the nucleon Higgs coupling. Using the estimate \nocite{Kaplan:1988ku} \cite{Fukugita:1994ba,Dong:1995ec,Lewis:2002ix}

\begin{equation}
<p| m_s \bar{s} s |p> \simeq 221 \pm 51 \; MeV <p|\bar{s}s|p>
\end{equation}

\noindent
gives the effective Higgs-nucleon coupling\index{Higgs!effective couplings} as

\begin{equation}
\begin{split}
<N|h|N> = \frac{1}{v_{EW}} & \left( \frac{2 N_H m_N}{3(11 - 2/3 N_L)  }  + \left( 1 - \frac{2 N_H}{3 (11 - 2/3 N_L)} \right)  \frac{<p|m_s \bar{s} s |p>}{<p|\bar{N} N|p>}  \right) \\  & \times <N|\bar{\psi}_N \psi_N |N> h 
\end{split}
\end{equation}

\noindent
The numerical value of this coupling for each type of Higgs will be given in the following sections. 

\par
In the following sections, the WIMP-nucleon scattering cross-section is calculated for each of the minimal model. For each model, this cross-section will then be compared to recent data from the CDMS\index{\ddme CDMS} and XENON10\index{\ddme XENON10} experiments\footnote{Data from DAMA is omitted, as the present bounds are weaker than the other two experiments for the models considered in this \thesis.}. Although there are several other dedicated dark matter experiments as well, these are the three which currently provide the most stringent bounds on the scattering cross-section \footnote{The constraints given in the following sections are the most stringent as of January 2008. Recently new results have been released by CDMS\index{\ddme CDMS} \cite{Ahmed:2008eu} which are slightly stronger for $m_{DM} \gtrsim 50 \;GeV$, however these new results do not significantly affect the results and have not been included in this dissertation.}.

\subsection{Model 1: Minimal Model of Dark Matter \label{Section::MDMrecoil}}\index{dark matter!Minimal Model|(}
\index{dark matter!scalar|(}

\par
For the scalar WIMPs\index{dark matter!WIMPs} in this model, the elastic scattering cross section depends on the single diagram in Figure \ref{Figure:FeynmanRecoil}. This cross-section depends on the Higgs-nucleon coupling, which is calculated using the low energy theorems outlined in the introduction to this section . For the Standard Model Higgs boson, the Higgs-nucleon coupling is approximated as\index{Higgs!effective couplings}

\begin{equation}
g_{hNN} \approx \frac{283 \pm 50 \; {\rm MeV}}{v_{EW}}
\end{equation}

\noindent
which gives

\begin{equation}
\sigma_{el} = \kappa^2 \left( \frac{50 \; {\rm GeV}}{m_S} \right)^2 ((0.87 \pm 0.04) \times 10^{-41} \; {\rm cm^2})
\end{equation}

\begin{figure}
\psfig{file=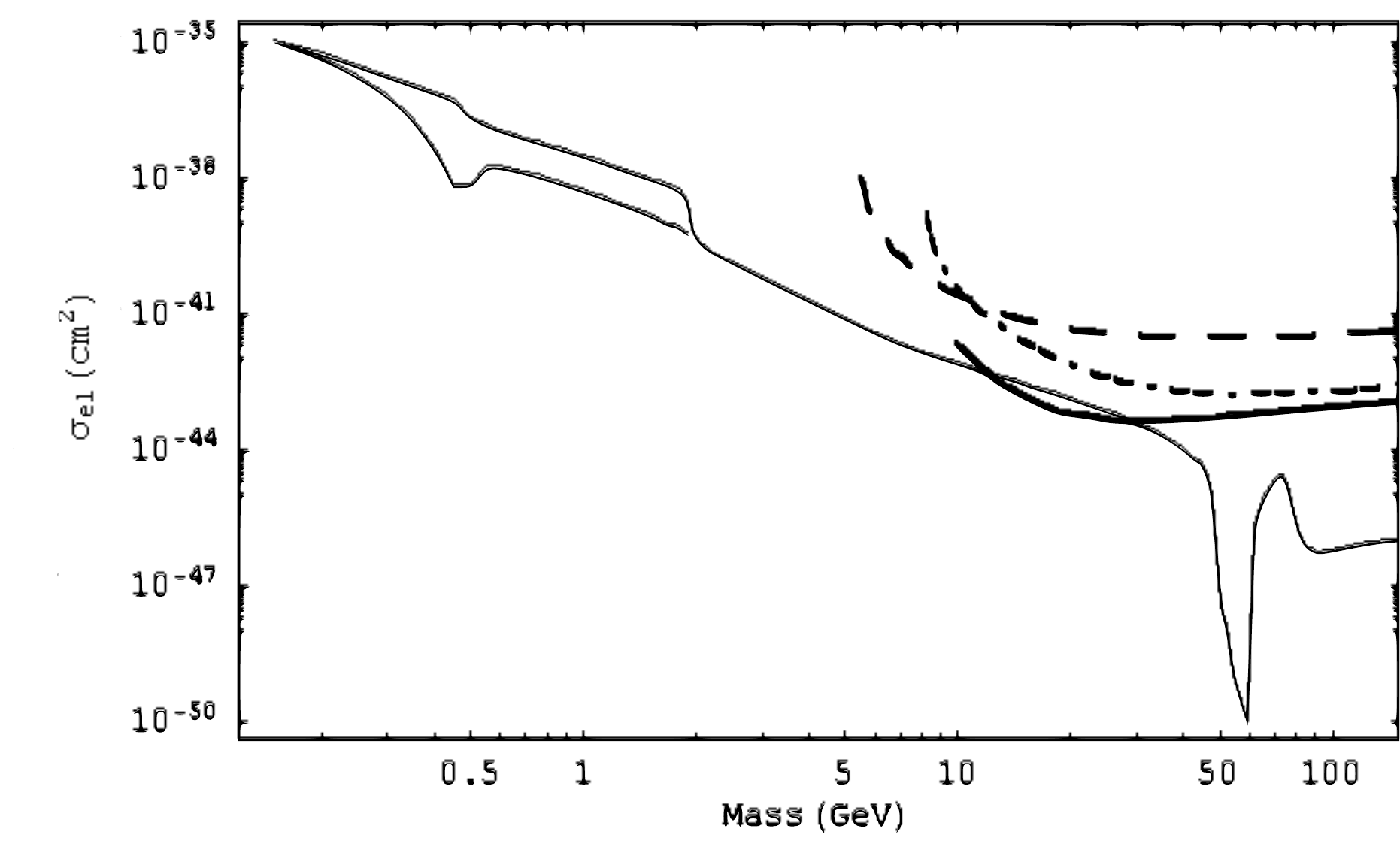,width=\textwidth,angle=0}
\caption{\label{Figure:DMLimits}Limits from dedicated dark matter searches on the Minimal Model of Dark Matter, using as an example $m_H = 120 \; {\rm GeV}$. The solid line represents the WIMP-nucleon scattering cross-section with the coupling constant determined by the abundance constraint. The dashed and dashed-dotted lines represent constraints from CDMS\index{\ddme CDMS} \cite{Akerib:2005kh} using the Silicon data and Germanium data respectively. The solid bold line represents the recently released constraints from the XENON10\index{\ddme XENON10} experiment\cite{Angle:2007uj}. 
}
\end{figure}

\noindent
Using the abundance constraints on $\kappa$ from Section \ref{Section::MDM-AC}, the elastic scattering cross-section is plotted in Figure \ref{Figure:DMLimits} with recent bounds from CDMS\index{\ddme CDMS}\cite{Akerib:2005kh} and XENON10\index{\ddme XENON10}\cite{Angle:2007uj} \footnote{Although there are several dedicated dark matter searches in operation, at present CDMS\index{\ddme CDMS} and XENON10\index{\ddme XENON10} have produced the strongest bounds.}. 
  \nocite{Ahmed:2008eu}
\par
In this figure it can be observed that these searches are insensitive to light scalars with masses below $\sim 10 \; GeV$. In this range, the low mass of the WIMPs\index{dark matter!WIMPs} results in small recoil velocities of the heavier nuclei (Germanium and Silicon in CDMS\index{\ddme CDMS} and gaseous Xenon in XENON10\index{\ddme XENON10}) in the detectors. It is expected that future experiments will be able to probe this range using lighter nuclei \cite{Giuliani:2007vg}, with several such experiments currently being planned \cite{Miuchi:2002zp,Cebrian:2005pc,Winkelmann:2006rg}\footnote{These experiments use lighter nuclei and are therefore more sensitive to light dark matter\index{dark matter!light}, however other factors in their design may still limit their ability to probe $O(1\; GeV)$ WIMPs\index{dark matter!WIMPs}}.

\par
In spite of this model being minimal, it avoids all prior bounds from dedicated searches with the only constraint arising from the XENON10\index{\ddme XENON10} data released in 2007. Full exclusion of this model will require several orders of magnitude improvement in the detector sensitivity, both for light WIMP mass\index{dark matter!light} for which current detectors are insensitive and for heavier WIMPs\index{dark matter!WIMPs} near the Higgs resonance, and is unlikely to happen with the next generation of experiments \cite{Brink:2005ej,Araujo:2006er,DMPlotter}.
\index{dark matter!Minimal Model|)}
\subsection{Model 2: Minimal Model of Dark Matter with 2HDM \label{Section::2HDM-ds}}

\par
The constraints from dedicated dark matter searches can be calculated in the same manner as in Section \ref{Section::MDMrecoil}, except that in the 2HDM model the Higgs-nucleon coupling is different due to the change in $v_{ew} \to v_u,v_d$, and the restriction that each of the higgs couples to only up-type or to down-type quarks and leptons.

\par
In the case of $\lambda_1$ dominant and $\lambda_3$ dominant, the scattering cross-section depends on the $H_d N N$ coupling. Using the same low-energy theories as in the Standard Model Higgs-nucleon coupling, the effective Higgs-nucleon coupling\index{Higgs!effective couplings} is 

\begin{equation}
L_{H_d N N} \approx \frac{220\pm 50 \; MeV}{v_d} H_d \bar{N} N 
\end{equation}

\noindent
In this case, there is no coupling of the Higgs to the gluons through the top-quark\index{top quark} loop, and $H_d$ couples to the nucleon predominantly through direct coupling to the virtual strange quarks within the nucleons. Since the quark content of the nucleon is not well known, the Higgs-nucleon coupling in this case contains a significant uncertainty\index{Higgs!effective couplings}.
\par
Using this effective coupling, the WIMP-nucleon scatting cross-section for the case $\lambda_1$ dominant is

\begin{equation}
\sigma_{el} = \kappa^2 \left( \frac{50 \; {\rm GeV}}{m_S} \right)^2 ((0.53 \pm 0.04)\times 10^{-41} \; {\rm cm^2})
\end{equation}

\noindent
and for the case $\lambda_3$ dominant is

\begin{equation}
\sigma_{el} = \kappa^2 \left(\frac{\tan \beta}{100} \right)^2\left( \frac{50 \; {\rm GeV}}{m_S} \right)^2 ((0.53 \pm 0.04)\times 10^{-37} \; {\rm cm^2})
\end{equation}

\begin{figure}
\begin{center}
\psfig{file=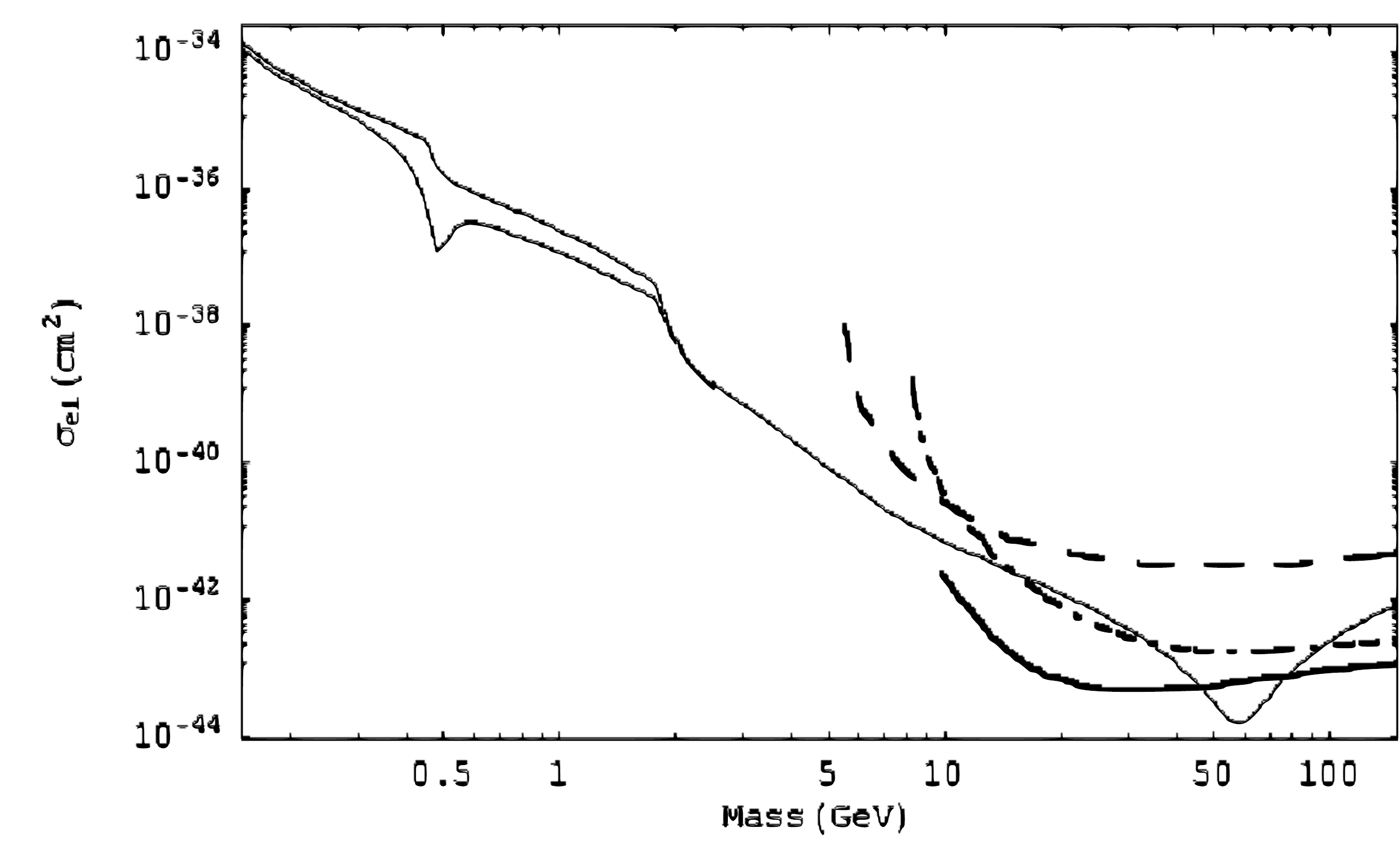,width=\textwidth,angle=0} 
\end{center}
\caption{\label{Figure::2HDM1-DS} Limits on the 2HDM+scalar model from dedicated dark matter searches. The bound is for the special cases of $\lambda_1$ dominant and $\lambda_3$ dominant. As in Figure \ref{Figure:DMLimits}, the dashed and dashed-dotted lines represent bounds from CDMS\index{\ddme CDMS}, while the solid bold line represents limits from XENON\index{\ddme XENON10}. }
\end{figure}

\noindent
From the abundance constraints on $\kappa$, the limits from dedicated searches can be derived. 
 The results are given in Figure \ref{Figure::2HDM1-DS}. With the exception of a small mass range near the Higgs resonance, data from the XENON10\index{\ddme XENON10} experiment can already exclude $m_{DM} \gtrsim 10 \; GeV$. 
 \par
 Since the annihilation cross-section and the scattering cross-section each contain a factor of $\tan \beta$ in the case of $\lambda_3$ dominant, these terms cancel out in the final result and the scattering cross-section is the same for $\lambda_1$ or $\lambda_3$ dominant.

\par
The case of $\lambda_2$ dominant is more difficult to detect in dedicated searches. Since $H_u$ is similar to the SM Higgs in the large $\tan \beta$ limit, the abundance constraints for this model are very similar to the abundance constraints on the minimal model of dark matter. However the SM Higgs coupling to the nucleon is dominated by a direct coupling to the strange quark content of the nucleon, and by heavy quark loops. In the case of $H_u$, there is no strange quark coupling and no coupling to bottom quark loops. Therefore the effective coupling is reduced to

\begin{equation}
L_{H_u N N} = \frac{82 \; MeV}{v_u} H_u \overline{N}N
\end{equation}

\noindent
In this case the Higgs couples to the nucleon predominantly through a top-quark loop, with no effects from direct coupling to the strange quark and therefore this coupling does not have the large uncertainty of the previous models.
\par 
Using this result, the scattering cross-section can then reduced to

\begin{equation}
\sigma_{el} = \kappa^2 \left( \frac{50 \; {\rm GeV}}{m_S} \right)^2 (0.75 \times 10^{-42} \; {\rm cm^2})
\end{equation}

\begin{figure}
\begin{center}
\psfig{file=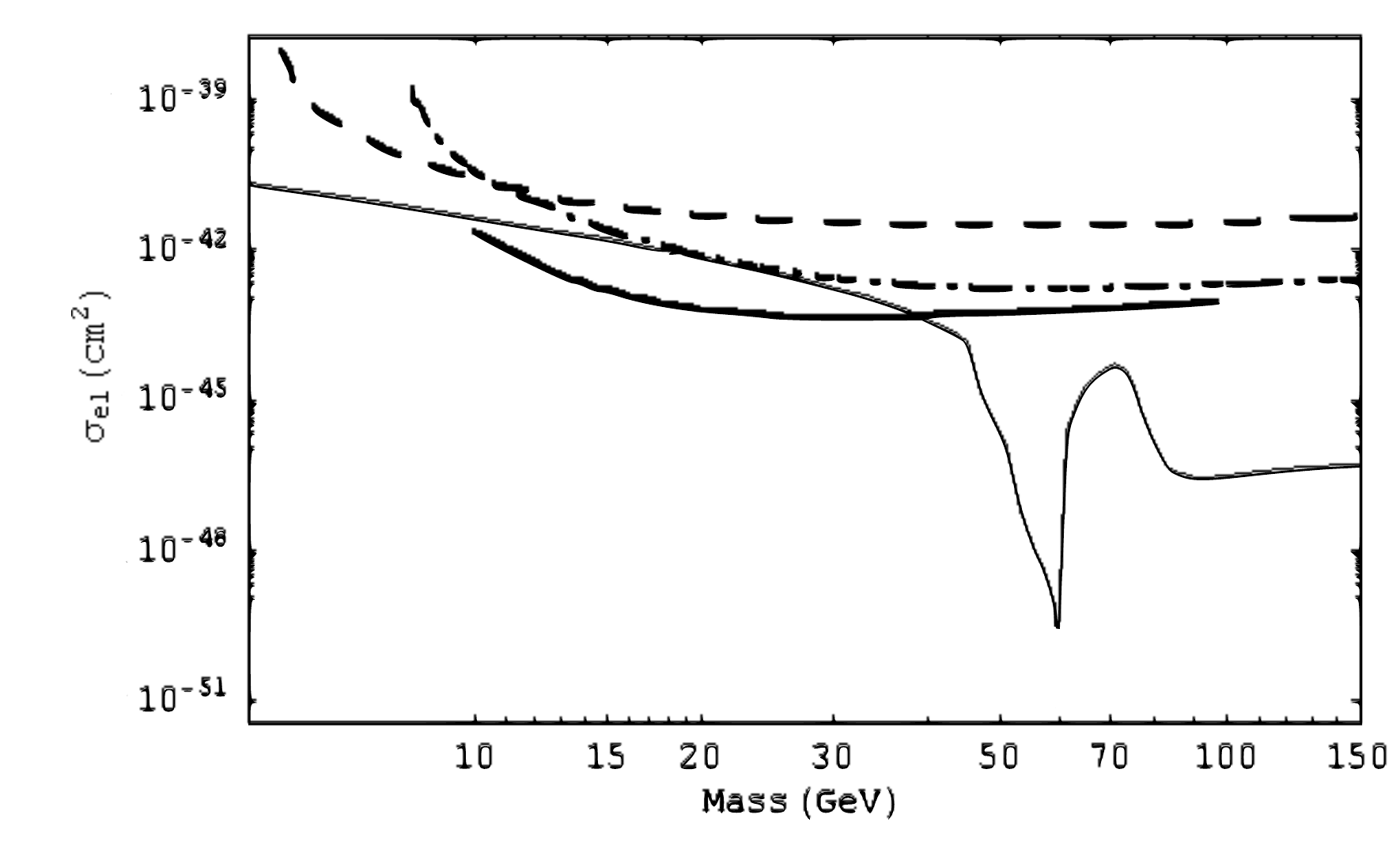,width=\textwidth,angle=0} 
\caption{\label{Figure::2HDM1-DS2} Limits on the 2HDM+scalar model from dedicated dark matter searches. In this plot, the special case of $\lambda_2$ dominant is plotted along with the usual experimental bounds.}
\end{center}
\end{figure}

\noindent
The dedicated search limits for this model are given in Figure \ref{Figure::2HDM1-DS2}. Although the Higgs-nucleon coupling\index{Higgs!effective couplings} is smaller for this model, the abundance constraints allow $\kappa$ to be slightly larger than in the minimal model of dark matter, and the scattering cross-section is similar for the two models. Existing data from XENON10\index{\ddme XENON10} can already exclude the range $10 \; Gev < m_S < 40 \; GeV$.

\par
In summary, all three of the special cases of this model can be probed with dedicated dark matter experiments. As with the minimal model of dark matter, the scalars in this model with $\lambda_1$ or $\lambda_3$ dominant can be as light as a few GeV and are invisible to present dedicated dark matter searches. However unlike the first model, heavier WIMPs\index{dark matter!WIMPs} have been almost completely excluded by XENON10\index{\ddme XENON10}. In the case of $\lambda_2$ dominant, sub-GeV WIMPs\index{dark matter!WIMPs} are forbidden by the abundance constraints and the requirement that $\kappa$ be perturbative, but heavier WIMPs\index{dark matter!WIMPs} can satisfy the abundance constraint without violating bounds set by nuclear recoil experiments.

\index{dark matter!scalar|)}

\subsection{Model 3: Minimal Model of Fermionic Dark Matter}
\index{dark matter!fermionic|(}\index{dark matter!Minimal Fermionic Model|(}

The nucleon scattering cross-section for the minimal model of fermionic dark matter is calculated in the same manner as the minimal model of scalar dark matter presented in Section \ref{Section::MDMrecoil}. Using the effective coupling, $\kappa$, and the Standard Model Higgs-nucleon coupling\index{Higgs!effective couplings} presented earlier, the scattering cross-section is

\begin{figure}
\begin{center}
\psfig{file=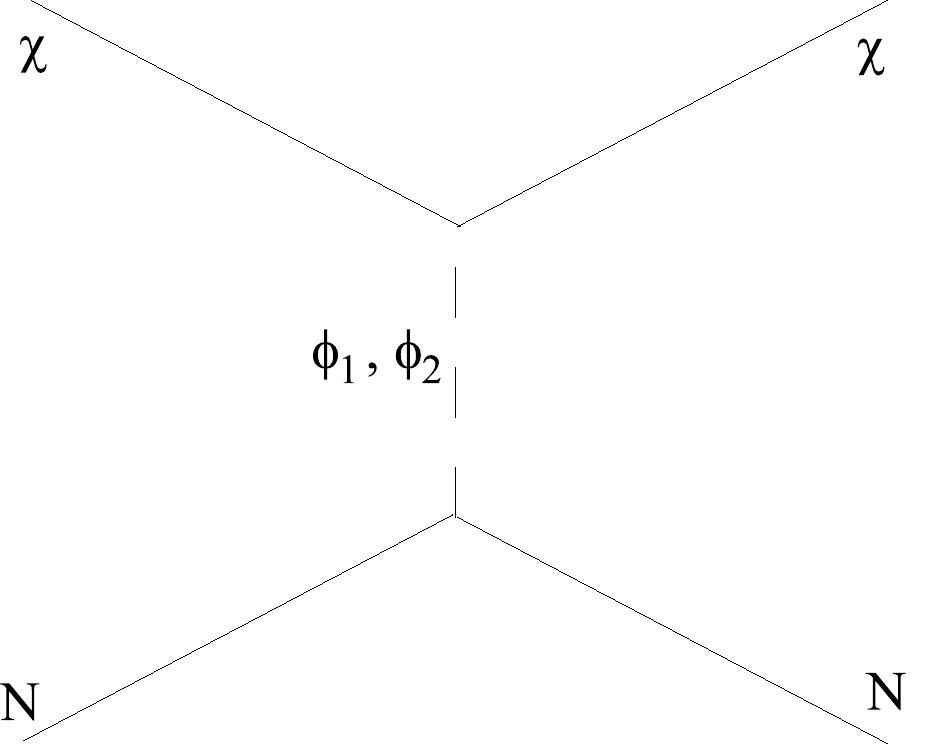,width=0.65\textwidth,angle=0}
\end{center}
\caption{\label{Figure::MFDM-nuclear} Feynman diagrams for the scattering of WIMPs\index{dark matter!WIMPs} and nucleons in the Minimal Model of Fermionic Dark Matter.}
\end{figure}

\begin{equation}
\sigma = 
 ((1.46 \pm 0.07 ) \times 10^{-42} cm^2) \kappa^2
\end{equation}

\begin{figure}
\begin{center}
\psfig{file=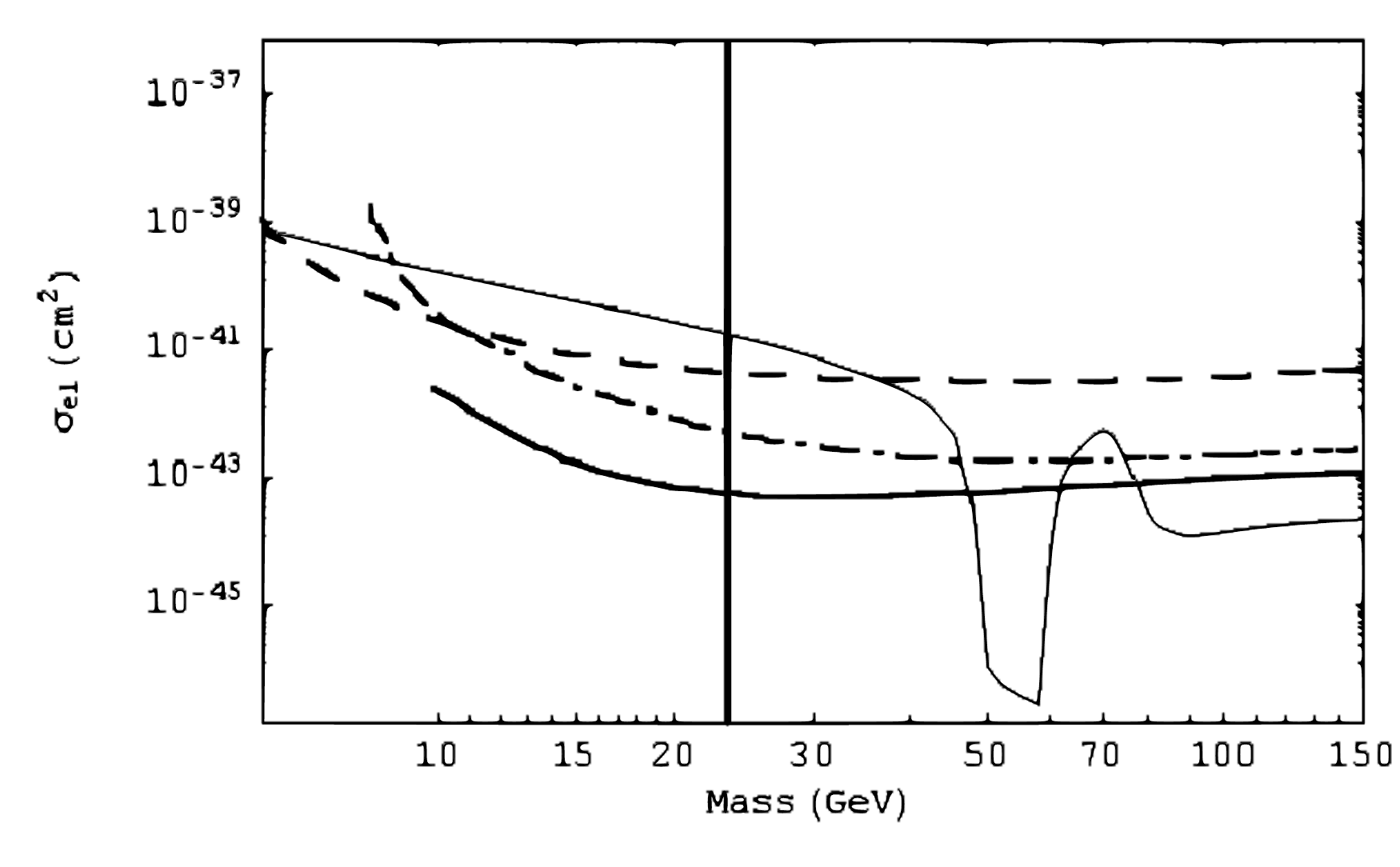,width=\textwidth,angle=0}
\end{center}
\caption{\label{Figure:MFDM-recoil} Expected WIMP-nucleon scattering cross section for the minimal model of fermionic dark matter (thin solid line), plotted as before with limits from CDMS (dashed and dashed-dotted lines) and XENON10 (thick solid line).The vertical line gives the cutoff at which lighter WIMPs\index{dark matter!WIMPs} require a non-perturbative coupling.}
\end{figure}

\noindent
This cross-section is enhanced relative to the minimal model of scalar dark matter. As shown in Eq \ref{EQ:MFDMenh}, the annihilation cross-section is suppressed for fermions relative to scalar WIMPs\index{dark matter!WIMPs}, resulting in larger values of $\kappa$ from the abundance bounds which also produces a larger scattering cross-section. The cross-section is plotted in Figure \ref{Figure:MFDM-recoil}, with the current limits from CDMS\index{\ddme CDMS} \cite{Akerib:2005kh} and XENON10\index{\ddme XENON10} \cite{Angle:2007uj}. 

\par
In contrast to the scalar models presented previously, the minimal model of fermionic dark matter cannot contain sub-GeV mass WIMPs\index{dark matter!WIMPs} and as such the entire parameter space can be explored by dedicated searches. For this model, data from the CDMS and XENON10 experiments have already excluded WIMPs\index{dark matter!WIMPs} with $m_{\chi} \lesssim 80 \; GeV$, with the only exception being a range of masses between $\sim 50 \; GeV$ and $60 \;GeV$\footnote{This range is calculated assuming a Higgs mass of $120 \; GeV$. If the Higgs is heavier than this, then this range will shift to allow for higher mass WIMPs\index{dark matter!WIMPs}.}. As in previous models, the mass range that is not excluded is due to the possibility that the WIMPs\index{dark matter!WIMPs} annihilate through a Higgs resonance, which enhances the cross-section and allows the WIMP couplings to be smaller, and which results in a suppressed nuclear scattering cross-section. 
 \index{dark matter!Minimal Fermionic Model|)}

\subsection{Model 4: Fermionic Dark Matter with 2HDM}

\begin{figure}
\psfig{file=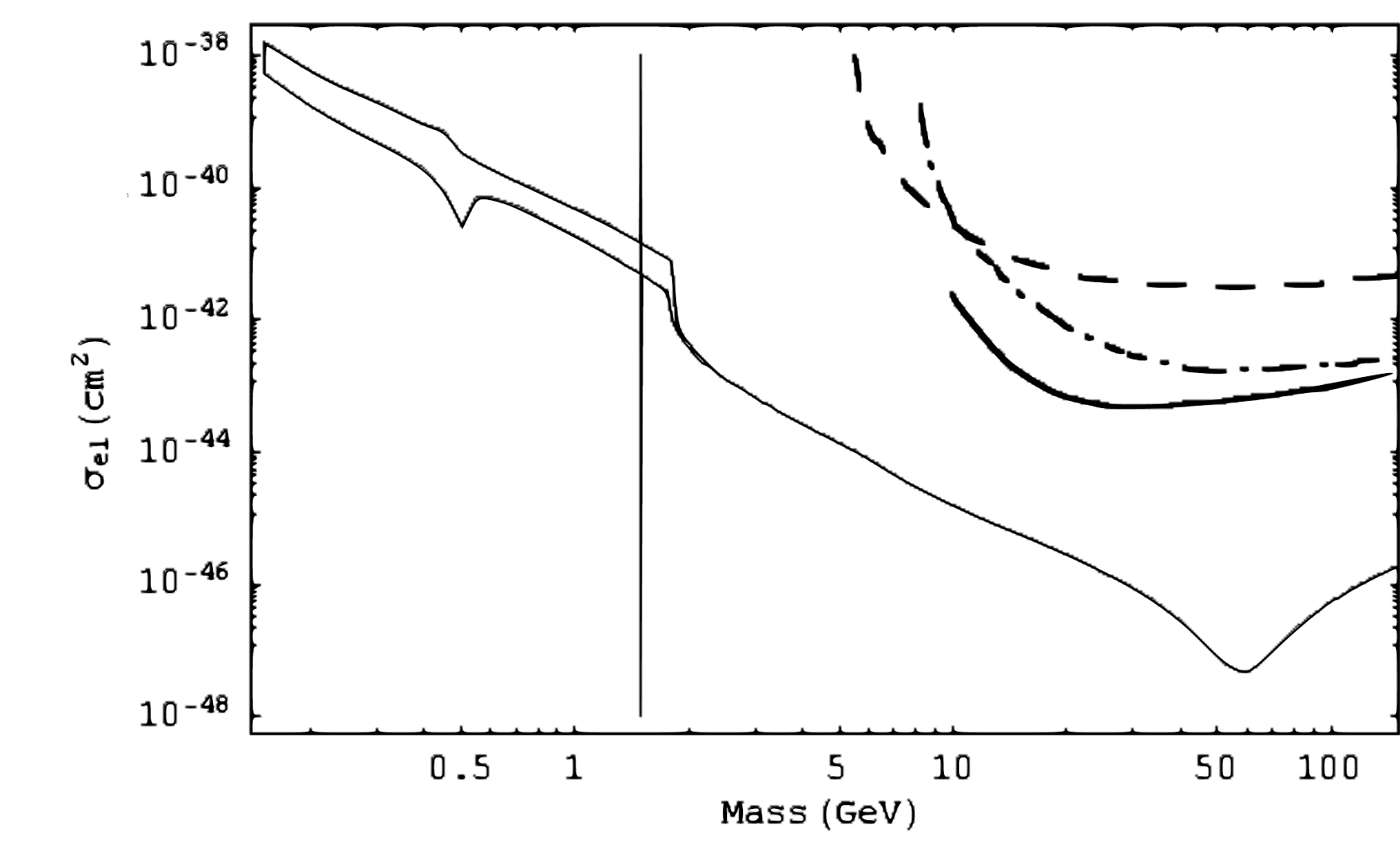,width=\textwidth,angle=0}
\caption{\label{Figure::2HDM+fermi} Dedicated Dark Matter search limits on the 2HDM+fermion model, with either $\lambda_1$ or $\lambda_3$ dominant. The experimental constraints from CDMS and XENON10 are as given before. The vertical line represents the cut-off in the $\lambda_1$ dominant case, where $\kappa$ becomes non-perturbative.}
\end{figure}

\par
As discussed in Section \ref{Section::2HDM+f}, the case of $\lambda_2$ dominant is similar to the minimal model of fermionic dark matter, and the bounds from dedicated dark matter searches are similiar. Therefore in this section, only the $\lambda_1$ and $\lambda_3$ dominant cases will be studied.

\par
The higgs-nucleon couplings\index{Higgs!effective couplings} in the 2HDM are as given in Section \ref{Section::2HDM-ds}, and the WIMP-nucleon cross-section is given by

\begin{equation}
\sigma_{el} 
\approx ((8.3\pm 0.6) \times 10^{-43} cm^2)\kappa^2
\end{equation}

\noindent
for the case of $\lambda_1$ dominant, and by

\begin{equation}
\sigma_{el} 
\approx ((8.3\pm 0.6) \times 10^{-39} cm^2)\left( \frac{\tan \beta}{100} \right)^2 \kappa^2
\end{equation}

\noindent
for the case of $\lambda_3$ dominant.
The scattering cross section and the corresponding experimental constraints are plotted in Figure \ref{Figure::2HDM+fermi}. As in the analogous scalar case, the scattering cross-section is the same for the $\lambda_1$ and $\lambda_3$ dominant cases. However in the $\lambda_1$ dominant case, the coupling constant becomes nonperturbative for light WIMPs\index{dark matter!WIMPs}\index{dark matter!light}. This is indicated in the figure by the solid vertical line.

\par
The difference between this model, in which there are no constraints from dedicated searches, and the minimal model of fermionic dark matter, in which a large range of masses is excluded, is that the large Higgs vev in the previous model leads to a smaller decay width for the Higgs while in this model the Higgs vev is $v_d \sim v_{EW}/\tan \beta$.  Because the decay width and the annihilation cross-section are larger in this model, the abundance constraints allows for a smaller $\kappa$. However the scattering cross-section does not get this same enhancement since it is assumed that the scattering occurs at lower energies away from the Higgs resonance, and so the overall effect is a suppression of the scattering cross-section. 

\par
Based on the results of this section, it is apparent that in the special case of $\lambda_1$ or $\lambda_3$ dominant, there are no constraints on the WIMP mass in this model, and even the next generation of experiments may not be able to probe it. In particular, the case of $\lambda_3$ dominant allows for sub-GeV WIMPs\index{dark matter!WIMPs}\index{dark matter!light} and yet has no constraints at any mass range from dedicated dark matter searches.
\index{dark matter!fermionic|)}

\subsection{Model 5: Dark Matter \& Warped Extra Dimensions}
\index{extra dimensions!Randall-Sundrum model|(}\index{dark matter!scalar|(}\index{dark matter!fermionic|(}
\par
In a similar manner to the higgs-nucleon coupling\index{Higgs!effective couplings}, the radion-nucleon coupling is given by the coupling of the radion\index{Kaluza-Klein gravitons!radions} to the trace of the energy momentum tensor,

\begin{equation}
L_{int} = \frac{\phi}{\Lambda_{\phi}} \Theta_{\mu}^{\mu}
\end{equation}

\noindent
where, at low energies,

\begin{equation}
<N| \Theta_{\mu}^{\mu} |N> = m_N <N| \bar{\psi}_N \psi_N |N>
\end{equation}

\noindent
This coupling is stronger than the corresponding Standard Model Higgs coupling, as the radion couples to the full energy-momentum tensor rather than just to the mass of the quarks and quark loops in the nucleon. This difference enhances the WIMP-nucleon scattering cross-section in this model by an order of magnitude compared to the Minimal Model of Dark Matter.

\begin{figure}
\psfig{file=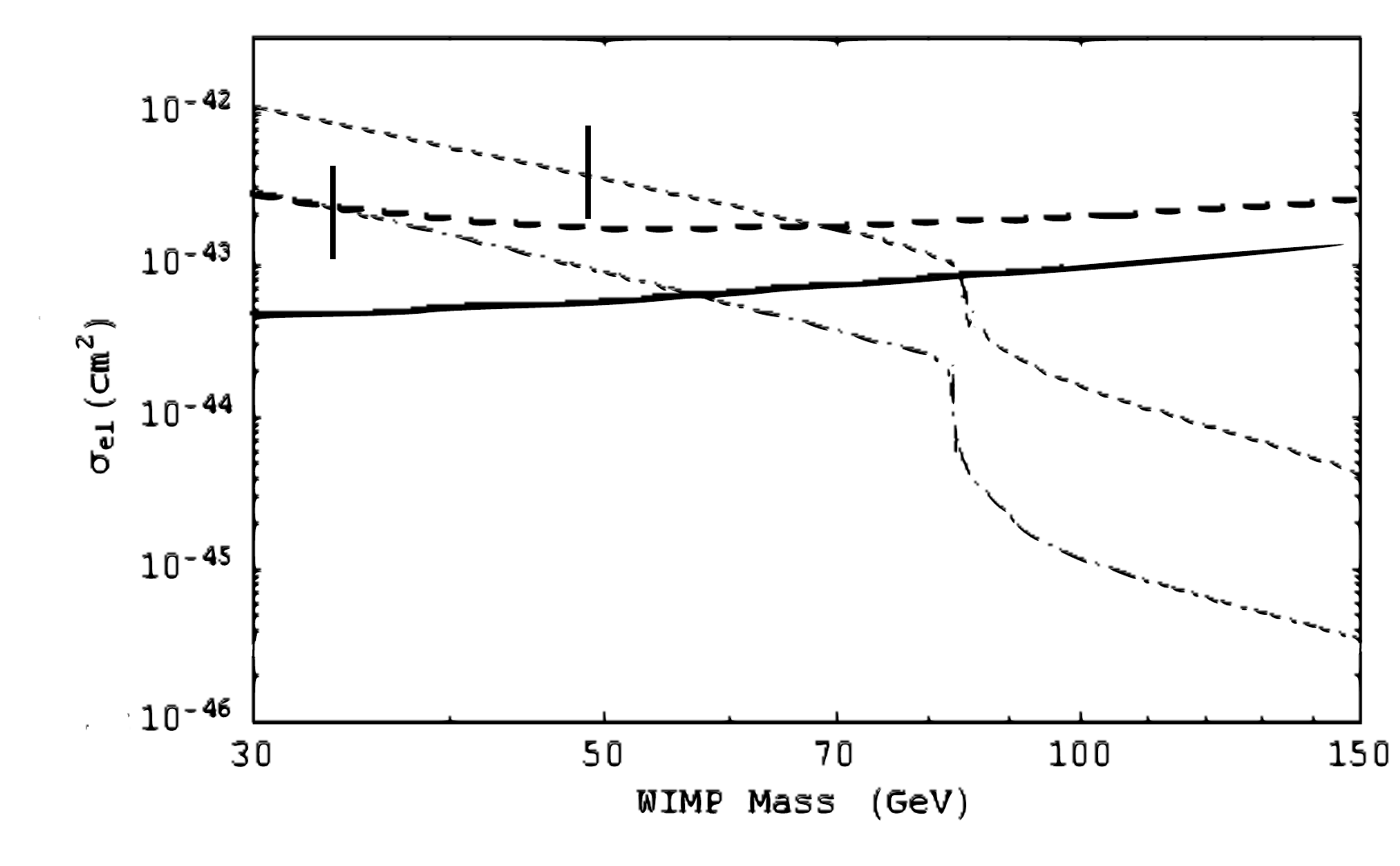,width=\textwidth,angle=0}
\caption{\label{Figure:RS-NC} WIMP-nucleon scattering cross-section for scalar WIMPs\index{dark matter!WIMPs} (dashed-dotted line) and fermion WIMPs\index{dark matter!WIMPs} (dashed line) with radion mediation. The current limits from CDMS\index{\ddme CDMS} and XENON10\index{\ddme XENON10} are indicated with the bold dashed and solid lines respectively. The short vertical lines represent the Lee-Weinberg bound on scalar and fermion WIMPs\index{dark matter!WIMPs}.}
\end{figure}

\par
For the scalar WIMPs\index{dark matter!WIMPs}, the WIMP-nucleon scattering cross section is

\begin{equation}
\sigma_{S} = \frac{m_N^4 m_S^2}{2 \pi \Lambda_{\phi}^4 m_{\phi}^4} = 1.3 \times 10^{-44} cm^2 \kappa^2
\end{equation}

\noindent
while for fermions the scattering cross-section is

\begin{equation}
\sigma_{f} = \frac{m_N^4 m_f^2}{\pi \Lambda_{\phi}^4 m_{\phi}^4} = 2.6 \times 10^{-44} cm^2 \kappa^2
\end{equation}

\noindent
Using the abundance constraint previously derived, the scattering cross-sections can be calculated, and the results are plotted in Figure \ref{Figure:RS-NC}.  

\par
The data from CDMS\index{\ddme CDMS} does not improve the constraints on this model, as the region it excludes requires a non-perturbative coupling for both scalars and fermions. However the recently reported bounds from the XENON10\index{\ddme XENON10} experiments improve the constraint, with scalars excluded for $m_{S} \lesssim 57 \; GeV$ and fermions excluded for $m_f \lesssim 85 \; GeV$. 

\index{extra dimensions!Randall-Sundrum model|)}\index{dark matter!scalar|)}\index{dark matter!fermionic|)}

\vspace{16pt}
\section{Collider Constraints \label{Section:ColliderS}}

\par
Another possibility is that dark matter will be detected in the next generation of high energy colliders, such as the LHC\index{Large Hadron Collider} or the ILC\index{International Linear Collider}. For the models presented in the section \ref{Section:DarkMatter}, the primary signal of WIMPs\index{dark matter!WIMPs}\footnote{As these channels require a real Higgs boson\index{Higgs} be produced, only WIMPs\index{dark matter!WIMPs} lighter than $m_h/2$ can be probed in this way.} will be an invisibly decaying Higgs boson\footnote{In the case of Model \EXDM, in which a radion\index{Kaluza-Klein gravitons!radions} is used as the mediator, the signal will be an invisibly decaying radion. However as demonstrated in Ref \cite{Bae:2001id}, the production and decay of a radion is very similar to the Standard Model Higgs boson. Therefore the arguments in this section also apply to invisible radion decays. }. Results from LEP-I\index{LEP-I} and LEP-II\index{LEP-II} have already excluded an invisible SM Higgs\footnote{These bounds assume a Standard Model Higgs boson with $BR(h \to \cancel{E}) = 100 \%$. If there are multiple Higgs fields, or if this branching ratio is less than $100 \%$, then these bounds are weaker. However a Higgs boson with a lower invisible branching ratio\index{Higgs!invisible} can be constrained by its visible decay modes } with mass below $114 \; GeV$ \cite{Abbiendi:2006gd,Abbiendi:2007jk,Nagai:2008zz} using the $e^+e^- \to Z^0 + h_{inv}$ channel, while the Tevatron\index{Tevatron} and the LHC\index{Large Hadron Collider} \nocite{Barger:2007im} will search for an invisible Higgs boson\index{Higgs!invisible} produced in either the $pp \to Z + h_{inv}$ channel\index{Higgstrahlung} \cite{Frederiksen:1994me,Davoudiasl:2004aj} or through weak boson fusion\index{weak boson fusion} \cite{Eboli:2000ze,Davoudiasl:2004aj}. 

\par
The discovery potential of the $pp \to Z(\to l^+ l^-) + h_{inv}$ channel was studied in Ref. \cite{Frederiksen:1994me,Davoudiasl:2004aj,Zhu:2005hv} and Ref. \cite{Martin:1999qf,Davoudiasl:2004aj} for the LHC\index{Large Hadron Collider} and Tevatron\index{Tevatron} respectively. The signal of an invisible Higgs\index{Higgs!invisible} in this channel is obscured by the $pp \to Z + jet$ background, where the jet could be either soft or otherwise undetected, and it was originally thought that this background would significantly weaken any signal detected in this channel \cite{Frederiksen:1994me}. However, as demonstrated in Ref \cite{Davoudiasl:2004aj}, the background can be reduced considerably by introducing specific cuts on the missing momentum, with the optimum results produced by requiring $\cancel{p}_T \gtrsim 75 \; GeV$. 
\par
The other significant backgrounds for this process are 

\begin{equation}
\begin{split}
p\bar{p} & \to Z(\to l^+ l^-) + Z (\to \bar{\nu} \nu) \\
p\bar{p} & \to W^+(\to l^+ \nu) + W^- (\to \bar{\nu} l^-) \\
p\bar{p} & \to Z(\to l^+ l^-) + W^{\pm} (\to l^{\pm} \nu) \\
\end{split}
\end{equation}

\noindent
where in the third channel, the charged lepton is not detected. The second channel is removed by requiring the invariant mass of the lepton pair to be close to the Z boson mass, with $|m_{l^+l^-} - m_Z| < 10 \; GeV$ and the lepton momenta should be in similar directions, 
which implies both leptons are produced by a single Z-boson rather than from two $W^{\pm}$ bosons, requiring a cut of $\phi_{l^+l^-} < 143^{\circ}$. The background from the first two channels can also be reduced by rejecting events with lower missing transverse momentum. As outlined in Ref \cite{Davoudiasl:2004aj}, the decays of Z and W bosons to missing energy tend to produce soft neutrinos and low momentum decay products, while invisibly decaying Higgs bosons\index{Higgs!invisible} are expected to generate larger values of $\cancel{p}_T$.  The third background channel, in which a third charged lepton is missed by the detector, is expected to be small due to the detector coverage at the LHC\index{Large Hadron Collider} \cite{Davoudiasl:2004aj}.

\begin{figure}
\begin{displaymath}
\begin{array}{cc}
\psfig{file=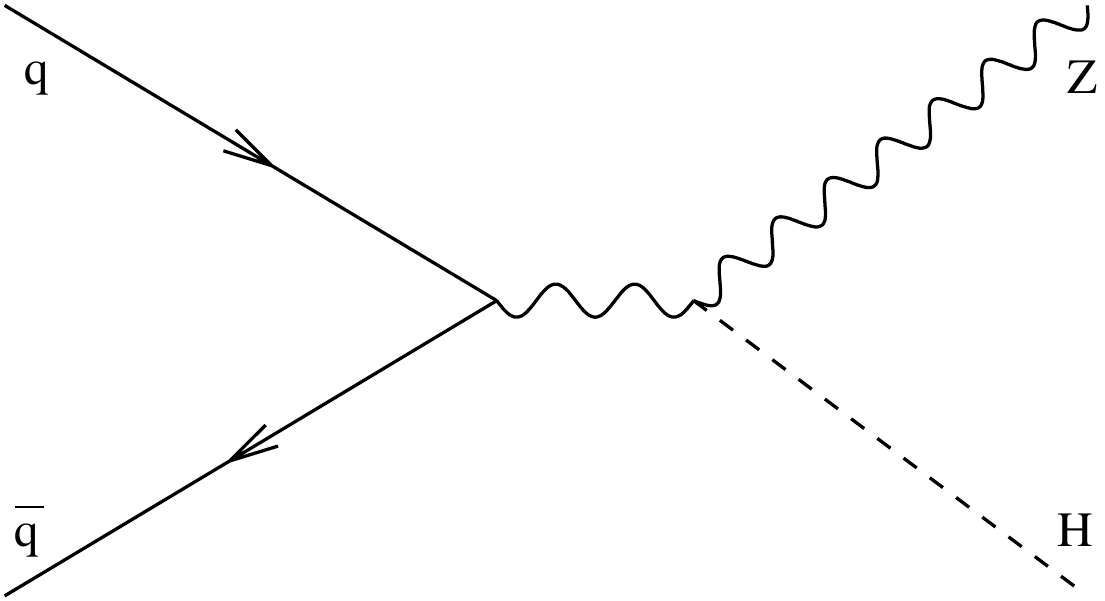,width=0.45\textwidth,angle=0} & \quad
\psfig{file=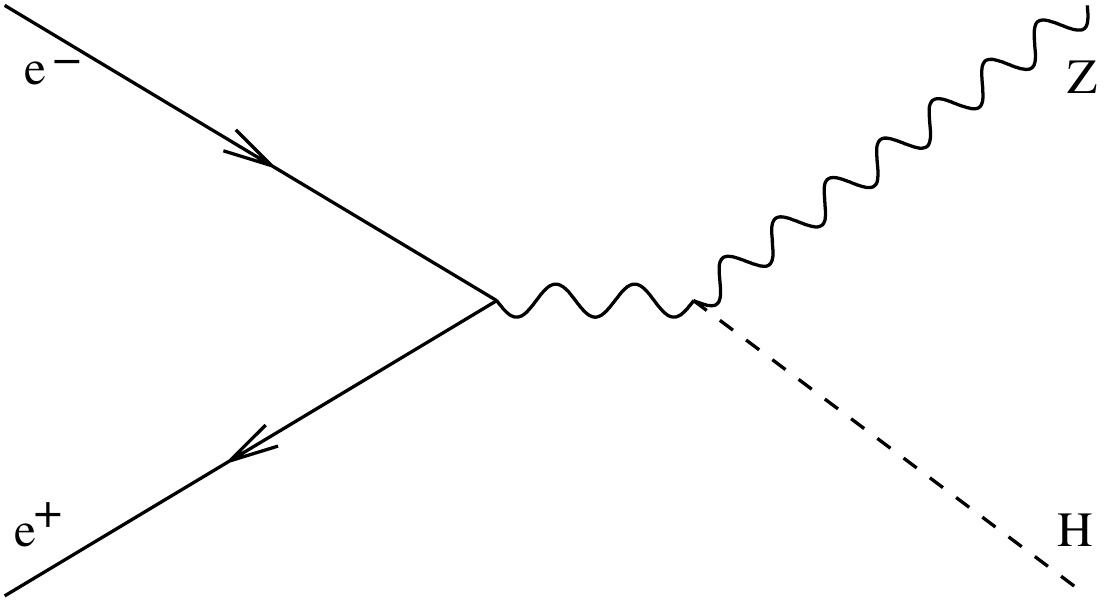,width=0.45\textwidth,angle=0}
\end{array}
\end{displaymath}
\caption{\label{Figure:ColliderDiagrams} Feynman diagram for the production of Higgs bosons through the $Z^0 + H$ channel\index{Higgstrahlung}, at hadron colliders (left) and electron-positron colliders (right)}
\end{figure}

\par
Using these cuts to reduce the background, it is expected that the LHC will be capable of detecting an invisibly decaying Standard Model Higgs\index{Higgs!invisible} with mass $m_h \lesssim 160 \; GeV$ through the $Z + h_{inv}$ channel. In comparison, the WBF channel\index{weak boson fusion}, which will be introduced later in this section, is expected to be capable of detecting an invisible Higgs as heavy as $m_h \lesssim 500 \; GeV$ with ${\mathcal L} = 10 \; fb^{-1}$ \cite{Eboli:2000ze}. 

\par
This channel can also be studied at ILC\index{International Linear Collider}, with the advantage of being able to accurately measure the mass of the Higgs, and to measure small SM branching ratios \cite{Richard:2007ru}. 

\par
The second channel which can be used for detection of an invisible Higgs is weak-boson fusion\index{weak boson fusion}, in which the Higgs boson couples to a virtual boson exchanged by two protons. The final signal in this channel will be $h_{inv} + 2 jets$, with the dominant contribution to the background being the $Z(\to \bar{\nu} \nu) + 2 jets$ channel and a lesser contribution to the background from the $W^{\pm} (\to l^{\pm} \nu) + 2 jets$ where as before the charged lepton is not detected. In the case of weak-boson fusion\index{weak boson fusion}, the accompanying jets are formed by high-energy partons in the $\bar{p}p$ collision, and are expected to have high momentum and a large rapidity gap. In contrast, the two background channels are expected to produce softer jets with lower rapidity gaps \footnote{The background event in which a Z-boson is produced in WBF\index{weak boson fusion}, and then decays to a neutrino-antineutrino pair, has very similar properties to the invisible Higgs decay\index{Higgs!invisible} and as such these cuts cannot reduce this particular channel}. Therefore the background can be reduced by requiring both the invariant mass of the jets and the rapidity gap to be large. For example, in Ref \cite{Davoudiasl:2004aj} the WBF\index{weak boson fusion} channel (at the Tevatron\index{Tevatron}) was studied using the cuts,

\begin{equation}
m_{jj} > 320-400 \; GeV \quad \quad \Delta \eta_{jj} > 2.8
\end{equation}

\noindent
with the strongest signal corresponding to $m_{jj} > 320 \; GeV$. In that study, it was demonstrated that although the WBF\index{weak boson fusion} channel is too weak to provide detection of an invisible Higgs\index{Higgs!invisible} at the Tevatron\index{Tevatron}, the combination of WBF\index{weak boson fusion} with the Higgstrahlung\index{Higgstrahlung} process described before would allow a $120 \; GeV$ higgs to be discovered for $ {\mathcal L} = 7\; fb^{-1}$ of data. The same channel has been studied for the LHC\index{Large Hadron Collider} \cite{Eboli:2000ze}, using

\begin{equation}
m_{jj} > 1200 \; GeV \quad \quad \phi_{jj} \lesssim 60^{\circ}
\end{equation}

\noindent
where $\phi_{jj}$ is the angle between the jets, and is restricted to count only forward jets. In addition, because the cross-section for the weak boson fusion\index{weak boson fusion} channel does not drop off as fast for larger Higgs masses, it can detect heavier Higgs bosons, with masses up to $m_H \sim 400 \; GeV$ for ${\mathcal L} = 30 \; fb^{-1}$ compared with a limit of $m_h \sim 175 \;GeV$ in the other channels \cite{Davoudiasl:2004aj}. 

\par
 This channel can be used for detection of an invisible Higgs at either LHC\index{Large Hadron Collider} or the Tevatron\index{Tevatron}, although the signal at the Tevatron is not expected to be strong unless several channels are combined \cite{Davoudiasl:2004aj}. 

\begin{figure}
\begin{center}
\psfig{file=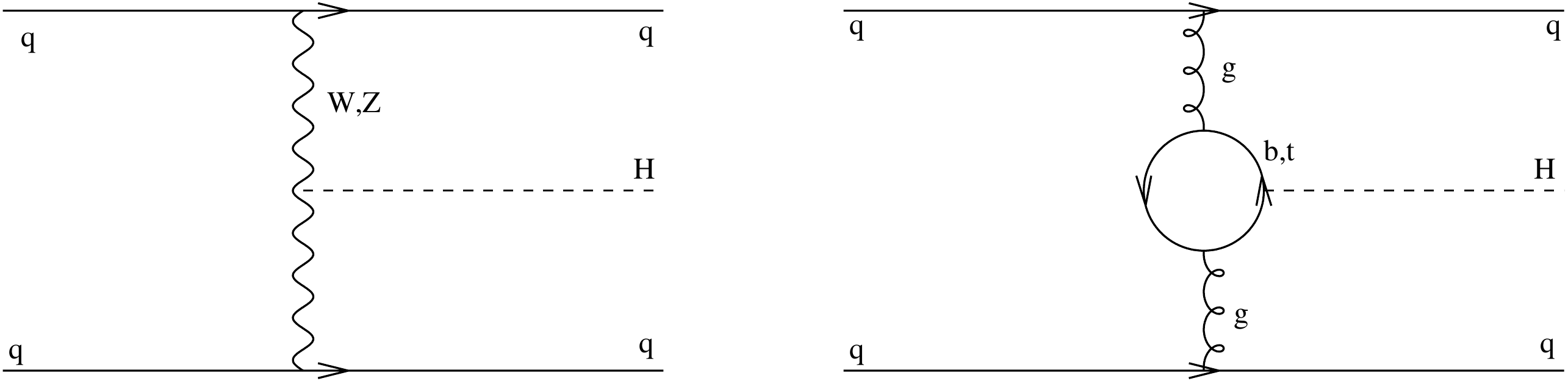,width=\textwidth,angle=0}
\end{center}
\caption{\label{Figure:Colliders} Feynman diagrams for the other two processes which could be used to probe invisible Higgs decays\index{Higgs!invisible} at hadron colliders, weak boson fusion\index{weak boson fusion} (left) and gluon fusion\index{gluon fusion} (right).}
\end{figure}

\par
Another possible channel which has been studied is the production of $h_{inv} + jet$ in gluon fusion\index{gluon fusion} \cite{Field:2003yy}. Gluon fusion is one of the main channels of Higgs production at both the LHC\index{Large Hadron Collider} and the Tevatron\index{Tevatron}. However as demonstrated in Ref \cite{Davoudiasl:2004aj}, for invisibly decaying Higgs bosons\index{Higgs!invisible} the background channel of $Z (\to \bar{\nu} \nu) + jet$ obscures this signal and the discovery potential of this channel is limited. 

\par
It has also been suggested that the production of top-quark\index{top quark} pairs with associated Higgs production could be used to probe invisible Higgs decays\index{Higgs!invisible} , however the analysis is significantly more difficult than the other processes and does not appear to provide a better signal \cite{Kersevan:2002zj}.

\par
The significance of the invisible Higgs\index{Higgs!invisible} signal depends on several factors, but for $m_h = 120 \; GeV$ the LHC\index{Large Hadron Collider} will be able to detect a $3\sigma$ signal for a branching ratio as small as 
\cite{Davoudiasl:2004aj}

\begin{equation}
BR_{h \to {\rm invisible}} \gtrsim \left( \begin{array}{ll} 0.41 & \quad {\mathcal L} = 10 \; {\rm fb^{-1}} \\ 0.24 & \quad {\mathcal L} = 30 \; {\rm fb^{-1}}  \end{array}\right)
\end{equation}

\noindent
At the Tevatron\index{Tevatron}, the analogous bounds are

\begin{equation}
BR_{h \to {\rm invisible}} \gtrsim \left( \begin{array}{ll} 0.90 & \quad {\mathcal L} = 10 \; {\rm fb^{-1}} \\ 0.52 & \quad {\mathcal L} = 30 \; {\rm fb^{-1}}  \end{array}\right)
\end{equation}

\par
It is also possible to search for invisible Higgs decays\index{Higgs!invisible} in $e^+e^-$ colliders. The primary search channel at electron-positron colliders in the 'Higgstrahlung'\index{Higgstrahlung} process, 

\begin{displaymath}
e^+e^- \to Z^0 + h_{inv}
\end{displaymath}
\noindent

measured by observing the decay of the Z-boson. The main background reactions for this reaction are

\begin{displaymath}
\begin{split}
e^+e^- &\to W^+W^- \to \ell \nu \bar{q}q \\
e^+e^- &\to Z^0 Z^0 \to  \bar{\nu} \nu \bar{q} q \\
e^+e^- & \to \bar{q} q + \gamma
\end{split}
\end{displaymath}

\noindent
The first reaction appears to produce missing energy if the lepton is hidden in a jet or is missed by the detectors, while the second reaction can mimic an invisible Higgs boson\index{Higgs!invisible} with a mass similar to that of the Z-boson. The third reaction can also contribute to the background if the photon is not detected. As with the searches at the LHC\index{Large Hadron Collider}, these backgrounds are reduced using a series of cuts outlined in \cite{Abbiendi:2006gd,Abbiendi:2007jk} for LEP\index{LEP-I}\index{LEP-II}, and in \cite{Richard:2007ru} for the ILC\index{International Linear Collider}.


\par
In this section, I will derive and present the expected sensitivity of collider experiments to each of the minimal models. For each model, it will be assumed that $m_h = 120 \; GeV$ for the purpose of demonstration. The general results are not expected to change significantly for different Higgs masses, with the exception the the location of the Higgs resonance in the abundance constraints and resulting drop in the invisible branching ratio for each model will shift to $m_{DM} \sim m_h/2$.

\subsection{Model 1: Minimal Model of Dark Matter \label{Section::MSDMColl}}\index{dark matter!Minimal Model|(}
\index{dark matter!scalar|(}

\begin{figure}
\psfig{file=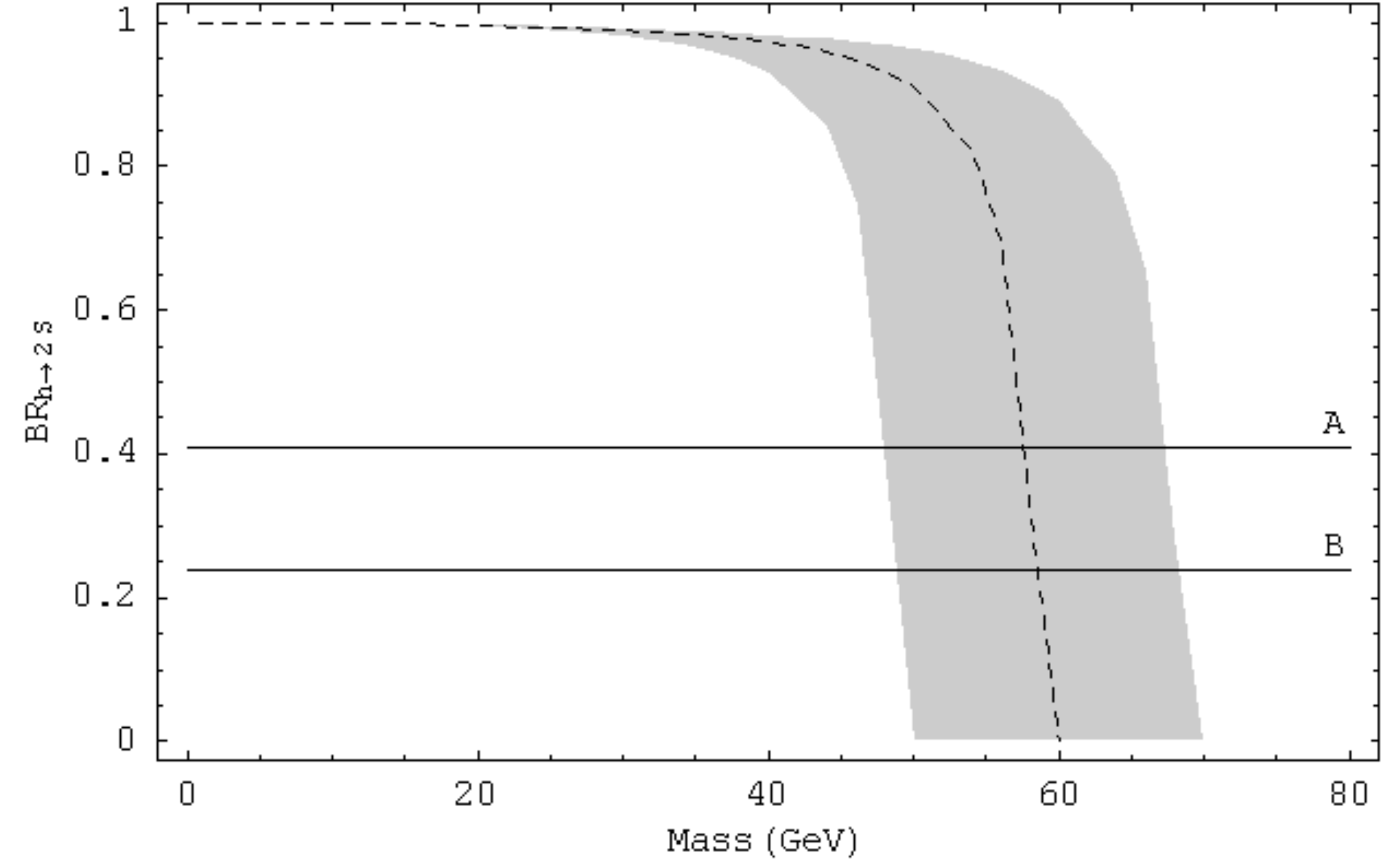,width=\textwidth,angle=0}
\caption{\label{Figure:LHCbounds} Invisible Higgs\index{Higgs!invisible} branching ratio, with the discovery potential for the LHC\index{Large Hadron Collider}. The central dashed line represents the invisible Higgs branching ratio for $m_h = 120 \; GeV$, with the grey region giving the branching ratio for the range $100\;GeV < m_h < 140 \; GeV$. The region above line A is detectable at the 3$\sigma$ level with $L = 10 \; fb^{-1}$ at LHC, while the region above line B can be detected with $L = 30 \; fb^{-1}$.}
\end{figure}

\par
In the minimal model of dark matter, the Higgs coupling to WIMPs\index{dark matter!WIMPs} is stronger than the coupling to Standard Model fields for light WIMPs\index{dark matter!WIMPs}, and for $m_s \lesssim m_h/2$ the branching ratio for invisible Higgs\index{Higgs!invisible} decays is almost $100 \%$. Using the search methods outlined in the introduction to this section, the LHC\index{Large Hadron Collider} will be able to search for WIMPs\index{dark matter!WIMPs} in this entire mass range, while higher mass WIMPs\index{dark matter!WIMPs} will produce a signal too weak to be detected. 
\par
The branching ratio for the minimal model of dark matter is plotted in Figure \ref{Figure:LHCbounds} for $100 \; GeV < m_h < 140 \; GeV$, along with the minimum branching ratio which can be detected by the LHC \footnote{It should be noted that the upper bounds on the invisible Higgs\index{Higgs!invisible} branching ratio is also dependent on the Higgs mass. However the variation over this range of masses is small.}. The Tevatron\index{Tevatron} can also probe this range, with a luminosity of $30 \; fb^{-1}$ being enough to detect $BR_{h_{inv}} \gtrsim 0.52$. Existing data from LEP\index{LEP} can already exclude $m_H < 114.4 \; GeV$ in this model when $BR(h_{inv}) = 100 \%$ \cite{Abbiendi:2006gd}.

\par
In summary, either the LHC\index{Large Hadron Collider} or the Tevatron\index{Tevatron} can detect scalar WIMPs\index{dark matter!WIMPs} in this model through an invisible Higgs decay\index{Higgs!invisible}. However as indicated in Figure \ref{Figure:LHCbounds} if $m_{DM} \approx m_h/2$ the invisible branching ratio is reduced and these bounds can be avoided. 
\index{dark matter!Minimal Model|)}
\subsection{Model 2: Minimal Model of Dark Matter with 2HDM \label{Section::2HDM-collider}}
\par
The special cases of $\lambda_1$ or $\lambda_3$ dominant are difficult to detect at colliders. In the previous model, the WIMPs\index{dark matter!WIMPs} could be detected through the invisible\index{Higgs boson!invisible} decay of a Standard Model Higgs boson which is produced through the $Z^0 + H$ channel or through weak boson fusion\index{weak boson fusion}. In the 2HDM\index{Higgs boson!two Higgs doublet} with large $\tan \beta$, only the up-type Higgs is produced in these reactions. The situation is the same for electron-positron colliders such as the ILC\index{International Linear Collider}, since the main channel for detecting an invisible Higgs is also $Z^0+H$ which does not occur in these two cases.
\par
The down-type Higgs can also be produced in the LHC\index{Large Hadron Collider} through a b-quark loop process, which results in a Higgs and a jet of other particles. However if the Higgs decays invisibly\index{Higgs!invisible}, this signal is difficult to separate from the background for a Standard Model Higgs boson.

Although the cross-section for down-type Higgs production is enhanced relative to the Standard Model by a factor of $\tan^2 \beta$, due to the smaller vacuum expectation value, it is also suppressed by the lack of a top-quark loop. From Ref \cite{Field:2003yy}, the production cross-section for the down-type Higgs through the b-quark loop is $ \sigma_h \sim O(100 \; pb)$ at the LHC, or $O(1 \; pb)$ for Tevatron\index{Tevatron}. However the background cross-section is $\sigma_b = 1.5 \times 10^5 \;pb$ ($300 pb$ for Tevatron) \cite{Davoudiasl:2004aj}.
\par
As a result, any collider constraints on these models are expected to be quite weak. With $30 \; fb^{-1}$ of data at LHC, the signal will still be well below $2\sigma$.

\par
For the purpose of collider searches, the third case of $\lambda_2$ dominant is identical to the minimal model of scalar dark matter. In this case, higgs production by gluon fusion\index{gluon fusion} involves only a top-quark\index{top quark} loop and not a bottom-quark loop, but this has only a small effect on the higgs production cross-section \cite{Field:2003yy}, and the gluon fusion channel is already difficult to detect when the higgs decays invisibly. 

\par
As mentioned in the introduction to this section, invisible Higgs\index{Higgs!invisible} decays can also be probed at electron-positron colliders through the $e^+ + e^- \to Z^0 + h_{inv}$  channel\index{Higgstrahlung}. However, as with the hadron colliders this channel only proceeds for the case of $\lambda_2$ dominant, in which case the results are the same as those for the minimal model of dark matter.
\index{dark matter!scalar|)}

\subsection{Model 3: Minimal Model of Fermionic Dark Matter \label{Section::MFDMColl}}
\index{dark matter!fermionic|(}\index{dark matter!Minimal Fermionic Model|(}

\begin{figure}
\psfig{file=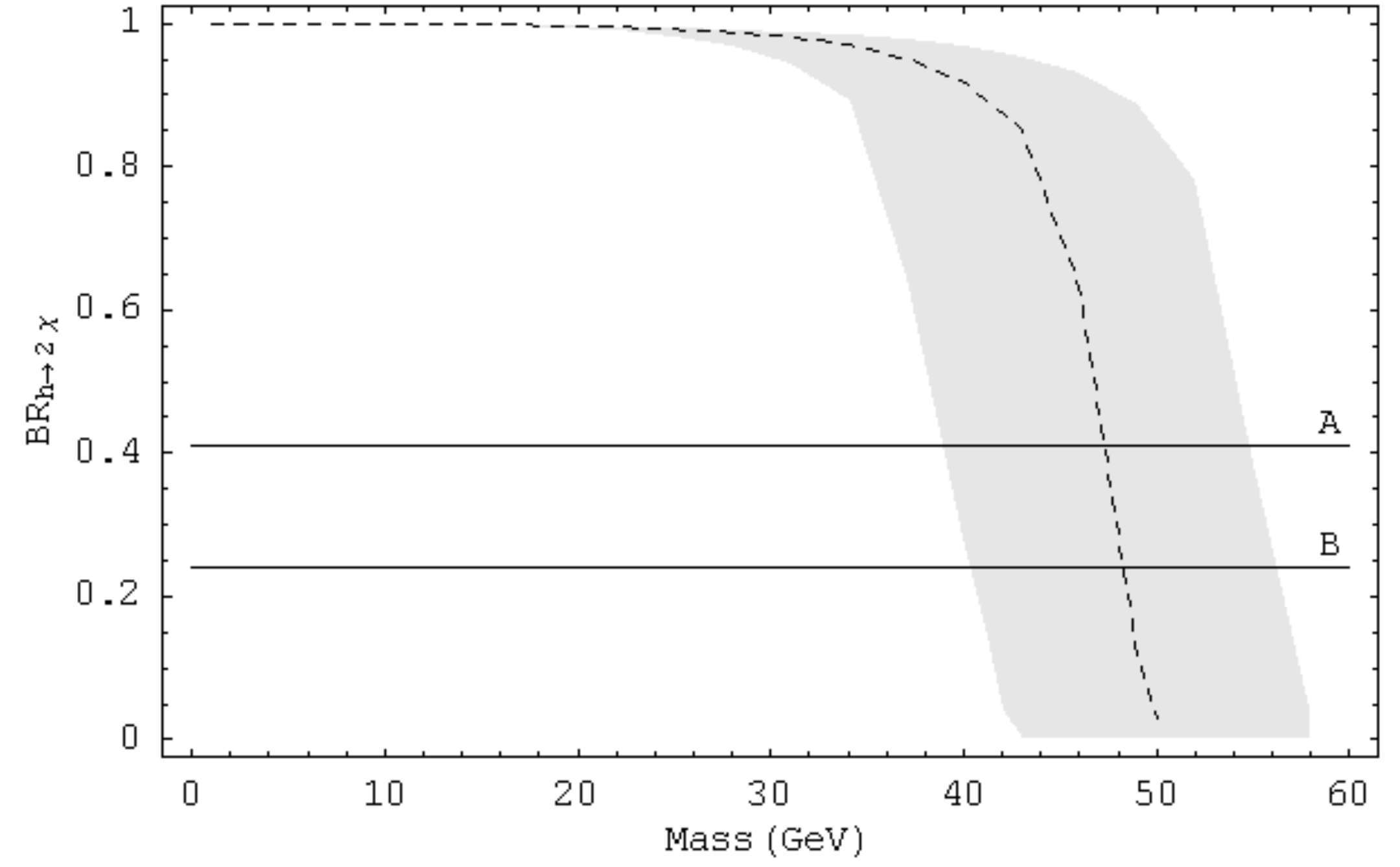,width=\textwidth,angle=0}
\caption{\label{Figure::LHCbounds} The branching ratio for $h \to \chi \chi$, with $m_h = 120\;Gev$ (dashed line) and in the range $100 \; GeV < m_h < 140 \; GeV$ (grey region). The horizontal lines represent the minimum branching ratio which would produce a $3 \sigma$ signal in the $h \to Z + \chi \chi$ channel at the LHC\index{Large Hadron Collider} for $L = 10 \; fb^{-1}$ and $L=30 \; fb^{-1}$. }
\end{figure}

\par
Using the abundance constraints on $\kappa$, and the three channels for the detection of an invisible Higgs\index{Higgs!invisible} given in the introduction to this section, the discovery potential of the LHC\index{Large Hadron Collider} and Tevatron\index{Tevatron} can be calculated for the minimal model of fermionic dark matter. 
\par
In this model, it is possible for the LHC to discover WIMPs\index{dark matter!WIMPs} as heavy as $~ 50 \; GeV$ (or slightly below $m_h/2$ for Higgs masses above $120 \;GeV$). However as with the minimal model of scalar dark matter, the rapid drop off in the Higgs branching ratio above $m_{\chi} \approx 50 \; GeV$ restricts detection of heavier WIMPs\index{dark matter!WIMPs}. As demonstrated in Figure \ref{Figure:MFDM-recoil}, the range of $m_{\chi} \lesssim 50 \; GeV$ has already been excluded by the dedicated dark matter searches, CDMS\index{\ddme CDMS} and XENON10\index{\ddme XENON10}. Therefore it is not expected that the LHC\index{Large Hadron Collider} will be able to detect signals from this model.
\index{dark matter!fermionic|)}\index{dark matter!Minimal Fermionic Model|)}

\subsection{Model 4: Fermionic Dark Matter with 2HDM}

\par
As in the model presented in Section \ref{Section::2HDM-collider}, an invisibly decaying down-type Higgs boson is difficult to detect in colliders, and therefore it is not expected that there will be strong constraints on either this model, or the higgsino model given in Section \ref{Section::Higgsino}, from any present or near-future collider. 
\par
As with the analogous model for scalar WIMPs\index{dark matter!WIMPs}, the cases of $\lambda_1$ or $\lambda_3$ dominant would be detected by an invisible $H_d$ decay. However in the LHC\index{Large Hadron Collider} and Tevatron\index{Tevatron}, the only channel capable of producing this Higgs boson is a b-quark loop, and the resulting signal is significantly smaller than the background.
\par
The third case, in which $\lambda_2$ is dominant, is not significantly different than the minimal model of fermionic dark matter. Therefore it is expected that either LHC\index{Large Hadron Collider} or the Tevatron\index{Tevatron} could probe this model up to $m_{DM} \sim 50 \; GeV$ if $m_H \approx 120 \; GeV$.

\subsection{Model 5: Dark Matter \& Warped Extra Dimensions}

\par
The radion used in this model has similar properties to the Standard Model Higgs boson, and so the collider signals and explorable region of parameter space for these models are similar to those presented in Section \ref{Section::MSDMColl} and Section \ref{Section::MFDMColl}. 

\begin{figure}
\psfig{file=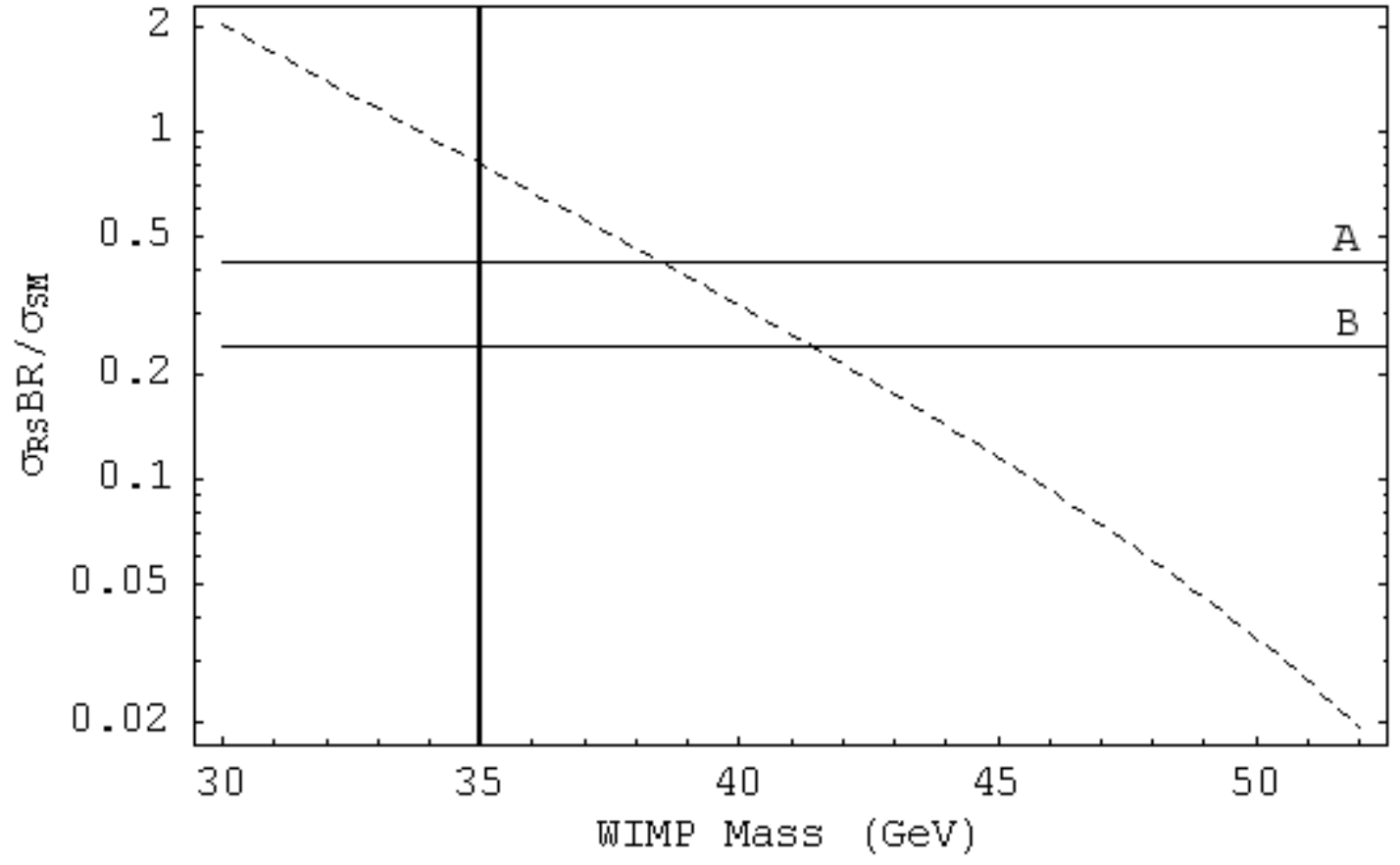,width=\textwidth,angle=0}
\caption{\label{Figure:RS-collider} Sensitivity of the LHC to scalar WIMPs\index{dark matter!WIMPs} through the invisible radion signal. Lines A and B represent the smallest branching ratio that can be detected at ${\mathcal L} = 10 \; fb^{-1}$ and ${\mathcal L} = 30 \; fb^{-1}$ respectively. The vertical line represents the WIMP mass at which the model becomes non-perturbative.}
\end{figure}

\par
The most significant difference is that, unlike in the Higgs models in which $v_{EW}$ is fixed, the radion vev $\Lambda_{\phi}$ is unknown. As a result, the cross-sections for radion production are enhanced (or suppressed) by a factor of $\sigma_{RS}/\sigma_{SM} \approx v_{EW}^2/\Lambda_{\phi}^2$. 

\par
The branching ratio for scalar WIMPs\index{dark matter!WIMPs} in this model is plotted in Figure \ref{Figure:RS-collider}, with a scaling factor added to account for the enhancement in the radion production cross-section.  For the purpose of comparison with previous models, it is assumed that $M_{\phi} = 120 \; GeV$ while $m_S$ and $\Lambda_{\phi}$ are varied to satisfy the abundance constraints. In addition, only scalar WIMPs\index{dark matter!WIMPs} are considered here, as the abundance constraints on the fermionic WIMPs\index{dark matter!WIMPs} in this model require non-perturbative couplings for the regions of parameter space which can be probed by the LHC\index{Large Hadron Collider} or Tevatron\index{Tevatron}.
\par
Using these results, the LHC\index{Large Hadron Collider} can probe scalar WIMPs\index{dark matter!WIMPs}\index{dark matter!scalar} in this model up to $m_S \lesssim 38 \;GeV$ with ${\mathcal L} = 10 \; fb^{-1}$ and up to $m_S \lesssim 42 \;GeV$ with ${\mathcal L} = 30 \; fb^{-1}$, while perturbative couplings require $m_S \gtrsim 35 \;GeV$.

\vspace{16pt}
\section{Light Dark Matter \label{Section::LDM}}\index{dark matter!light|(}

\par
In the previous sections, WIMPs\index{dark matter!WIMPs} as heavy as $m_{DM} \sim O(100 \;GeV)$ have been studied, while even heavier WIMPs\index{dark matter!WIMPs} can be used to explain dark matter. However there are also many reasons to study models which contain WIMPs\index{dark matter!WIMPs} with masses in the $O(1 \; GeV)$ range. \nocite{Pospelov:2008jk}

\par
Light dark matter\index{dark matter!light} has several benefits. The 511 keV $\gamma$-ray flux\index{$\gamma$-rays!511 keV} from the galactic core\index{galactic positron excess}, which is believed to be caused by a non-localized source of positrons \cite{Knodlseder:2003sv}, can be explained by the annihilation \cite{Boehm:2003bt}  or decay \cite{Picciotto:2004rp}\nocite{Pospelov:2007xh} of O(100 MeV) WIMPs\index{dark matter!WIMPs} that have been captured in the galactic core. It has also been demonstrated that this does not result in an overproduction of other galactic $\gamma$-rays\index{$\gamma$-rays} \cite{Beacom:2004pe,Beacom:2005qv}. Light dark matter\index{dark matter!light} may also be able to explain the observed flux of $\gamma$-rays in the 1 MeV to 20 MeV range \cite{Ahn:2005ck} without conflicting with measurements of the extragalactic $\gamma$-ray background \cite{Rasera:2005sa}. 

\nocite{Milne:1999ri}\nocite{Schanne:2004gy}\nocite{Bertone:2004ek}

\par
There is also a known discrepancy between results from different dedicated dark matter experiments. The data from DAMA\index{\ddme DAMA} indicates the existence of a WIMP consistent with $m_{DM} \sim 60\; GeV$, while CDMS\index{\ddme CDMS} and other experiments have excluded this region of parameter space. Both of these experiments search for dark matter by detecting the recoil of nuclei scattered by WIMPs\index{dark matter!WIMPs}. However the experiments use different nuclei, and the Na atoms in the DAMA\index{\ddme DAMA} detector are more sensitive to light WIMPs\index{dark matter!WIMPs} than the Ge atoms in the CDMS\index{\ddme CDMS} detector, due to the difference in the WIMP velocity that is required to cause a detectable recoil. As demonstrated in Ref \cite{Gondolo:2005hh}, this difference in the detection of light dark matter\index{dark matter!light} can explain both the positive signal at DAMA\index{\ddme DAMA} and the negative at all other detectors. \nocite{Aalseth:2008rx}

\par
The abundance constraints on sub-GeV WIMPs\index{dark matter!WIMPs}\index{dark matter!light} were given for each of the minimal models in Section \ref{Section:Abundance}. 
In this section, constraints on light dark matter\index{dark matter!light} from B-meson\index{B mesons} decays are calculated and presented for each of the models
which permits light WIMPs\index{dark matter!WIMPs}. The work in this section was originally published by the author and collaborators in Ref \cite{Bird:2004ts} and Ref \cite{Bird:2006jd}.

\subsection{Constraints on Light Dark Matter from B-decays}\index{B mesons}\index{dark matter!light}

\par
At present the experimental constraints on light dark matter\index{dark matter!light} are weak. As discussed in previous sections, light WIMPs\index{dark matter!WIMPs}\index{dark matter!light} are difficult to detect in experiments relying on nuclear recoil, as the heavier nuclei do not have a measurable recoil from lighter WIMPs\index{dark matter!WIMPs}\index{dark matter!light} unless they have a large velocity. Experiments which study the $\gamma$-ray\index{$\gamma$-rays} flux from both the galaxy and from extragalactic sources can provide better constraints on such models, but are still limited. One alternative is to use the existing B-physics experiments, in which large numbers of B mesons are produced and accurate decay rates can be measured.

\par
In general, WIMPs\index{dark matter!WIMPs} with masses of a few GeV or less will be produced in the decay of heavier mesons such as the B-meson\index{B mesons}. The requirement that the light dark matter\index{dark matter!light} abundance correspond to the observed dark matter abundance places a constraint on the WIMP annihilation cross section, and therefore a lower bound on the coupling constants. This also implies a lower bound on the decay widths for the invisible decays of heavy Standard Model particles. 
\par
By measuring the B meson\index{B mesons} branching ratios involving missing energy, the presence of light WIMPs\index{dark matter!WIMPs}\index{dark matter!light} could be inferred \cite{Bird:2004ts,Bird:2006jd}. As will be demonstrated in the next section, this method allows B-physics experiments such as BaBar\index{BaBar} and BELLE\index{BELLE} to search for sub-GeV WIMPs\index{dark matter!WIMPs}, and for many models this provides the strongest constraints on their existence and properties. It should also be noted that we demonstrated the possibility of detecting dark matter\index{dark matter} in the decays of B-mesons\index{B mesons} prior to either BELLE\index{BELLE} or BaBar\index{BaBar} placing stringent bounds on the invisible decay widths, and that this idea has now been invoked as a motivation for building future B-meson\index{B mesons}\index{SuperB} experiments \cite{Hewett:2004tv,Browder:2008em,Hitlin:2008gf} capable of further probing invisible decays.

\par
It should be noted that in this section only the decay $B \to K + \cancel{E}$ is studied. The decays $B \to \cancel{E}$ ,$B \to \gamma + \cancel{E}$, and $B \to \pi + \cancel{E}$ will also produce WIMP pairs in these experiments, however neither of the first two decays can be easily detected, while the third decay is suppressed by a factor of $\sim | V_{ts} /V_{td}|^2 \sim 0.04 $ relative to the kaon decay\index{kaons} and is also not as well constrained from experiments \cite{PDG}. It is also possible to produce WIMP pairs in the decays of heavier mesons, such as $\Upsilon \to \gamma + \cancel{E}$. However in general there is less data available for these decays, while some specific decays like this one have large widths in the Standard Model and therefore will have very small branching ratios for WIMP production. 

\subsection{B-Meson Experiments} 

\par
Although the decay of $B^+ \to K^+ + \cancel{E}$ can be detected through the detection of the kaon, it is not possible to determine the missing energy without first knowing the energy of the initial B-meson\index{B mesons}. This can be achieved by considering the system in which there are two B-mesons produced, and reconstructing one decay completely. Such a system can be studied effectively by considering the threshold production of two B-mesons in the decay of the $\Upsilon(4S)$\index{$\Upsilon(4S)$} state.

\par
For the reconstructed B-meson\index{B mesons}, either hadronic decays or semileptonic decays are allowed and the decay products of this B-meson are accounted for and removed from consideration. If the only remaining particle in the detector is a single kaon \footnote{In practice, many of these decays also have some background particles or spurious effects of the decay products of the reconstructed B-meson\index{B mesons}. As such, it is common to use weaker selection criteria which suppress the background without adversely affecting the efficiency of the missing energy searches.}, then it could represent a missing energy signal. 
\par
When the kaon momentum is small, there also exists a large background which can overwhelm the signal. The most common of these are normal B-meson decays\index{B mesons} in which some particles are undetected, such as $B \to (D \to K + \ell \bar{\nu}) + \ell \bar{\nu} $, and in decays to the long lived neutral kaon, $K_L^0$ which can pass through the detectors without depositing any significant energy. As such only decays in which the kaon is above a predetermined minimum are counted as a signal. For the BaBar\index{BaBar} and BELLE\index{BELLE} experiments, this restriction is ${\bf p}_{K} > 1.2 \; GeV$, which restricts the range of WIMP masses that can be probed to be $m_{DM} < 1.9 \; GeV$. For the CLEO\index{CLEO} experiment, ${\bf p}_{K} > 0.7 \; GeV$ and can probe up to $m_{DM} < 2.1 \; GeV$.

\par
At present, BaBar\index{BaBar} has examined $8.8\times 10^7$ $\Upsilon(4S) \to \bar{B} B$ decays with both hadronic and semileptonic decays of the reconstructed B-meson\index{B mesons}, and restricted the invisible branching ratio to \cite{Aubert:2004ws} 

\begin{equation}
BR(B^+ \to K^+ + \cancel{E}) < 5.2 \times 10^{-5}
\end{equation}

\noindent
at a $90\%$ c.l. In contrast, BELLE\index{BELLE} examined $53.5\times10^7$ $\Upsilon(4S) \to \bar{B} B$\index{$\Upsilon(4S)$} decays with only hadronic B-mesons reconstructed, and set a limit of \cite{Chen:2007zk}

\begin{equation}
BR(B^+ \to K^+ + \cancel{E}) < 1.4 \times 10^{-5}
\end{equation}

\noindent
at a $90 \%$ c.l. 
\par
In principle it is also possible to search for dark matter in the decays of the neutral B-meson\index{B mesons}, through the reaction $B^0 \to K^0 + \cancel{E}$, however it more both more difficult to reconstruct $B^0$ and less probable that the resulting $K^0$\index{kaons} can be detected and as such this reaction does not significantly enhance the constraints on missing energy. 
\par
Another possible channel is the decay of B-meson to an excited state of the kaon, $B \to K^* + \cancel{E}$. The sensitivity from this decay mode is comparable to that of $B^+ \to K^+ + \cancel{E}$ and can probe WIMPs\index{dark matter!WIMPs} only slightly lighter than that case. 

\par
As more data is collected from B-meson factories, it is expected that this sensitivity to light dark matter\index{dark matter!light} will improve. In particular, increased luminosity will allow these experiments to probe smaller branching ratios, while possible improvements in separating signal from background may allow the experiments to detect lower energy kaons and therefore heavier WIMPs\index{dark matter!WIMPs}.

\subsection{Model 1: Minimal Model of Dark Matter}
\index{dark matter!Minimal Model|(}
\index{dark matter!scalar|(}
\par
As discussed in Section \ref{Section::LDM}, light dark matter can be detected in B meson decays\index{B mesons}. In particular, the decay $B \to K + missing \; energy$ provides a strong probe of the sub-GeV range of the minimal model of dark matter\index{dark matter!light}\index{dark matter!scalar}.

\begin{figure}
\psfig{file=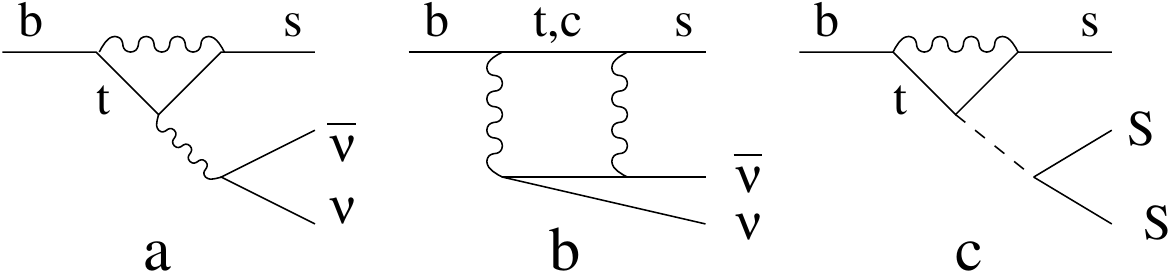,width=0.95\textwidth,angle=0}
\caption{\label{diagram:MSMgraphs}Feynman diagrams which contribute to B-decay with missing energy in the minimal scalar model of dark matter.\index{dark matter!scalar}}
\end{figure}

\par
The decay $B \to K + missing \; energy$ can occur in two ways. The Standard Model predicts this decay will occur as $B \to K + \overline{\nu} \nu$ \cite{Buchalla:2000sk}, with 

\begin{displaymath}
Br(B^+ \to K^+ + \overline{\nu} \nu) \simeq (4 \pm 1) \times 10^{-6}
\end{displaymath}

\noindent
As demonstrated in Ref \cite{Bird:2004ts,Bird:2006jd}, this decay can also proceed as a decay to dark matter\index{dark matter} which would not be seen in the present detectors.  The diagrams which contribute to the Standard Model decay and to the decay in the Minimal Model of Dark Matter\index{dark matter!Minimal Model} are given in Figure \ref{diagram:MSMgraphs}. 

\par
The loop process $b \to s + h$, which is required for the decay to dark matter scalars\index{dark matter!scalar}, can be calculated as the derivative of the $b \to s$ self-energy term with respect to $v_{EW}$ (see for example Ref \cite{Willey:1982mc}), resulting in an effective vertex 

\begin{equation}
L_{bsh} = \left( \frac{3 g_W^2 m_b m_t^2 V_{ts}^* V_{tb}}{64 \pi^2 M_W^2 v_{EW}} \right) \overline{s}_L b_R h + h.c.
\end{equation}

\noindent
Since the Higgs is constrained to be significantly heavier than the other particles involved in this decay, it can be integrated out to form an effective Lagrangian for b-decays.

\begin{equation} \label{Equation:BDML}
L_{bsE} = \frac{1}{2} C_{DM} m_b \overline{s}_L b_R S^2 - C_{\nu} \overline{s}_L \gamma^{\mu} b_R \overline{\nu}  \gamma_{\mu} \nu + h.c.
\end{equation}

\noindent
At leading order, the coefficients are 

\begin{equation}
\begin{split}
&C_{DM} = \frac{\lambda}{m_h^2} \frac{3 g_W^2 V_{ts}^* V_{tb}}{32 \pi^2 } x_t\\
&C_{\nu} = \frac{g_W^2 }{M_W^2} ~\frac{g_W^2V_{ts}^* V_{tb}}{16\pi^2}
\left[\frac{x_t^2+ 2 x_t}{8(x_t-1)}
+\frac{3x_t^2-6x_t}{8(x_t-1)^2}\ln x_t \right]
\end{split}
\end{equation}

\noindent
where $x_t = m_t^2/M_W^2$. From these effective interactions and the hadronic form factors for B and K mesons\index{B mesons}\index{kaons} \cite{Ali:1999mm,Bobeth:2001sq,Ball:2004ye}, the decay width for the process $B \to K + SS$ can be calculated:

\begin{equation}
\frac{d\Gamma_{B^+\to K^+SS}}{d\hat s} = \frac{x_t^2 C_{DM}^2 f_0(\hat s)^2}{512 \pi^3}  
\frac{I(\hat s,m_S)m_b^2 (M_B^2-M_K^2)^2}{M_B^3(m_b-m_s)^2},
\end{equation}

\noindent
where $\hat s = (p_B - p_K)^2$,$I(\hat s,m_S)$ reflects the available phase space, 

\begin{equation}
I(\hat s,m_S)= [\hat s^2 - 2\hat s(M_B^2 +M_K^2)+(M_B^2-M_K^2)^2]^{\frac{1}{2}}
[1-4m_S^2/\hat s]^{\frac{1}{2}}.
\end{equation}

\noindent 
and $f_0(\hat s) \simeq 0.33 exp(0.63 \hat s/M_B^2 - 0.095 {\hat s}^2/M_B^4 + 0.591 {\hat s}^3/M_B^6) $ encodes the internal structure of the B-meson\index{B mesons} and kaon\index{kaons} \footnote{In Ref \cite{Ball:2004ye}, the form factor is given a different parameterization. However for the purpose of this calculation, the two parameterizations give nearly identical results.}.

\par
From these results, the branching ratio for invisible B decays is calculated to be 

\begin{equation}\label{Eq:MDM-BR}
BR_{B^+ \to K^+ + E} =  4 \times 10^{-6} + 2.8 \times 10^{-4} \kappa^2 F(m_S)
\end{equation}

\begin{figure}
\begin{center}
\psfig{file=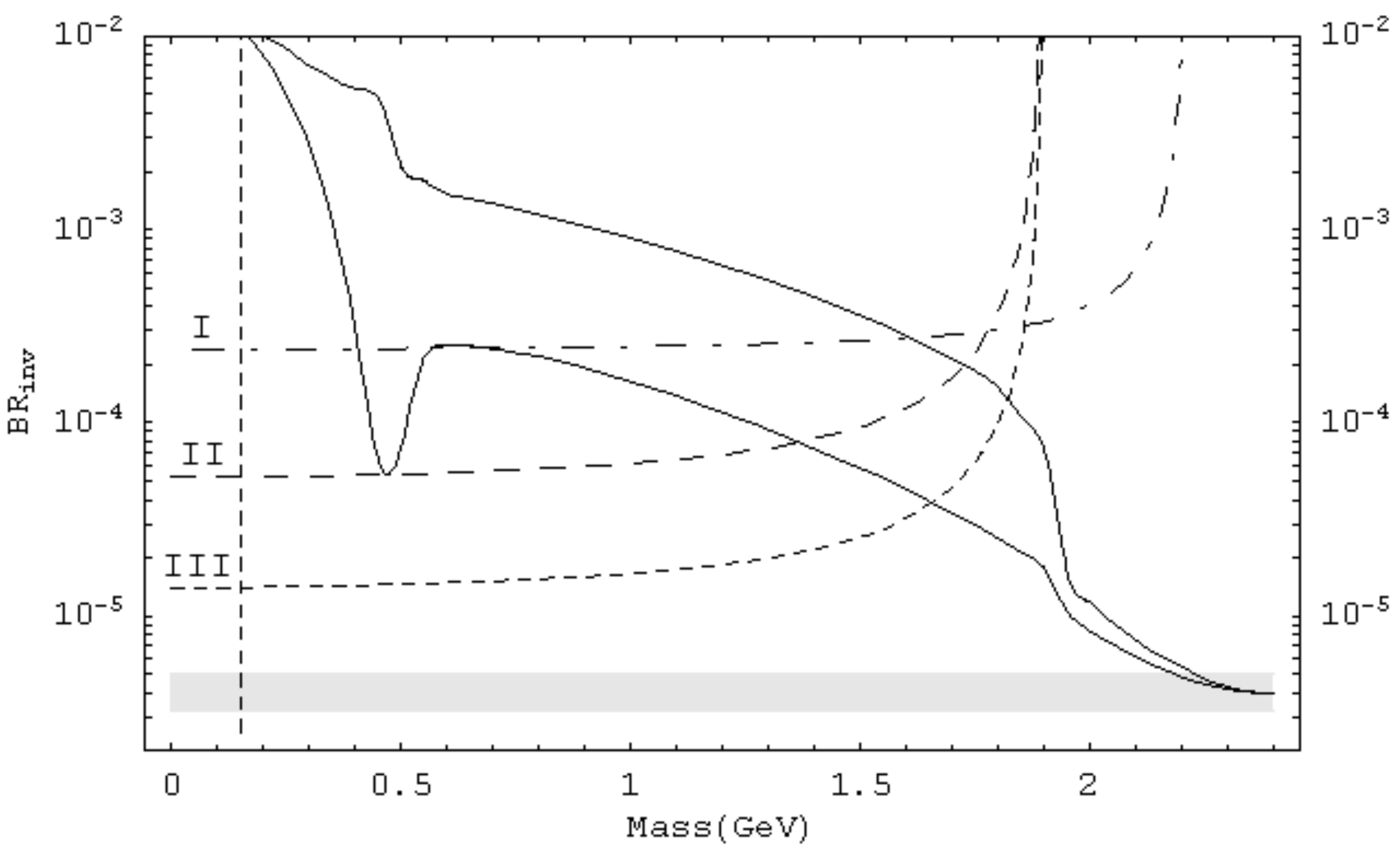,width=\textwidth,angle=0}
\end{center}
\caption{ \label{figure:BranchRatio}
Predicted branching ratios for the decay $B^+ \to K^+ +$ {\it missing energy}, with $\kappa$ determined by the abundance constraints, and with current limits from  
CLEO\index{CLEO} (I) \cite{Browder:2000qr}, $\babar$  (II) \cite{Aubert:2003yh} , and BELLE\index{BELLE} (III) \cite{Chen:2007zk}. The grey bar shows the expected $B\to K\nu\bar \nu$ signal. The parameter space to the left of the vertical dashed line can also be probed with  $K^+ \to \pi^+ + missing ~energy $.}
\end{figure}

\noindent
where as before,

\begin{equation}
\kappa \equiv \lambda \left( \frac{100 \; GeV}{m_h} \right)^2
\end{equation}

\noindent
and $F(m_S)$ represents the available phase space,

\begin{equation}
F(m_S) = \int_{\hat s_{min}}^{\hat s_{max}} f_0(\hat s)^2 I(\hat s,m_S) d{\hat s} \left[ \int_{\hat s_{min}}^{\hat s_{max}} f_0(\hat s)^2 I(\hat s,0) d{\hat s} \right]^{-1}
\end{equation}

\noindent
The first term in the branching ratio represents the Standard Model result, and the second term represents the additional effects of dark matter. The branching ratio is plotted in Figure \ref{figure:BranchRatio} with recent limits from BaBar and BELLE\index{BELLE}. 
In this figure, the region between the two solid lines gives the correct abundance for dark matter, with the region above the dashed lines excluded by B-physics\index{B mesons} experiments. The region to the right of these dashed lines corresponds to WIMP masses for which this decay produces a kaon\index{kaons} with momentum less than the experimental cut-off. For very light WIMPs\index{dark matter!WIMPs}\index{dark matter!light} it is also possible to produce dark matter in the kaon decay\index{kaons}, $K^+ \to \pi^+ + \cancel{E}$, and measurements of this decay width can already exclude the kinematically allowed region of $m_S \lesssim 150 \;MeV$ \cite{Adler:2001xv}. 
\par
From the abundance calculation, it is clear that $\kappa \sim O(1)$ for most kinematically allowed scalars, resulting in a branching ratio at least one order of magnitude larger than predicted by the Standard Model. As discussed in the introduction, recent results from BaBar\index{BaBar} \cite{Aubert:2004ws} have set a limit of $Br(B^+ \to K^+ + E) < 5.2 \times 10^{-5}$ at $90\%$ confidence level, while preliminary results from BELLE\index{BELLE} \cite{Chen:2007zk} report a limit of $Br(B^+ \to K^+ + E) < 1.4 \times 10^{-5}$. 

\par
Based on the results of these experiments, the mass of scalar WIMPs\index{dark matter!WIMPs} in the minimal model can be constrained to be $m_S \gtrsim 1.7 \; GeV$.

\index{dark matter!Minimal Model|)}
\subsection{Model 2: Minimal Model of Dark Matter with 2HDM}

\begin{figure}
\begin{center}
\psfig{file=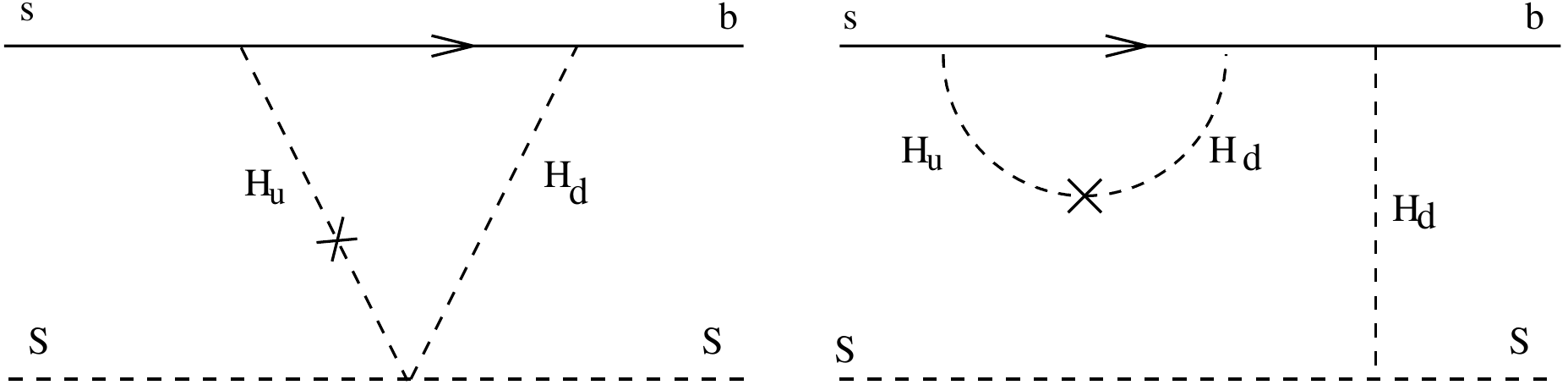,width=0.95\textwidth,angle=0}
\end{center}
\caption{\label{figure:2HDMdiagrams}Diagrams contributing to the decay $b \to s +SS$ in   the 2HDM plus 
scalar dark matter model when $\lambda_1$ is dominant and $\tan\beta$ is large. 
Inside the loops, $H_u$ and $H_d$ denote the two charged Higgs bosons\index{Higgs!two higgs doublets}, 
with the mixing of the two doublets denoted by a cross.}
\end{figure}

\par
As before, \we consider the three special cases in which a single $\lambda_i$ is taken to be dominant. For the first case, $\lambda_1 >> \lambda_2,\lambda_3$ the diagrams which contribute are given in Figure \ref{figure:2HDMdiagrams} and, using the same effective Lagrangian given in Eq \ref{Equation:BDML}, the Wilson coefficient is

\begin{equation}
C_{DM} = \frac{\lambda_1}{M_{H_d}^2} \frac{g_W^2 V_{ts}^* V_{tb} x_t}{32 \pi^2} \left( \frac{1-a_t + a_t \ln a_t}{(1-a_t)^2} \right)
\end{equation}

\noindent
where $a_t = m_t^2/M_{H}^2$. The corresponding branching ratio is

\begin{equation}
BR_{B \to K + E} = 4.0 \times 10^{-6} + 3.2 \times 10^{-5} \kappa^2  \left( \frac{1-a_t + a_t \ln a_t}{(1-a_t)^2} \right)^2 F(m_S)
\end{equation}

\begin{figure}
\begin{center}
\psfig{file=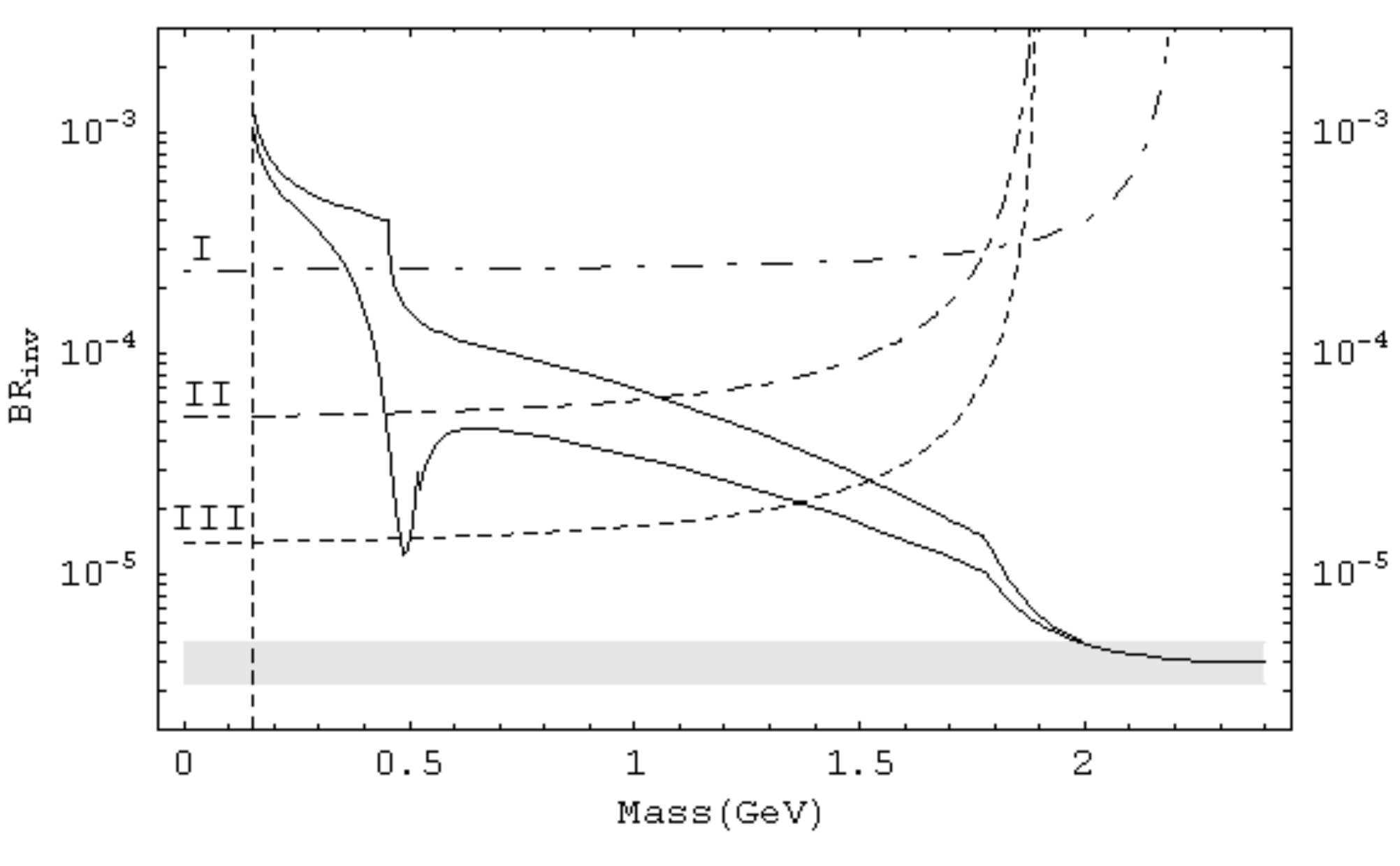,width=\textwidth,angle=0} \\
\end{center}
\caption{\label{2HDM:BR1}Branching Ratios for $B \to K+missing \; energy$ in the two higgs doublet model, with scalar WIMPs\index{dark matter!WIMPs} coupled
 primarily to $H_d$.  The labeling of current limits from CLEO\index{CLEO} (I),\babar\ (II), and BELLE\index{BELLE} (III)
 is the same as in Figure \ref{figure:BranchRatio}. }
\end{figure}

\noindent
where 

\begin{equation}
\kappa = \lambda_1 \left( \frac{100 \; GeV}{M_H} \right)^2
\end{equation}

\noindent
and as for the minimal model of dark matter,

\begin{equation}
F(m_S) = \int_{\hat s_{min}}^{\hat s_{max}} f_0(\hat s)^2 I(\hat s,m_S) d{\hat s} \left[ \int_{\hat s_{min}}^{\hat s_{max}} f_0(\hat s)^2 I(\hat s,0) d{\hat s} \right]^{-1}
\end{equation}

\noindent
The branching ratio is plotted against current experimental limits in Figure \ref{2HDM:BR1}. 

\par
The second case is $\lambda_2$ dominant, corresponding to a scalar WIMP\index{dark matter!WIMPs}\index{dark matter!scalar} which is coupled to the $H_u$ boson. In the limit of large $\tan \beta$, $v_u \approx v_{ew}$ and the model becomes very similar to the minimal model of dark matter\index{dark matter!Minimal Model} presented in Section \ref{Section::MDM}. However because the WIMPs\index{dark matter!WIMPs} cannot annihilate to down or strange-type quarks, the Lee-Weinberg\index{dark matter!Lee-Weinberg limit} limit excludes a larger region of the parameter space.  

\begin{figure}
\begin{center}
\psfig{file=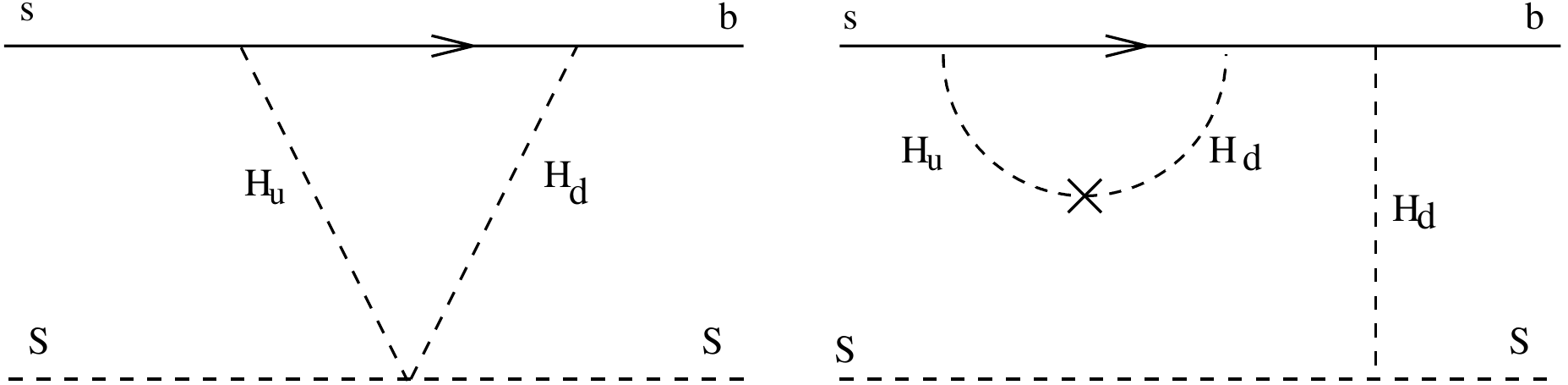,width=0.95\textwidth,angle=0}
\end{center}
\caption{\label{figure:l3diagrams}Diagrams contributing to the decay $b \to s +SS$ in   the 2HDM plus 
scalar dark matter model when $\lambda_3$ is dominant and $\tan\beta$ is large. }
\end{figure}

\par
The final case is the one in which $\lambda_3 \gg \lambda_1,\lambda_2$. The diagrams for this model are given in Figure \ref{figure:l3diagrams}. As noted in Section \ref{Section:MDM2HDM-AC}, the annihilation cross-section in this model is enhanced relative to the case of $\lambda_1$ dominant, resulting in a suppression of $\lambda_3$ by a factor of $\tan \beta$. However in this model the leading terms in the two B-decay diagrams (ie. the terms which are proportional to $\tan \beta$) cancel, resulting in a suppressed coupling $C_{DM} \sim O(\tan^{0} \beta)$. The overall result is that $BR(B \to K + SS)$ is reduced by $\tan^{-2} \beta \sim 10^{-4}$, and so

\begin{equation}
BR(B \to K + SS) \ll BR(B \to K + \overline{\nu} \nu)
\end{equation}

\noindent
and there are no constraints on this model. In the case with both $\lambda_1$ and $\lambda_3$ significant, the B-decay constraints can exclude certain regions of parameter space but cannot exclude WIMPs\index{dark matter!WIMPs} at any range of masses.

\par
For the case of $\lambda_1$ dominant, the results are similar to those plotted in Figure \ref{figure:BranchRatio}, and as with the minimal model the experiments exclude WIMPs\index{dark matter!WIMPs} lighter than $\sim 1.7 \; GeV$. For the case of $\lambda_2$ dominant, the results are similar and the data from BELLE\index{BELLE} excludes WIMPs\index{dark matter!WIMPs} lighter than $\sim 1.5 \; GeV$. The case of $\lambda_3$ dominant is quite different, in that the decays of B-mesons\index{B mesons} to WIMP pairs\index{dark matter!WIMPs} is suppressed by a factor of $\tan^2 \beta$ relative to the other two cases. As a result, B physics experiments cannot provide constraints on this model. 
\par
For the most general case, in which the three coupling constants are comparable in magnitude, the experimental limits are weaker. In the annihilation cross-section, the $\lambda_3$ terms dominate and are therefore $\lambda_1$ and $\lambda_2$ are not well constrained. However the invisible branching ratio is dominated by $\lambda_1$ and $\lambda_2$ terms, with $\lambda_3$ terms representing an $O(\tan^2 \beta)$ correction. As a result, it is more difficult to place constraints on this model than in the special cases, as unlike those models $BR(B \to K + SS)$ is not related to $\Omega_{DM}$. 
\index{dark matter!scalar|)}

\subsection{Model 3: Minimal Model of Fermionic Dark Matter}\index{dark matter!fermionic}\index{dark matter!Minimal Fermionic Model|(}

\par
As demonstrated in Section \ref{Section:MFDM-AC}, the minimal model of fermionic dark matter cannot describe sub-GeV WIMPs\index{dark matter!WIMPs} and maintain perturbative couplings\footnote{As discussed in Section \ref{Section::MFDM}, it may be possible for this model to contain light fermionic WIMPs\index{dark matter!WIMPs} if the mediator particle is also light. However such models are constrained by existing experiments, and are not considered in this dissertation.}. Therefore it is not possible to search for this type of dark matter in B-physics experiments.

\index{dark matter!Minimal Fermionic Model|)}

\subsection{Model 4: Fermionic Dark Matter with 2HDM}

\par
As was demonstrated in Section \ref{Section:2HDMf-ab}, it is possible to have sub-GeV fermionic WIMPs\index{dark matter!WIMPs}\index{dark matter!light}, without violating the abundance constraints, if they are coupled to the Higgs bosons in a two-higgs doublet\index{Higgs!two higgs doublet}. In particular, in the special case in which the 
\begin{displaymath}
\lambda_3 \overline{\chi} \chi v_u H_d
\end{displaymath}

\noindent
term dominates all other WIMP-Higgs interactions, there is a $\tan^2 \beta$ enhancement of the WIMPs\index{dark matter!WIMPs} annihilation cross-section and therefore $\lambda_3$ can be $O(\tan \beta)$ smaller, making it perturbative for light WIMPs\index{dark matter!WIMPs}\index{dark matter!light}.
\par
\begin{figure}
\begin{center}
\psfig{file=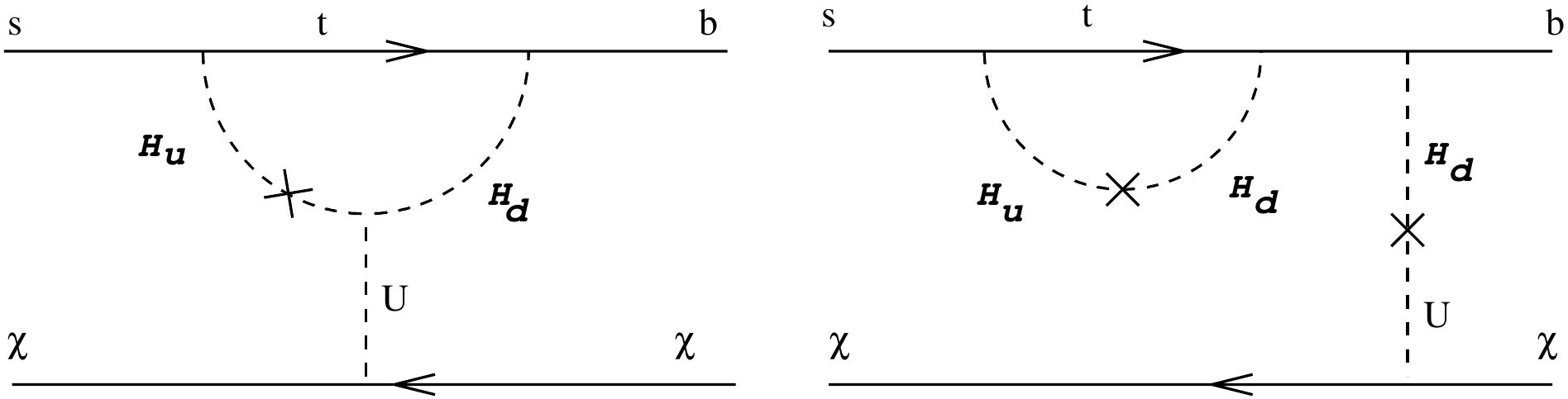,width=0.95\textwidth,angle=0}
\end{center}
\caption{\label{diagrams:Fermion2} The Feynman diagrams which contribute to 
$b \to s + E\!\!\!\!/$ in the 2HDM plus fermionic WIMP model for the special case of $\lambda_3$-dominant. }
\end{figure}

However for this special case, the diagrams which contribute to the decay $B \to K + \chi \chi$, which are given in Figure \ref{diagrams:Fermion2}, cancel at leading order in $\tan \beta$. As a result, 

\begin{displaymath}
BR(B \to K + \chi \chi) \ll BR(B \to K + \overline{\nu} \nu)
\end{displaymath}

\noindent
and therefore this model is not constrained by B-meson decays\index{B mesons}. 

\subsection{Model 4b: Higgs-Higgsino Model \label{Section::LightHiggsino}}\index{dark matter!higgsino}\index{higgsino}

\par
\begin{figure}
\begin{center}
\psfig{file=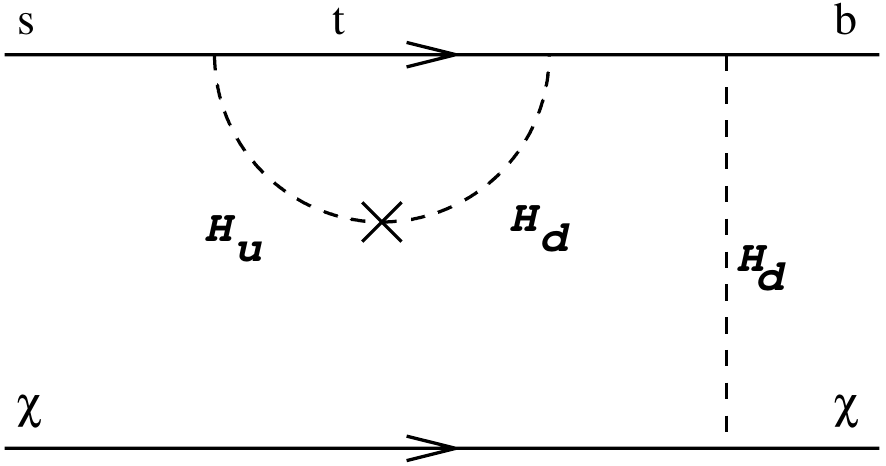,width=0.5\textwidth,angle=0}
\caption{\label{Figure::btos-higgsino} The Feynman diagram for the leading order decay of $b \to s + \chi \chi$ in the Higgs-Higgsino model.}
\end{center}
\end{figure}

In this model, the decay of the b-quark to dark matter is given by a single diagram, given in Figure \ref{Figure::btos-higgsino}. The effective Lagrangian for this process is as given in Eq \ref{Equation:BDML}, with the Wilson coefficient 

\begin{equation}
C_{DM} = \frac{V_{ts}^* V_{tb} \tan \beta}{32 \pi^2 v_{sm}^3} \left( \frac{\lambda_d \lambda_u v_u \mu}{\lambda_u^2 v_u^2 + \mu^2}  \right) \frac{a_t \ln a_t}{1 - a_t}
\end{equation}

\noindent
where, as before, $a_t \equiv m_t^2/M_H^2$. Using the parameter $\kappa$ from Eq \ref{Eq:higgsino-kappa}, and the phase space integral for fermions,

\begin{equation}
F(m_{\chi}) = \int_{\hat{s}_{min}}^{\hat{s}_{max}} f_0(\hat{s})^2 (\hat{s} -  2 m_{\chi}^2/M_B^2) I(\hat{s},m_{\chi}) d\hat{s} \left[ \int_{\hat{s}_{min}}^{\hat{s}_{max}} f_0(\hat{s})^2 \hat{s}  I(\hat{s},0) d\hat{s}  \right]^{-1}
\end{equation}

\noindent
the branching ratio for this decay is

\begin{equation}
BR_{B  \to K + \slash{E}} = 4.0 \times 10^{-6} + 9.8 \times 10^{-5} \kappa^2 \left( \frac{a_t \ln a_t}{1-a_t}\right)^2 F(m_1)
\end{equation}

\noindent
The first term represent the Standard Model decay to neutrinos\index{neutrinos}, and the second term represents the contribution from decays to light dark matter\index{dark matter!light}. 

\begin{figure}
\psfig{file=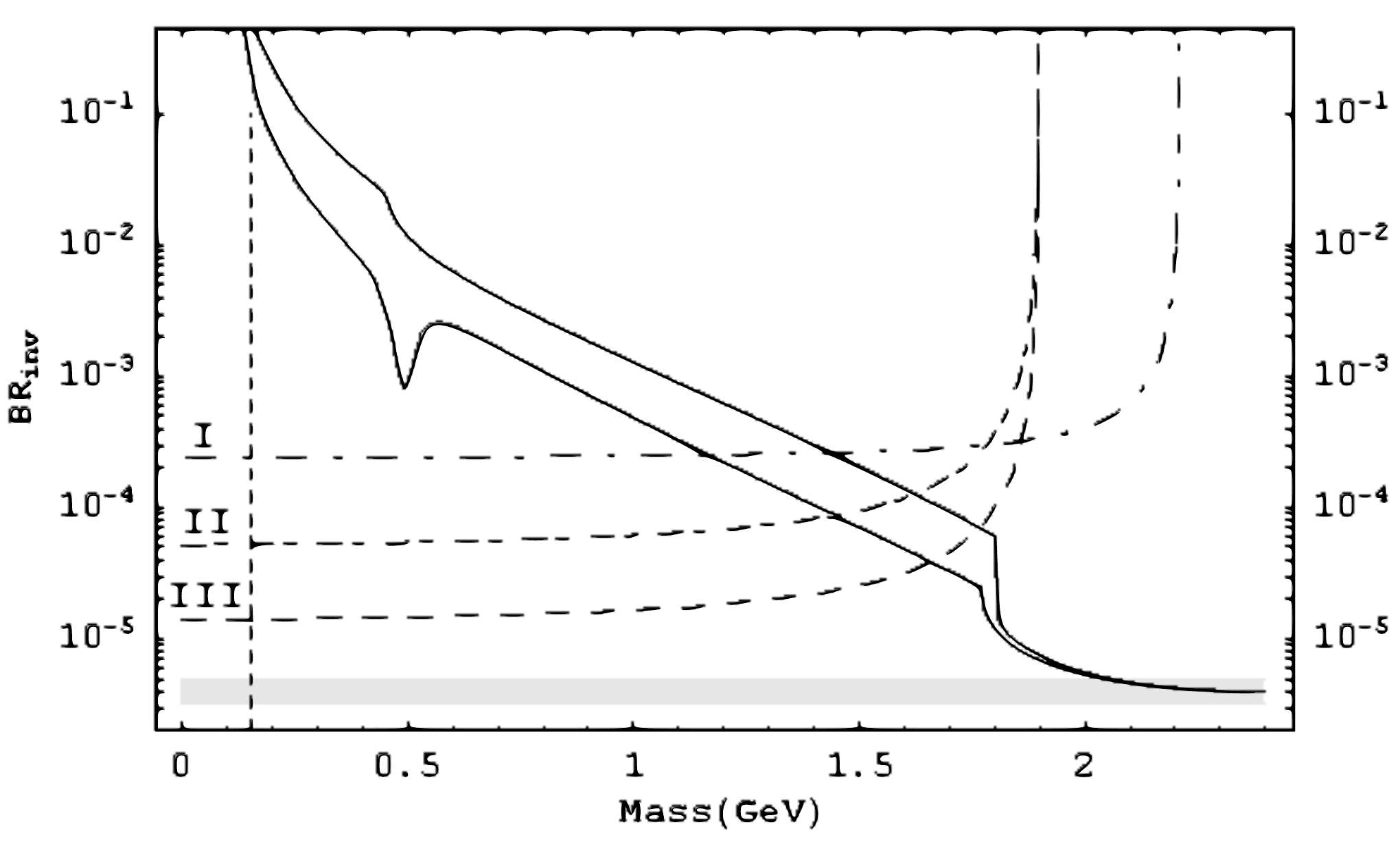,width=\textwidth,angle=0}
\caption{\label{Figure::B-higgsino} Branching ratio in the Higgs-Higgsino model}
\end{figure}

\par
The bounds on this model from B-decays\index{B mesons} are given in Figure \ref{Figure::B-higgsino}. Using the most recent data from BELLE\index{BELLE} \cite{Chen:2007zk}, neutralinos\index{neutralinos} lighter than $\sim 1.7 \;GeV$ can be excluded \footnote{As outline in Section \ref{Section::Higgsino}, this model does not include all of the complications of supersymmetry\index{supersymmetry}, and it may be possible to avoid these bounds in certain supersymmetric models.}. Furthermore, WIMPs\index{dark matter!WIMPs} heavier than $\sim 2 \; GeV$ produce a signal smaller than the uncertainty in the predictions of the Standard Model, and as such the effect of decays to WIMP pairs will not be detectable. 

\subsection{Model 5: Dark Matter \& Warped Extra Dimensions}

\par
As outlined in the previous section, the radion\index{Kaluza-Klein gravitons!radions} has similar properties to the Standard Model Higgs boson and as such existing experimental data can place a lower bound on its mass of $m_{\phi} \gtrsim 100 \; GeV$. From the abundance constraints in Section \ref{Section:RSDM}, this bound on the radion mass implies that the WIMP mass satisfies $m_{DM} \gtrsim 30 \; GeV$ which is too heavy to be produced in B-meson decays.\index{dark matter!light|)}

\section{Conclusion}

\par
The nature of dark matter\index{dark matter} has remained a mystery for several decades, in spite of increasing evidence for its existence. Measurements of the cosmic microwave background\index{cosmic microwave background} and other astrophysics experiments suggest an abundance of $\Omega_{DM} \sim 0.11$, which for WIMP\index{dark matter!WIMPs} dark matter corresponds to a total annihilation cross-section of $<\sigma_{ann} v> \approx 0.7 \; pb$, while dedicated dark matter searches have constrained the nuclear recoil cross-section. 
However most of the properties of dark matter are still unknown.

\par
In this chapter, I have presented seven models of dark matter in which only a minimal amount of new physics is introduced. For each model, I have calculated and presented the dark matter abundance and used this result to constrain the model. Each of these models was able to reproduce the observed abundance using natural values of the parameters, and perturbative couplings. 
\par
In Section \ref{Section:DDMS}, I calculated the WIMP-nucleon scattering cross-section for each model using the abundance constraints, and used this result and existing data from dedicated dark matter searches to constrain the parameter space.
I have also demonstrated how each model can be explored at collider experiments, such as the LHC\index{Large Hadron Collider} and the Tevatron\index{Tevatron}, with the primary signals being invisible Higgs decays through the channel $pp \to Z^0 + h_{inv}$\index{Higgstrahlung} and through weak-boson fusion\index{weak boson fusion}. These experiments should be able to detect WIMPs\index{dark matter!WIMPs} lighter than $m_{DM} \lesssim M_h/2$, while heavier WIMPs\index{dark matter!WIMPs} can avoid detection.

\par
In Section \ref{Section::LDM}, I outlined the motivations for light dark matter and the existing bounds. I then proved that light dark matter\index{dark matter!light} could be detected in the decay of B-mesons\index{B mesons}, as originally published by myself and collaborators in Refs \cite{Bird:2004ts} and \cite{Bird:2006jd}. It is expected that this result is generic, with most models of sub-GeV dark matter\index{dark matter!light} producing an observable signal in B-meson decays\index{B mesons}.
\par
Throughout this chapter I have shown how each of the minimal models, while defined in a similar manner, have very different properties. The Minimal Model of Dark Matter is capable of reproducing the observed dark matter abundance for the most of mass range considered, with the Lee-Weinberg limit\index{dark matter!Lee-Weinberg limit} only requiring $m_S \gtrsim 300 \;MeV$. Dedicated dark matter searches have only recently been able to probe this model, with XENON10\index{\ddme XENON10} data released in 2007 excluding a range of $10 \lesssim m_S \lesssim 30\;GeV$. However it is expected that the LHC\index{Large Hadron Collider} will be able to further probe this model for WIMP masses as large as half the Higgs mass, or $m_S \gtrsim 55\; GeV$ for $m_H \approx 120 \;GeV$. Furthermore, using
existing data from BELLE\index{BELLE} and \babar\ can already exclude WIMPs\index{dark matter!WIMPs} lighter than $\sim 1.6 - 1.8 \; GeV$, although in this model these light WIMPs\index{dark matter!WIMPs}\index{dark matter!light} require a significant fine-tuning of the parameters to provide $O(1 \; GeV)$ masses.

\par
In Model 2, the scalar WIMP\index{dark matter!scalar} was coupled to two different Higgs fields in the Type-II two Higgs-doublet model\index{Higgs!two higgs doublet}. For the special cases of this model, WIMPs\index{dark matter!WIMPs} could have masses below $O(100 \; MeV)$ without fine-tuning or violating the Lee-Weinberg limit\index{dark matter!Lee-Weinberg limit} due to $\tan \beta$ enhancements of the coupling constants. However these same models can be excluded by nuclear recoil experiments for $m_S \gtrsim 10\;GeV$, with the exception of a small range at $50\; GeV \lesssim m_S \lesssim 70 \;GeV$ due to the Higgs resonance. Since these special cases do not include strong couplings to the weak gauge bosons, they are not expected to be probed well by colliders, which search for Higgs bosons in the $Z^0 + h$\index{Higgstrahlung} channel or from weak boson fusion\index{weak boson fusion}. From B-meson experiments\index{B mesons}, the case of $\lambda_1$ dominant is excluded for $m_S \lesssim 1.4 - 1.6 \; GeV$, while the $\lambda_3$ dominant case cannot be probed in these experiments.

\par
The minimal model of fermionic dark matter\index{dark matter!fermionic} is similar to the MDM, but with fermion WIMPs\index{dark matter!WIMPs} replacing the scalars. While this model can produce the correct abundance of dark matter, the Lee-Weinberg\index{dark matter!Lee-Weinberg limit} bound is $m_{\chi} \sim 25 \;GeV$ and nuclear recoil experiments exclude $m_{\chi} \lesssim 50 \;GeV$ (when $m_H \sim 120 \;GeV$). The LHC\index{Large Hadron Collider} and Tevatron\index{Tevatron} are not expected to be able to probe WIMPs\index{dark matter!WIMPs} heavier than this bound.

\par
When the SM Higgs in the MFDM is replaced with two higgs doublets\index{Higgs!two higgs doublet}, the results can be quite different. For the special case of $\lambda_3$ dominant, this model can produce sub-GeV fermionic WIMPs\index{dark matter!WIMPs}\index{dark matter!fermionic} without violating the Lee-Weinberg limit\index{dark matter!Lee-Weinberg limit}, due to $\tan^2 \beta$ enhancements to the annihilation cross-section, while the case of $\lambda_1$ dominant can contain WIMPs\index{dark matter!WIMPs} as light\index{dark matter!light} as $m_{\chi} \sim 1.6 \;GeV$. Furthermore, these two cases are not presently constrained by data from nuclear recoil experiments. When the WIMPs\index{dark matter!WIMPs} are light, B-meson decays can exclude $m_{\chi} \lesssim 1.6 - 1.8 \;GeV $ in the case of $\lambda_1$ dominant, while the WIMPs\index{dark matter!WIMPs} in the case of $\lambda_3$ dominant cannot be probed by these experiments. The Higgsino\index{dark matter!higgsino}\index{higgsino} model has similar properties to this model with $\lambda_1$ dominant.

\par
The final model presented is dark matter in the presence of warped extra dimensions\index{extra dimensions!warped}. Although the WIMPs\index{dark matter!WIMPs} in this model have no non-gravitational interactions, the radion\index{Kaluza-Klein gravitons!radions} field generated by the extra dimensions allow them to annihilate efficiently. The result is that this model can provide the correct abundance, with a Lee-Weinberg limit\index{dark matter!Lee-Weinberg limit} of $m_S \gtrsim 35 \; GeV$ for scalars\index{dark matter!scalar} and $m_f \gtrsim 50 \; GeV$ for fermions\index{dark matter!fermionic}. Dedicated dark matter searches\index{\ddme} further restrict this to $m_S \gtrsim 60 \; GeV$ and $m_f \gtrsim 80 \; GeV$. Furthermore, it is expected that both the LHC\index{Large Hadron Collider} and Tevatron\index{Tevatron} will be able to probe this model further, with the precise limit of their sensitivity depending on the radion mass and vev. As with the minimal model of fermionic dark matter, this model cannot contain light WIMPs\index{dark matter!WIMPs}\index{dark matter!light} and therefore is not constrained by B-meson experiments.

\par
Although the models presented are minimal, their properties are similar to more complicated models. From the figures on dedicated dark matter searches and collider searches, it is clear that the next generation of experiments will be able to probe most of the parameter space of these models and possibly detect dark matter\index{dark matter|)}.

%% file: chapter1f.tex
\index{dark matter!scalar|)}

\subsection{Model 3: Minimal Model of Fermionic Dark Matter \label{Section::MFDM}}\index{dark matter!fermionic|(}\index{dark matter!Minimal Fermionic Model|(}

\par
The models discussed previously have used scalar dark matter. However there are no observed scalars in nature, and many candidates for dark matter are fermionic. For this reason, minimal models containing fermion WIMPs\index{dark matter!fermionic} also need to be considered.
\par
As with the minimal model of scalar dark matter presented in Section \ref{Section::MDM}, it is possible to construct a minimal model of fermionic dark matter \cite{Bird:2006jd}. However in this case a new scalar must be introduced as well to mediate the interaction between the WIMP\index{dark matter!WIMPs} and the Higgs\index{Higgs} \footnote{Although it is possible to construct a minimal model without this additional scalar field, the resulting model in non-renormalizable.}. The Lagrangian for this model is

\begin{equation} \label{Lag0}
\begin{split}
L = & \frac{1}{2} \partial_{\mu} \Phi \partial^{\mu} \Phi - \frac{M_2^2}{2} \Phi^2 + \frac{1}{2} \chi i \cancel{\partial} \chi - \frac{M_3}{2} \chi^2 - \frac{\lambda_1}{2} \Phi \chi^2 \\& - \lambda_2 \Phi H^{\dag} H - \frac{\lambda_3}{2} \Phi^2 H^{\dag} H - \frac{\lambda_4}{6} \Phi^3 - \frac{\lambda_5}{24} \Phi^4 
\end{split}
\end{equation}

\par
The first constraint imposed on this model is the requirement that it have a stable vacuum state. If it does not contain a stable vacuum, it cannot be a realistic model. The potential for this model in the unitary gauge, $\sqrt{2} H^{\dag} = (h,0)$, is

\begin{equation}
\begin{split}
V = & \frac{M_2^2}{2} \Phi^2 + \frac{M_3}{2} \chi^2 + \frac{\lambda_1}{2} \Phi \chi^2  + \frac{\lambda_2}{2} \Phi h^2 \\& + \frac{\lambda_3}{4} \Phi^2 h^2 + \frac{\lambda_4}{6} \Phi^3 + \frac{\lambda_5}{24} \Phi^4 + \frac{\lambda_h}{4}(h^2-v_0^2)^2
\end{split}
\end{equation}

\noindent
where the final term is the usual potential for the Higgs boson\index{Higgs!potential}, but with $v_0$ an arbitrary parameter instead of $v_{ew}$. This potential is bounded from below if  

\begin{equation}
\lambda_5,\lambda_h  > 0 \quad \quad \quad \lambda_h \lambda_5 > 6 \lambda_3^2
\end{equation}

\noindent
or if 

\begin{equation}
\lambda_5 = \lambda_4 = 0 \quad \quad  \lambda_h > 0 \quad \quad \lambda_3 + 4 M_2^2 > 0
\end{equation}

\noindent
The minimum 
of this potential is

\begin{equation}
<\chi> = 0 \quad <h> = v_{ew} = \pm v_{0} \sqrt{1 - \frac{  2 \lambda_2 w + \lambda_3 w^2 }{\lambda_h}} \quad <\Phi> \equiv w 
\end{equation}

\noindent
where w is the solution of the cubic equation

\begin{equation}
\lambda_2 <h>^2 + (M_2^2 + \lambda_3 <h>^2) w + \frac{\lambda_4}{2} w^2 + \frac{\lambda_5}{6} w^3 = 0
\end{equation}

\noindent
and where $<h> = v_{ew}$, as is required in the Standard Model.



\par
Because of the mixing terms, h and $\phi$ do not represent physical fields. Instead the physical particles are linear combinations of the two states, which we will denote by $\sca = h \cos \theta + \phi \sin \theta$ and $\scb = \phi \cos \theta - h \sin \theta $, and the Lagrangian is of the form

\begin{equation} \label{Lag}
\begin{split}
L = & \frac{1}{2} \partial_{\mu} \sca \partial^{\mu} \sca -\frac{m_1^2}{2} \sca^2 + \frac{1}{2} \partial_{\mu} \scb \partial^{\mu} \scb 
- \frac{m_{2}^2}{2} \scb^2 + \frac{1}{2} \chi i \cancel{\partial} \chi - \frac{m_\chi}{2} \chi^2 
\\& - \frac{\eta_1 \sin \theta}{2} \sca \chi^2 
- \frac{\eta_1 \cos \theta}{2} \scb \chi^2 - \frac{\eta_3}{2} \sca^2 \scb - \frac{\eta_4}{2} \scb^2 \sca - \frac{\eta_5}{4} \sca^2 \scb^2 \\& - \frac{\eta_6}{6} \sca^3 - \frac{\eta_7}{6} \scb^3 - \frac{\eta_8}{24} \sca^4- \frac{\eta_9}{24} \sca^4
\end{split}
\end{equation}

\noindent
The couplings to the Standard Model are taken to be the usual Higgs couplings, with $h = \sca \cos \theta - \scb \sin \theta$. 
\par
For the remainder of this section, it will be assumed that $m_2 \gg m_1,m_{\chi}$. As a result, the last three terms in the Lagrangian will not contribute to the annihilation or scattering cross-sections at tree level, and can be omitted. This requirement, although not required for the model, ensures that only the fermion contributes significantly to the dark matter abundance, and that existing experimental bounds on new forces below the electroweak scale are not violated. \nocite{Pospelov:2007mp}

%% file: chapter2.tex
\chapter{Charged Relics \label{Chapter:ChargedRelics}}\index{charged relics|(}\index{CHAMPs|(}

\par
In addition to dark matter, it is also possible that the early Universe contained a small abundance of exotic charged massive particles (referred to as CHAMPs). The presence of these relics in the early universe can have many effects on the standard cosmology and on astrophysical processes, depending on their properties.

\par
Heavy charged particles are predicted to exist in several modern theories, with the most commonly cited examples being in supersymmetry and in models with extra dimensions. In supersymmetry\index{supersymmetry}, every known charged particles would have a supersymmetric partner, which would also be charged, and which would be present in the early Universe. These models often also include candidates for dark matter, in the form of the lightest stable particle, with the next-to-lightest (quasi)stable particles being charged. For example, both the Constrained Minimal Supersymmetric Standard Model (CMSSM) and Minimal Supergravity (mSUGRA)\index{supergravity!minimal} theories, contain gravitino dark matter\index{dark matter!gravitino} accompanied by a stau NLSP which is long lived, massive, and charged. 


\par
 The properties of CHAMPs\index{CHAMPs}\index{charged relics} are already constrained by experiments at LEP \cite{LEP-CHAMP} and the Tevatron\index{Tevatron} \cite{Acosta:2002ju}, which are currently able to probe up to $m_{X} \sim O(100 \; GeV)$. However these constraints tend to be model dependent, and the existing experiments cannot effectively probe higher masses. 
\par
In addition to accelerator based searches for charged particles, CHAMPs which survive to the present age of the universe could be detected using heavy water experiments\index{heavy water experiments} \cite{Dimopoulos:1989hk, Kudo:2001ie,Byrne:2002ri} as well as in cosmic ray\index{cosmic rays} and $\gamma$-ray detectors \cite{Dimopoulos:1989hk} and neutrino detectors \cite{Ando:2007ds}. A detailed review of the bounds can be found in Ref \cite{Perl:2001xi}. However these searches are insensitive to relics which are long-lived, with lifetimes in the range of $O(100 s - 1000 s)$ or longer, but which are not stable. Since such particles decay in the early Universe, new methods are required to search for the existence.

\par
The focus of this chapter is on the recently proposed model of catalyzed big bang nucleosynthesis\index{Big Bang Nucleosynthesis!catalyzed}\index{CBBN}, in which the primordial abundances of light nuclei are altered by the formation of bound states with charged massive particles. In such models, metastable CHAMPs can be detected through their effects on the early universe without violating the existing bounds on stable charged massive particles. By comparing the predictions of CBBN\index{Big Bang Nucleosynthesis!catalyzed} to observed abundances of primordial elements, the properties of CHAMPs can be better constrained.

\section{Catalyzed Big Bang Nucleosynthesis}\index{Big Bang Nucleosynthesis!catalyzed|(}\index{CBBN|(}\index{Big Bang Nucleosynthesis|(}

\par
Big Bang Nucleosynthesis (BBN) is the process by which the light nuclei in the Universe are formed, and is also a powerful tool for constraining new physics (for a review, see eg. Ref \cite{Sarkar:1995dd}) . The theory relies on particle physics, nuclear physics, and general relativity to predict the abundances, and the results depend on a single parameter, the baryon-to-photon ratio\index{baryon-to-photon ratio}, which can be measured from the cosmic microwave background \cite{Spergel:2003cb}. The predictions have proven to be accurate when compared to the observed abundances in older, metal-poor stars. Therefore the effects of new physics on nucleosynthesis also provide strong constraints on proposed models.

\par
If there exist charged particles which survive to the era of nucleosynthesis, then they can affect the standard processes in several ways. For the purposes of this section, there are two classes of model in which the charged particle can have a naturally long lifetime. One possibility is that all relevant coupling constants in the theory are small, which results in a low decay rate, and which we refer to as Type I. The other possibility is that the decay releases very little energy while having natural coupling constants. We refer to these models as Type II. 

\par
It is already well known that the presence of massive relics can affect Big Bang nucleosynthesis\index{Big Bang Nucleosynthesis}\index{nucleosynthesis}. The most common mechanism is through the decay of a massive relic during the epoch of BBN, with the decays producing more energy and entropy, which in turn alter the predictions of the primordial abundances of light elements \cite{Cyburt:2002uv,Jedamzik:2006xz,Jedamzik:2005dh,Jedamzik:2004er,Jedamzik:2005dh}. In addition the decay of the relic can produce energetic photons, which will photo-dissociate the light nuclei \cite{Ellis:2005ii,Kusakabe:2006hc}, and additional hadrons which can convert protons to neutrons leading to a higher neutron-to-proton ration and an overabundance of \hef\  \cite{Reno:1987qw,Kawasaki:2004qu,Jedamzik:2006xz,Jedamzik:2006xz} as well as non-thermally producing other light nuclei \cite{Dimopoulos:1987fz,Dimopoulos:1988ue,Kawasaki:2004qu,Jedamzik:2006xz,Jedamzik:2006xz}.  

\par
However it is also known that any charged particles present during nucleosynthesis will form bound states with light nuclei. These heavy bound states result in a reduced Coulomb repulsion of other nuclei, as well as changing the rates of the nuclear reactions. This results in a shift in the relative abundances of light nuclei \cite{Pospelov:2006sc,Kohri:2006cn,Kaplinghat:2006qr,Pospelov:2007js}.
 
\par
In this Chapter, two results of catalyzed BBN\index{Big Bang Nucleosynthesis!catalyzed} will be studied. In Section \ref{Section:L6} it will be demonstrated that the presence of a charged massive relic can increase the production rate of \lisx , and by comparison with observations of the primordial \lisx\ abundance this result is used to constrain the properties of CHAMPs. In Section \ref{Section:L7}, I demonstrate how the same charged relics can suppress the \lisv\ abundance, and using observations of the primordial \lisv\ abundance I derive limits on the lifetime and abundance of the charged relics\index{CHAMPs}\index{charged relics}. This work was originally published by the author and collaborators in Ref \cite{Bird:2007ge}.
 
\subsection{Bound States of Nuclei} 
 
\par 
The predictions of CBBN\index{Big Bang Nucleosynthesis!catalyzed}\index{CBBN} rely on the properties of bound states of nuclei with heavy charged particles. However the exact properties of these states are not well known, and depend on the internal structure of each nuclei. 
\par
For a first approximation, the bound state can be treated as a hydrogen-like state, with the $X^-$ treated as a massive point particle and the light nucleus in a bound state with the naive Bohr energies. However it is common for the Bohr radius in these bound states to be comparable to or even smaller than the nuclear radius, leading to inconsistencies in the calculations. Furthermore, the nuclei have more complicated charged distributions than a point-particle, and as such the binding energies and wavefunction are also significantly altered. Therefore the Bohr approximation fails. 
\par
Assuming that the internal structure of the nuclei is unaffected, ground states energies for these bound states can be calculated using experimental data for the nuclear radii and variational methods in quantum mechanics to determine the ground state wavefunctions and energies. In each case the charge distribution is assumed to be:

\begin{equation}
\rho_c(r) = \rho_0 e^{-(r/r_0)^2} \quad \quad \rho_0 = Z \alpha / (\sqrt{\pi} r_0 )^{1/3}
\end{equation} 

\noindent
as this distribution has been previously found to be a good approximation to the actual charge distribution of the light nuclei \cite{Karol:1975zz,nucCharge}. The value of $r_0$ is chosen such that the rms of this distribution matches the experimental rms radius, or $r_0 = \sqrt{2/3} R_N$.
\par
For the calculations that follow, each wavefunction was calculated using both a trial wavefunction which is tuned to minimize the ground state energy and a numerical solution of the Schrodinger equation, although these two methods have been found to give identical results.
\par
The energies for the common light elements involved in CBBN\index{Big Bang Nucleosynthesis!catalyzed}\footnote{It should be noted here that the properties of $(^4HeX^{--})$ are included in this table, but as yet have not been included in detailed calculations of CBBN\index{Big Bang Nucleosynthesis!catalyzed}} are given in Table \ref{Table:BoundProp}. In this table,$|E_b^0|$ and  $a_0$ are the Bohr energy and radius for a point-like nucleus,  $R_N^{sc}$ and $|E_b(R_N^{sc})|$ are the nuclear radius and bound state binding energy for assuming a uniform charge distribution and $R_N = 1.22 A^{1/3}$, and  $R_{Nc}$ and $|E_b(R_{Nc})|$ represent the nuclear radius and bound state binding energy using experimental determinations of the nuclear radius, $R_N = (5/3)^{1/3} R_c$ where $R_c$ is the measured charge radius. $T_0$ represents the ionization energy of the bound state. 

\begin{table}
\begin{center}
\begin{tabular}{|c|c|c|c|c|c|c|c|}
\hline
bound st. &     $|E_b^0|$&  $a_0$ & $R_N^{sc}$ & $|E_b(R_N^{sc})|$ & $R_{Nc}$  & $|E_b(R_{Nc})|$ &~$T_0 (keV)$~ \\ \hline\hline
\hef$X^-$&          397  &   3.63 &    1.94    &       352         &    2.16   &         346     &  8.2 \\ \hline
\lisx$X^-$&        1343  &   1.61 &    2.22    &       930         &    3.29   &         780     &  19  \\ \hline
\lisv$X^-$&        1566  &   1.38 &    2.33    &       990         &    3.09   &         870     &  21  \\ \hline
\bes$X^-$&         2787  &   1.03 &    2.33    &       1540        &      3    &         1350    &  32  \\ \hline
\beet$X^-$&        3178  &   0.91 &    2.44    &       1600        &      3    &         1430    &  34  \\ \hline
\hef $X^{--} $ &   1589  &   1.81 &    1.94    &       1200        &      2.16 &         1150    &  28  \\ \hline\hline
D$X^-$&              50  &    14  &      -     &       49          &      2.13 &           49    & 1.2  \\ \hline
p$X^-$&              25  &    29  &      -     &       25          &      0.85 &           25    & 0.6  \\ \hline
\end{tabular}
\end{center}
\caption{\label{Table:BoundProp}Properties of bound states for charged relics and light nuclei. In this table, $|E_b^0|$,$|E_b(R_N^{sc})|$, and $|E_b(R_{Nc})|$ are given in units of keV, while $a_0$,$R_N^{sc}$, and $R_{Nc}$ are given in fm.\index{deuterium}\index{helium!$^4He$}\index{helium!$^3He$} }
\end{table}

\par
When nucleosynthesis starts, the temperature of the Universe is too high to form bound states. Although some states will form, ambient high energy photons will immediately break-up the bound states. As the Universe cools, nucleosynthesis proceeds as in the standard model of BBN until the temperature drops below the binding energy for each state. At that time, the charged relics will begin to capture light nuclei which have already been formed. The abundance of bound states can be calculated in the usual manner using the Boltzmann equation,

\begin{equation}
- H T \frac{dY_{(NX^-)}}{dT} = <\sigma_{rec} v> n_{N} Y_{X} - <\sigma_{ph} v> n_{\gamma} Y_{(NX^-)}
\end{equation}

\noindent
where $Y_{(NX^-)},Y_{X},n_{N}$ are the abundances of bound states, free relics, and free nuclei respectively, and the cross-sections represent recombination and photodisintegration respectively. In previous papers \cite{Hamaguchi:2007mp,Kawasaki:2007xb,Jittoh:2007fr} it has been common to use a Saha equation, 

\begin{equation}
Y_{(NX^-)}(T) = \frac{Y_X e^{-T_{\tau}^2/T^2}}{1 + n_{He}^{-1} (m_{\alpha} T)^{3/2} e^{-E_b/T} }
\end{equation}

\noindent
with $T_{\tau} = T/\sqrt{2 \tau H(T)}$ and which assumes a thermal distribution of bound and unbound relics, as an approximation to the solution of the Boltzmann equation. However this assumes that the recombination and photodisintegration reactions are in equilibrium and that there exists a thermal distribution of bound and unbound systems throughout the epoch of nucleosynthesis. As demonstrated in Ref \cite{kkmsc,Kawasaki:2008qe}, this approximation is not valid and can result in order of magnitude errors in the final abundances of the light nuclei. 

\par
Throughout this section, bound states of charged relics are denoted by $(NX^-)$, where N can be any nuclei. The S-factors are defined in the usual way, as $S(E) = E \sigma/G$ where $\sigma$ is the reaction cross section and G is the Gamow factor, which measures the probability that two particles can overcome Coulomb repulsion during a fusion reaction.

\subsection{Overproduction of \lisx \label{Section:L6}}

\par
The effects of catalyzed Big Bang nucleosynthesis are most noticeable in the rarer elements such as Lithium-6 and Lithium-7. The theory of CBBN\index{Big Bang Nucleosynthesis!catalyzed} was introduced in Ref \cite{Pospelov:2006sc}, where it was demonstrated that the presence of any significant abundance of charged relics with $\tau \gtrsim O(10^4 \; {\rm s})$ would produce additional \lisx\, in excess of observations. As a result, experimental constraints on the primordial abundance of \lisx\ can be used to constrain theories which include metastable charged relics.
\par
It should also be noted that the observed abundance of \lisx\ is known to be several orders of magnitude larger than predicted in SBBN. This excess is believed to be produced by more modern sources, such as in $\alpha-\alpha$ or $p-\alpha$ fusion by cosmic rays \cite{Prantzos:2005mh,Rollinde:2006zx}. However the excess is still present in metal-poor regions in which cosmic ray reactions should not have produced such large abundances of \lisx, and furthermore there appears to exist a \lisx\ plateau analogous to the Spite plateau of \lisv\ in which the \lisx\ abundance appears to be constant over large regions of space and is independent of metallicity. This suggests that the excess of \lisx\ is produced in the early Universe (for a review of the \lisx\ problem, see Ref \cite{Asplund:2005yt,Cyburt:2006uv}). 

\par
In the Standard BBN scenario, \lisx\ is produced in small quantities by the reaction,

\begin{equation} \label{Eq:SBBN-LI6}
^4He + D \to ^6Li + \gamma
\end{equation}

\noindent
The cross-section for this reaction is suppressed due to the presence the photon in the final state. As outlined in Ref \cite{Pospelov:2006sc}, the \lisx\ nucleus can be modeled as a $^4He-D$ bound state, and the E1 transition would normally be expected to dominate this reaction. However the two terms in the amplitude for this transition cancel due to the almost equal charge-to-mass ratio of the $^4He$\index{helium!$^4He$} and D nuclei\index{deuterium}. As a result, the reaction in Eq \ref{Eq:SBBN-LI6} proceeds by an E2 transition in which the quadrupole moment of the $^4He-D$  system couples to the photon. The result is that this cross-section is proportional to the inverse fifth power of the photon wavelength, leading to a strong suppression relative to other BBN reactions.

\par
If there exist charged particles\index{CHAMPs}\index{charged relics} during BBN then they can form bound states with $^4He$\index{helium!$^4He$}, which has a relatively large primordial abundance. These bound states can then produce additional \lisx\, primarily through the reaction 

\begin{equation}
(^4HeX^-)+ D \to ^6Li + X^- 
\end{equation}

\begin{figure}
\begin{center}
\psfig{file=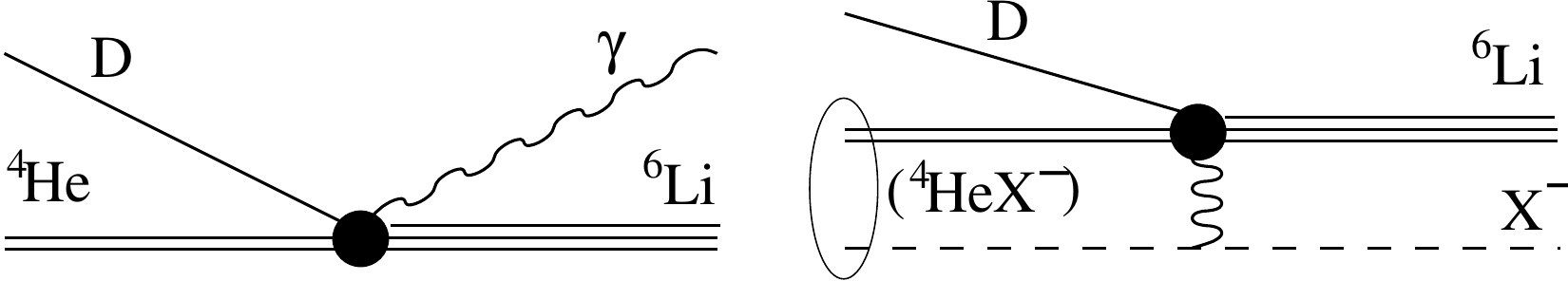,width=\textwidth,angle=0}
\caption{\label{Figure:Li6Production} Diagrams contributing to the production of \lisx\ , in the standard BBN (left) and in CBBN\index{Big Bang Nucleosynthesis!catalyzed} (right).}
\end{center}
\end{figure}

\noindent
The cross section for this reaction does not have the same suppression as the reaction Eq \ref{Eq:SBBN-LI6}. For this reaction, the photon emitted by the \lisx\ nucleus is virtual and can a much smaller wavelength. This results in a cross-section that is approximately eight orders of magnitude larger than the primary channel in the Standard BBN\index{Big Bang Nucleosynthesis}\index{nucleosynthesis}. The diagrams for the BBN and CBBN\index{Big Bang Nucleosynthesis!catalyzed} reactions are given in Figure \ref{Figure:Li6Production}, and the relationship between the S-factors for the two processes can be approximated as

\begin{equation}
S_{CBBN} = S_{BBN} \times \frac{8}{3 \pi^2} \frac{p_f a_0}{(\omega a_0)^5} \left( 1 + \frac{m_D}{m_{^4He}}\right)^2
\end{equation}

\noindent
where $a_0$ is the Bohr radius of the $(^4HeX^-)$ system, $p_f = \sqrt{2 m_{^6Li} (Q_{CBBN} + E}$ is the momentum of the \lisx\ nucleus in the CBBN\index{Big Bang Nucleosynthesis!catalyzed} reaction (with $Q_{CBBN} = 1.13 \; MeV$) and $\omega = Q_{BBN} + E$ is the energy of the photon in the SBBN\index{Big Bang Nucleosynthesis}\index{nucleosynthesis} reaction 
(with $Q_{BBN} = 1.47 \; MeV$). A more detailed calculation of the S-factor has been completed using a ground state wavefunction for $(^4HeX^-)$ obtained by variational methods rather than the Bohr approximation, and Coulomb wavefunctions for the final state rather 
than plane waves, however the results are similar to this approximation. Using the results of Ref \cite{Kiener:1991zz} interpolated to the relevant energies, the S-factor is found to be $S_{CBBN}\simeq 0.3 \; MeV \; bn$.

\par
The presence of a charged relic in the bound state with $^4He$ also affects the Gamow factor in the reaction. Since the relic carries a negative charge, it screens the Coulomb repulsion between the $^4He$ and D nuclei and reduces the Gamow suppression. Because the relic is also much more massive than the nuclei, the change in the reduced mass of the system will also affect the Gamow factor. The Gamow energy for the system is changed as

\begin{equation}
E_{SBBN}^{Gamov} = 5249 \; keV \to E_{CBBN}^{Gamov} = 1973 \; keV
\end{equation}

\noindent
Combining the new S-factor and the reduced Gamov energy gives the \lisx\ production cross-section as 

\begin{equation}
<\sigma_{CBBN} v> \simeq 1.8 \times 10^9 \times T_9^{-2/3} exp(-5.37 T_9^{-1/3})
\end{equation}

\noindent
where the units are those common to BBN calculations, $N_A^{-1} cm^3 s^{-1} g^{-1}$, with $T_9$ the temperature in $10^9 K$. It should be noted that this assumes that all nuclear distances are smaller than the Bohr radius for the bound state. In Ref \cite{Hamaguchi:2007mp}  a solution of the three-body Schrodinger equation was utilized to improve on the cross-section, with the new results being

\begin{equation}
<\sigma_{CBBN} v > \simeq 2.4 \times 10^8 (1 - 0.34 T_9) T_9^{-2/3} e^{- 5.33 T_9^{-1/3}}
\end{equation}

\noindent
 For the purpose of comparison, the cross-section for the reaction in Eq \ref{Eq:SBBN-LI6} is

\begin{equation}
<\sigma_{SBBN} v> \simeq 30 \times T_9^{-2/3} exp(-7.435 T_9^{-1/3})
\end{equation}

\noindent
From these equations, it is apparent that the presence of a charged relic can increase the rate of \lisx\ production by several orders of magnitude.
 
\par
The predicted abundance of \lisx\ in this model will also depend on the initial abundance of $(^4HeX^-)$ bound states. For the purpose of this calculation, it is assumed that the relic abundance is small, and that by the time the \lisx\ producing reactions occurs all $X^-$ particles are either in bound states with $^4He$ nuclei or are unbound \footnote{A full treatment of CBBN\index{Big Bang Nucleosynthesis!catalyzed} would require a fraction of the relics to be bound to other nuclei, however these fractions are quite small and are not expected to significantly alter these results.}. 


\noindent
As discussed in the previous section, the abundance of bound states must be calculated using the full Boltzmann

\begin{equation}
- H T \frac{dY_{(NX^-)}}{dT} = <\sigma_{rec} v> n_{N} Y_{X} - <\sigma_{ph} v> n_{\gamma} Y_{(NX^-)}
\end{equation}

\noindent
where $n_N$ and $Y_X$ represent the abundance of unbound nuclei and relics, and $Y_{(NX^-)}$ is the abundance of bound states.

\par
Using the CBBN\index{Big Bang Nucleosynthesis!catalyzed} cross-section and the abundance of bound states, the evolution equation for the abundance of \lisx\ is 

\begin{equation}
- H T \frac{d^6Li}{dT} = D (n_{BS} <\sigma_{CBBN}v> + n_{He} <\sigma_{SBBN}v>)  - ^6Li \; n_p< \sigma_p v>
\end{equation}

\noindent
where \lisx\ and D\index{deuterium} are used to denote the (hydrogen normalized) abundance of each element, and $\sigma_p$ is the destruction cross-section for the reaction

\begin{equation}
^6Li + p \to ^3He + ^4He
\end{equation}

\begin{figure}
\begin{center}
\psfig{file=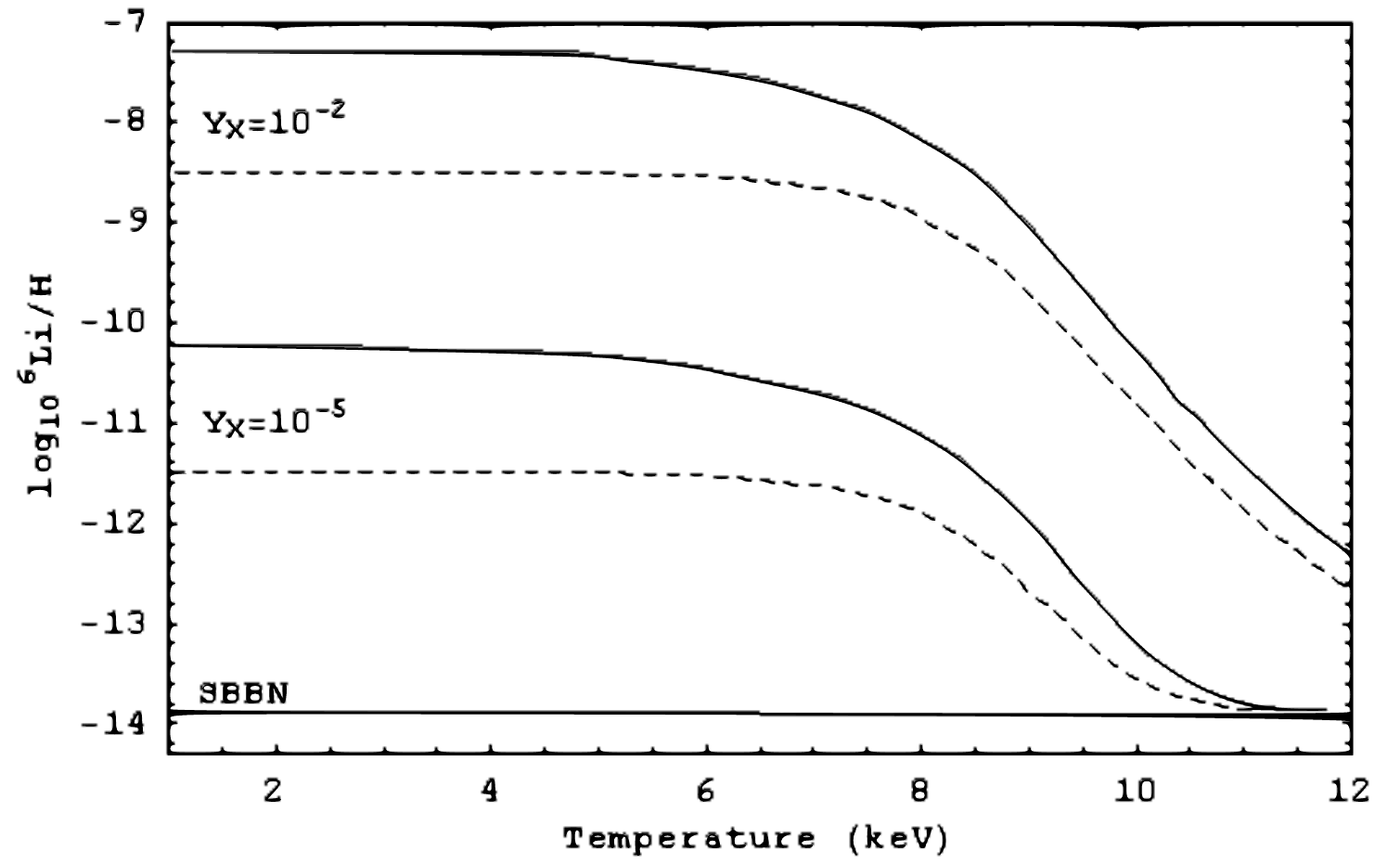,width=\textwidth,angle=0}
\end{center}
\caption{\label{Figure:Li6Ab} Abundance of \lisx\ as a function of temperature (in keV) in the CBBN\index{Big Bang Nucleosynthesis!catalyzed} model. The solid lines correspond to stable relics, while the dashed lines correspond to  $\tau = 10^4 \;s$. The \lisx\ abundance is given for two abundances of relics,  $Y_X = 10^{-2}$ and $Y_X = 10^{-5}$. The predicted abundance from SBBN is given at the bottom of the figure.}
\end{figure}

\noindent
The solution of this equation is given in Figure \ref{Figure:Li6Ab} for stable and for long-lived ($\tau = 10^4 s$) relics, with abundances of  $Y_X = 10^{-2}$ and $Y_X = 10^{-5}$. As an example of the strength of this method of constraining the properties of CHAMPs, it can be noted that for $\tau \sim 10^5 \;s$ and a \lisx\ abundance of $Y_{^6Li} \approx 2 \times 10^{-11}$ \cite{Cyburt:2002uv}, the abundance of $X^-$ must satisfy

\begin{equation}
Y_{X^-} \lesssim 8 \times 10^{-8}  \quad {\rm or}\quad \frac{n_{X^-}}{s} \lesssim 10^{-17}
\end{equation}

\noindent
This method of constraining CHAMPs\index{CHAMPs}\index{charged relics} with nucleosynthesis\index{Big Bang Nucleosynthesis}\index{nucleosynthesis} is also independent of the nature of the decay of the relics, whereas other constraints are based primarily on the effects of those decays \cite{Cyburt:2002uv,Jedamzik:2006xz,Jedamzik:2005dh}. 
\par
As outlined in Ref \cite{Pospelov:2006sc}, a scan over the entire parameter space of relic lifetimes and abundances yields a constraint (for reasonable choices of the relic annihilation cross-section) of $\tau \lesssim 4 \times 10^3 s$ using only the \lisx\ abundance. It is possible that including other reactions involving rare nuclei in the early Universe will strengthen this bound.

\subsection{Suppression of \lisv\ \label{Section:L7}}

\par
In addition to the overproduction of \lisx\, in Ref \cite{Bird:2007ge} my collaborators and I demonstrated that there is a region of parameter space which satisfies the constraints from \lisx\ production while also reducing the \lisv\ abundance. The calculations and primary results of the \lisv\ suppression by catalyzed BBN\index{Big Bang Nucleosynthesis!catalyzed} processes are repeated and reviewed in this section.

\par
The abundance of \lisv\ is one of the few problems with the standard BBN. Using the value for the baryon-to-photon ratio derived from the CMB anisotropy, which leads to consistent predictions for the other abundances of light elements, the relative abundance of \lisv\ is predicted to be $Y_{^7Li} = 5.24^{+0.71}_{-0.67} \times 10^{-10}$ \nocite{Coc:2003ce} \cite{Cyburt:2008kw}. 
However observations of low metallicity stars gives a value of $Y_{^7Li} = 1.23^{+0.68}_{-0.32} \times 10^{-10}$ \cite{Asplund:2005yt,Cyburt:2003fe} while observations of globular clusters have given values of $Y_{^7Li} = 2.19^{+0.30}_{-0.26} \times 10^{-10}$ \cite{BoniLi} and $Y_{^7Li} = 2.34^{+0.35}_{-0.30} \times 10^{-10}$ \cite{Melendez:2004ni}, which is a factor of a few smaller than predicted. 

\par
There have been several proposed solutions for the \lisv\ problem. If the relevant nuclear reactions had been measured incorrectly, or if there were different rates in the early Universe, the predictions would be altered. However the possible corrections do not produce a significant change in the \lisv\ abundance   \cite{Coc:2003ce,Angulo:2005mi}. There is a possibility that there is an unknown mechanism in which stars burn \lisv\ faster than current models predict \cite{Asplund:2005yt,Korn:2006tv,Tatischeff:2006tw}, as well as changes to the predictions from new developments in understanding how heavier elements such as \lisv\ are diffused and destroyed in stars \cite{Korn:2006tv,Tatischeff:2006tw}, both of which could affect the inferred primordial abundances. 
\par
It is also possible that an unknown physical effect, or a new particle, in the early Universe could affect the BBN predictions. In Ref \cite{Jedamzik:2004er,Jedamzik:2005dh} it was demonstrated that a decaying relic with  $\tau \sim 1000-3000 {\rm s}$ could release enough energy to lower the \lisv\ abundances. As will be demonstrated in the section, yet another possibility is that charged metastable particles can catalyze lithium burning reactions.

\vspace{16pt}
\subsubsection{Recombination of \bes\ and $X^-$}

\par
In the standard BBN scenario, \lisv\ is produced through the destruction of \bes\ by neutrons, through the reaction

\begin{equation}
^7Be + n \to ^7Li + p
\end{equation}

\noindent
and after the epoch of nucleosynthesis through electron capture. From these two processes, all of the primordial \bes\ is converted to \lisv\ (or other lighter nuclei) by the present time. Therefore depletion of the \lisv\ abundance can be achieved by first  depleting the \bes\ abundance. In the CBBN\index{Big Bang Nucleosynthesis!catalyzed} scenario, similar reactions
  using the bound state \bex can destroy \bes\ which further depletes the \lisv\ abundance relative to the SBBN predictions. It should also be noted that in the SBBN, the \bes\ abundance is frozen out at about $T \sim 40 \; keV$, while the formation of bound states occurs at $T \sim 33\; keV$. Therefore the \bes\ abundance 
  will be constant throughout the timespan in which CBBN\index{Big Bang Nucleosynthesis!catalyzed} becomes important, with the only changes arising from CBBN\index{Big Bang Nucleosynthesis!catalyzed} processes.
\par
Therefore the most important aspect of the \lisv\ suppression via CBBN\index{Big Bang Nucleosynthesis!catalyzed} is the recombination of \bes\ with the charged relic, $X^-$. If the capture rate is too low, then the density of bound states will be too low for the catalyzed reactions to affect the primordial \lisv\ abundance. As was demonstrated in Ref \cite{Bird:2007ge}, the formation of these bound states is the critical stage in the series of CBBN\index{Big Bang Nucleosynthesis!catalyzed} reactions which determines the primordial \lisv\ abundance.

\begin{figure}
\psfig{file=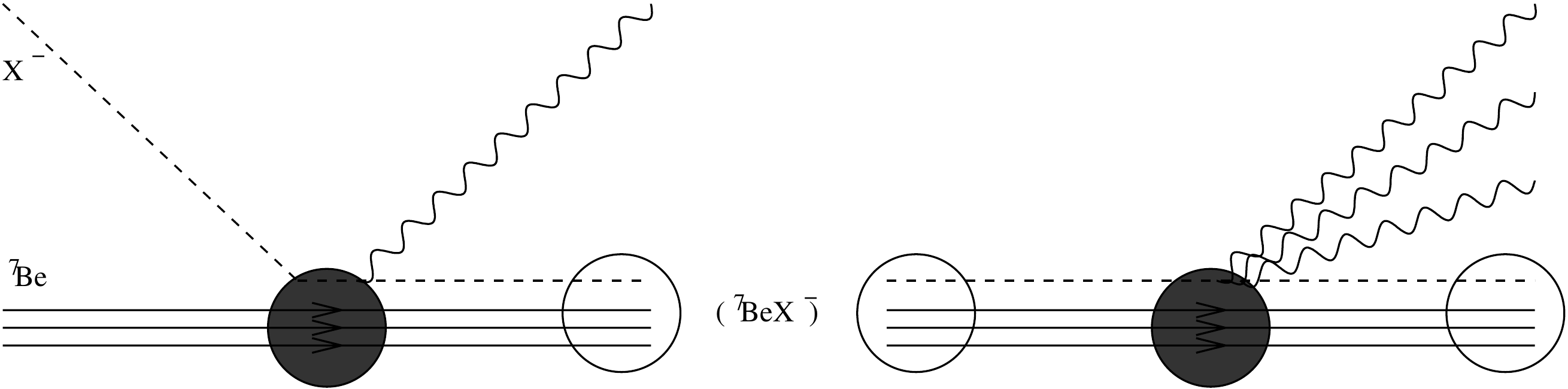,width=\textwidth,angle=0}
\caption{\label{Figure:BeRecomb} Recombination of \bes\ and $X^-$. The relic is captured to an excited state (left) and then radiates a number of photons to reach the ground state (right).}
\end{figure}

\par
The recombination rate also depends strongly on the properties of the bound state, \bex\ . The bound state energy could be approximated with the \bes\ nucleus represented as a point particle, in which case the ground state energy is $E_g^{(0)} = -2785 \; keV$. However in this approximation the Bohr radius is $a_B = 1.03 \;fm$, while the actual \bes\ radius is $<r^2>^{1/2} = 2.50 \pm 0.04 \; fm$ \cite{radii}. Assuming a Gaussian charge distribution, and a finite nuclear radius, the ground state energy is approximated as

\begin{equation}
E_g = - 1330 \pm 20 \; keV
\end{equation}

\noindent
The recombination rate will also depend on the internal structure of the \bes\ nucleus. The first excited state of \bes\, with spin 1/2, has an energy of only $429 \;keV$ above the ground state and can also form bound states with $X^-$. Therefore it is possible to form \bex\ through intermediate states. The most important of these are

\begin{eqnarray}
(^7{\rm Be}_{1/2} X^-), ~~ n=3,~~l=0,~~E_R=(-239+429){\rm keV= 190 keV};\nonumber\\
\phantom{(^7{\rm Be}_{1/2} X^-),}~~ n=3,~~l=1,~~E_R=(-290+429){\rm keV= 140 keV};\\
\phantom{(^7{\rm Be}_{1/2} X^-),}~~ n=3,~~l=2,~~E_R=(-308+429){\rm keV= 121 keV}.\nonumber
\end{eqnarray}
\noindent
where $E_R$ is the energy level of the state relative to 
$\besm_{3/2}+X^-$ . The recombination cross-section is

\begin{equation} \label{Eq:RecA}
< \sigma_{rec} v > = \frac{6 \times 10^3}{T_9^{1/2}} + \frac{1.9 \times 10^4}{T_9^{3/2}} exp(-1.40/T_9) + \frac{1.5 \times 10^4}{T_9^{3/2}} exp(-1.62/T_9)
\end{equation}

\noindent
where $T_9 \equiv T/10^9 {\rm K}$ is the temperature, and the rate is given in the standard BBN units of $N_A^{-1} cm^3 s^{-1} g^{-1}$. The first term corresponds to the non-resonant reactions

\begin{displaymath}
\begin{split}
&^7Be_{3/2} + X^- \to (^7Be_{3/2} X^-) + \gamma \to  (^7Be X^-) + k \gamma \\
&^7Be_{3/2} + X^- \to (^7Be_{1/2} X^-) + \gamma \to  (^7Be X^-) + k \gamma
\end{split}
\end{displaymath}

\noindent
where k is a number of photons,while the other two terms correspond to the resonances and captures to the excited states of \bes . In addition to these reactions, it is possible that the recombination rate is increased by effects from the 2s state via the reaction

\begin{displaymath}
^7Be_{3/2} + X^- \to (^7Be_{1/2} X^-, n=2,l=0) \to (^7Be_{1/2} X^-, n=2,l=1) + \gamma \to (^7BeX^-) + 3 \gamma
\end{displaymath}

\noindent
However the position of this resonance is unknown, and this reaction introduces additional uncertainty into the recombination rate. Therefore two special cases will be considered, corresponding to either no effect from this reaction or to a resonance located at +10 keV. In the second case, the recombination cross-section contains an additional term,

\begin{equation} \label{Eq:RecB}
<\sigma_{rec} v> \to <\sigma_{rec} v> + \frac{4 \times 10^3}{T_9^{3/2}} exp(-0.12/T_9)
\end{equation}

\noindent
This additional term can increase the total recombination rate by a factor of a few, and therefore is the source of a significant uncertainty.

\par
It is also important that the bound states are not destroyed by thermal photons. The rate for this process is related to the recombination rate via the detailed balance relation \cite{Angulo:1999zz},

\begin{equation}
\Gamma_{photo} = \int_{E > |E_g|} \sigma_{photo} dn_{\gamma}(E) = <\sigma_{rec} v> \times \left( \frac{m_{Be} T}{2 \pi} \right)^{3/2} exp(- |E_g|/T)
\end{equation}


\noindent
or

\begin{equation}
\Gamma_{photo}= <\sigma_{rec} v> \times \frac{5.5 \times 10^6}{T_9^{3/2}} exp(-15.42/T_9) n_{\gamma}
\end{equation}

\noindent
where $n_{\gamma}(T) = 0.24 T^3$ is the total photon number density.
Using both the recombination rate and photodisintegration rates, the Boltzmann equation for the density of bound states is

\begin{equation}
-H T \frac{dY_{B}}{dT} = n_{B} Y_X(1-Y_B)<\sigma_{rec} v> - \Gamma_{photo} Y_{B}
\end{equation}

\noindent
with the assumption that $n_X \gg n_B$. 

\par
It should also be noted again that in this calculation the abundance of \bes\ nuclei is taken to be constant. In the SBBN scenario, \bes\ freezes out at a temperature of $T \sim 40 \; keV$, while the bound states \bex\ form at temperatures of $T \sim 33 \; keV$. Therefore in the temperature range at which CBBN\index{Big Bang Nucleosynthesis!catalyzed} reactions become important, the abundance of \bes\ is approximately constant.

\vspace{16pt}
\subsubsection{Catalyzed Reactions}

\par
Once the bound states of \bes\ have formed, there are several reactions which destroy the \bes\ nuclei and decrease the abundance of \lisv\ + \bes. Some of these reactions are enhanced relative to the SBBN case, while other reactions are not possible in SBBN.

\par
The first reaction to be considered is the destruction of \bes\ by protons. The reaction

\begin{displaymath}
^7Be + p \to ^8B + \gamma
\end{displaymath}

\noindent
is significantly enhanced in the catalyzed form, $(^7BeX^-) + p \to ^8B + X^-$, but the threshold energy is too large for this reaction to occur at any significant rate. It is also possible to catalyze this reaction as\footnote{In Ref \cite{Kusakabe:2007fu,Kusakabe:2007fv}, it was also suggested that this reaction could proceed through a resonant excited state of $^8B$. However the threshold energy for this reaction is much larger and the reaction is unimportant. }

\begin{displaymath}
(^7BeX^-) + p \to (^8BX^-) + \gamma
\end{displaymath}

\noindent
In this reaction, there is a reduction in the Gamow suppression due to the Coulomb shielding of the \bes\ nucleus by $X^-$, as well as an increase in the rate due to the emitted photon having approximately five times more energy than in the non-catalyzed reaction. The non-catalyzed reaction has been well studied (see for example Ref. \cite{Baye:2000gi} for the properties of the S-factor), and is known to depend on the third power of the photon energy through the S-factor,

\begin{equation}
S_{SBBN}(0) = \frac{5 \pi}{9} \alpha \left( \frac{Z_1 A_2 - Z_2 A_1}{A} \right)^2 \omega_{SBBN}^3 I^2(0)
\end{equation}

\noindent
In this equation, $A_i$ and $Z_i$ represent the mass and charge of the initial nuclei in this reaction and $I(0)$ is the radial nuclear integral. Therefore, at first approximation the S-factor is taken to be

\begin{equation}
S_{CBBN}(0) = S_{SBBN}(0) \left( \frac{\omega_{CBBN}}{\omega_{SBBN}} \right)^3 \frac{1}{0.37^2} \sim 15 \; {\rm keV \; bn}
\end{equation}

\noindent
where \we\ have taken $S_{SBBN}(0)=21 \; {\rm eV \; bn}$ and the factor of represents the effects of the mass and charge of $X^-$.

\par
This reaction also receives a significant contribution from the bound state resonances, with the most important being

\begin{displaymath}
(^7BeX^-) + p \to (^8BX^-,n=2,l=1) \to (^8BX^-) + \gamma \quad \quad E_R = 167 \; keV
\end{displaymath}

\noindent
where the energy is given relative to the $(^7BeX^-) + p $ unbound system. The rate for this reaction is 

\begin{equation}
<\sigma v> \simeq 1.6 \times 10^8 T_9^{-2/3} exp(-8.86/T_9^{1/3}) + 1.6 \times 10^6 T_9^{-3/2} exp(-1.94/T_9)
\end{equation}

\noindent
The non-resonant contribution is small relative to the resonant rate, and the bound states, $(^7BeX^-)$, are destroyed by p-burning until $T_9 \approx 0.2$.
\par
The bound states, $(^8BX^-)$, will $\beta$-decay into $(^8BeX^-)$ which remains stable for the lifetime of the charged relic. Once it decays, the remaining $^8Be$\index{beryllium!$^8Be$} decays into two $\alpha$-particles, but the amount of $^4He$ produced in this way does not affect the primordial abundance of Helium. It may also be possible for the reaction 
\begin{displaymath}
(^8BeX^-) + n \to ^9Be + X^-
\end{displaymath}
 \noindent
 to occur, resulting in a small abundance of primordial $^9Be$\index{beryllium!$^9Be$}.

\par
The second reaction to be considered is destruction by neutrons. In the Standard BBN, the main  channel for \bes\ depletion is 

\begin{displaymath}
^7Be + n \to ^7Li + p
\end{displaymath}

\noindent
However as outlined in Ref \cite{Bird:2007ge}, the corresponding bound state reaction

\begin{displaymath}
(^7BeX^-) + n \to (^7LiX^-) + p
\end{displaymath}

\noindent
is not enhanced, as the widths of the $^8Be$ resonances which mediate the reaction are too large to be affected by the binding energies in any significant way.




\par
From the relative magnitude of the proton and neutron destruction of \bes\ , as well as the relative abundances of protons and neutrons at the appropriate temperatures, it is clear the in CBBN\index{Big Bang Nucleosynthesis!catalyzed} p-burning is the dominate channel for suppressing the \bes\ + \lisv\ abundance.

\vspace{16pt}
\subsubsection{Internal Decays of $X^-$}

\par
Another possible destruction mechanism is through the decay of the charged relic while in the bound state. The three main processes for this are:

\begin{list}{}
\item A: $(^7BeX^-) \to ^7Be + {\rm decay \; products} + {\rm hard} \; \gamma \to ^3He + ^4He + ...$ 
\item B: $(^7BeX^-) \to ^7Be(E \gg T) + {\rm background \; particles} + ... \to ^3He + ^4He + ... $
\item C: $(^7Be^*X^-) \to^3He + ^4He + ...$
\end{list}

\noindent
In the first case, the decaying $X^-$ produces a high energy photon which destroys the \bes\ nucleus. In this case, the decaying $X^-$ can release energies of $O(100 \; GeV)$ while the photodisintegration threshold for \bes\ is $E_{thr} \simeq 1.59 \; MeV$. In the second case, the decaying $X^-$ does not directly affect the nucleus, but instead release the \bes\ nucleus back into the background of thermal nuclei. Because the orbital energies of the \bes\ are of order $O(1\;MeV)$ in the bound state, the recoiling nucleus is sufficiently energetic that it may be broken up by collisions with other nuclei and background particles. In the third case, the effect of being bound to a charged particle polarizes the \bes\ nucleus, effectively forming a superposition of the ground state with several excited states of the free nucleus. When the $X^-$ decays, this polarization is removed and the \bes\ excited states can decay into $^3He+^4He$.

\par
The efficiency of case A can be estimated as follows. Assuming a Type I model (to provide sufficient energy in the decay) and that $X^-$ decays to a single charged particle with an energy of $E_{max} \gg 1.59 \; MeV$, and approximating the flux of high energy photons colliding with the \bes\ nucleus, the probability of radiative break-up is

\begin{equation}
P_{rad\;br} \simeq \int_0^{\infty} dr |\psi(r)|^2 \int_{E_{thr}}^{E_{max}} \sigma_{\gamma}(E) dn_{\gamma}(E)
\end{equation}

\noindent
where $dn_{\gamma}(E)$ is the number of photons with $E<E_{max}$ and $\sigma_{\gamma}(E)$ is the cross-section for radiative breakup of the \bes\ nucleus. Using the measured cross-section\footnote{In the calculation of the probability of break-up for the \bes\ nuclei, the cross-sections are given in different units than usual for BBN, as this calculation applies to a single nuclei with a certain energy rather than a thermal distribution, and is not required in the BBN calculation.} for $^7Be + \gamma \to ^3He + ^4He$ \cite{Cyburt:2002uv}, 

\begin{equation}
\begin{split}
\sigma_{\gamma} = 0.504 \; mb & \left( \frac{2371 \; MeV}{\omega}\right)^2 exp \left( -5.19 /\sqrt{E_{cm}} - 0.548 E_{cm}\right) \\
& \times (1 - 0.428 E_{cm}^2 + 0.534 E_{cm}^3 - 0.115 E_{cm}^4)
\end{split}
\end{equation}

\noindent
where $\omega$ is the energy of the virtual photon, $E_{thr} = 1.587 \; MeV$ is the threshold at which this reaction can occur, and $E_{cm} \equiv \omega - E_{thr}$ is the released energy in MeV. The probability is given by

\begin{equation}
P_{rad\;br} \approx 1.0 \times 10^{-4} \quad \quad E_{max} = 100 \; {\rm GeV}
\end{equation}

\noindent
The other channel which can destroy the bound \bes\ is $^7Be + \gamma \to ^3He + ^4He$,  with cross-section

\begin{equation}
\sigma_{\gamma} = 32.6 \; mb \frac{E_{thr}^{10} E_{cm}^2}{\omega^{12}} + 2.27 \times 10^6 \; mb \frac{E_{thr}^{8.83} E_{cm}^{13}}{\omega^{21.83}}
\end{equation}

\noindent
where $E_{thr} = 5.61 \; MeV$. As with the previous channel, the probability is

\begin{equation}
P_{rad\;br} \approx 1.3 \times 10^{-4} \quad \quad E_{max} = 100 \; {\rm GeV}
\end{equation}

\noindent
This indicates that this mechanism is not efficient, with $\lesssim 0.1 \%$ of the bound \bes\ nuclei being destroyed in this manner, and does not constitute a significant channel for suppression of the \bes\ + \lisv\ abundance.

\par
In case B, the decay of the relic $X^-$ leaves the \bes\ nucleus unbound with O(MeV) kinetic energy. In effect, the released nucleus is bombarded with a flux of background photons and other particles which serve to destroy the nucleus. 
\par
The exact destruction rate depends on the relative rates of the slowing of the \bes\ nucleus and the reaction $^7Be +  p \to ^8B+\gamma$. The slowing of the \bes\ nucleus occurs via elastic collisions, in the form of Coulomb scattering from electrons and positrons\footnote{The nuclei can also be slowed by Compton scattering of thermal background photons, however the cross-section for this process is several orders of magnitude smaller than the cross-section for Coulomb scattering \cite{Reno:1987qw}}. In Ref \cite{Bird:2007ge}, \we\ have estimated that an $O(MeV)$ nucleus will thermalize in this manner in $\sim 10^{-6} s$. The rate for \bes\ destruction is significantly smaller, with the rate of destruction by the reactions $^7Be + \gamma \to ^4He + ^3He$ and $^7Be + p \to ^9B + \gamma$ estimated as $\Gamma_{\gamma} \sim 10^{-10} s^{-1}$ and $\Gamma_{p} \sim 10^{-5} s^{-1}$ respectively \cite{kkmsc}. Therefore the probability of the \bes\ nucleus being destroyed before thermalization is $\lesssim 10^{-11}$, and therefore this mechanism is also not significant. 

\par
In case C, the rate of destruction depends on the part of the bound state \bes\ wavefunction which corresponds to an excited nucleus. This can be approximated as

\begin{equation}
\sum \left| \frac{{\bf d\cdot E}_{0i}}{E_0 - E_i} \right|^2
\end{equation}

\noindent
where ${\bf d\cdot E_{0i}}$ is of order 1 MeV, and the energy difference is $(E_i - E_0) \geq 10 \; MeV $. From these typical values, it is clear that this mechanism is not efficient either, with less than $1 \%$ of recoiling nuclei decaying from excited states. 

\par
These three channels do not contribute significantly to the destruction of \bes\ nuclei, and therefore the effects of internal decays of the charged relic can be ignored.

\vspace{16pt}
\subsubsection{Internal Conversion Channels}

\par
Another possibility is that there is a second relic state, $X^0$, which has a similar mass to $X^-$, and which can be a decay product of $X^- \to X^0 + {\rm charged \; particles}$. The most obvious choice is a weak-coupling, $X^- \to X^0 + W^-$, and in this case the $W^-$ boson can then hit the \bes\ nucleus in the bound state,$(^7BeX^-)$, and convert it into a \lisv\ nucleus,

\begin{displaymath}
(^7BeX^-) \to ^7Li + X^0
\end{displaymath}

\begin{figure}
\begin{center}
\psfig{file=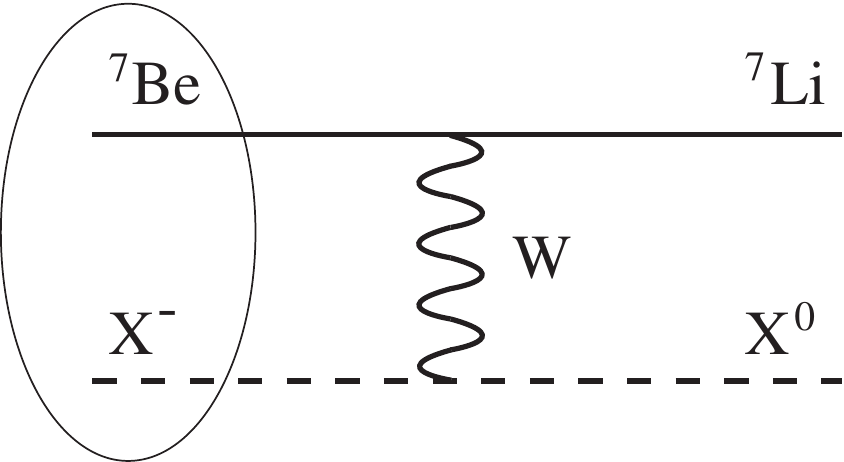,width=0.65\textwidth,angle=0}
\caption{\label{Figure::LiBeConversion} The internal conversion of \bes\ to \lisv\ via a weak-boson exchange with $X^-$.}
\end{center}
\end{figure}

\noindent
This reaction is shown in Figure \ref{Figure::LiBeConversion}. As the \lisv\ nucleus is more fragile, it is subsequently destroyed through the SBBN channels\index{Big Bang Nucleosynthesis}\index{nucleosynthesis}.

\vspace{16pt}
\subsubsection{Energy Injection From $X^+X^-$ Annihilation}

\par
In addition to the catalysis of nuclear reactions, the presence of charged relics can also affect the primordial abundances of the light elements through the injection of energy from $X^+X^-$ annihilations and decays. In particular, when the temperature of the Universe is low enough the relic can form a bound state with its antiparticle, and subsequently annihilate through a positronium-like state. 

\par
The rate of energy released by $X^+X^-$ annihilations is

\begin{equation}
\Gamma_{ann} = <\sigma_{X^+X^-} v_{rel}> n_{X^+}
\end{equation}

\noindent
where the cross-section for bound state formation is 

\begin{equation}
<\sigma_{X^+X^-} v_{rel}> = \frac{2^{10}\pi^{3/2} \alpha^3}{3 e^4 m_X^{3/2} T^{1/2}}
\end{equation}

\noindent
It should be noted that this cross-section is sufficiently small that this mechanism does not have a significant effect on the abundance of $X^-$ during CBBN. 

\par
As is common in calculations of this nature, the effects of this mechanism are quantified using the energy released per photon, normalized on 1 GeV. For annihilating particles, this takes the form

\begin{equation}
\xi = \frac{2 m_X}{1\; GeV} Y_X^2 \eta_B^2 \times \int_{T_1}^{T_2} \frac{<\sigma_{X^+X^-} v_{rel}> n_{\gamma}}{T H} dT
\end{equation}

\noindent
where $T_1$ to $T_2$ represents the range of temperatures at which these effects are important. In this calculation, this range is between $\sim 10\; keV$ and $\sim 40 \;keV$. At higher temperatures the effects of the annihilating particles will be overwhelmed by the thermal nuclear reactions.

\par
Using the cross-section given above, $\xi$ is given by

\begin{equation}
\xi = 2.2 \times 10^{-12} \left(\frac{500 \; GeV}{m_X} \right)^{1/2} \left( \frac{Y_X}{0.02}\right)^2
\end{equation}

\noindent
At this scale, the energy released in the form of photon pairs produced in $X^+X^-$ annihilations is not sufficient to affect the nuclear abundances. However if the annihilations produce hadrons with any significant branching ratio, then this energy release can suppress the \lisv\ abundance by a factor of a few\cite{Jedamzik:2005dh}. It should also be noted that the $m_X^{3/2}$ scaling of $\xi$, which is a result of the expected linear dependence of $Y_X$ on $m_X$, means that heavier CHAMPs will produce a stronger suppression of the \lisv\ abundance due to hadronic annihilations. As a result, the predicted abundance of \lisv\ in the CBBN model depends on both the catalyzed reaction rates, calculated in previous sections, and on the effects of energy injected into the system by both the annihilations and decays of the CHAMP being considered.

\vspace{16pt}
\subsubsection{Abundance}

\par
In the Standard BBN\index{Big Bang Nucleosynthesis}\index{nucleosynthesis}, the dominant processes which determine the \lisv\ abundance are:

\begin{equation} {\rm SBBN}: ~~
\hetm+\hefm\ \to \besm +\gamma; ~
\besm + n \to \lisvm + p; ~ \lisvm+p\to 2\hefm {\rm ~or~ D} +\lisxm.
\label{Equation::SBBNreac}
\end{equation}

\noindent
In addition to these reactions, the abundance is also affected by secondary reactions:

\begin{equation} {\rm SBBN}: ~~~~
^3H+\hefm\ \to \lisvm +\gamma; ~
\besm+p\to {\rm ^8B} +\gamma; \besm+{\rm D}\to p+2\hefm;
\label{Equation::SBBNreac2}
\end{equation}

\noindent
These processes have been studied extensively in relation to BBN \cite{Coc:2003ce,Cyburt:2004cq}.

\par
As demonstrated in the previous sections, the presence of charged relics can add additional destruction channels:


\begin{equation} \label{Equation::CBBN}
\begin{split}
{\rm CBBN}: ~~~~ &\besm+X^-\to(\besm X^-)+\gamma;~~ (\besm X^-)+\gamma\to \besm+X^-;\\ &
 \lisvm+X^-\to(\lisvm X^-)+\gamma;~~ (\lisvm X^-)+\gamma\to \lisvm+X^-.
 \end{split}
\end{equation}
\begin{displaymath}
\begin{array}{ll}
{\rm Type ~ I~ and ~ II}: & (\besm X^-)+p\leftrightarrow ({\rm ^8B}X^-) +\gamma; ~~~
({\rm ^8B}X^-)\to ({\rm ^8Be}X^-).\\
&(\besm X^-)+n\to (\lisvm X^-)+p; ~(\lisvm X^-)+p\to X^-+2\hefm;\\&(\lisvm X^-)+p\to X^-+{\rm D}+ \lisxm.\\ & \\
{\rm Type ~  II ~ only }:& (\besm X^-) \to \lisvm+X^0; ~~({\rm ^8B} X^-) \to \beetm+X^0.
\end{array}
\end{displaymath}

\noindent
In addition to these primary channels, destruction of the \bex\ can occur by D-burning. In $(\lisvm X^-)+p$ 
and \bex+D reactions the only change implemented relative to the SBBN\index{Big Bang Nucleosynthesis}\index{nucleosynthesis} rate is in the Coulomb penetration factor. In Type II models, it is also possible to produce \lisv\ through the internal conversion process $\besm + X^- \to (\besm X^-) + \gamma \to \lisvm + X^0 + \gamma$, as discussed in the previous section.


\begin{figure}
\psfig{file=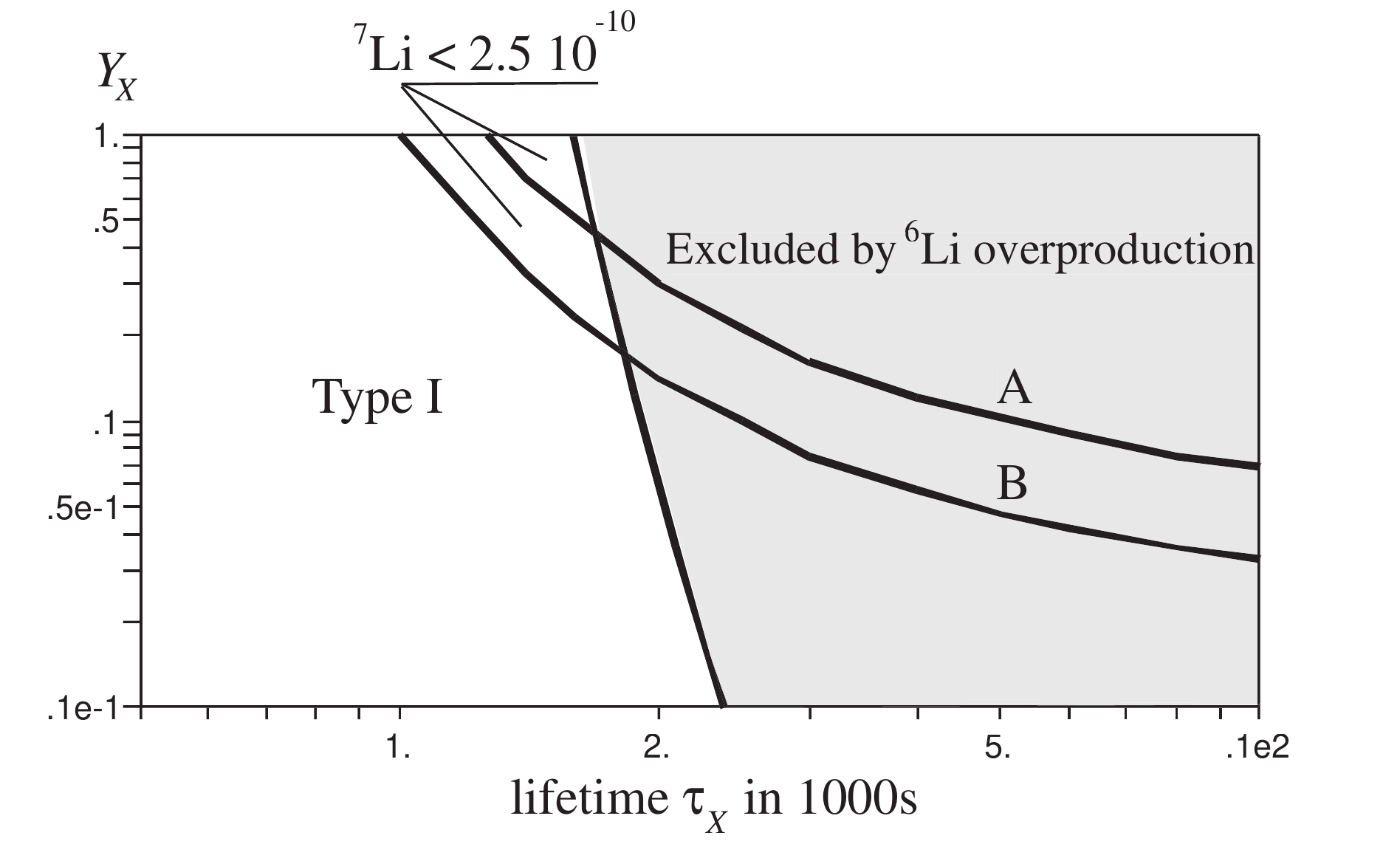,width=0.88 \textwidth,angle=0} \\ \begin{center}(a) \end{center} 
\psfig{file=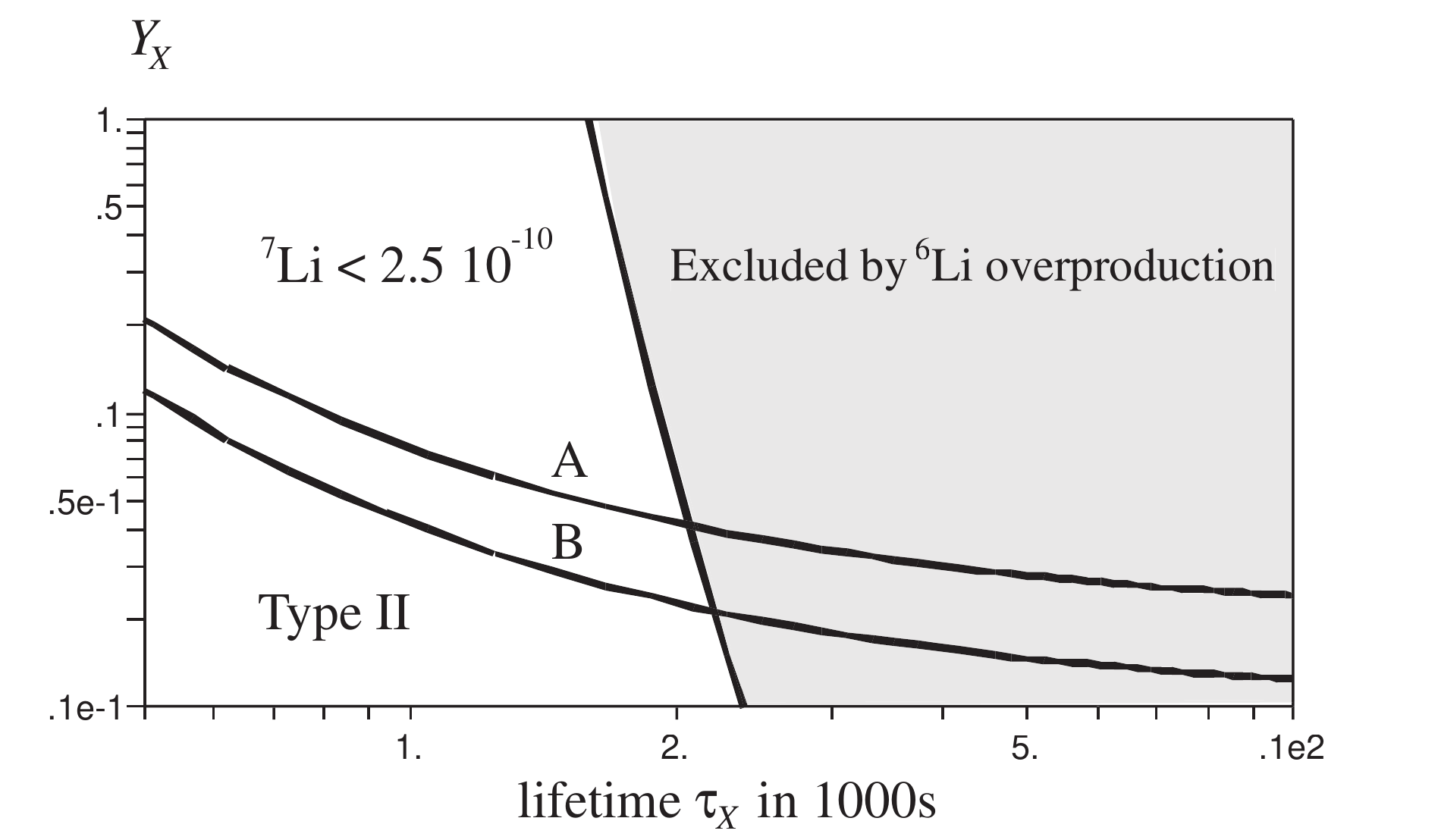,width=0.88 \textwidth,angle=0} \\ \begin{center}(b) \end{center}  
\caption{\label{Figure:Li7abundance} 
The combined constraints on the abundance and lifetime of charged relics from CBBN\index{Big Bang Nucleosynthesis!catalyzed} in (a) Type-I and (b) Type-II models. The gray region is excluded due to an overproduction of \lisx\, while the lines represent the region that can explain the observed \lisv\ suppression. Curve A corresponds to the abundance when Eq \ref{Eq:RecA} is used, while curve B corresponds to the abundance when Eq \ref{Eq:RecB} is used.}
\end{figure}

\par
Although a full treatment of CBBN\index{Big Bang Nucleosynthesis!catalyzed} would require the inclusion of several hundred catalyzed reactions, the calculation of the \lisv\ abundance can be performed to reasonable precision using only the Li-Be reactions \footnote{During the writing of this \thesis, two more calculations of the \lisx\ and \lisv\ abundance in CBBN\index{Big Bang Nucleosynthesis!catalyzed} were published \cite{Kusakabe:2007fu,Kusakabe:2007fv}, in which additional reaction rates were estimated and included. However as indicated in Table \ref{Table:CHAMPs}, these additional reactions do not significantly alter the constraints given here.} since the final abundances of these elements are too small for a backreaction to significantly affect the more abundant elements such as \hef\ , \het\ , and D. Using the predicted temperature dependent abundances D(T),\het\ (T),etc. from the full SBBN code, and the portions of the code corresponding to the reactions in Eq \ref{Equation::SBBNreac} and Eq \ref{Equation::SBBNreac2}, the prediction for the \lisv\ + \bes\ abundance is

\begin{equation}
(\lisvm_{tot})_{SBBN} \equiv (\besm + \lisvm)_{SBBN} = 4.1 \times 10^{-10}
\end{equation}

\noindent
which agrees with more detailed calculations\footnote{It should be noted that after this calculation was completed, the reaction rates were updated and the SBBN prediction for $(\lisvm_{tot})_{SBBN}$ has increased \cite{Cyburt:2008kw}. However these updated reaction rates are not expected to significantly alter the conclusions of this chapter.} of the SBBN \cite{Coc:2003ce,Cyburt:2003fe,Cyburt:2004cq}.

\par
When the reactions of Eq \ref{Equation::CBBN} are included, the \lisv\ abundance is reduced. The abundance depends on the mass and lifetime of the charged relic\index{charged relic}\index{CHAMPs}, and is given in Figure \ref{Figure:Li7abundance} along with the constraints from \lisx\ abundances. For relics along the solid line, the \lisv\ abundance is $2.5 \times 10^{-10}$, which is consistent with the observed abundance of primordial \lisv\ . From these results, a charged relic\index{charged relic}\index{CHAMPs} with

\begin{displaymath}
{\rm few} \times 100 s < \tau_X < 2000 s
\end{displaymath}

\noindent
could reduce \lisv\ to observed levels without overproducing \lisx\ . 

\vspace{16pt}
\section{Conclusions}

\par
The existence of massive charged relics is well motivated in theories of supersymmetry\index{supersymmetry} and in other theories of new physics. In the past the properties of such particles has been constrained by both collider experiments and heavy water experiments\index{heavy water experiments}, as well as indirect constraints from the $\gamma$-ray background\index{$\gamma$-ray background}. However collider experiments have thus far only been capable of excluding $m_X \lesssim O(100 \; GeV)$ in very model dependent ways, and heavy water experiments cannot constrain heavy metastable CHAMPs\index{CHAMPs}\index{charged relics} with low abundance. 

\begin{table}
\begin{center}
\begin{tabular}{|l|c|c|c|} \hline
& nuclei& $\tau_X (s)$ & $Y_X$ \\ \hline
Pospelov \cite{Pospelov:2006sc} & \lisx\ &$\lesssim 4000$&  $\lesssim 10^{-7}$ 
 \\
Hamagutchi \cite{Hamaguchi:2007mp} & \lisx\ &3300-3700 & $\sim 10^{-3} - 10^{-4}$\\
Koopmans \cite{kkmsc} & \lisx\ & 1600 - 7000 & $\sim 5 \times 10^{-4} - 0.07$ \\ \hline
Bird et. al. \cite{Bird:2007ge} & \lisv\ &1000-2000& $\gtrsim 0.04$  \\ 
Jittoh et. al. \cite{Jittoh:2007fr} & \lisv\ & 730-1736 & $\gtrsim 1.2 \times 10^{-10}$ \\

Kusakabe et. al. \cite{Kusakabe:2007fu,Kusakabe:2007fv} & \lisx\ , \lisx\ & 1600-2800 & 0.09 - 0.6
\\ Kusakabe et. al. \cite{Kusakabe:2007fu,Kusakabe:2007fv}  &\lisx\ , \lisv\ &1400-2700& 0.04 - 0.2 
\\ \quad (with $X^-$ internal decays) && &
\\ \hline Pospelov \cite{Pospelov:2007js} & $^9Be$& $ \lesssim few \times 10^3$& -
\\ \hline
\end{tabular}
\end{center}
\caption{\label{Table:CHAMPs} Preferred parameters for CHAMPs\index{CHAMPs}\index{charged relics} from various calculations of CBBN\index{Big Bang Nucleosynthesis!catalyzed}}
\end{table}

\par
In this chapter, I have demonstrated how the presence of charged massive particles\index{CHAMPs}\index{charged relics} in the early Universe could affect the predictions of Big Bang nucleosynthesis\index{Big Bang Nucleosynthesis}\index{nucleosynthesis} by catalyzing nuclear reactions. In particular, the formation of bound states of charged relics with Lithium\index{lithium} and Beryllium\index{beryllium} nuclei can lead to a suppression of the primordial \lisv\ abundance. As the \lisv\ abundance is one of the few discrepancies between BBN predictions and observation, with the observed abundance a factor of a few lower than predicted, this result can be viewed as another motivation for the existence of charged massive particles in the early universe. 

\par
The primary constraint on CHAMPs\index{CHAMPs}\index{charged relics} in CBBN\index{Big Bang Nucleosynthesis!catalyzed} is from \lisx\ production. This bound and the \lisv\ bounds are reasonably model independent, with the main assumption being that the relic has unit charge and is massive.

\par
As originally published in Ref \cite{Bird:2007ge}, for CHAMPs with lifetimes in the range of $\tau \sim 1000-2000 \; s$ and abundances in the range of $Y_{X} \gtrsim 0.04$, the predicted \lisv\ abundance is consistent with observations without violating observed limits on the \lisx\ abundance. At present, the region of concordance between all CBBN\index{Big Bang Nucleosynthesis!catalyzed} bounds gives the preferred CHAMP\index{CHAMPs}\index{charged relics} properties as $\tau \sim 1500 \;s $ with $Y_X \sim O(0.1)$. 

\par
In the last year, the predictions of catalyzed Big Bang nucleosynthesis\index{Big Bang Nucleosynthesis!catalyzed} have been studied extensively \cite{Pospelov:2006sc,Kohri:2006cn,Kaplinghat:2006qr, Bird:2007ge,Hamaguchi:2007mp,Kawasaki:2007xb,Jittoh:2007fr,Kersten:2007ab,Pradler:2007is,Pradler:2007ar,Spanos:2007cz,Kusakabe:2007fu,Kusakabe:2007fv,Pospelov:2008ta,Kamimura:2008fx} and will be studied in more detail in the future. A summary of the constraints on CHAMPs\index{CHAMPs}\index{charged relics} in each study is given in Table \ref{Table:CHAMPs}. The first three studies \cite{Pospelov:2006sc,Hamaguchi:2007mp,kkmsc} used the production of \lisx\ as a restriction, with different estimates of the relevant production rates and the fractional abundance of bound states of $(\hefm X^-)$ , and constrain generic properties of the CHAMPs\index{CHAMPs}\index{charged relics} based on the observed small but non-zero abundance of primordial \lisx\ . The \lisv\ calculations \cite{Bird:2007ge,Jittoh:2007fr} assume that the CHAMP\index{CHAMPs}\index{charged relics} is responsible for the observed suppression of primordial \lisv\ , although these constraints could be avoided if \lisv\ is suppressed by other mechanisms as well. The two papers which simultaneously treat \lisx\ and \lisv\ also estimate additional reaction rates in CBBN\index{Big Bang Nucleosynthesis!catalyzed} which slightly alter the constraints. Furthermore, these two papers give different constraints, as the first result is independent of the CHAMP\index{CHAMPs}\index{charged relics} decay mode while the second result includes the effects of the decay $(NX^-) \to N + X^0 + ...$ \cite{Kusakabe:2007fu,Kusakabe:2007fv}. 
\nocite{Kusakabe:2008kf}\nocite{Kohri:2008cf}
\par
In addition, CBBN\index{Big Bang Nucleosynthesis!catalyzed} predicts a fairly model independent abundance of $^9Be$\index{beryllium!$^9Be$} relative to the \lisx\ abundance \cite{Pospelov:2007js}, which should be measurable in future astrophysics experiments. It has also been suggested that the region of parameter space which resolves the \lisv\ discrepancy should be thoroughly probed in future collider experiments \cite{Ibe:2007km,Cakir:2007xa}. 

\par
Based on these results, it is clear that CBBN\index{Big Bang Nucleosynthesis!catalyzed|)} provides a strong constraint on the properties of charged relics\index{CHAMPs|)}\index{charged relics|)} in the Universe as well as providing a possible explanation for the higher abundance of \lisx\ and the suppression of \lisv\ in the early Universe. \index{Big Bang Nucleosynthesis|)}

\index{CBBN|)}

%% file: chapter3.tex
\chapter{Extra Dimensions \label{Chapter:ExtraDimensions}}\index{extra dimensions}

\section{Overview}

\par
The possibility that there may exist hidden dimensions in our Universe has been considered for several centuries. The first serious research into extra dimensions occurred in 1915 with the four-dimensional general theory of relativity\index{general relativity}, and the five and six dimensional theories of Nordstrom \cite{Nordstrom}, Kaluza \cite{Kaluza}, and Klein \cite{Klein}. These theories were largely ignored until the 1970's, with the realization that certain models of supersymmetry\index{supersymmetry} and superstring\index{string theory} theory were only consistent in higher dimensional spacetimes.

\par
In 1998, the idea that the Universe could contain extra dimensions was revitalized with the realization that higher dimensions could solve several problems with Standard Model. It was discovered that if the Standard Model fields were confined to a small region of the extra dimension \cite{Arkani-Hamed:1998rs}, or confined to a three-dimensional membrane \cite{Randall:1999ee,Randall:1999vf}, then gravity could be made as strong as the other three forces in the Standard Model without violating existing experimental bounds\index{hierarchy problem}, resolving the {\it hierarchy  problem}. In addition, these models allowed extra dimensions as large as $R \sim 1 \; mm$ which made it possible to search for extra dimensions in modern gravity experiments. Later models have also been used to explain the fermion hierarchy \cite{Arkani-Hamed:1999za,Huber:2000ie}, number of spacetime dimensions\cite{Chodos:1979vk}, dark matter\index{dark matter} \cite{Servant:2002aq,Cheng:2002ej,Cembranos:2003mr}, dark energy\index{dark energy} \cite{Dvali:2003rk}, inflation\index{inflation} \cite{Levin:1994yw}, the low energy scale of the cosmological constant\index{cosmological constant} \cite{Chen:2000at,Aghababaie:2003wz}, and non-singular alternatives to inflation\index{inflation} \cite{Khoury:2001wf}.

\par
The original models introduced by Kaluza and Klein included no difference between higher dimensions and the usual dimensions of space, with the exception of compactification. At present there are several models of extra dimensions\index{extra dimensions}, with no clear indication of which if any is correct. Many of the models are known to be unstable with respect to small perturbations. However each model predicts unique phenomena which could be detected. The three main classes of extra dimensional models are:
\vspace{16pt}
\begin{list}{}
\item {\it Universal Extra Dimensions (UED)}\index{extra dimensions!universal} -  In the UED model, all Standard Model fields are allowed to propagate through all dimensions. This also results in very strong constraints on the size of the extra dimensions, with colliders restricting $R\lesssim 10^{-16} \; cm $. Since the SM fields can propagate in the higher dimensions, these models have been used to explain the three generations of the Standard Model \cite{Arkani-Hamed:1999za,Huber:2000ie}, dark matter\index{dark matter} \cite{Servant:2002aq,Cheng:2002ej,Kakizaki:2006dz}, but may also predict a decaying proton\index{proton decay} \cite{Mohapatra:2002ug,Hsieh:2006fg}.
\vspace{16pt}

\item {\it Nonwarped Braneworlds} - The second class of extra dimensional models involves 4D branes embedded in higher dimensional spacetimes. In such models, the extra dimensions are not warped, and are usually given the topology of a d-dimensional torus. In these models, only gravity can propagate in all dimensions, while the Standard Model fields are restricted to the brane. Gravity experiments still require these extra dimensions to be compact, but their sizes can be of O(1 mm) \cite{Arkani-Hamed:1998rs} which is observable in modern gravity experiments \cite{Adelberger:2003zx}. 
\par One of the features of these models is the presence of Kaluza-Klein gravitons\index{Kaluza-Klein gravitons}. The gravitational field, like any function with periodic boundary conditions, can be described by discrete modes. Each of these modes satisfies the equation of motion for either a spin-2 particle (KK graviton), a vector field (graviphoton)\index{Kaluza-Klein gravitons!graviphotons}, or a scalar (radions or graviscalars)\index{Kaluza-Klein gravitons!radions} \cite{Han:1998sg}. These effective particles are known as Kaluza-Klein modes, and have a mass spectrum of the form

\begin{equation}
m_{\stackrel{\rightarrow}{n}}^2 = m_0^2 + \frac{\stackrel{\rightarrow}{n}^2}{R^2}
\end{equation}

\noindent
where $\stackrel{\rightarrow}{n}$ represents a set of d-integers. Nonwarped brane models also have a reduced Planck mass, which is given by

\begin{equation}
M_* ^{d+2} = M_{PL}^2/ R^d
\end{equation}

\noindent
and which can be as low as a few TeV\cite{Arkani-Hamed:1998rs}. 
\vspace{16pt}
\item {\it Warped Extra Dimensions} - In models with warped extra dimensions, the Standard Model fields are still trapped on a 4D brane while gravity can propagate in higher dimensions. However unlike the previous model, the strong warping of the extra dimensions allows for large or even infinite dimensions as the warping limits the range over which gravity can operate.
\par
The most common example of warped extra dimensions are the Randall-Sundrum models \cite{Randall:1999ee,Randall:1999vf}. In the Randall-Sundrum model\index{extra dimensions!Randall-Sundrum model}, the metric of spacetime is given by

\begin{equation}
ds^2 = e^{-2 k \phi(x^{\mu})|y|} \eta_{\mu \nu} dx^{\mu} dx^{\nu} - \phi(x^{\mu})^2 dy^2
\end{equation}

\noindent
As outlined in Section \ref{Section::EXDM}, the field $\phi(x^{\mu})$ has the same properties as a scalar field trapped on the 4D brane, and is referred to as the {\it radion}\index{Kaluza-Klein gravitons!radions}. As with the nonwarped models, the Randall-Sundrum\index{extra dimensions!Randall-Sundrum model} has a reduced Planck mass given by

\begin{equation}
M_* \approx M_{PL} e^{- k \pi r_c}
\end{equation}

\noindent
where $r_c \equiv <\phi>$. In this model, $M_*$ can be as small as 1 TeV.
\end{list}

\noindent
In this section I review the existing constraints on the ADD model \cite{Arkani-Hamed:1998rs} of nonwarped extra dimensions, and then calculate two new sets constraints arising from Big Bang nucleosynthesis\index{Big Bang Nucleosynthesis} in Section \ref{Section:BBNEX} and from positron production\index{galactic positron excess} in the galactic core in Section \ref{Section:PosEX}. Although each constraint assumes the extra dimensions form a d-dimensional torus with all compact dimensions having the same size, the methods used are generic and could also be applied to more general models to provide additional constraints.

\section{Previous Constraints \& Searches}

\par
 The most important constraint on such models is that they not violate current limits from gravitation experiments \cite{Adelberger:2003zx}. In the ADD model\index{extra dimensions!ADD model}, a single extra dimension cannot resolve the hierarchy problem unless $R \simeq 1000 \; km$. However a dimension this large would lead to a $1/r^3$ gravitational force at the scale of the solar system, and so is already excluded. For higher dimensions, the present limit from gravitational force measurements is $R \lesssim 0.13 \;mm$ \cite{Hoyle:2004cw}. For comparison with other bounds, the gravitational bound on two extra dimensions limits the Planck mass to be $M_* \gtrsim 1.9 \; TeV$.

\subsection{Astrophysics Constraints}

\par
In the early Universe, large numbers of Kaluza-Klein gravitons would have been created in both thermal processes and non-thermal processes. For thermal production, the dominant production channels are $\gamma \gamma \to KK, \bar{\nu} \nu \to KK$ and $e^+e^- \to KK$ \cite{Han:1998sg,Hall:1999mk}. 
By assuming that the abundance of Kaluza-Klein modes is smaller than the abundances of the Standard Model fields, which avoids backreactions of the KK modes, the Boltzmann equation can be written as

\begin{equation}
\dot{n}_m = \sum_{i = \gamma, \nu, e^+} <\sigma_{ann} v>_i n_i^2 -3 H n_m
\end{equation}

\noindent
where H is the Hubble constant and $n_i$ is the number abundance of the fields. In this equation, it is assumed that the Kaluza-Klein modes\index{Kaluza-Klein gravitons} produced in this way will have small mass, which results in a long life \cite{Han:1998sg} and as such the decay term in the Boltzmann equation has be omitted. Following Ref \cite{Hall:1999mk}, this equation can be reduced to 

\begin{equation}
\dot{n}_m = - 3 H n_m + \frac{11 m^5 T}{128 \pi^3 M_P^2} K_1(m/T)
\end{equation}

\noindent
where m is the mass of the Kaluza-Klein mode, $M_P$ is the reduced Planck mass (or the Planck mass as viewed on the brane) and $K_1$ is the modified Bessel function of the second kind.
Taking in consideration the subsequent KK mode decays\index{Kaluza-Klein gravitons}, gives the present abundance of a specific mode as

\begin{equation}
n_m \simeq \frac{19 T_0^3 }{64 \pi^3 \sqrt{g_{*,RH}}} \frac{m}{M_P} e^{- \Gamma_{dec}(m) t_0}  \int_{m/T_{RH}}^{\infty} q^3 K_1(q) dq
\end{equation}

\noindent
or for all modes with mass m,

\begin{equation} \label{Eq:KK}
n_m 
 \simeq (1.9 \times 10^{-22} \; GeV^4)S_{d-1} \frac{T_{RH}^{d+1} }{M_*^{d+2}} \left( \frac{m}{T_{RH}}\right)^{d} e^{- \Gamma_{dec}(m) t_0}  \int_{m/T_{RH}}^{\infty} q^3 K_1(q) dq
\end{equation}

\noindent
where $T_{RH}$ is the temperature at which the Universe becomes radiation dominated\index{reheating temperature}, $S_{d-1} = 2 \pi^{d/2} /\Gamma(d/2)$ is the area of a d-dimensional sphere. It is assumed that the modes are so closely
 spaced and that R is sufficiently large that the sum over indices, which for a given mass m must satisfy $\sum_{i=1..d} \left(\frac{k_i}{R} \right)^2 = m^2$ , can be replaced with an integral over the surface of a sphere.
 
\par
The second production mechanism in the early Universe is non-thermal production during reheating. It is believed that the Universe underwent a period of inflationary expansion\index{inflation}, generated by a massive field known as the {\it inflaton}. At the end of this expansion, the inflatons decayed into other fields, which in effect reheated the Universe and began the period of thermal interactions which produced the usual Standard Model fields and dark matter. Since gravity is coupled to energy, these inflaton decays would necessarily be accompanied by production of Kaluza-Klein gravitons.

\par
The abundance of Kaluza-Klein modes\index{Kaluza-Klein gravitons} produced by reheating can be calculated in a similar manner to the abundance of thermally produced modes (see eg. Ref \cite{Hannestad:2001nq}). The final result is an abundance of

\begin{equation} \label{Eq:NTKK}
n_m 
 \simeq (1.9 \times 10^{-22} \; GeV^4)S_{d-1} \frac{T_{RH}^{d+1} }{M_*^{d+2}} \left( \frac{m}{T_{RH}}\right)^{d-7} e^{- \Gamma_{dec}(m) t_0}  \int_{m/(\alpha T_{RH})}^{m/T_{RH}} q^{10} K_1(q) dq
\end{equation}

\noindent
where $\alpha \equiv T_{MAX}/T_{RH}$ is a measure of the maximum temperature produced during reheating. 
\par
The total (number) abundance of Kaluza-Klein gravitons is given by the sum of the thermal and non-thermal abundance, given in Eq. \ref{Eq:KK} and Eq. \ref{Eq:NTKK} respectively, summed over all modes. The total (energy) abundance is similarly given by 

\begin{equation}
\rho_{KK} = \int_{0}^{\infty} dm \; m (n_{m,thermal} + n_{m,reheat})
\end{equation}

As before, it is assumed that the modes are so closely spaced that the sum over modes with the same mass can be replaced with an integral over the surface of a sphere, and the sum over modes with different masses can be replaced with an integral over the mass.

\subsubsection{Diffuse $\gamma$-ray Background}\index{$\gamma$-ray background}
\par
Since these gravitons can have sufficiently long lifetimes that they still exist in the Universe at present, the decay of these modes can be detected in satellite experiments. Some of these gravitons will decay to photons, with the decay rate \cite{Han:1998sg}

\begin{equation}
\Gamma_{dec} (m) = \frac{m^3}{80 \pi M_{PL}^2}
\end{equation}

\noindent
producing a flux of extragalactic $\gamma$-rays\index{$\gamma$-rays!extragalactic} which can be observed by several satellite based experiments. 

\par
Using the measurements from EGRET\index{EGRET} \cite{egret} and COMPTEL\index{COMPTEL} \cite{comptel}, a lower limit on the Planck mass, $M_*$, was calculated\footnote{
These results assume that $T_{MAX} \gtrsim 1 \; GeV$, while $T_{RH} \sim 0.7 \; MeV$, although it is possible that the difference between these two temperatures could be higher resulting in stronger constraints. It has also been argued that if $T_{MAX}$ is too large, the KK modes produced would be heavier and therefore decay too early to be detected at present \cite{Macesanu:2004gf}.} \cite{Hannestad:2001nq}. The results are given in Table \ref{Table::ExtraDimensionBounds}. This also places an upper bound on the size of the extra dimensions from the relation,

\begin{equation}
\frac{R}{1 \; mm} \lesssim 2 \times 10^{31/d - 16}  \left( \frac{1 \; TeV}{M_*}\right)^{1+2/d}
\end{equation}

\noindent
From the table, it can be seen that the constraints on the size of extra dimensions from this method are already several orders of magnitude better than from gravity experiments\index{gravity experiments}. As before, it should also be noted that these bounds assume that the extra dimensions form a torus, whereas other topologies and geometries are also possible and could avoid these bounds.

\begin{table}[h]
\begin{center}
\begin{tabular}{|c|c|c|c|c|c|}
\hline
& $d=2$ & $d=3$ &$d=4$&$d=5$& $d=6$ \\ \hline
 $M_*$ & 167 TeV & 21.7 TeV & 4.75 TeV & 1.55 TeV 
& $< 1$  TeV$$
\\ \hline
R& 22 nm & $2.5 \times 10^{-2}$ nm & 1.1 pm & 0.17 pm & $>$ 0.029 pm 
\\ \hline \hline
 $M_*$ & 700  TeV & 25.5  TeV & 2.8 TeV 
& 0.57 TeV & 0.17 TeV \\
\hline R &$1.3 $ nm &$1.9 \times 10^{-2}$ nm &2.4  pm & 0.70 pm &$0.16$ pm
\\ \hline
\end{tabular}
\end{center}
\caption{\label{Table::ExtraDimensionBounds} Bounds on the size of nonwarped extra dimensions and the reduced Planck mass from astrophysical experiments, with the first two lines representing bounds from the $\gamma$-ray background\index{$\gamma$-ray background} and the second two lines representing bounds from neutron stars\index{neutron stars}. The size of the extra dimensions is given as an upper bound, while the Planck mass given is a lower bound. 
}
\end{table}

\subsubsection{Neutron Stars  \& Supernovae}\index{neutron stars}\index{supernovae}
\par
In addition to the Kaluza-Klein gravitons produced in the early Universe, it is also expected that high energy astrophysical processes will produce a significant abundance of Kaluza-Klein gravitons. In particular, it is expected that a fraction of the energy released in a supernova\index{supernovae} will be in the form of KK modes. The immediately imposes a constraint on the size of the extra dimensions, since the energy loss will affect the flux of neutrinos and other particles emitted by the supernova\index{supernovae}. For SN 1987A\index{supernovae!SN 1987A}, observations of the neutrino flux restrict the fraction of energy lost to KK modes to be $f_{KK} \lesssim 0.5$ \cite{Cullen:1999hc,Barger:1999jf,Hanhart:2001fx}, which corresponds to $M_* \gtrsim 8.9 \; TeV$ for\footnote{For the purpose of clarity, only the constraints on the case of $d=2$ are presented. The results for higher dimensions are given in Ref \cite{Hannestad:2003yd}} $d=2$. The bounds can be strengthened further by using the EGRET data to search for sources of $\gamma$-rays created in the decay or annihilation of the KK gravitons surrounding all cosmic supernovae\index{supernovae}, with $f_{KK} \lesssim 0.5 \times 10^{-2}$ or $M_* \gtrsim 28 \; TeV$ for $d=2$ \cite{Hannestad:2001jv,Hannestad:2003yd}.

\par
It is also possible to search for the effects of Kaluza-Klein gravitons in nearby neutron stars\index{neutron stars}. During a supernova\index{supernovae}, a large number of KK modes\index{Kaluza-Klein gravitons} are produced but they tend to have small kinetic energies, and as such are trapped in the gravitational potential of the supernova\index{supernovae} core. The final result is a neutron star\index{neutron stars} surrounded by a halo of KK gravitons\index{Kaluza-Klein gravitons}. Since these modes are decaying and annihilating, this also results in a flux of $\sim O(100 \; MeV)$  neutrinos, electrons/positrons, and $\gamma$-rays. Therefore experiments designed to search for $\gamma$-ray fluxes from localized sources, such as GLAST\index{GLAST}, will be able to detect the effects of this halo, with the strongest bounds arising from nearby neutron stars\index{neutron stars}. Existing data from EGRET\index{EGRET} can already be used is this manner to restrict $M_* \gtrsim 54 \; TeV$ for $d=2$ \cite{Hannestad:2001xi,Hannestad:2003yd}.

\par
The Kaluza-Klein graviton\index{Kaluza-Klein gravitons} halo also affects the neutron star\index{neutron stars} itself. The same decays and annihilations which produce an observable flux at Earth will also produce a flux of particles and $\gamma$-rays\index{$\gamma$-rays} that reheat the neutron star\index{neutron stars} and prevent it from cooling. By observing the thermal emissions of pulsars, this effect can be used to place strong bounds on the ADD\index{extra dimensions!ADD model} model of $M_* \gtrsim 700 \; TeV$ for $d=2$  \cite{Hannestad:2001xi,Hannestad:2003yd}. The bounds for higher numbers of dimensions and the corresponding upper bounds on the size of the extra dimensions are given in Table \ref{Table::ExtraDimensionBounds}.\footnote{It should be noted that there is some uncertainty in the production of KK modes in supernovae, and various authors have produced different constraints, as summarized in Ref. \cite{Kumar:2007gb}}

\par
From Table \ref{Table::ExtraDimensionBounds}, it can be observed that the constraints from neutron stars\index{neutron stars}, in which the KK-modes are produced through neutron-neutron scattering, are better than those from the $\gamma$-ray background\index{$\gamma$-ray background},in which the KK-modes are thermally produced\index{Kaluza-Klein gravitons!thermally produced}, for $d=2$ and $d=3$ while for higher dimensions the $\gamma$-ray background\index{$\gamma$-ray background} is stronger. This difference is due to the fact that $\gamma$-rays are produced by thermal KK-modes while neutron stars\index{neutron stars} produce modes via neutron-neutron scattering which have different  dependencies on the number of dimensions.

\subsection{Collider Constraints}
\subsubsection{Existing Experiments}
\par
Aside from astrophysics experiments, Kaluza-Klein modes\index{Kaluza-Klein gravitons} can also be produced in collider experiments. 
In the ADD\index{extra dimensions!ADD model} model all particle processes can emit gravitons, however due to the weak gravitation couplings the rates of graviton emission are proportional to inverse powers of the Planck energy $M_{PL} = 1.2 \times 10^{19} \; GeV$. As such it is not possible to search for events with a specific number of missing gravitons.

\par
However it is possible to search for the missing energy due to the emission of large numbers of gravitons. Although the probability of emitting a single graviton is small, in the ADD\index{extra dimensions!ADD model} model the number of graviton modes with energy less than E is $\sim (R E)^d$. For reasonable energies, this number is quite large and as such the probability of producing any gravitons is significant. At the Tevatron\index{Tevatron}, searches have been conducted for the reaction $p \overline{p} \to {\rm jet} + \cancel{E}$ \cite{Abulencia:2006kk} while LEP searched for the reactions $e^+e^ \to \gamma+\cancel{E}\cite{ADLOLEWG2004}, Z^0 + \cancel{E}$ \cite{Acciarri:1997im}. The combined results from LEP give the limits in Table \ref{Table:XDcollider} at $95\%$ c.l. while the limits from the Tevatron\index{Tevatron} are given in Table \ref{Table:CDFbound}.

\begin{table}[h]
\begin{center}
\begin{tabular}{|c|c|c|c|c|c|}
\hline
d & 2 & 3& 4 & 5 & 6 \\
\hline
$M_*$ & 1.60  TeV &1.20 TeV& 0.94 TeV& 0.77 TeV& 0.66 TeV \\ \hline 
R & 250 $\mu m$ & 3.2 nm & 12 pm & 0.45 \; pm & 50 \; fm \\ \hline
\end{tabular}
\end{center}
\caption{\label{Table:XDcollider} Lower bounds on $M_*$ and upper bounds on R, due to missing energy experiments at LEP\index{LEP} through the reaction $e^+e^- \to \gamma + \cancel{E}$. \cite{ADLOLEWG2004} }
\end{table}

\begin{table}[h]
\begin{center}
\begin{tabular}{|c|c|c|c|c|c|}
\hline
d & 2 & 3& 4 & 5 & 6 \\
\hline
$M_*$ & 1.18 TeV & 0.99 TeV& 0.91 TeV & 0.86 TeV& 0.83 TeV \\ \hline 
R &  350 \; $\mu m$ & 3.6 \; nm & 11 \;pm & 0.35 \; pm & 34 \; fm\\ \hline
\end{tabular}
\end{center}
\caption{\label{Table:CDFbound}Lower bounds on $M_*$ and upper bounds on R, due to missing energy experiments at Tevatron\index{Tevatron!CDF} through the reaction $p\bar{p} \to jet + \cancel{E}$. \cite{Abulencia:2006kk} }
\end{table}

\par
The constraints on the ADD\index{extra dimensions!ADD model} model are stronger at LEP\index{LEP} for $d=2,3,4$, and stronger at Tevatron for $d=5,6$, due to different dependences on the number of dimensions for the two processes studied. However both of these collider results are weaker than the astrophysics constraints on the ADD\index{extra dimensions!ADD model} model, with the exception of $d=6$ for which the CDF\index{Tevatron!CDF} at the Tevatron\index{Tevatron} provides the strongest constraints.

\subsubsection{Future Experiments}

\par
The ADD\index{extra dimensions!ADD model} model will also be probed at the LHC\index{Large Hadron Collider} (and the ILC\index{International Linear Collider}) using several channels. Since the Standard Model Higgs field couples to the Kaluza-Klein gravitons, it may be possible for the ADD\index{extra dimensions!ADD model} model to be probed using the invisible Higgs decays outlined in Section \ref{Section:ColliderS} \cite{Battaglia:2004js,Battaglia:2005rn}. As shown in Ref \cite{Battaglia:2004js}, it is also possible to probe the ADD\index{extra dimensions!ADD model} model using the mixing of a Higgs boson and a Kaluza-Klein graviscalar\index{Kaluza-Klein gravitons!radions}. However the constraints from these channels at future colliders are still expected to be weaker than those from existing satellite experiments, particularly for $d=2,3$ in which the scale of gravitation interactions,$M_*$, is expected to be above the energies probed at these colliders. 
\par
Another channel which can be used at the LHC\index{Large Hadron Collider} is dimuon production, in which virtual graviton exchange produce an observable effect in the production of muon pairs\index{muons} with large invariant mass. This channel can provide the limits given in Table \ref{Table:LHC-ADD} and Table \ref{Table:LHC-ADD2} \cite{Golutvin:2005ev}. Using this channel, the LHC\index{Large Hadron Collider} bounds for $d=2,3,4$ are weaker than those from the diffuse $\gamma$-ray background, as given in Table \ref{Table::ExtraDimensionBounds}, even for very high luminosity. However the LHC\index{Large Hadron Collider} could probe for extra dimensions with $d=5,6$ with luminosity as low as ${\mathcal L} = 10 \; fb^{-1}$, as the existing astrophysics bounds are weak for higher numbers of dimensions.

\begin{table}
\begin{center}
\begin{displaymath}
\begin{array}{|c|c|c|c|c|}
\hline
& \quad {\mathcal L} = 10 \; fb^{-1} \quad & \quad {\mathcal L} = 30 \; fb^{-1} \quad & \quad {\mathcal L} = 100 \; fb^{-1} \quad & \quad {\mathcal L} = 300 \; fb^{-1} \quad \\ \hline \hline 
d=3 & 5.0 \; TeV & 6.1 \; TeV & 7.5 \; TeV & 8.6 \; TeV \\ \hline 
d=6 & 4.2 \; TeV & 4.8 \; TeV & 5.8 \; TeV & 6.7 \; TeV \\
\hline
\end{array}
\end{displaymath}
\end{center}
\caption{\label{Table:LHC-ADD} The values of $M_*$ which provide $5\sigma$ signal from dimuon production at the LHC\index{Large Hadron Collider}, for $d=3$ and $d=6$.}
\end{table}

\begin{table}
\begin{center}
\begin{displaymath}\begin{array}{|c|c|c|c|c|}
\hline
& \quad {\mathcal L} = 10 \; fb^{-1} \quad & \quad {\mathcal L} = 30 \; fb^{-1} \quad & \quad {\mathcal L} = 100 \; fb^{-1} \quad & \quad {\mathcal L} = 300 \; fb^{-1} \quad \\ \hline \hline 
d=3 & 0.29 \; nm & 0.21 \; nm & 0.15 \; nm & 0.12 \; nm \\ \hline 
d=6 & 4.3 \; fm & 3.6 \; fm & 2.8 \; fm & 2.3 \; fm \\
\hline
\end{array}\end{displaymath}
\end{center}
\caption{\label{Table:LHC-ADD2} The values of R which provide $5\sigma$ signal from dimuon production at the LHC\index{Large Hadron Collider}, for $d=3$ and $d=6$.}
\end{table}

\par
It has also been suggested that colliders could search for extra dimensions through the production of black holes\index{black holes} (for a review, see eg Ref. \cite{Cavaglia:2002si,Kanti:2004nr,Cavaglia:2007ir}). The production of sub-atomic black holes\index{black holes} is not well understood, however it is believed that they can form when the energy in the collider exceeds the Planck scale. Although this will not happen for $M_{PL} \simeq 1.2 \times 10^{16} \; TeV$ in the normal four-dimensional spacetime, the ADD\index{extra dimensions!ADD model} model reduces the Planck mass to the TeV scale, which is accessible.   
\par
Although the production of black holes\index{black holes} in colliders is not well understood, or even certain to occur, the energy scales which can be probed by colliders is limited. In order to produce black holes\index{black holes} at the LHC\index{Large Hadron Collider}, the Planck mass would have to be $M_* \lesssim O(10 \; TeV)$, which from astrophysics constraints on the ADD\index{extra dimensions!ADD model} model means that only $d=4,5,6$ could be probed in this way.

\par
The next generation of collider experiments has the potential to probe an interesting region of parameter space in the ADD\index{extra dimensions!ADD model} model. For low numbers of dimensions, $d=2,3$ the range of Planck masses which can be studied is already excluded by existing astrophysics bounds. For $d=4,5$ the LHC\index{Large Hadron Collider} can explore regions which are allowed by the neutron star\index{neutron stars} bounds, however as will be shown in Section \ref{Section:XDnew}, this range is excluded by nucleosynthesis and by galactic positron\index{galactic positrons} production \footnote{The bounds from these two mechanisms could be weakened sufficiently to allow LHC\index{Large Hadron Collider} to detect extra dimensions, however this requires some tuning of the models.}. For $d=6$, the LHC\index{Large Hadron Collider} is expected to be able to probe further than any existing astrophysics bounds.

\input{chapter3b.tex}

%% file: chapter3b.tex
\section{New Constraints \label{Section:XDnew}}
\vspace{16pt}
\subsection{Nucleosynthesis Constraints \label{Section:BBNEX}}

\par
The presence of extra dimensions, and in particular the decay of the Kaluza-Klein modes\index{Kaluza-Klein gravitons}, will also affect the relative abundance of the light nuclei predicted by BBN\index{Big Bang Nucleosynthesis}\index{nucleosynthesis}. By comparing the observed abundances with the abundances predicted by BBN\index{Big Bang Nucleosynthesis} in the presence of KK gravitons\index{Kaluza-Klein gravitons}, the ADD model\index{extra dimensions!ADD model} can be further constraints. In this section, I derive and present the resulting limits, as previously published by the author and collaborators in Ref \cite{Allahverdi:2003aq}. 

\par
In the standard cosmology, it is believed that the early universe experienced a period of rapid expansion known as inflation\index{inflation}. This expansion is generated by the presence of a field, referred to as the {\it inflaton}\index{inflation}, which then decays into the Standard Model fields and reheats the Universe. 

\par 
However if the Universe also contains extra dimensions, then some fraction of the energy released by inflaton decay will be in the form of Kaluza-Klein gravitons\index{Kaluza-Klein gravitons}. Since the KK modes interact through gravity, they couple to the mass and energy of the inflaton, and as such are relatively insensitive to the precise details of the inflaton model\index{inflation}. This also requires that every decay channel of the inflation, $\phi \to \psi_i \psi_j$, be paired with a KK-mode producing channel, $\phi \to \psi_i \psi_j + KK$.
\par
The KK modes\index{Kaluza-Klein gravitons} produced in inflaton decays\index{inflation} are expected to have lifetimes of several years, which results in their decay occurring after all light elements have been generated by Big Bang Nucleosynthesis\index{Big Bang Nucleosynthesis}\index{nucleosynthesis}. As a result of this delay in the release of some of the inflaton\index{inflation} energy, the relative abundances of the light elements will be distorted compared to models without extra dimensions, and observed abundances can be then used to restrict the nature of the higher dimensions. 

\par
\begin{figure}[t]
\begin{center}
\epsfig{file=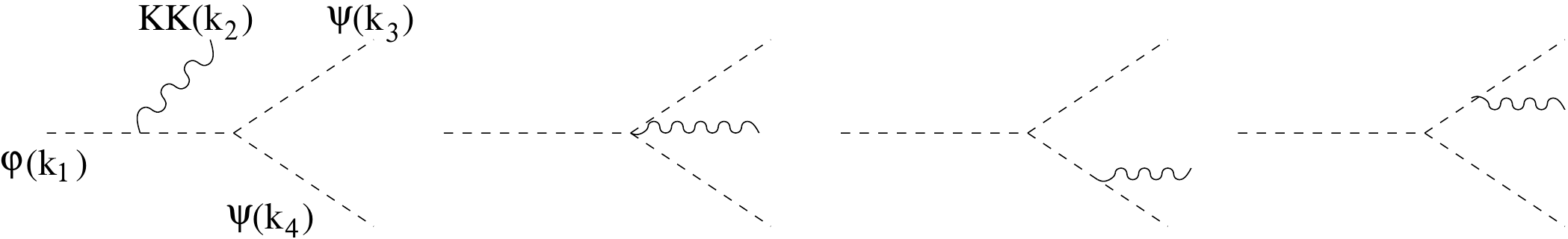, width=\textwidth, angle=0}
\end{center}
\vskip .1in
\caption{\label{fig:3body} 
The Feynman diagram for emission of KK modes\index{Kaluza-Klein gravitons} in the decay of the inflaton into a pair of the scalars $\psi$. }
\end{figure}

For the purpose of this calculation, it will be assumed that the inflaton\index{inflation} decays primarily to Higgs bosons\index{Higgs}. The Feynman diagrams which contribute to this decay are given in Figure \ref{fig:3body}. As I will demonstrate in this section, the constraints on extra dimensions which result from this calculation depend on the energy released rather than the coupling constants. Therefore these constraints are relatively insensitive to which decays dominate. 

\par
Although there is a large range of masses and lifetimes for the KK-modes which are produced, any neutral relics which decay after recombination\index{recombination} $(\tau_{KK} \gtrsim 10^{12} \;sec)$ will leave a signature in the $\gamma$-ray background\index{$\gamma$-ray background}, which provides a very strong bound \cite{Ellis:1990nb,Kribs:1996ac}. For neutral relics with shorter lifetimes, it is expected that constraints from the dissociation of light nuclei will provide a stronger bound \cite{Ellis:1990nb,Cyburt:2002uv}. From Ref. \cite{Cyburt:2002uv} the bound on the energy released by these decays is

\begin{equation}
\frac{\sum n_{KK} m_{KK}}{s \cdot } \approx 2\times 10^{-12} \; GeV
\end{equation}

\noindent
for $\tau_{KK} \gtrsim 10^{8} s$, while for $10^8 s > \tau_{KK} > 10^6 s$ the bound is

\begin{equation} \label{Eq:BBNbound}
\frac{\sum n_{KK} m_{KK}}{s \cdot } \approx 2\times 10^{-9} \; GeV
\end{equation}

\begin{figure}[t]
\begin{center}
\epsfig{file=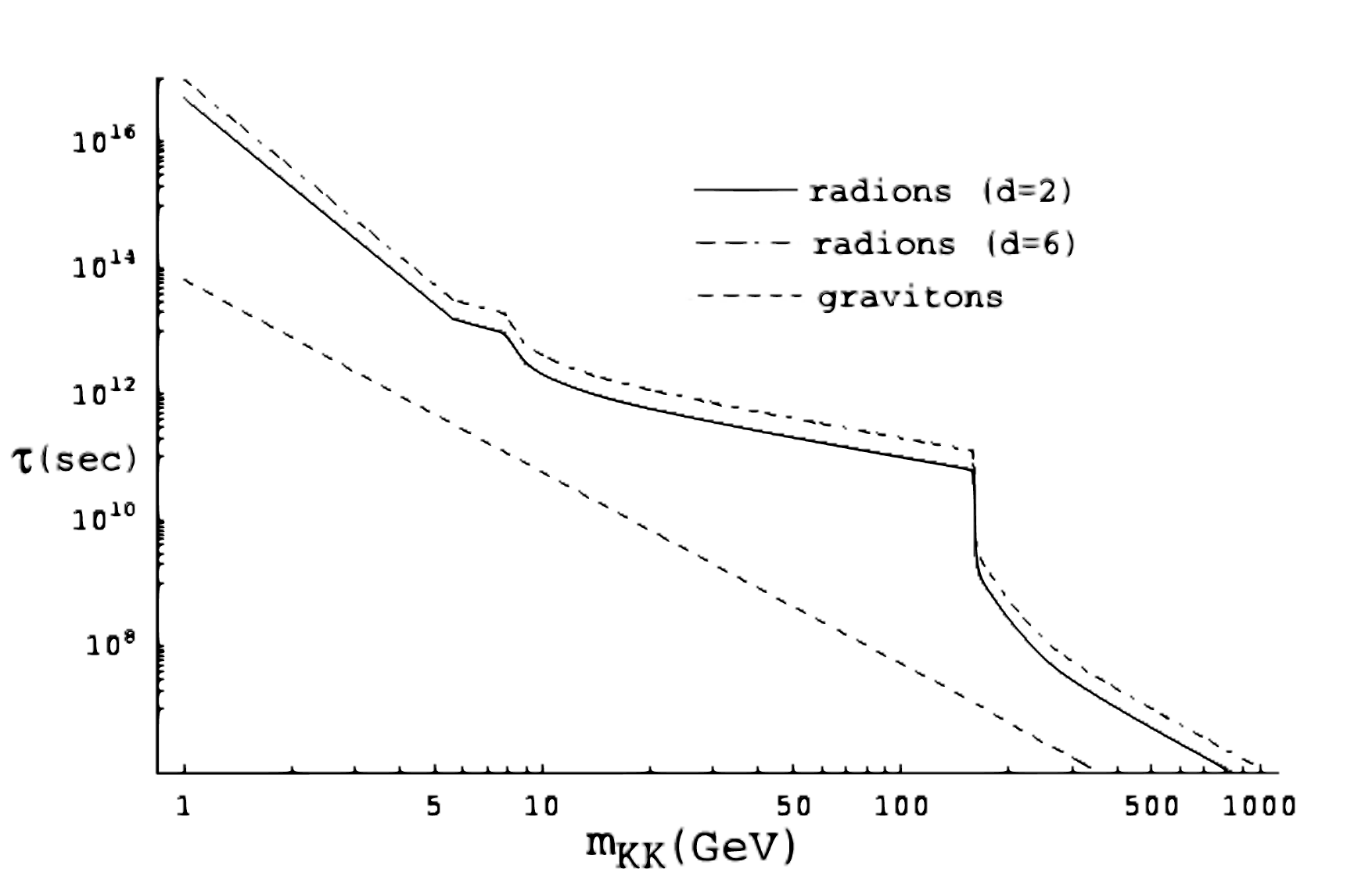, width=\textwidth, angle=0}
\end{center}
\caption{\label{fig:lifetimes} 
The lifetime of the KK gravitons\index{Kaluza-Klein gravitons} as a function of mass is
indicated by the dashed line. The lifetime of the KK
radions\index{Kaluza-Klein gravitons!radions} is shown for $d=2$ (solid line) and for $d=6$ (dashed-dotted line). 
} 
\end{figure}

\noindent
For KK modes\index{Kaluza-Klein gravitons} with lifetimes in this range, the masses are in the electroweak energy scale, and the decays are to Standard Model particles. The lifetime of the graviton modes is \cite{Han:1998sg}

\begin{equation}
\tau_{gr} \simeq 5 \times 10^{4} \;{\rm sec} \left( \frac{1 \; {\rm TeV}}{m_{KK}} \right)^3
\end{equation}

\noindent
while for the radions the lifetime is

\begin{equation}
\tau_{rad} \simeq 2(d+2) \times 10^{5} \; {\rm sec} \left( \frac{1\; {\rm TeV}}{m_{KK}}\right)^3
\end{equation}

\noindent
The precise lifetimes are plotted in Figure \ref{fig:lifetimes}.

\par
By using the requirement that $\tau \lesssim 10^8 \; sec$ for the KK mode to have a significant effect on BBN\index{Big Bang Nucleosynthesis}\index{nucleosynthesis}, a maximum mass can be calculated

\begin{equation}
m_{max}^{\rm gr} \simeq 80 {\rm GeV}; 
\qquad 
\begin{array}{l c c c c c c l }
m_{max}^{\rm rad}  & \simeq &  200, & 210, & 225, & 240, & 250 &
{\rm GeV},  
\\[.3ex] 
\text{for}\, ~ d & = & 2, & 3, & 4, & 5, & 6, & 
\end{array}
\label{mmax}
\end{equation}

\begin{figure}[t]
\begin{center}
\epsfig{file=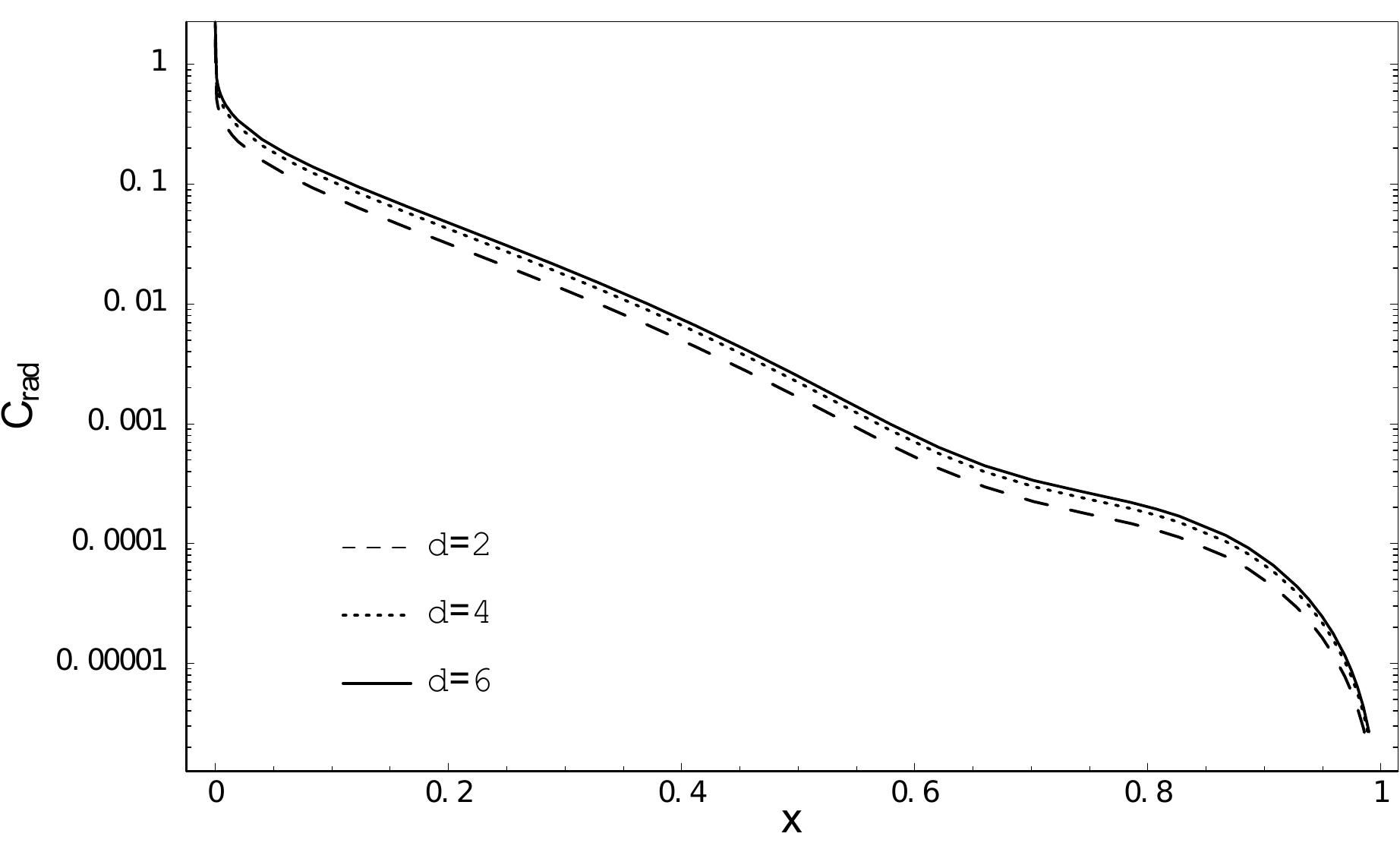, width=\textwidth, angle=0}
\end{center}
\caption{\label{fig:LogC} 
The dependence of $C_{{\rm rad}}(x)$, as defined
in \eqref{decaykk}, on the ratio $x = m_{KK}^{\rm rad}/m_{\phi}$. Separate plots are given for the number of extra
dimensions $d= 3,4,$ and $6$. }
\end{figure}

\par
The density of KK modes\index{Kaluza-Klein gravitons} is calculated by first calculating the ratio of emitted KK modes to inflatons\index{inflation}. This ratio is parameterized as

\begin{equation}\label{decaykk}
\frac{n^i_{KK}}{n_{\phi}} = C_i(x) \frac{m_{\phi}^2}{M_{PL}^2} \quad \quad x = \frac{m^{i}_{KK}}{m_{\phi}} \quad \quad i = {\rm gr,rad}
\end{equation}

\noindent
and is equal to the three body decay width involving a single KK mode divided by the total decay width of the inflaton. For the purpose of this calculation, it is assumed that the inflaton decay width is dominated by the two-body decays and that the ratio using Standard Model particles is similar to the ratio using only scalar fields in the decay products. The diagrams for the decay of the inflaton are given in Figure \ref{fig:3body}. The functions $C_i(x)$ which are calculated in this manner are plotted in Figure \ref{fig:LogC}. 

\par
The entropy released by inflaton\index{inflation} decay satisfies the relation

\begin{equation}
\frac{n_{\phi}}{s} \simeq \frac{3 T}{m_{\phi}}
\end{equation}

\noindent
which can be combined with the results of \eqref{decaykk} to give the energy density of the KK modes\index{Kaluza-Klein gravitons},

\begin{equation}
\sum_{KK}\frac{m^{i}_{KK} n_{KK}^{i}}{s} \simeq  3 B_i(x) \frac{3 T m_{\phi}}{M_*}  \left(  \frac{m^i_{max}}{M_*}\right)^{d+1}, \quad \quad x = \frac{m_{max}^i}{m_{\phi}}
\end{equation}

\begin{figure}[t]
\begin{center}
\epsfig{file=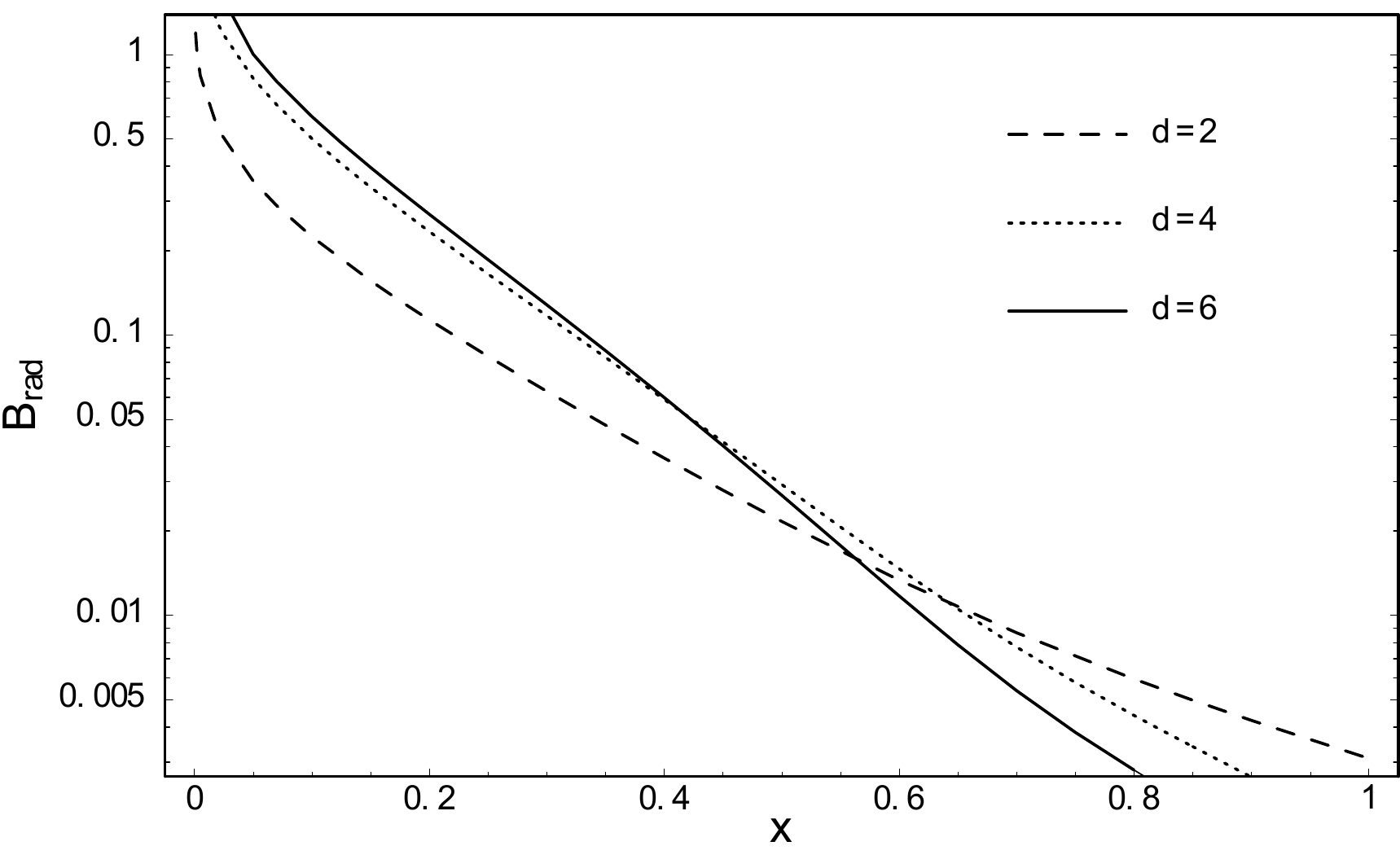, width=\textwidth, angle=0}
\end{center}
\caption{\label{fig:LogB} 
The dependence of $B_{{\rm rad}}(x)$ on the ratio $x = m_{max}^{\rm rad}/m_{\phi}$. Separate plots are given for the number of extra
dimensions $d= 3,4,$ and $6$. }
\end{figure}

\noindent
where $B_i(x) = \sum_{KK} m^i_{KK} C_i(m_{KK}^i/m_{\phi}) $. Since the KK mode masses are very close together, this sum can be replaced with an integral over $m_{KK}^i$,  

\begin{equation}
B_i(x) = \frac{S_d}{x^{d+1}} \int_0^x dy C_i(y) y^d 
\end{equation}

\noindent
where $S_d = 2 \pi^{d/2} /\Gamma(d/2)$ is the surface area of a d-dimensional sphere. The function $B_{rad}(x)$ is plotted in Figure \ref{fig:LogB}.

\begin{table}[h]
\begin{center}
\renewcommand{\arraystretch}{1.5}
\begin{tabular}{|c||c|c|c|c|c|}
\hline 
$m_\phi$ & $d=2$ & $d=3$ &$d=4$&$d=5$& $d=6$
\\ \hline \hline
1 TeV & 35 TeV  & 13 TeV & 7.1 TeV & 4.5 TeV & 2.8 TeV
\\ 
2 TeV & 47 TeV  & 17 TeV & 9.1  TeV & 5.7 TeV & 3.4 TeV
\\
$M_*$ & 220 TeV  & 42 TeV &  15 TeV &
7.9 TeV &  4.0 TeV
\\ \hline
\end{tabular}
\caption{\label{Table::EDBBN1} Lower bounds on the reduced Planck mass from nucleosynthesis\index{nucleosynthesis}\index{Big Bang Nucleosynthesis} constraints as a function of $m_{\phi}$.}
\end{center}
\end{table}

\begin{table}[h]
\begin{center}
\renewcommand{\arraystretch}{1.5}
\begin{tabular}{|c||c|c|c|c|c|}
\hline 
$m_\phi$ & $d=2$ & $d=3$ &$d=4$&$d=5$& $d=6$
\\ \hline \hline
1 TeV & $0.52 \mu m$  & 60 pm & 0.59 pm & 39 fm & 7.4 fm
\\ 
2 TeV & $0.29 \mu m$  & 38 pm & 0.41 pm & 28 fm & 5.7 fm
\\
$M_*$ & $0.013 \mu m$ & 8.5 pm & 0.19 pm &
18 fm &  4.6 fm
\\ \hline
\end{tabular}
\caption{\label{Table::EDBBN2} Upper bounds on the size of nonwarped extra dimensions in the ADD model from nucleosynthesis\index{nucleosynthesis}\index{Big Bang Nucleosynthesis} constraints as a function of $m_{\phi}$.}
\end{center}
\end{table}

\begin{figure}
\begin{center}
\psfig{file=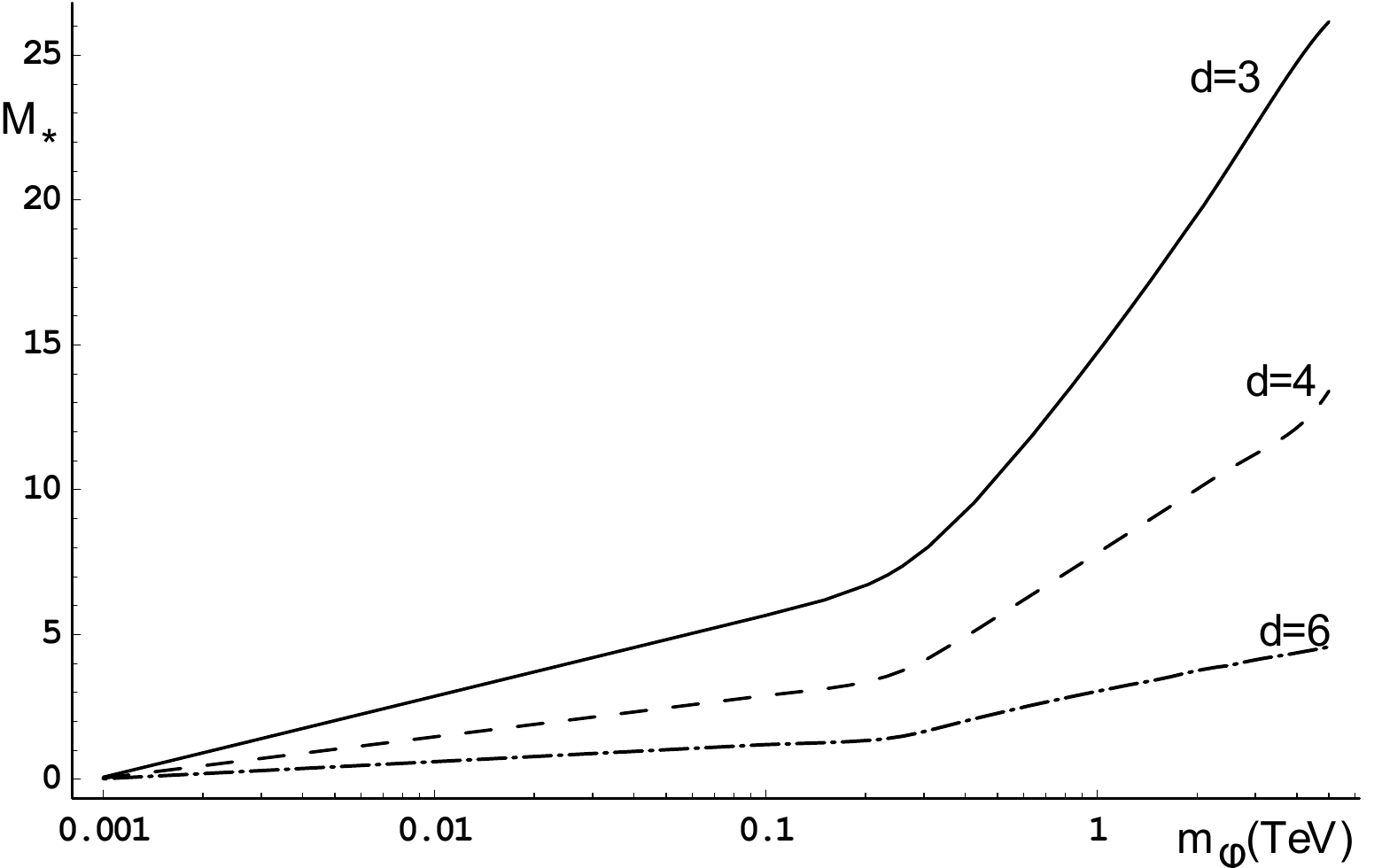, width=0.95\textwidth,angle=0}
\end{center}
\caption{\label{Figure:ADD-inflaton} Lower bounds on $M_*$ for d=3,4,6 and for a range of inflaton\index{inflation} masses.}
\end{figure}

\par
In the limit of $m_{max} \ll m_{\phi}$, the energy density of KK modes can be approximated as

\begin{equation}
\frac{\sum_{KK} m_{KK}^i n_{KK}^i}{s} \simeq \frac{2 S_d}{(d+1)\pi} \frac{T_R m_{\phi}}{M_*} \left[ \frac{d}{d+2} \left( \frac{m_{max}^{rad}}{M_*} \right)^{d+1} + 2 \left(\frac{m_{max}^{gr}}{M_*} \right)^{d+1} \right] \ln \left( \frac{m_{\phi}}{m_{max}} \right)
\end{equation}

\noindent
where \we take $\ln m^{gr}_{max} \approx \ln m^{rad}_{max}$, and $T_R \simeq 0.7 \; MeV$ is the reheat temperature\footnote{Recently it has been suggested that any reasonable model of inflation would require $T_R \geq 4 \; MeV$\cite{Hannestad:2004px}, which would further strengthen these bounds}\index{reheat temperature}. The constraint in Eq \ref{Eq:BBNbound} then places a constraint on $M_*$, and therefore on the scale of the extra dimensions. These limits are given in Table \ref{Table::EDBBN1} and Table \ref{Table::EDBBN2}. In addition, limits on $M_*$ for d=3,4,6 and for a range of inflaton masses are given in Figure \ref{Figure:ADD-inflaton}.

\par
These constraints are stronger than the previous constraints for $d=4,5,6$. However unlike the previous constraints, these results are model dependent. For heavier inflatons\index{inflation}, the constraints are expected to be even stronger.

\vspace{16pt}
\input{chapter3c-2.tex}

\section{Conclusions}

\par
The possible existence of extra dimensions has been studied for almost a century, with several models proposed to explain a variety of phenomena both within the Standard Model and beyond the Standard Model. The simplest of these is the addition of a number of non-warped space-like dimensions with the topology of a torus, referred to as the ADD model\index{extra dimensions!ADD model}.

\par
In this chapter, I have reviewed the existing bounds on the ADD model from collider searches, from the $\gamma$-ray background, and from the properties of astrophysical objects such as supernovae\index{supernovae} and neutron stars\index{neutron stars}. For $d=2,3,4$ the bounds from neutron stars\index{neutron stars} are the strongest constraints on the ADD model\index{extra dimensions!ADD model}. For higher dimensions of $d=5,6$, astrophysics experiments have provided constraints, but the next generation of high energy colliders such as the LHC\index{Large Hadron Collider} are expected to be able to probe regions of parameter space which are not yet constrained.

\par
I then demonstrated how the existing bounds could be improved upon using Big Bang nucleosynthesis\index{nucleosynthesis}\index{Big Bang Nucleosynthesis}, as originally published by myself and collaborators in Ref \cite{Allahverdi:2003aq}, and using the $511 \; keV$ $\gamma$-ray\index{$\gamma$-rays!511 keV} flux from the galactic core, which has not been previously published.

\par
In the first case, decaying inflatons\index{inflation} lose a fraction of their energy to the production of Kaluza-Klein gravitons\index{Kaluza-Klein gravitons}. The energy stored in these gravitons is then released later in the evolution of the Universe, which effectively reheats the Universe. For KK gravitons\index{Kaluza-Klein gravitons} with masses in the range of $m_{gr} \sim 80 \; GeV$, this reheating occurs before the completion of Big Bang nucleosynthesis\index{nucleosynthesis}\index{Big Bang Nucleosynthesis} and alters the predicted abundance of light nuclei. Since these abundances are tightly constrained by observations, the properties of the extra dimensions are also constrained, although these constraints are dependent on the mass of the inflaton \index{inflation}. For $d\geq4$ these bounds are stronger than previous constraints. However it is expected that the LHC\index{Large Hadron Collider} will be able to probe the case of $d=6$ up to $M_* \sim 6 \; TeV$, which could improve upon the nucleosynthesis bounds of $M_* \gtrsim 3 \; TeV$.

\par
In the second case, the effects of Kaluza-Klein\index{Kaluza-Klein gravitons} gravitons in the modern Universe were studied. These particles are produced thermally\index{Kaluza-Klein gravitons!thermally produced} in the early Universe, and survive long enough to get trapped in the gravitational well of the galaxy. They then decay to $\gamma$-rays and to electron-positron pairs, which subsequently annihilate to $511 \; keV$ $\gamma$-rays\index{$\gamma$-rays!511 keV}, producing a flux of $\gamma$-rays at our solar system. Using existing measurements of this flux from the INTEGRAL\index{INTEGRAL} experiment, the properties of the KK gravitons\index{Kaluza-Klein gravitons} can be constrained, and in turn the properties of the extra dimensions can be constrained. If the reheat temperature\index{reheat temperature} is $T_{RH} \sim 1 \; MeV$, then the resulting bounds are comparable to the bounds derived from other astrophysics experiments. In models with a higher reheat temperature, the constraints from the $511 \; keV$ $\gamma$-ray flux\index{$\gamma$-rays!511 keV} are expected to be stronger than other astrophysics constraints for $d=2$ and are expected to be comparable to the constraints from the $\gamma$-ray background\index{$\gamma$-ray background} for $d \geq 3$. However these bounds are dependent on the energy loss of positrons in the galactic core\index{galactic positrons}, and therefore contain a level of uncertainty.
\par
Although these two methods of probing extra dimensions have been applied to a specific model, it is expected that they can be generalized to other models. It is also clear from these results and from the previous constraints that astrophysics experiments are able to constrain models of extra dimensions better than colliders and other terrestrial experiments for a large extra dimensions with $d=2,3,4$,  and are expected to be competitive with colliders for $d=5,6$.

%% file: chapter3c-2.tex
\subsection{Galactic Positron Constraints \label{Section:PosEX}}

\par
In the previous sections,the effects of Kaluza-Klein excitations\index{Kaluza-Klein gravitons} on the early Universe were considered. However it is also possible that observable effects could be created in the modern Universe by quasi-stable modes,which were produced in the early Universe, and which the ADD model\index{extra dimensions!ADD model} predicts would still exist in the present. 

\par
As outlined in Section \ref{Section::LDM}, measurements by the SPI spectrometer\index{SPI spectrometer} on the INTEGRAL\index{INTEGRAL} satellite \cite{Jean:2003ci,Knodlseder:2005yq} have confirmed previous observations of a flux of $511 \; {\rm keV}$ photons\index{$\gamma$-rays!511 keV} from the galactic center \cite{oldflux1,oldflux2,oldflux3}. These experiments have also determined that the $\gamma$-rays are most likely produced by a diffuse source rather than by a few point sources, which is consistent with a galactic halo composed of dark matter\index{dark matter!galactic} or KK-modes\index{Kaluza-Klein gravitons}. 

\par
In this section I consider the possibility that a significant density of KK gravitons\index{Kaluza-Klein gravitons} could be trapped in the gravitational potential of the galaxy. 
Although these modes are quasi-stable, some will decay 
into electron-positron pairs and into $\gamma$-rays\index{$\gamma$-rays}. This should result in an observable flux of $\gamma$-rays both from direct decay of the KK-modes\index{Kaluza-Klein gravitons} and from subsequent electron-positron annihilation. The observed flux can then be used to constrain the nature and size of the extra dimensions by comparing the predicted flux with the observed flux in the solar system\footnote{During the preparation of this dissertation, \we\ became aware of another paper addressing this issue \cite{Kasuya:2006kj}. However in that paper, only a single modulus field was considered rather than the complete spectrum of KK-modes}.  

\par
Although it is possible to derive constraints from the entire spectrum of galactic $\gamma$-rays, the complexity of such a calculation places it beyond the scope of this dissertation. Therefore only the $511 \; keV$ $\gamma$-ray flux, which is produced by positron production\index{galactic positron excess} in the galaxy, will be considered.
It should also be noted that the constraint considered in this section results from the production of low-energy positrons which lose energy as they travel through the interstellar medium\index{interstellar medium}. It is assumed that the positrons become non-relativistic\footnote{It is also possible to detect relativistic positrons produced by dark matter annihilations or KK-mode\index{Kaluza-Klein gravitons} decays in astrophysics experiments \cite{Baltz:1998xv,Hooper:2004bq,Hooper:2004yc}, with some experiments indicating an unexplained excess of high energy positrons \cite{Barwick:1997ig}. However there are also a number of uncertainties inherent to modeling positrons flux through the galaxy (see eg. Ref \cite{Delahaye:2007fr}), and as such these bounds will not be considered in this \thesis.} within a short distance and annihilate in the galactic bulge. However the rate of energy loss by these positrons and the cut-off energy above which the positrons can no longer contribute to the observed flux are not well understood, and could potentially cause some uncertainty in the final limits. 


\par
For the purpose of this calculation, it can be assumed that the $\gamma$-ray flux\index{$\gamma$-rays} which results from KK decays depends only on the partial decay width and galactic abundance of the modes. It will also be assumed that all positrons produced in the decay of KK-modes\index{Kaluza-Klein gravitons} with mass below a certain cut-off, denoted $m_{max}$, will become non-relativistic and annihilate within the galactic bulge. 
The decay widths for graviton decay to electron-positron pairs 
is given in Ref. \cite{Han:1998sg}, 

\begin{equation}\label{Eq:KKwidth}
\Gamma_{e^+e^-} (m_{KK}) = \frac{m_{KK}^3}{80 M_{PL}^2} 
\end{equation}

\noindent
while the abundance of gravitons\index{Kaluza-Klein gravitons} in the Universe was previously calculated in Ref. \cite{Hall:1999mk} and Ref. \cite{Macesanu:2004gf}. At cosmic scales, the number density of each KK-mode as a function of mass is

\begin{equation}
\begin{split}
n_0(m_{KK}) \simeq & \frac{19 T_0^3}{64 \pi^3 \sqrt{g_*}}  \frac{m_{KK}}{M_{PL}} e^{-t_0/\tau_{KK}} \\ & \times \left( \int_{m_{KK}/T_{RH}}^{\infty} q^3 K_1(q) dq +2 \left(\frac{m_{KK}}{T_{RH}} \right)^{-7} \int_{m_{KK}/T_{MAX}}^{m_{KK}/T_{RH}} q^{10} K_1(q) dq  \right)
\end{split}
\end{equation}


\noindent
where $m_{KK}$ is the mass of the Kaluza-Klein mode,  $T_0$ is the present (neutrino) temperature of the Universe, and $T_{RH}$ is the temperature at which the Universe becomes dominated by radiation.  In this equation, the first term represents KK production by thermal processes which occur during this radiation dominated epoch , while the second term represents KK production during an earlier period of reheating\index{reheating temperature}. The second integral also depends on $T_{MAX}$, which is the maximum temperature at which KK modes are produced during the reheating phase\index{reheating temperature}. In this section, only the two special cases of $T_{MAX} \sim T_{RH}$ and $T_{MAX} \gg T_{RH}$ will be considered.

\par
The total energy density of Kaluza-Klein modes\index{Kaluza-Klein gravitons} is obtained by summing over all masses. However as before, at the relevant energy scales the difference between the masses of neighbouring KK modes\index{Kaluza-Klein gravitons} is small and as such the sum can be replaced with an integral over a d-dimensional sphere. The resulting energy density is

\begin{equation}\label{Eq:KKabundance}
\begin{split}
\rho_G = \sum_{all\;modes} m_{KK}n_0(m_{KK}) \simeq & (1.9\times 10^{22} \; GeV^4) S_{d-1}  \left( \frac{T_{RH}}{M_*} \right)^{d+2} I^{(1)}_d (T_{RH}/T_{max})
\end{split}
\end{equation}

\noindent
where

\begin{equation} \label{Eq:SpInt}
\begin{split}
I_d^{(\sigma)} (\beta ) = \int_{0}^{\infty} dz e^{-t_0/\tau_{KK}} & \left(  \; z^{d+\sigma}   \int_{z}^{\infty} dq \; q^3 K_1(q) \right. \left. + 2 \; z^{d+\sigma-7} \int_{\beta z}^{z} dq \; q^{10} K_1(q) \right)
\end{split}
\end{equation}

\noindent
represents the integral over all modes. 

\par
The bound derived in this section assumes that these KK modes\index{Kaluza-Klein gravitons}, which were formed in the early Universe, have become trapped in the gravitational potential of the galaxy. Using the cosmological KK-mode abundance, 
the distribution of KK modes\index{Kaluza-Klein gravitons} in the galaxy can then be approximated as

\begin{equation}\label{Eq:KKdist}
\rho_{KK}(r) = \frac{\rho_G}{\rho_{DM,cosmic}} \rho_{DM}(r)
\end{equation}

\noindent
where $\rho_{DM,cosmic}$ is the total cosmic dark matter abundance, and

\begin{equation}
\rho_{DM} (r) = \rho_0 exp \left( -\frac{2}{\alpha} \left[ \left(\frac{r}{r_0} \right)^{\alpha} - 1 \right] \right)
\end{equation}

\noindent
with $r_0 = 20h^{-1} kpc$ and $0.1 < \alpha < 0.2$,is the galactic\index{dark matter!galactic} dark matter distribution \cite{Navarro:2003ew}. The constant $\rho_0$ is determined by requiring $\rho_{DM} (8.5 kpc) = 0.3 \; GeV/cm^3$, which is the accepted value of the dark matter density at the Solar system\index{dark matter!Solar System}. In this calculation it is assumed that the KK modes\index{Kaluza-Klein gravitons} are distributed in the halo with the same mass distribution as in the early Universe. In practice the lightest modes are too light to be captured in the halo, however the effect of these missing modes is small. 

\par
\begin{figure}
\begin{center}
\psfig{file=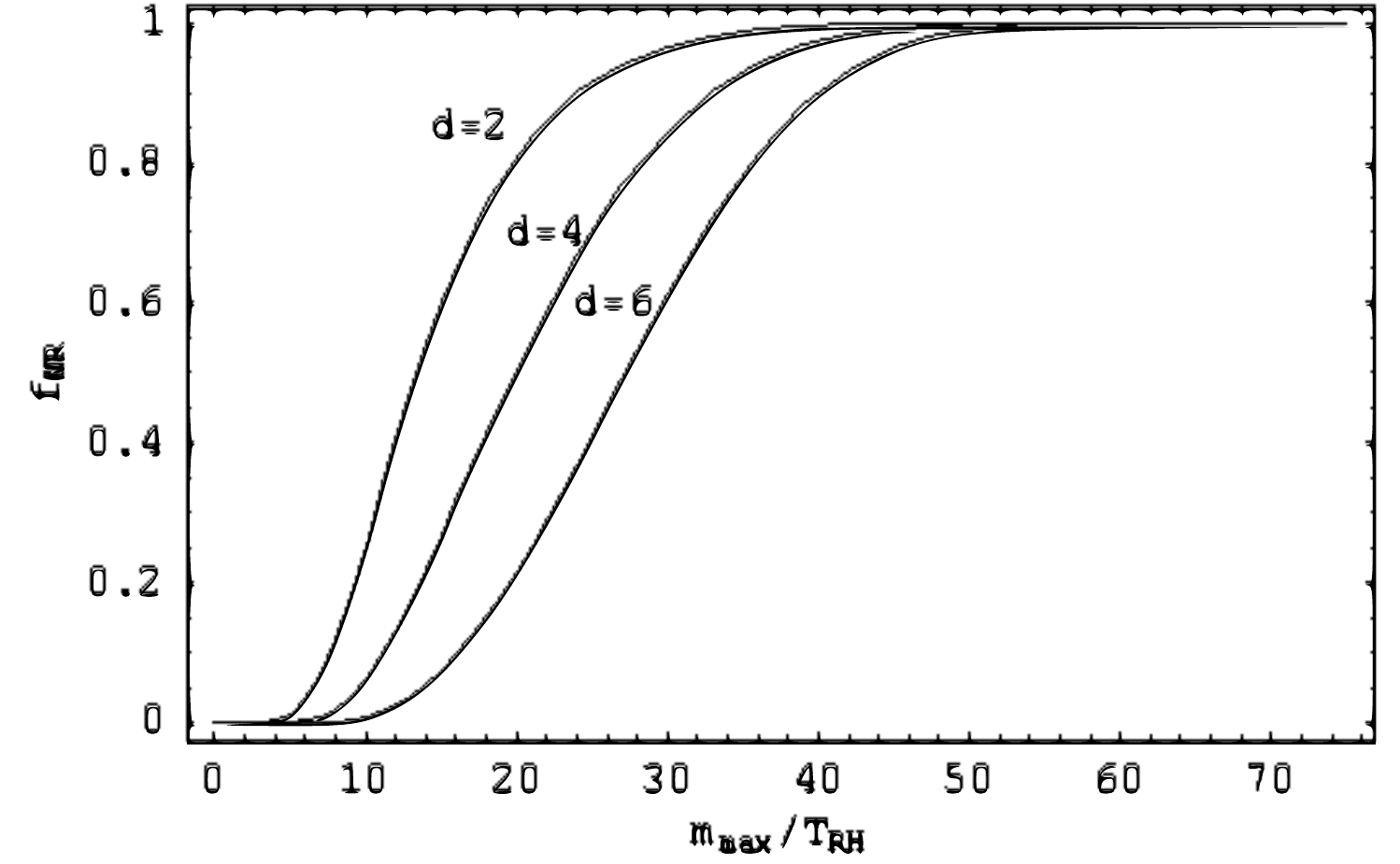,width=\textwidth,angle=0} \\ (a) \\
\psfig{file=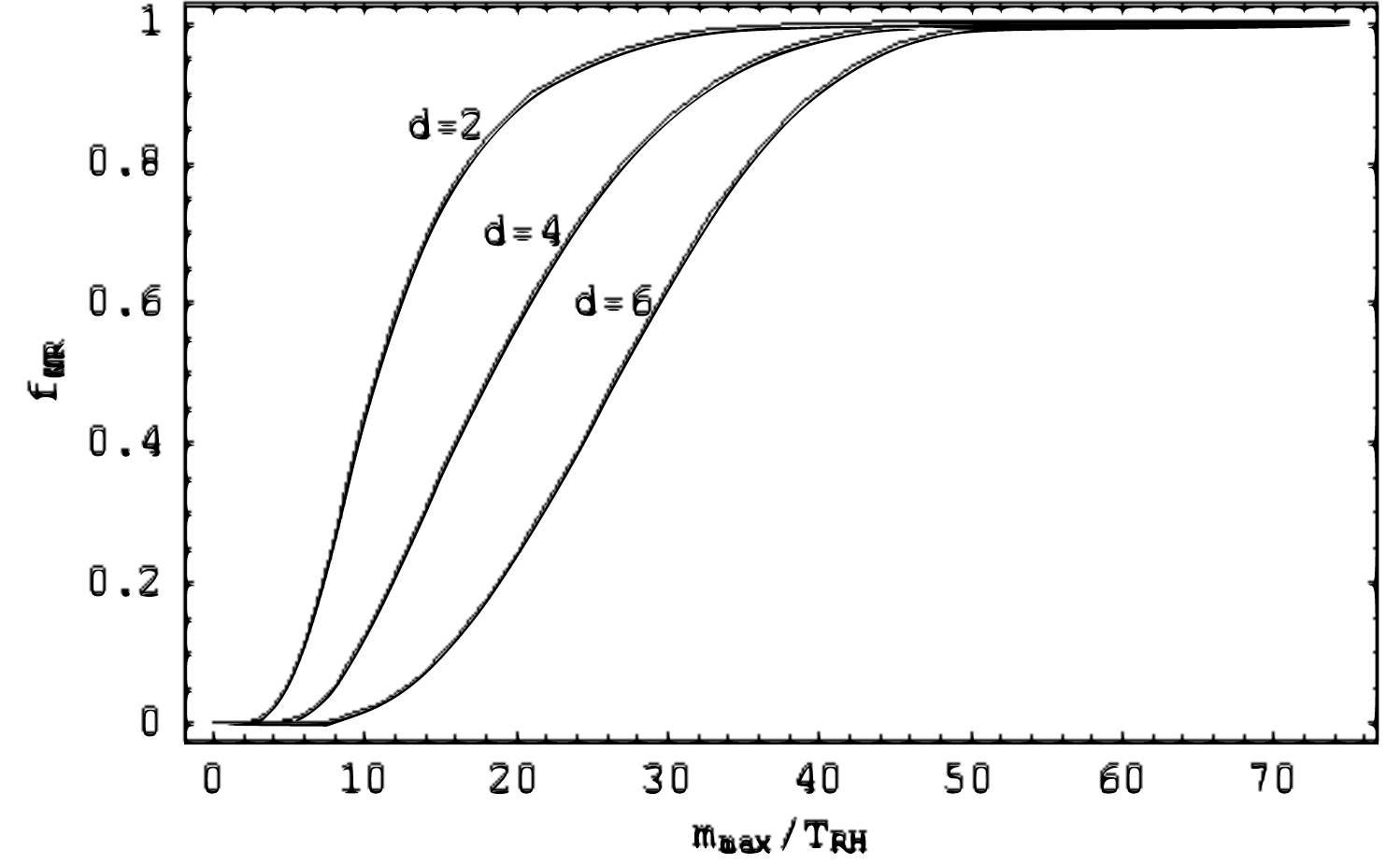,width=\textwidth,angle=0} \\ (b)
\end{center}
\caption{\label{Fig:PosFrac}The fraction of Kaluza-Klein modes with mass below $m_{max}$ for the cases of (a) $T_{MAX} \sim T_{RH}$ and (b) $T_{MAX} \gg T_{RH}$.}
\end{figure}

Using similar methods, the rate of positron production in the Universe can be determined by summing over the partial width for the decay $KK \to e^+e^-$ of each KK mode, 

\begin{equation}
\begin{split}
N_{e^+,cosmic}& =\sum_{all \; modes} \Gamma_{e^+e^-}(m_{KK})n_0(m_{KK}) \\& \simeq  (1.9\times 10^{22} \; GeV^4) S_{d-1}  \left( \frac{T_{RH}}{M_*} \right)^{d+2} \frac{T_{RH}^2}{80 M_{PL}^2} I^{(3)}_d (T_{RH}/T_{max})
\end{split}
\end{equation}



\noindent
and the number density of non-relativistic positrons produced per unit time in the galaxy\index{galactic positron excess} follows from Eq \ref{Eq:KKdist},

\begin{equation} \label{Eq:PosDen}
\begin{split}
N_{511 keV} (r)  & \simeq N_{e^+,cosmic} \frac{ \rho_{DM}(r)}{\rho_{DM,cosmic}}\\&\simeq  (2.57 \times 10^{-3d - 4} \; cm^{-3} s^{-1} )(0.85 + 1.49 \alpha) S_{d-1}  \left( \frac{1000 \; T_{RH}}{M_*} \right)^{d+2}  \\ & \times \left(\frac{T_{RH}}{1 \; MeV}\right)^2 I_d^{(3)}(T_{RH}/T_{MAX}) f_{NR}(m_{max}/T_{RH}) exp \left( -\frac{2}{\alpha} \left[ \left(\frac{r}{r_0} \right)^{\alpha} - 1 \right] \right)
\end{split}
\end{equation}


\noindent
where the integral $I_d^{(3)}$ is given in Eq \ref{Eq:SpInt}, and $f_{NR}(m_{max}/T_{RH})$ is the fraction of Kaluza-Klein modes\index{Kaluza-Klein gravitons} which are lighter than $m_{max}$, and which are assumed to decay to non-relativistic positrons. This fraction is plotted in Figure \ref{Fig:PosFrac} for the cases of $T_{MAX} \sim T_{RH}$ and $T_{MAX} \gg T_{RH}$.

\par
As outlined in Ref \cite{Picciotto:2004rp}, the resulting flux of $511 \; keV$ $\gamma$-rays at the Solar system can be derived by calculating the line-of-sight integral of this density. However it should be noted that not all positrons contribute to the $511 \; keV$ $\gamma$-ray flux. As outlined in Ref \cite{positronium1,positronium2,Beacom:2005qv} positrons produced in the galaxy can form positronium states with electrons before annihilating. The positronium\index{positronium} states only produce $511 \; keV$ photons in approximately $25 \%$ of the annihilations, while the other states decay to three photons each with lower energies. By measuring the flux of $511 \; keV$ photons relative to lower energy photons, the fraction of positrons bound into positronium\index{positronium} states in the galactic core is determined to be $f = 0.967 \pm 0.022$ \cite{Jean:2005af}. As a result, on average each positron produced contributes $2(1-0.75 f) = 0.55$ photons to the $511 \; keV$ flux. Therefore the rate of positron production is expected to be a factor of $\sim 3.6$ larger than would be expected based solely on the $511 \; keV$ flux.

\par
As discussed in the introduction to this section, the $511 \; keV$ flux\index{$\gamma$-rays!511 keV} has been measured by several experiments. The SPI spectrometer on the INTEGRAL\index{INTEGRAL} $\gamma$-ray observatory measured
 an azimuthally symmetric distribution of $511\; keV$ photons with FWHM
 of $\sim 8^\circ \pm 1^\circ$, and total flux $\Phi_{obs} = (1.05\pm 0.06) \times 10^{-3} \; ph \; cm^{-2} \; s^{-1}$ \cite{Jean:2003ci,Knodlseder:2005yq}. Integrating the density in Eq \ref{Eq:PosDen} over
 this solid angle with the line of sight integral gives a total predicted flux of

\begin{equation}
\begin{split}
\Phi_{th}&= (0.86 \times 10^{20-3d} \; cm^{-2} s^{-1} )
(1.4 - 3.8 \alpha) S_{d-1}  \left( \frac{1000 \; T_{RH}}{M_*} \right)^{d+2}  \\ & \times \left(\frac{T_{RH}}{1 \; MeV}\right)^2 f_{NR}(m_{max}/T_{RH}) I_d^{(3)}(T_{RH}/T_{MAX}) 
\end{split}
\end{equation}


\noindent
Comparing this result with the observed $511 \;keV$ $\gamma$-ray flux constrains the Planck mass,

\begin{equation}\label{DMC}
\begin{split}
\left( \frac{M_*}{1000 \; T_{RH}} \right)^{d+2} \gtrsim (0.43 \pm 0.03) \times 10^{23-3d} &  
(1.4 - 3.8 \alpha) S_{d-1} \left(\frac{T_{RH}}{1 \; MeV}\right)^2 \\ & \times f_{NR}(m_{max}/T_{RH})I_d^{(3)}(T_{RH}/T_{MAX}) 
\end{split}
\end{equation}


\noindent
where as before $S_d = 2 \pi^{d/2} / \Gamma(d/2)$, and where it is assumed that the contribution from direct production of $511 \; keV$ photons is negligible. Although it is possible to produce a flux of $\gamma$-rays in this energy range through direct production, the effect is expected to be small and will not significantly alter the constraints given.



\begin{table}
\begin{center}
\begin{tabular}{|c|c|c|c|c|c|}
\hline &d=2&d=3&d=4&d=5&d=6\\ \hline 
$T_{MAX} \gg T_{RH}$ &&&&& \\ \hline
$\alpha=0.1$ & 580 \; TeV & 20 \; TeV & 2.1 \; TeV & 0.43 \; TeV & 0.13 \; TeV \\ 
$\alpha=0.2$ & 520 \; TeV & 18 \; TeV & 1.9 \; TeV & 0.39 \; TeV & 0.12 \; TeV \\ \hline
$T_{MAX} \sim T_{RH}$ &&&&& \\ \hline
$\alpha=0.1$ & 340 \; TeV & 12 \; TeV & 1.2 \; TeV & 0.17 \; TeV & 0.069 \; TeV \\
 $\alpha=0.2$ & 300 \; TeV & 11 \; TeV & 1.1 \; TeV & 0.16 \; TeV & 0.066 \; TeV \\ \hline
\end{tabular}
\caption{\label{Table::PosBound}Limits of sensitivity on $M_*$ (TeV) for different values of $\alpha$ and different dimensions, assuming $T_{RH} = 1 \; MeV$. For larger values of $M_*$, the positron flux from decaying KK modes is lower than the observed $511 \; keV$ $\gamma$-ray flux. }
\end{center}
\end{table}

\begin{table}
\begin{center}
\begin{tabular}{|c|c|c|c|c|c|}
\hline &d=2&d=3&d=4&d=5&d=6\\ \hline 
$T_{MAX} \gg T_{RH}$ &&&&& \\ \hline
$\alpha=0.1$ & 1.9 \; nm & 30 \; pm & 3.7 \; pm & 1.0 \; pm & 0.45 \; pm\\ 
$\alpha=0.2$ & 2.3 \; nm & 35 \; pm& 4.3 \; pm& 1.2 \; pm& 0.50 \; pm\\ \hline
$T_{MAX} \sim T_{RH}$ &&&&& \\ \hline
$\alpha=0.1$ & 5.5 \; nm & 69 \; pm & 8.6 \; pm & 3.8 \; pm & 1.0 \; pm\\
 $\alpha=0.2$ & 7.0 \; nm& 79 \; pm & 9.7 \; pm& 4.1 \; pm& 1.1 \; pm\\ \hline
\end{tabular}
\caption{\label{Table:PosBound2} Limits of sensitivity on the size of extra dimensions, R  for different values of $\alpha$ and different dimensions and assuming $T_{RH} = 1 \; MeV$. For smaller values of R, the positron flux from decaying KK modes is lower than the observed $511 \; keV$ $\gamma$-ray flux.}
\end{center}
\end{table}

\par
 The corresponding bounds on the Planck mass and the size of the extra dimensions are given in Table \ref{Table::PosBound} and Table \ref{Table:PosBound2} respectively. In each bound, it is assumed that $T_{RH} \sim 1 \; MeV$, although recent data from the CMB and from large scale structure suggest $T_{RH} > 4 \; MeV$ for most models \cite{Hannestad:2004px}.  
\nocite{Kawasaki:1999na,Kawasaki:2000en} If this stronger temperature bound is used, then each of the constraints on $M_*$ is improved by a factor of $4^{(d+4)/(d+2)}$.  The corresponding bounds\footnote{It should be noted that for $T_{RH} \geq 4\; MeV$, a significant fraction of the produced positrons have energies of order $O(100 \; MeV)$, and therefore may remain relativistic. This introduces a level of uncertainty into these constraints. } in this case are given in Table \ref{Table::PosBound3} and Table \ref{Table:PosBound4} respectively.

\begin{table}
\begin{center}
\begin{tabular}{|c|c|c|c|c|c|}
\hline &d=2&d=3&d=4&d=5&d=6\\ \hline 
$T_{MAX} \gg T_{RH}$ &&&&& \\ \hline
$\alpha=0.1$ & 4600 \; TeV & 140 \; TeV & 13 \; TeV & 2.6 \; TeV & 0.74 \; TeV\\ 
$\alpha=0.2$ & 4200 \; TeV & 130 \; TeV & 12 \; TeV & 2.3 \; TeV & 0.68 \; TeV \\ \hline
$T_{MAX} \sim T_{RH}$ &&&&& \\ \hline
$\alpha=0.1$ & 2700 \; TeV & 83 \; TeV & 7.8 \; TeV & 1.0 \; TeV & 0.39 \; TeV \\
 $\alpha=0.2$ & 2400 \; TeV & 77 \; TeV & 7.0 \; TeV & 0.95 \; TeV & 0.37 \; TeV \\ \hline
\end{tabular}
\caption{\label{Table::PosBound3}Limits of sensitivity on $M_*$ (TeV) for different values of $\alpha$ and different dimensions, assuming $T_{RH} \sim 4 \; MeV$ and $f_{NR} \approx 1$.  }
\end{center}
\end{table}

\begin{table}
\begin{center}
\begin{tabular}{|c|c|c|c|c|c|}
\hline &d=2&d=3&d=4&d=5&d=6\\ \hline 
$T_{MAX} \gg T_{RH}$ &&&&& \\ \hline
$\alpha=0.1$ & 30 \; pm & 1.2 \; pm & 0.23 \; pm & 0.082 \; pm & 0.045 \; pm\\ 
$\alpha=0.2$ & 36 \; pm & 1.4 \; pm& 0.27 \; pm& 0.098 \; pm& 0.050 \; pm\\ \hline
$T_{MAX} \sim T_{RH}$ &&&&& \\ \hline
$\alpha=0.1$ & 86 \; pm & 2.7 \; pm & 0.54 \; pm & 0.31 \; pm & 0.10 \; pm\\
 $\alpha=0.2$ & 110 \; pm& 3.1 \; pm & 0.61 \; pm& 0.34 \; pm& 0.11 \; pm\\ \hline
\end{tabular}
\caption{\label{Table:PosBound4} Limits of sensitivity on the size of extra dimensions, R  for different values of $\alpha$ and different dimensions assuming $T_{RH} \sim 4 \; MeV$ and $f_{NR} \approx 1$. }
\end{center}
\end{table}

\par
It is also assumed that the mass at which decaying modes no longer produce non-relativistic positrons is sufficiently high as to not affect this bound. The production of nonrelativistic positrons requires $m_{max} \lesssim O(200 \; MeV)$ \cite{Boehm:2003bt}. Limits on $\gamma$-ray production by bremsstrahlung processes suggest $m_{max} \lesssim 40 \; MeV$ \cite{Beacom:2004pe} or $m_{max} \lesssim 6 \; MeV$ \cite{Beacom:2005qv}. If these bounds are used as a mass cut-off, the resulting constraints on the ADD model are weaker than from other astrophysics experiments. However it should be noted that these bounds assume all positrons are created with uniform energy. In this model, the positrons have a range of energies corresponding to the range of KK-mode masses, and as such $m_{max}$ can be larger without violating these bounds. As indicated in Figure \ref{Fig:PosFrac}, $m_{max}/T_{RH}$ can be as low as $\sim 40$ without significantly weakening the constraints.

%

\par
It should also be noted that these bounds are based on several assumptions. First, there is an assumption that the relative abundance of KK modes\index{Kaluza-Klein gravitons} of different masses is the same in the galaxy as in the early Universe. There are also uncertainties in the dark matter halo\index{dark matter!galactic} profile of the galaxy, which can affect the observed flux, although this variation is small for most reasonable profiles and parameters. It is also assumed that the positrons annihilate within a short distance compared to the scale of the galactic center. This diffusion process is not completely understood, and if the positrons travel further then the observed flux would have a wider angular distribution which could improve these bounds \cite{Beacom:2004pe}. It is also assumed that all annihilating positrons produce $511 \; keV$ photons (or lower energy $\gamma$-rays from the decays of positronium states.). However if a significant fraction of the positrons annihilate at higher energies, these photons would not be counted in the observed $511 \; keV$ flux. This would result in less stringent constraints on the number of positrons, and would weaken the bounds given in Table \ref{Table::PosBound} and Table \ref{Table:PosBound2}, while improving bounds obtained from considering the entire $\gamma$-ray spectrum.


\par
 From Table \ref{Table::PosBound}, it is apparent that the accumulated KK modes in the galactic core provide a strong bound on the size of the extra dimensions in the ADD model\footnote{Although the ADD model is considered in this section, it is expected that other models of extra dimensions can be constrained in this manner as well.},as well as providing an explanation for the source of the observed population of galactic positrons. When these modes do decay, they inject photons and positrons into the galaxy which could be observed by existing experiments. As an example, I derived the non-relativistic positron density in the galaxy which is produced by the decay of light KK-modes. By requiring the flux of $511 \; keV$ photons produced by the subsequent annihilations of the positrons to be lower than the observed flux, constraints can be placed on the properties of extra dimensions.

\par
In summary, Kaluza-Klein gravitons which were produced in the early Universe can still exist in the present. Furthermore, these particles can accumulate in the galactic halo, leading to a high density in the galactic core. If KK modes heavier than $\sim 40 \; MeV$ can exist in the galaxy with a significant abundance and can produce nonrelativistic positrons, then the bounds from galactic positron production are significantly stronger than the bounds from collider experiments for $d \leq 4$, and are comparable to the bounds from the extragalactic $\gamma$-ray background and other astrophysics experiments for all dimensions.


%% file: conclusions.tex
\chapter{Conclusions}

\par
In spite of its many successes, it is clear that the Standard Model cannot describe all aspects of nature. Aside from the lack of a valid theory of quantum gravity\index{quantum gravity}, astrophysics experiments have determined that $\sim 96 \%$ of  the energy content of the Universe is unexplained. There also remain questions about why the Standard Model forces are several orders of magnitude stronger than gravity,\index{hierarchy problem} why the cosmological constant is so small, and several other unexplained phenomena. Terrestrial experiments have searched for signs of new physics, but as yet have been unsuccessful.

\par
In this \thesis, I have demonstrated how new theories can be probed using remnants of the early Universe. In the first few minutes, the Universe contained ultra-high energy particles and fields, and therefore any new theories are expected to contribute to its evolution. It is expected that some of these effects left signatures that could be detected in the present. Some examples have been reviewed in this dissertation, and in several cases the constraints imposed by examining the early Universe have been shown to be stronger than the corresponding constraints from terrestrial experiments.

\par
In the Chapter \ref{Chapter:DarkMatter}, I reviewed the motivations for introducing dark matter as well as the current experimental limits on dark matter\index{dark matter}. In particular, I introduced a series of minimal models in which only a minimum amount of new physics is introduced. In each model, I applied existing experimental constraints from dedicated searches, from high energy colliders, and from astrophysical experiments, to demonstrate their properties. These minimal models are generic, and therefore these results can be applied to a larger class of dark matter models. 
\par
In addition to existing constraints, I have shown that for sufficiently light particles\index{dark matter!light}, which can exist in several of the minimal models, it is possible to produce the correct dark matter abundance while also providing observable effects in B-meson decays\index{B mesons}. I also showed that light dark matter models can exist which are not constrained by any present experiments, including B-meson\index{B mesons} decays. These results are general, and can be used to probe or constrain most models of light dark matter\index{dark matter!light}.

\par
In Chapter \ref{Chapter:ChargedRelics}, the possibility of long lived charged relics was presented. In particular, I have shown that charged particles present during Big Bang nucleosynthesis\index{nucleosynthesis}\index{Big Bang Nucleosynthesis} could form bound states and catalyze the standard nuclear reactions. This catalysis serves to create \lisx\ while destroying \lisv\ , and in a certain region of parameter space can explain both the observed \lisx\ abundance and the observed suppression of the primordial \lisv\ abundance relative to the standard BBN model. Through the use of CBBN\index{Big Bang Nucleosynthesis!catalyzed}, bounds can be placed on models of heavy charged relics. 

\par
In the Chapter \ref{Chapter:ExtraDimensions}, the motivations for introducing additional spacetime dimensions were reviewed along with the existing experimental constraints from colliders and from astrophysical experiments. I demonstrated that the effects of extra dimensions could affect nucleosynthesis\index{nucleosynthesis}\index{Big Bang Nucleosynthesis} in the early universe and derived constraints on both the size of the higher dimensions and on the higher dimensional Planck mass. For the case of $d \geq 4$ these bounds are stronger than the previous astrophysics and collider bounds on the ADD model\index{extra dimensions!ADD model}.
Also in this chapter, I showed how the presence of excited Kaluza-Klein states\index{Kaluza-Klein gravitons} in the galactic core and their subsequent decays could be used to explain the $511 \; keV$ $\gamma$-ray\index{$\gamma$-rays!511 keV} line. By requiring the flux from KK-graviton decay not exceed the observed flux, I was able to add additional constraints the ADD model\index{extra dimensions!ADD model}.

\par
In conclusion, in this dissertation I have demonstrated how three different types of new physical theories -neutral particles, charged particles\index{CHAMPs}\index{charged relics}, and extra dimensions\index{extra dimensions} - can be studied and constrained using their effects on the early Universe. For each model, existing experimental data was used to apply new constraints and in most cases the limits from astrophysics and cosmology were comparable to or stronger than the limits from terrestrial experiments, thus proving that the early Universe provides a viable probe of new physics.